\documentclass{aa}
\usepackage{graphicx}
\usepackage{txfonts}
\usepackage{natbib}
\usepackage{xcolor}
\usepackage{float}

\usepackage{supertabular}
\usepackage{subfig}
\usepackage{rotating}
\DeclareCaptionFormat{cont}{#1 (cont.)#2#3\par}
\usepackage{upgreek}
\usepackage{pifont}

\usepackage[colorlinks=true,linkcolor=blue,citecolor=blue,urlcolor=blue]{hyperref}

\def\as {\ifmmode {\rlap.}$\,$''$\,$\! \else ${\rlap.}$\,$''$\,$\!$\fi}

\begin{document} 

	\title{Physical and chemical complexity in high-mass star-forming regions with ALMA.}

	\subtitle{I. Overview and evolutionary trends of physical properties}

	\author{C. Gieser\inst{1,2}
	\and
	H. Beuther\inst{1}
	\and
	D. Semenov\inst{1,3}
	\and
	A. Ahmadi\inst{4}
	\and
	Th. Henning\inst{1}
	\and
	M.~R.~A. Wells\inst{1}
	}

	\institute{
	Max Planck Institute for Astronomy, Königstuhl 17, 69117 Heidelberg, Germany
	\and
	Max-Planck-Institut für extraterrestrische Physik, Giessenbachstrasse 1, 85748 Garching, Germany,\\
	\email{gieser@mpe.mpg.de}
	\and
	Department of Chemistry, Ludwig Maximilian University, Butenandtstr. 5-13, 81377 Munich, Germany
	\and
	Leiden Observatory, Leiden University, P.O. Box 9513, 2300 RA Leiden, the Netherlands
	}

	\date{Received...; accepted...}
 
	\abstract
	{High-mass star formation is a hierarchical process from cloud ($>$1\,pc), to clump (0.1-1\,pc) to core scales ($<$0.1\,pc). Modern interferometers achieving high angular resolutions at mm wavelengths allow us to probe the physical and chemical properties of the gas and dust of protostellar cores in the earliest evolutionary formation phases.}
	{In this study, we investigate how physical properties, such as the density and temperature profiles, evolve on core scales through the evolutionary sequence during high-mass star formation ranging from protostars in cold infrared dark clouds to evolved UCH{\sc ii} regions.}
	{We observed 11 high-mass star-forming regions with the Atacama Large Millimeter/submillimeter Array (ALMA) at 3\,mm wavelengths. Based on the 3\,mm continuum morphology and H(40)$\alpha$ recombination line emission, tracing locations with free-free (ff) emission, the fragmented cores analyzed in this study are classified into either ``dust'' or ``dust+ff'' cores. In addition, we resolve three cometary UCH{\sc ii} regions with extended 3\,mm emission that is dominated by free-free emission. The temperature structure and radial profiles ($T \sim r^{-q}$) are determined by modeling molecular emission of CH$_{3}$CN and CH$_{3}^{13}$CN with \texttt{XCLASS} and by using the HCN-to-HNC intensity ratio as probes for the gas kinetic temperature. The density profiles ($n \sim r^{-p}$) are estimated from the 3\,mm continuum visibility profiles. The masses $M$ and H$_{2}$ column densities $N$(H$_{2}$) are then calculated from the 3\,mm dust continuum emission.}
	{We find a large spread in mass and peak H$_{2}$ column density in the detected sources ranging from 0.1 - 150\,$M_\odot$ and 10$^{23}$ - 10$^{26}$\,cm$^{-2}$, respectively. Including the results of the CORE and CORE-extension studies to increase the sample size, we find evolutionary trends on core scales for the temperature power-law index $q$ increasing from 0.1 to 0.7 from infrared dark clouds to UCH{\sc ii} regions, while for the the density power-law index $p$ on core scales, we do not find strong evidence for an evolutionary trend. However, we find that on the larger clump scales throughout these evolutionary phases the density profile flattens from $p \approx 2.2$ to $p \approx 1.2$.}
	{By characterizing a large statistical sample of individual fragmented cores, we find that the physical properties, such as the temperature on core scales and density profile on clump scales, evolve even during the earliest evolutionary phases in high-mass star-forming regions. These findings provide observational constraint for theoretical models describing the formation of massive stars. In follow-up studies we aim to further characterize the chemical properties of the regions by analyzing the large amount of molecular lines detected with ALMA in order to investigate how the chemical properties of the molecular gas evolve during the formation of massive stars.}


	\maketitle

\section{Introduction}\label{sec:ALMAintroduction}

	Current research on high-mass star formation (HMSF) is centered around how gas accretion can efficiently be funneled from large scales toward individual protostars and what processes set the fragmentation within a single high-mass star-forming region (HMSFR). For dedicated reviews regarding observations and theories of HMSF we refer to \citet{Beuther2007}, \citet{Bonnell2007}, \citet{Zinnecker2007}, \citet{Smith2009}, \citet{Tan2014}, \citet{Krumholz2015}, \citet{Schilke2015}, \citet{ Motte2018}, and \citet{Rosen2020}. Since the formation of a massive star is fast compared to its low-mass analogues and the distances are typically at a few kiloparsec, they can be best studied in their earliest phases of their formation with interferometric observations at (sub)mm wavelengths allowing us to probe the dust and molecular gas on core scales ($<$0.1\,pc).
	
	Massive stars form in the densest regions of molecular clouds with typical formation time scales on the order of $10^{5}$\,yr \citep{McKee2002, McKee2003, Mottram2011, Kuiper2018}. The evolutionary stages of high-mass protostars can be empirically categorized into four different stages based on observed properties that are further explained in the following:
	\begin{itemize}
	\setlength\itemsep{0em}
	\item infrared dark cloud (IRDC)
	\item high-mass protostellar object (HMPO)
	\item hot molecular core (HMC)
	\item ultra-compact H{\sc ii} (UCH{\sc ii}) region
	\end{itemize}
Considering the underlying physical properties during HMSF \citep[e.g.,][]{Beuther2007,Zinnecker2007,Gerner2014}, these evolutionary stages can be related to the high-mass star formation scenario proposed by \citet{Motte2018}.
	
	Due to their high extinctions at near-infrared (NIR) and mid-infrared (MIR) wavelengths, IRDCs were initially identified as absorption features against the bright Galactic MIR background, while at far-infrared (FIR) wavelengths they appear in emission \citep[e.g.,][]{Rathborne2006,Henning2010}. IRDCs with sizes of $1-10$\,pc have a typical density of $n \gtrsim 10^{4}$\,cm$^{-3}$ and temperature of $T < 20$\,K, and are the birth places of stars \citep[e.g.,][]{Carey1998,Pillai2006a,Rathborne2006,Zhang2015}. In local high-density regions, star formation with MIR-bright protostellar cores takes place, while in other parts of the cloud prestellar cores can be found that are only detected at FIR and mm wavelengths. In this stage, low- and intermediate-mass protostars and potentially high-mass starless cores exist \citep{Motte2018}.
	
	Dense cores that undergo gravitational collapse can harbor high-mass protostars that have high luminosities, $L>10^{3}\,L_\odot$, and high accretion rates, $\dot M \gtrsim 10^{-4}\,M_\odot$\,yr$^{-1}$, inferred from large bipolar molecular outflows \citep[e.g.,][]{Beuther2002B,DuarteCabral2013}. These are referred to as HMPOs \citep[e.g.,][]{Beuther2002,Williams2004,Motte2007,Beuther2010}. The temperature in the envelope increases due to the central heating of the protostar and since outflows are commonly observed, disks should also be present, but due to their small sizes ($\lesssim$1\,000\,au) they remain challenging to observe \citep[e.g.,][]{SanchezMonge2013, Beltran2016}, while they are commonly found around low-mass protostars \citep[e.g.,][]{ALMA2015,Andrews2018,Avenhaus2018,Oeberg2021}. HMPOs are bright at mm wavelengths, but have no or only weak emission at cm wavelengths.
	
	As the protostar heats up the envelope to $\geq$100\,K, molecules that have resided and/or formed on dust grains evaporate into the gas-phase revealing line rich emission spectra, being classified as HMCs or ``hot cores'' \citep{Cesaroni1997,Osorio1999,Belloche2013,SanchezMonge2017,Beltran2018}. In HMCs, complex organic molecules \citep[COMs, consisting six or more atoms, following the definition by][]{Herbst2009} are very abundant, for example, methanol (CH$_{3}$OH), acetone (CH$_{3}$COCH$_{3}$), methyl formate (CH$_{3}$OCHO), and ethyl cyanide (CH$_{3}$CH$_{2}$CN) as revealed by spectral line surveys \citep[e.g.,][]{Belloche2013}. Due to the ionizing radiation of the protostar, in the central region of the HMC, a hyper-compact (HC) H{\sc ii} region might already be present.
	
	The strong protostellar radiation causes an expansion and further ionization of the surrounding envelope that is eventually disrupted revealing the massive star. The region is then classified to be a UCH{\sc ii} region \citep[e.g.,][]{Wood1989,Hatchell1998,Garay1999,Kurtz2000,Churchwell2002,Palau2007, Qin2008,SanchezMonge2013b,Klaassen2018}. UCH{\sc ii} regions can be studied at cm wavelengths due to free-free emission from scattered electrons \citep{Churchwell2002,Peters2010b} and show diverse spatial morphologies from compact cores to cometary halos \citep{Churchwell2002}. Typical sizes of UCH{\sc ii} regions are $R<0.1$\,pc, while HCH{\sc ii} regions are even more compact with $R<0.01$\,pc. At NIR and MIR wavelengths, the massive star might be detectable depending on the extinction in the cloud.

	During the formation of high-mass stars, the physical and chemical properties on core scales are diverse \citep{Gieser2019,Gieser2021,Gieser2022} as revealed in the CORE \citep{Beuther2018} and CORE-extension HMSFRs \citep{Beuther2021} using sub-arcsecond interferometric NOEMA observations at 1\,mm wavelengths. Since HMSFRs are typically located at distances of a few kpc, the linear resolution is on the order of a few thousand au. This allows us to resolve the envelope of protostellar cores, for which the temperature profile can be described by a power-law profile in the form of 
	\begin{equation}
	\label{eq:temperatureprofile}
	T(r) = T_{\mathrm{in}} \times \bigg(\frac{r}{r_{\mathrm{in}}}\bigg)^{-q}
	\end{equation}
	and the radial density profile in the form of
	\begin{equation}
	\label{eq:densityprofile}
	n(r) = n_{\mathrm{in}} \times \bigg(\frac{r}{r_{\mathrm{in}}}\bigg)^{-p}.
	\end{equation}
	
	In \citet{Gieser2021,Gieser2022} the temperature and density profiles of a sample of HMSFRs were derived as well as mass, $M$, and H$_{2}$ column density, $N$(H$_{2}$), estimates. The regions in the CORE sample \citep{Gieser2021} are classified to be roughly in the HMPO and HMC stages, while the regions in the CORE-extension sample \citep{Gieser2022} were classified to be younger and colder regions. In these studies, a mean temperature and density power-law index of $q \approx 0.4$ and $p \approx 2.0$, respectively, is derived. With a spread in $q$ and $p$ around the mean values, it was not clear whether a spread is real or due to uncertainties and a small number of analyzed cores.
	
	To increase the sample size of HMSFRs and to investigate evolutionary trends of both physical and chemical properties, we carried out ALMA observations in Cycle 6 targeting in total 11 HMSFRs in Band 3 covering the full evolutionary sequence during HMSF - from protostars in cold IRDCs to evolved UCH{\sc ii} regions. The ALMA continuum and spectral line observations with an angular resolution of $\approx$1$''$ allow us to study the fragmentation and physical properties on core scales and characterize the molecular content of the cores and in the surrounding structures they are embedded in. 
	
	The 3\,mm spectral setup (Table \ref{tab:almaspwsummary}) covers a variety of molecular lines targeting, for example, high- and low-density and temperature regimes, photochemistry, shocks and outflows, and C-/O-/N-/S- bearing species. Since molecule formation and destruction depends heavily on the underlying physical conditions, with this sample covering multiple evolutionary stages we are able to probe the evolution of physical properties and molecular abundances with time.
	
	In this first study, we analyze the fragmentation properties based on the ALMA 3\,mm continuum and the temperature and density profiles. Furthermore, we investigate how the properties change along the evolutionary sequence taking into account the results from the CORE and CORE-extension regions \citep{Gieser2021,Gieser2022}. Future studies will be dedicated to a detailed chemical analysis targeting a variety of processes, for example the formation and spatial morphology of COMs, isotopologues and isomers, shocks, and outflows.

\section{Sample}\label{sec:ALMAsample}

\begin{table*}[!htb]
\caption{Overview of the sample.}
\label{tab:ALMA_regions}
\centering
\renewcommand{\arraystretch}{1.1}
\begin{tabular}{lrrrrrrr}
\hline\hline
Region & \multicolumn{2}{c}{Phase center} & & \multicolumn{4}{c}{ATLASGAL clump properties$^{(*)}$}\\ \cline{2-3}\cline{5-8}
& & & Velocity & Distance & Dust temperature & Mass & Luminosity\\
 & $\alpha$ & $\delta$ & $\varv_\mathrm{LSR}$ & $d$ & $T_\mathrm{dust}$ & log $M$ & log $L$\\
 & (J2000) & (J2000) & (km\,s$^{-1}$) & (kpc) & (K) & (log $M_\odot$) & (log $L_\odot$)\\
\hline
IRDC\,G11.11$-$4 & 18:10:28.30 & $-$19:22:31.5 & $+29.2$ & $2.9$ & 16 & 3.0 & 2.8\\ 
IRDC\,18223$-$3 & 18:25:08.40 & $-$12:45:15.5 & $+45.3$ & $3.4$ & 13 & 3.1 & 2.7\\ 
IRDC\,18310$-$4 & 18:33:39.42 & $-$08:21:10.4 & $+86.5$ & $5.9$ & 13 & 3.0 & 2.5\\ 
HMPO\,IRAS\,18089 & 18:11:51.52 & $-$17:31:28.9 & $+33.8$ & $2.5$ & 23 & 3.1 & 4.3\\ 
HMPO\,IRAS\,18182 & 18:21:09.21 & $-$14:31:45.5 & $+59.1$ & $4.7$ & 25 & 3.1 & 4.3\\ 
HMPO\,IRAS\,18264 & 18:29:14.68 & $-$11:50:24.0 & $+43.6$ & $3.3$ & 20 & 3.2 & 3.9\\ 
HMC\,G9.62$+$0.19 & 18:06:14.92 & $-$20:31:39.2 & $+4.4$ & $5.2$ & 32 & 3.5 & 5.4\\ 
HMC\,G10.47$+$0.03 & 18:08:38.20 & $-$19:51:50.1 & $+67.8$ & $8.6$ & 25 & 4.4 & 5.7\\ 
HMC\,G34.26$+$0.15 & 18:53:18.54 & $+$01:14:57.9 & $+58.0$ & $1.6$ & 29 & 3.2 & 4.8\\ 
UCH{\sc ii}\,G10.30$-$0.15 & 18:08:55.92 & $-$20:05:54.6 & $+13.5$ & $3.6$ & 30 & 3.3 & 5.2\\ 
UCH{\sc ii}\,G13.87$+$0.28 & 18:14:35.95 & $-$16:45:36.5 & $+48.3$ & $3.9$ & 34 & 3.1 & 5.1\\ 
\hline 
\end{tabular}
\tablefoot{$^{(*)}$ taken from \citet{Urquhart2018}.}
\end{table*}

	The 11 target regions were selected from the chemical study by \citet{Gerner2014,Gerner2015} targeting a total of 59 HMSFRs at all evolutionary phases during HMSF from IRDCs to UCH{\sc ii} regions. Their analysis was based on observations with single-dish telescopes (e.g., with the IRAM 30m telescope) that can not resolve individual fragmented cores at the typical distances of HMSFRs but only trace the larger-scale clump structures. In addition, protostars in various evolutionary phases can be present within one clump as revealed by the CORE \citep{Beuther2018,Gieser2021} and CORE-extension NOEMA observations \citep{Beuther2021,Gieser2022}. We therefore observed 11 HMSFRs of the sample by \citet{Gerner2014,Gerner2015} at an angular resolution of $\approx$1$''$ with ALMA in Cycle 6 at 3\,mm wavelengths. 
	
	In Table \ref{tab:ALMA_regions} the region properties such as the coordinates of the phase center and velocity $\varv_\mathrm{LSR}$ are summarized with additional information of the larger scale clump properties (distance $d$, dust temperature $T_\mathrm{dust}$, mass $M$, and luminosity $L$) taken from the APEX Telescope Large Area Survey of the Galaxy \citep[ATLASGAL,][]{Urquhart2018}. A short description of each region is given in Appendix \ref{app:sampleoverview}. Within this study, we consistently use the clump properties ($d$, $L$, $M$, $T_\mathrm{dust}$) taken from the ATLASGAL survey listed in Table \ref{tab:ALMA_regions}. Archival mid-infrared (MIR) and far-infrared (FIR), and cm observations with the Atacama Pathfinder EXperiment (APEX), \textit{Herschel}, and \textit{Spitzer} telescopes, and the \textit{Karl G. Jansky} Very Large Array (VLA) are presented in Appendix \ref{app:sampleoverview} in Figs. \ref{fig:overview_IRDC_G1111} $-$ \ref{fig:overview_UCHII_G1387} for a multi-wavelength and multi-scale overview of each region. The \textit{Herschel} 250\,$\upmu$m data, for example, reveal that, except for IRDC\,18310$-$4, all regions are embedded in dense clump structures while being surrounded by filamentary structures that might feed the central hub with new material \citep{Kumar2020}. 
	
	The regions are located at distances in the range of 1.6 kpc to 8.6\,kpc (Table \ref{tab:ALMA_regions}). Since HMSFRs are less common compared to star-forming regions in the low-mass only regime, it is hardly possible to find a large sample of HMSFRs at different evolutionary stages and at similar distances. However, in Sect. \ref{sec:ALMAdiscussionclump} we show that the overall clump properties are unbiased by distance in our selected sample.

\section{Observations}\label{sec:ALMAallobs}

\begin{table*}[!htb]
\caption[Overview of the ALMA 3\,mm continuum data products.]{Overview of the ALMA 3\,mm continuum data products.}
\label{tab:ALMAcontinuumdataproducts}
\centering
\renewcommand{\arraystretch}{1.1}
\begin{tabular}{lrrrrrr}
\hline \hline
Region & \multicolumn{2}{c}{Beam} & Noise & Peak intensity & Flux density & Field\\
\cline{2-3}
 & $\theta_\mathrm{maj}\times\theta_\mathrm{min}$ & PA & $\sigma_\mathrm{cont}$ & $I^\mathrm{region}_\mathrm{3mm}$ & $F^\mathrm{region}_\mathrm{3mm}$ & \\
 & ($''\times''$) & ($^\circ$) & (mJy\,beam$^{-1}$) & (mJy\,beam$^{-1}$) & (mJy) & \\
\hline
IRDC\,G11.11$-$4 & 1.0$\times$0.8 & 104 & 0.033 & 1.4 & 15 & Field 2\\ 
IRDC\,18223$-$3 & 1.0$\times$0.7 & 108 & 0.022 & 2.9 & 23 & Field 3\\ 
IRDC\,18310$-$4 & 1.0$\times$0.7 & 111 & 0.022 & 1.2 & 4.0 & Field 3\\ 
HMPO\,IRAS\,18089 & 0.8$\times$0.7 & 101 & 0.057 & 24 & 130 & Field 2\\ 
HMPO\,IRAS\,18182 & 1.1$\times$0.7 & 107 & 0.029 & 13 & 90 & Field 3\\ 
HMPO\,IRAS\,18264 & 1.0$\times$0.7 & 108 & 0.031 & 9.9 & 130 & Field 3\\ 
HMC\,G9.62$+$0.19 & 0.9$\times$0.7 & 104 & 0.049 & 58 & 640 & Field 2\\ 
HMC\,G10.47$+$0.03 & 0.8$\times$0.7 & 100 & 0.25 & 340 & 830 & Field 2\\ 
HMC\,G34.26$+$0.15 & 0.8$\times$0.7 & 118 & 1.8 & 2\,500 & 7\,100 & Field 1\\ 
UCH{\sc ii}\,G10.30$-$0.15 & 1.0$\times$0.8 & 104 & 0.041 & 18 & 1\,000 & Field 2\\ 
UCH{\sc ii}\,G13.87$+$0.28 & 1.0$\times$0.8 & 105 & 0.065 & 7.0 & 1\,800 & Field 2\\ 
\hline
\end{tabular}
\tablefoot{The regions are grouped into three fields (Field 1, Field 2, and Field 3) and were observed together during the ALMA observations to reduce off-source calibration time. The flux density $F^\mathrm{region}_\mathrm{3mm}$ is computed considering only areas with $S$/$N > 5$ within the full ALMA FOV.}
\end{table*}

	In this study, we use the ALMA 3\,mm observations toward the sample in combination with archival MIR, FIR, and cm data with the \textit{Spitzer}, \textit{Herschel}, and APEX telescopes, and the VLA. The ALMA data calibration and imaging procedure is explained in Sect. \ref{sec:ALMAobs} and an overview of the archival data is given in Sect. \ref{sec:ALMAarchivdata}.

\subsection{ALMA}\label{sec:ALMAobs}

	The ALMA observations were carried out during Cycle 6 with the project code 2018.1.00424.S (PI: Caroline Gieser). In total, 12 science targets were observed in Band 3 covering a (non-continuous) spectral range (SPR) from 86$-$110\,GHz in three different spectral range setups (referred to as SPR1, SPR2, and SPR3) and in total 39 spectral windows (spws), summarized in Table \ref{tab:almaspwsummary}. In order to reduce off-source time, the regions were grouped into three fields (referred to as Field 1, Field 2 and Field 3, last column in Table \ref{tab:ALMAcontinuumdataproducts}). One of the observed science targets turned out to be a misclassified UCH{\sc ii} region \citep{Wood1989} and is in fact a planetary nebula \citep{Walsh2003,Thompson2006} and is not further discussed in the following analysis, however the ALMA data will be presented in Moraga et al. (in preparation).
	
	The initial strategy of the observations was to observe all regions with three array configurations, two with the ``12m-array'' (C43-4 and C43-1) and one with the Atacama Compact Array (ACA, also referred to as the ``7m-array'') covering spatial scales from 1$''$ to 60$''$. Since not all observations could be carried out during Cycle 6, the intermediate C43-1 configuration is missing for the most part, however, the $uv$ coverage of the C43-4 and ACA configurations have a sufficient overlap in order to successfully combine the data. For Field 2/SPR3, the C43-4 configuration is missing, however, these regions were observed with the C43-1 configuration instead. For the remaining fields and spectral setups, observations in the C43-1 configuration are missing, but data in the C43-4 configuration is available. Therefore, the angular resolution of Field 2/SPR3 is lower ($\approx$3$''$) compared to the SPR3 line data products of Field 1 and Field 3 ($\approx$1$''$).
	
	A summary of all ALMA Cycle 6 observations - including the array configuration, observation date, precipitable water vapor (PWV) and minimum and maximum baseline - is presented in Table \ref{tab:almaobssummary}. During the observations from October 2018 until May 2019, the PWV ranged from $1-6$\,mm sufficient for observations at 3\,mm wavelengths. The baselines cover 9$-$1\,400\,m. We do not have complementary observations with the total power array to recover missing short-spacing information, we therefore filter out spatial scales $>$60$''$ in both the continuum and spectral line data.
	
	The covered frequency ranges and spectral resolution of each of the 39 spws is summarized in Table \ref{tab:almaspwsummary}. In total, there are 36 high-resolution spws with a channel width of 244\,kHz ($\approx$0.8\,km\,s$^{-1}$ at 3\,mm) covering 0.12\,GHz each and 3 low-resolution spws with a channel width of 1\,MHz ($\approx$3\,km\,s$^{-1}$ at 3\,mm) covering 1.9\,GHz each. The channel width of the three low-resolution spws is not sufficient to resolve typical line widths toward IRDCs ($\lesssim$3\,km\,s$^{-1}$), however, for the remaining regions, the line emission can be spectrally resolved.
	
\subsubsection{Calibration}

	The data of each observation block (Table \ref{tab:almaobssummary}) were calibrated using the \texttt{CASA} pipeline (\texttt{CASA} version 5.4.0). When multiple observations were carried out for each SPR, field and array configuration, the data were merged using the \texttt{concat} task. Science targets were extracted from the calibrated $uv$ tables using the \texttt{split} task.
	
	In order to extract the 3\,mm continuum, we computed an average spectrum for each region, array configuration, and spw in order to determine all line-free channels. All line-free channels were merged using the \texttt{concat} task to create a continuum table containing all SPRs and array configurations. To increase the $S$/$N$, the continuum visibilities were averaged over 30\,s at a central frequency of 98.3\,GHz (corresponding to $\approx$3\,mm). 
	
	The continuum is subtracted from the spectral line data using the \texttt{uvcontsub} task by fitting a first-order polynomial to the line-free channels. Using the \texttt{concat} task, all array configurations were merged for each SPR. With the \texttt{split} task, for each spw a $uv$ table of the merged spectral line data is created.
	
\subsubsection{Self-calibration and imaging}
	
	The ALMA continuum and spectral line data are imaged using the \texttt{tclean} task. All data are \texttt{CLEAN}ed using the Hogbom algorithm \citep{Hogbom1974} with Briggs weighting \citep{Briggs1995} using a robust parameter of 0.5. 
	
\begin{figure*}
\includegraphics[width=0.98\textwidth]{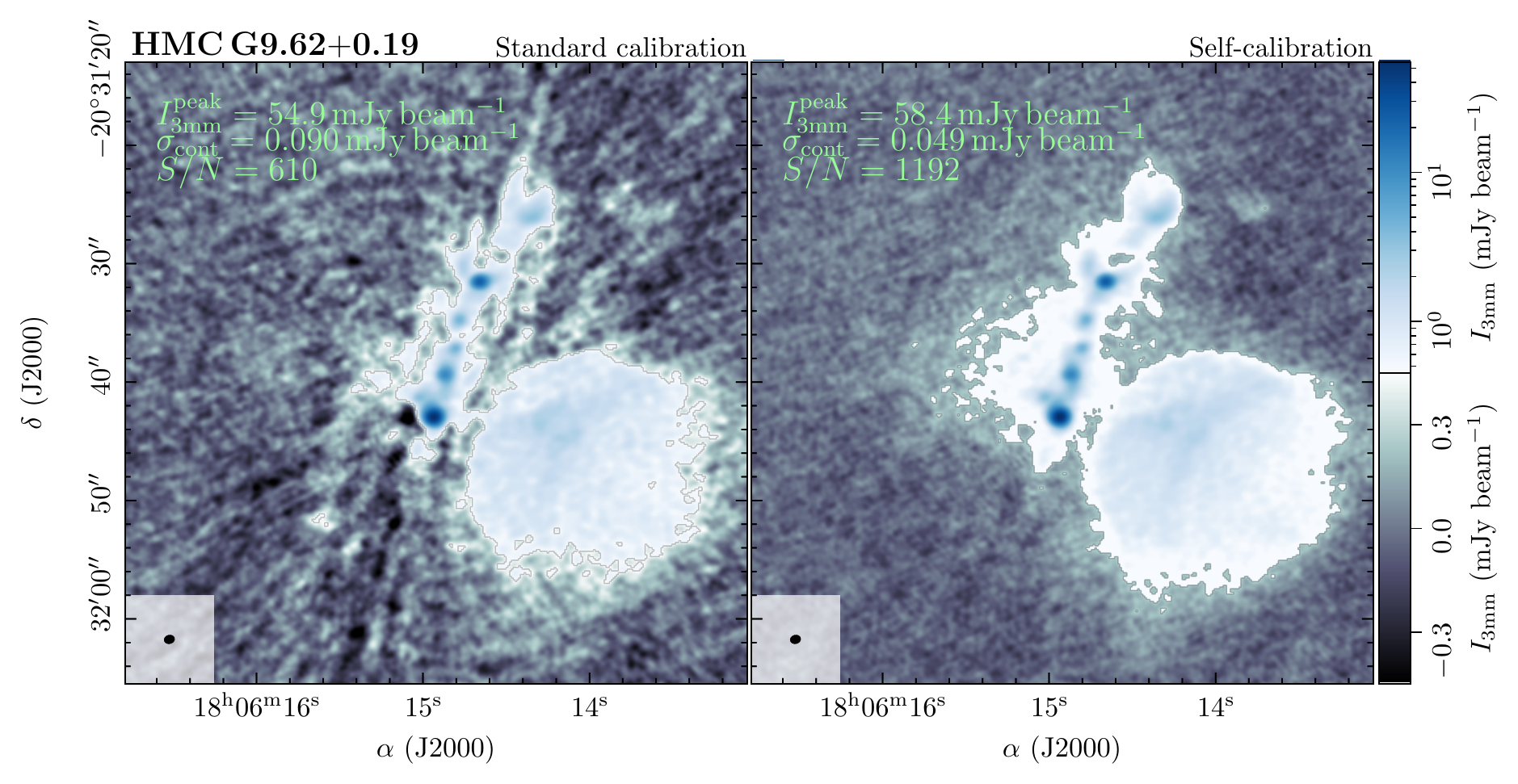}
\caption{Comparison of \texttt{CASA} standard calibrated (\textit{left panel}) and \texttt{CASA} self-calibrated (\textit{right panel}) HMC\,G9.62$+$0.19 ALMA continuum data. In each panel, the beam size is shown in the bottom left corner. The linear grey-scale and logarithmic blue-scale color map highlight the continuum intensity between $-5\sigma_\mathrm{cont}$ to $+5\sigma_\mathrm{cont}$ and between $+5\sigma_\mathrm{cont}$ and peak intensity of the standard calibrated data, respectively.}
\label{fig:ALMAselfcal}
\end{figure*}
	
	To increase the $S$/$N$, we perform phase self-calibration on the continuum data using a similar method adopted for the CORE NOEMA data \citep{Gieser2021} and apply the solutions to the spectral line data. First, a shallow model of the continuum data is created using \texttt{tclean} with a stopping criterion of either 2\,000 \texttt{CLEAN} components or a peak residual intensity of 0.1\,mJy\,beam$^{-1}$. In a first self-calibration loop, the gains based on the source model are determined using the \texttt{gaincal} task with a solution interval of 120\,s. The gain solutions are applied with the \texttt{applycal} task using the ``calonly'' mode, that does not flag baselines for which the $S$/$N$ is too small for self-calibration ($S$/$N < 3$), therefore long baselines with low $S$/$N$ are not self-calibrated but kept in the $uv$ table in order to maintain a high angular resolution. In the second and third self-calibration loop, the number of iterations increases to 4\,000 and 8\,000, the peak residual intensity decreases to 0.05\,mJy\,beam$^{-1}$ and 0.025\,mJy\,beam$^{-1}$, and the solution interval decreases to 60\,s and 30\,s, respectively. 

	The final self-calibrated continuum data are then imaged with a stopping criterion of either 8\,000 \texttt{CLEAN} components or a peak residual intensity of 0.025\,mJy\,beam$^{-1}$ with a pixel size of 0\as15. An overview of the self-calibrated continuum data products for each region is presented in Table \ref{tab:ALMAcontinuumdataproducts}. The synthesized beam size is $\approx$1$''$. Table \ref{tab:ALMAcontinuumdataproducts} also shows the continuum peak intensity $I^\mathrm{region}_\mathrm{3mm}$ and region-integrated flux density $F^\mathrm{region}_\mathrm{3mm}$ considering only areas with $S$/$N>5$. The continuum noise $\sigma_\mathrm{cont}$ in the regions is smaller than 0.07\,mJy\,beam$^{-1}$, except for HMC\,G10.47$+$0.03 and G34.26$+$0.15. These two regions are bright continuum sources and their spectra do not have many line-free emission channels, which results in a higher continuum noise compared to regions with fainter continuum emission and more line-free channels.
	
		A comparison between the standard-calibrated and self-calibrated continuum data of the HMC\,G9.62$+$0.19 region is presented in Fig. \ref{fig:ALMAselfcal}. This region has a similar complex morphology as the W3\,IRS4 region of the CORE sample \citep[Fig. 1 in ][]{Gieser2021}. Strong noise artifacts throughout the field-of-view (FOV) due to a high phase noise are clearly corrected by the phase self-calibration procedure decreasing the continuum noise $\sigma_\mathrm{cont}$ by a factor of two. The peak intensity is slightly enhanced as well. Overall, the $S$/$N$ is increased by a factor of two from $\approx$610 to $\approx$1\,200 by applying phase self-calibration in the HMC\,G9.62$+$0.19 region.

\begin{table}
\caption{Transition properties of lines analyzed in this work.}
\label{tab:ALMAline}
\centering
\setlength{\tabcolsep}{5pt}
\renewcommand{\arraystretch}{1.1}
\begin{tabular}{lrrrr}
\hline \hline
Line & $\nu_0$ & log $A_{\mathrm{ul}}$ & $E_{\mathrm{u}}/k_{\mathrm{B}}$ & SPR spw\\
 & (GHz) & (log s$^{-1}$) & (K) & \\
\hline
H(40)$\alpha$ & 99.023 & $\ldots$ & \ldots & SPR2 spw6\\ 
HCN $1-0$ & 88.630 & $-4.6$ & 4 & SPR1 spw5\\ 
HNC $1-0$ & 90.664 & $-4.6$ & 4 & SPR3 spw9\\ 
CH$_{3}$CN $5_{4}-4_{4}$ & 91.959 & $-4.8$ & 128 & SPR3 spw13\\ 
CH$_{3}$CN $5_{3}-4_{3}$ & 91.971 & $-4.6$ & 78 & SPR3 spw13\\ 
CH$_{3}$CN $5_{2}-4_{2}$ & 91.980 & $-4.4$ & 42 & SPR3 spw13\\ 
CH$_{3}$CN $5_{1}-4_{1}$ & 91.985 & $-4.4$ & 20 & SPR3 spw13\\ 
CH$_{3}$CN $5_{0}-4_{0}$ & 91.987 & $-4.4$ & 13 & SPR3 spw13\\ 
CH$_{3}^{13}$CN $5_{4}-4_{4}$ & 91.913 & $-4.6$ & 128 & SPR3 spw11\\ 
CH$_{3}^{13}$CN $5_{3}-4_{3}$ & 91.926 & $-4.4$ & 78 & SPR3 spw11\\ 
CH$_{3}^{13}$CN $5_{2}-4_{2}$ & 91.935 & $-4.3$ & 42 & SPR3 spw11\\ 
CH$_{3}^{13}$CN $5_{1}-4_{1}$ & 91.940 & $-4.2$ & 20 & SPR3 spw11\\ 
CH$_{3}^{13}$CN $5_{0}-4_{0}$ & 91.942 & $-4.2$ & 13 & SPR3 spw11\\ 
\hline
\end{tabular}
\tablefoot{The line properties are taken from the Cologne Database for Molecular Spectroscopy \citep[CDMS,][]{CDMS} and the Jet Propulsion Laboratory \citep[JPL,][]{JPL} database. The last column refers to the corresponding line data product (SPR and spw) that covers the transition in the ALMA observations (Table \ref{tab:almaspwsummary}).}
\end{table}

\begin{table*}
\caption[Overview of the ALMA spectral line data products.]{Overview of the ALMA spectral line data products used in this work.}
\label{tab:ALMAlinedataproducts}
\centering
\renewcommand{\arraystretch}{1.1}
\begin{tabular}{lrrr|rrr|rrr}
\hline \hline
Line & \multicolumn{3}{c}{Field 1} & \multicolumn{3}{c}{Field 2} & \multicolumn{3}{c}{Field 3}\\ 
 & \multicolumn{2}{c}{Beam} & Noise & \multicolumn{2}{c}{Beam} & Noise & \multicolumn{2}{c}{Beam} & Noise\\ \cline{2-3}\cline{5-6}\cline{8-9}
& $\theta_\mathrm{maj}\times\theta_\mathrm{min}$ & PA & $\sigma_\mathrm{line}$ & $\theta_\mathrm{maj}\times\theta_\mathrm{min}$ & PA & $\sigma_\mathrm{line}$ & $\theta_\mathrm{maj}\times\theta_\mathrm{min}$ & PA & $\sigma_\mathrm{line}$\\
 & ($''\times''$) & ($^\circ$) & (K) & ($''\times''$) & ($^\circ$) & (K) & ($''\times''$) & ($^\circ$) & (K)\\
\hline
H(40)$\alpha$ & 1.5$\times$1.2 & 122 & 0.12 & 1.9$\times$1.0 & 105 & 0.12 & 1.5$\times$1.0 & 111 & 0.11\\ 
HCN $1-0$ & 0.9$\times$0.8 & 97 & 0.73 & 4.2$\times$2.5 & 104 & 0.02 & 0.9$\times$0.7 & 102 & 0.52\\ 
HNC $1-0$ & 0.8$\times$0.7 & 80 & 0.84 & 4.2$\times$2.5 & 104 & 0.08 & 1.0$\times$0.7 & 104 & 0.66\\ 
CH$_{3}$CN $J$=$5-4$,$K$=0-4 & 0.8$\times$0.7 & 72 & 0.85 & 4.2$\times$2.5 & 104 & 0.09 & 1.1$\times$0.7 & 98 & 0.59\\ 
CH$_{3}^{13}$CN $J$=$5-4$,$K$=0-4 & 0.8$\times$0.7 & 75 & 0.78 & 4.2$\times$2.5 & 104 & 0.09 & 0.9$\times$0.7 & 102 & 0.66\\ 
\hline
\end{tabular}
\tablefoot{The transition properties are summarized in Table \ref{tab:ALMAline}. The line noise $\sigma_\mathrm{line}$ is a mean value taken from the regions grouped within the three fields. Table \ref{tab:ALMAcontinuumdataproducts} lists which region is grouped into which field.}
\end{table*}

	The gain solutions of the phase self-calibrated continuum data are applied to all 39 spws using the \texttt{applycal} task, keeping non self-calibrated visibilities with $S$/$N<3$ in order to not flag long baseline data. The spectral line data are imaged with a pixel size of 0\as15 (except for Field 2/SPR3 with a pixel size of 0\as5). The \texttt{CLEAN} stopping criterion of the 36 high-resolution spws is a peak residual intensity of 20\,mJy\,beam$^{-1}$ (except for Field 2/SPR3 where the threshold is 28\,mJy\,beam$^{-1}$). The 3 low-resolution spws are \texttt{CLEAN}ed with a stopping criterion set to 10\,mJy\,beam$^{-1}$ of the peak residual intensity.
	
	In this work, we focus on the physical properties of fragmented objects within the regions, we therefore only utilize a few emission lines. H(40)$\alpha$ recombination line emission is used to estimate for which regions and fragments there is a contribution of free-free emission to the 3\,mm continuum emission aside from dust emission (Sect. \ref{sec:ALMAffemission}). Molecular line emission of HCN, HNC, CH$_{3}$CN, and CH$_{3}^{13}$CN is used to estimate the temperature structure in the regions (Sect. \ref{sec:ALMAtemperature}). The line properties, such as rest frequency $\nu_0$ and upper energy level $E_{\mathrm{u}}/k_{\mathrm{B}}$, are summarized in Table \ref{tab:ALMAline}.
	
	The spectral line data products of the lines analyzed in this work are summarized in Table \ref{tab:ALMAlinedataproducts}. For each field, a mean value for the beam size and line noise are shown. The line noise $\sigma_\mathrm{line}$ in the low-resolution spw containing the H(40)$\alpha$ emission is $\approx$0.1\,K. The line noise in the high-resolution spws containing the HCN, HNC, CH$_{3}$CN, and CH$_{3}^{13}$CN is $\approx$0.7\,K (except for Field 2/SPR3, $\approx$0.1\,K). For the purpose of this work, for all regions in Field 2, the HCN spectral line data are imaged using the same baselines, pixel size, and stopping criterion as the HNC observations have (Table \ref{tab:almaobssummary}). Since in this work we use the HCN-to-HNC intensity ratio \citep{Hacar2020} to estimate the temperature in the regions (Sect. \ref{sec:TmapHCNHNC}) that requires the same angular resolution for both lines for a reliable comparison.

\subsection{Archival data}\label{sec:ALMAarchivdata}

	We use archival MIR, FIR, and cm data in order to create a multi-wavelength picture of each region in Figs. \ref{fig:overview_IRDC_G1111} - \ref{fig:overview_UCHII_G1387}. Science-ready \textit{Spitzer} IRAC 4.5\,$\upmu$m and MIPS 24\,$\upmu$m data are taken from the \textit{Spitzer} Heritage Archive and science-ready \textit{Herschel} PACS 160\,$\upmu$m and SPIRE 250\,$\upmu$m data products are taken from the \textit{Herschel} Science Archive.
	
	The clump properties of the 11 target regions are taken from ATLASGAL results. The 870\,$\upmu$m observations with APEX using the Large APEX BOlometer CAmera (LABOCA) have an angular resolution of 19\as2 \citep{Schuller2009}. The clump properties used in this study are taken from \citet{Urquhart2018} and listed in Table \ref{tab:ALMA_regions}. In Figs. \ref{fig:overview_IRDC_G1111} - \ref{fig:overview_UCHII_G1387} it can be clearly seen that an ATLASGAL clump roughly covers the ALMA FOV that is highlighted by a grey circle.
	
	Radio continuum observations at 5\,GHz (6\,cm) were taken with the VLA as part of the COordinated Radio aNd Infrared Survey for High-mass star formation (CORNISH) project \citep{Hoare2012} for all regions in our sample except for HMC\,G9.62$+$0.19 (Sect. \ref{sec:ALMAffemission}). The CORNISH project is a 5\,GHz (6\,cm) radio continuum survey with the VLA covering the Galactic plane at $10^\circ < l < 65^\circ$ and $|b| < 1^\circ$ at an angular resolution of 1\as5 and sensitivity of $\approx$0.3\,mJy\,beam$^{-1}$ with the primary aim to study UCH{\sc ii} regions. The CORNISH data give a complementary overview for which sources the 3\,mm ALMA emission is contaminated by free-free emission (Sect. \ref{sec:ALMAffemission}).
	
	\section{Continuum}\label{sec:ALMAcont}

\begin{figure*}
\includegraphics[width=0.94\textwidth]{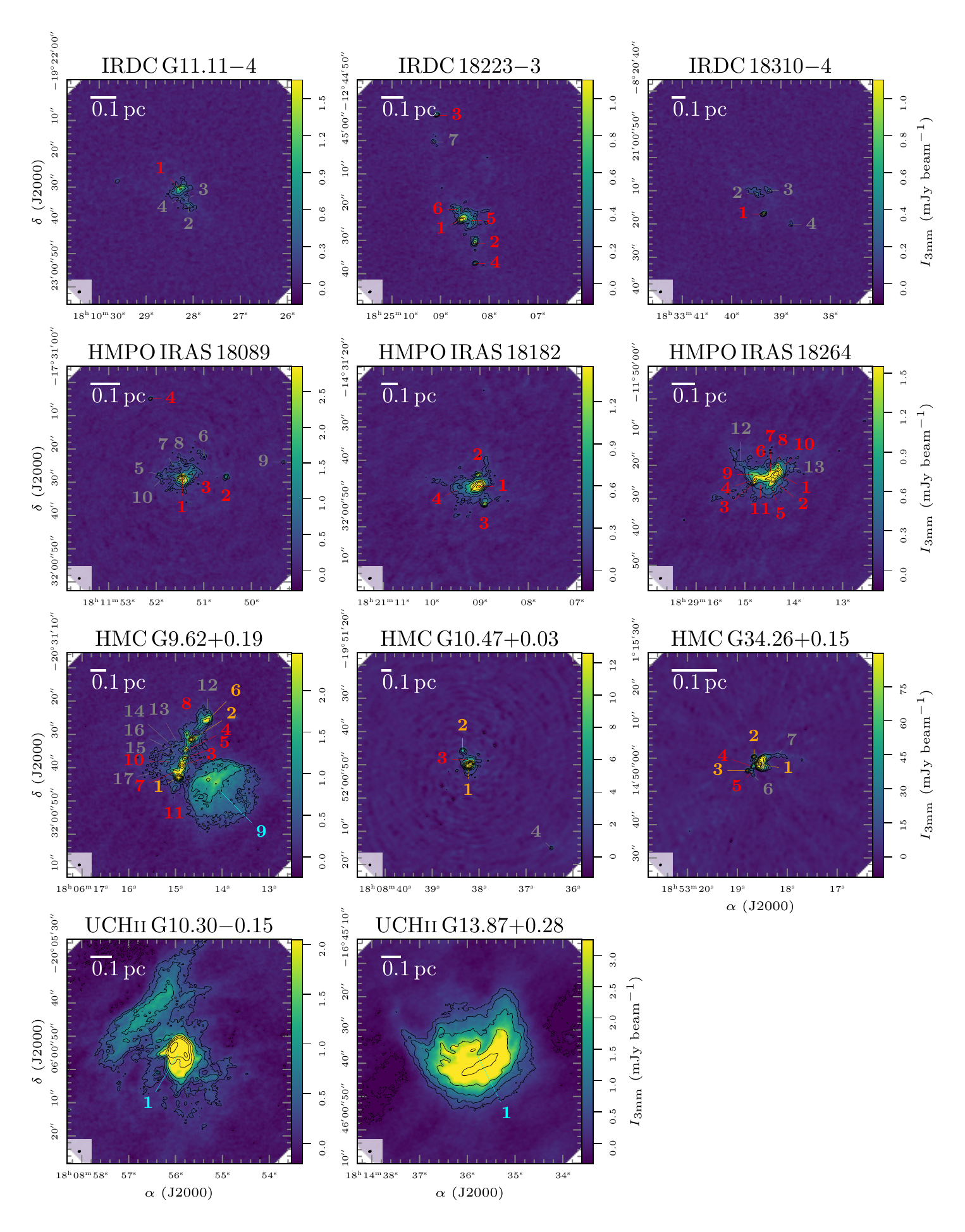}
\caption[ALMA 3\,mm continuum images of the sample.]{ALMA 3\,mm continuum. In each panel, the continuum data of the region is shown in color and black contours. The dotted black contour marks the $-5\sigma_\mathrm{cont}$ level. The solid black contours start at $5\sigma_\mathrm{cont}$ and contour steps increase by a factor of 2 (e.g., 5, 10, 20, $40\sigma_\mathrm{cont}$). The synthesized beam size is shown in the bottom left corner. The bar in the top left corner indicates a linear spatial scale of 0.1\,pc. The continuum noise and synthesized beam size are listed in Table \ref{tab:ALMAcontinuumdataproducts}. The continuum fragments are classified into dust cores (red), dust+ff cores (orange), cometary UCH{\sc ii} regions (cyan), further explained in Sect. \ref{sec:ALMAfrag}. Fragments with $S$/$N < 15$ are not analyzed in this study and are labeled in grey.}
\label{fig:ALMAcontinuum}
\end{figure*}

	In this section, we analyze the fragmentation properties in the regions and classify the fragmented objects (Sect. \ref{sec:ALMAfrag}). The free-free contribution in the 3\,mm continuum data for evolved cores and UCH{\sc ii} regions is estimated using the H(40)$\alpha$ recombination line (Sect. \ref{sec:ALMAffemission}).

	The ALMA 3\,mm continuum data is presented in Fig. \ref{fig:ALMAcontinuum} which consists not only of dust emission but also of free-free emission in regions containing evolved protostars. All regions show some level of fragmentation, with some being dominated by a single core (e.g., IRDC\,G11.11$-$4 and HMPO\,IRAS\,18089), while others have a number of bright cores (e.g., HMPO\,IRAS\,18182 and HMPO\,IRAS\,18264). In IRDC\,18223$-$3 and HMC\,G9.62$+$0.19 the cores are aligned in an elongated filamentary structure. In HMC\,G34.26$+$0.15, three UCH{\sc ii} regions \citep[referred to as A, B, and C in the literature, e.g.,][corresponding to source 3, 2, and 1 in Fig. \ref{fig:ALMAcontinuum}]{Mookerjea2007} are clearly resolved and detected. The UCH{\sc ii} regions G10.30$-$0.15 and G13.87$+$0.28 show extended 3\,mm emission in the shape of cometary halos. In HMC\,G9.62$+$0.19 there is an extended cometary UCH{\sc ii} region toward the south-west of a filament with embedded cores.

\subsection{Fragmentation and classification}\label{sec:ALMAfrag}

	In order to quantify the fragmentation in the continuum data, we use the \texttt{clumpfind} algorithm \citep{clumpfind}. The algorithm groups the continuum emission into individual ``clumps'' and we choose a starting level of $5\sigma_\mathrm{cont}$ and contour spacing of $5\sigma_\mathrm{cont}$. Although the algorithm is called \texttt{\textit{clump}find}, we may not refer to the structures as clumps, but refer to them in the following as cores since with the ALMA observations we trace the core scales with fragments typically smaller than 0.1\,pc. The minimum number of pixels in a core is set to 20, which covers a slightly smaller area of one synthesized beam.
	
	The properties (coordinates, relative position, peak intensity, integrated flux, radius) of all cores extracted by the \texttt{clumpfind} algorithm are listed in Table \ref{tab:ALMApositions}. The relative positions are given with respect to the phase center (Table \ref{tab:ALMA_regions}). The fragments are sorted by peak intensity and assigned with increasing number at decreasing peak intensity. The radius is calculated from the area $A$ covered by each identified core by $r_\mathrm{out} = \sqrt{\frac{A}{\pi}}$. 
	
	Since the three cometary UCH{\sc ii} regions (UCH{\sc ii}\,G10.30$-$0.15 1, UCH{\sc ii}\,G13.87$+$0.28 1, and HMC\,G9.62$+$0.19 9, Fig. \ref{fig:ALMAcontinuum}) have extended clumpy morphology, we exclude the area covered by these cometary UCH{\sc ii} regions from the \texttt{clumpfind} analysis and treat them as individual structures in our analysis. The peak intensity and integrated flux are computed over the area with $S$/$N$ $> 5$ covered by the extended cometary UCH{\sc ii} regions. 
	
	The 3\,mm continuum emission reveals that cores are embedded in extended envelopes which can themselves be clumpy. In the following analysis we therefore only focus on the cores identified by the \texttt{clumpfind} algorithm with a $S$/$N > 15$. With additional 5\,GHz data of the CORNISH project and observed H(40)$\alpha$ recombination line revealing regions where the 3\,mm emission is a composite of dust and free-free emission, we can further classify the fragmented sources. Cores with no detected H(40)$\alpha$ recombination line emission are referred to as ``dust cores'', while those with detected H(40)$\alpha$ and 5\,GHz emission (Sect. \ref{sec:ALMAffemission}) are classified as ``dust+ff cores''. The three extended UCH{\sc ii} regions are classified as ``cometary UCH{\sc ii} regions''. The classification for each fragment is listed in the last column in Table \ref{tab:ALMApositions}.
	
	In the following, ``protostellar sources'' refer to all fragments classified as dust cores, dust+ff cores, and cometary UCH{\sc ii} regions. While the structures with $S$/$N < 15$ might also contain protostellar objects, we refrain from analyzing their properties in this work due to insufficient sensitivity and angular resolution, since faint cores are typically unresolved in our data. Since these are minor substructures in our data and the goal of this work is to study radial density and temperature profiles, we exclude these cores in our analysis.
	
	In total, we extract 48 protostellar sources: 37 dust cores, 8 dust+ff cores, 3 cometary UCH{\sc ii} regions. In total 24 cores extracted using \texttt{clumpfind} have $S$/$N < 15$ and are not further analyzed in this work. Excluding these cores, the 3\,mm peak intensity and integrated flux of the fragments cover 5 orders of magnitudes from 0.44 to 2\,500\,mJy\,beam$^{-1}$ and 0.75 to 6\,600\,mJy, respectively and the radii range from 1\,100 to 66\,000\,au.
	
	In the following analysis, we derive the temperature and density profiles ($T \sim r^{-q}$ and $n \sim r^{-p}$), H$_{2}$ column densities $N$(H$_{2}$), and masses $M$ of the protostellar sources. In order to reliably estimate the H$_{2}$ column density and mass (Sect. \ref{sec:ALMANH2}), free-free emission that can have a significant contribution at 3\,mm wavelengths has to be subtracted first from the ALMA 3\,mm continuum in order to determine the 3\,mm dust emission (Sect. \ref{sec:ALMAffemission}).
	
\subsection{Free-free emission}\label{sec:ALMAffemission}

\begin{table*}[!htb]
\caption[Free-free contribution at 3\,mm in the ALMA regions.]{Free-free contribution at 3\,mm in the ALMA regions.}
\label{tab:ffemission}
\centering
\begin{tabular}{lrrrrl}
\hline \hline
 & $F_\mathrm{3mm}$ & $F_\mathrm{ff,3mm}$ & $F_\mathrm{dust,3mm}$ & $F_\mathrm{ff,3mm}$/$F_\mathrm{3mm}$ & classification \\
 \cline{2-4}
 & \multicolumn{3}{c}{(mJy)} & (\%) & \\
\hline
HMC\,G9.62$+$0.19 1 & $76.38$ & $46.46$ & $29.85$ & $60.84$ & dust+ff core\\ 
HMC\,G9.62$+$0.19 2 & $14.23$ & $2.92$ & $11.31$ & $20.51$ & dust+ff core\\ 
HMC\,G9.62$+$0.19 6 & $9.78$ & $9.37$ & $0.41$ & $95.84$ & dust+ff core\\ 
HMC\,G9.62$+$0.19 9 & $177.23$ & $158.8$ & $18.21$ & $89.6$ & cometary UCH{\sc ii} region\\ 
HMC\,G10.47$+$0.03 1 & $709.92$ & $189.83$ & $519.3$ & $26.74$ & dust+ff core\\ 
HMC\,G10.47$+$0.03 2 & $8.98$ & $8.21$ & $0.62$ & $91.34$ & dust+ff core\\ 
HMC\,G34.26$+$0.15 1 & $5276.23$ & $3636.25$ & $1639.98$ & $68.92$ & dust+ff core\\ 
HMC\,G34.26$+$0.15 2 & $81.24$ & $63.25$ & $17.99$ & $77.86$ & dust+ff core\\ 
HMC\,G34.26$+$0.15 3 & $43.44$ & $38.69$ & $4.75$ & $89.07$ & dust+ff core\\ 
UCH{\sc ii}\,G10.30$-$0.15 1 & $452.35$ & $389.58$ & $59.7$ & $86.12$ & cometary UCH{\sc ii} region\\ 
UCH{\sc ii}\,G13.87$+$0.28 1 & $481.71$ & $355.29$ & $112.71$ & $73.76$ & cometary UCH{\sc ii} region\\ 
\hline
\end{tabular}
\tablefoot{The free-free emission is estimated using the H(40)$\alpha$ recombination line (Fig. \ref{fig:freefree}).}
\end{table*}

	The continuum emission at mm wavelengths toward young and cold protostars typically arises from optically thin dust emission. An additional contribution from free-free emission can be present for more evolved protostars toward the HMC and UCH{\sc ii} regions. Free-free emission arises from free electrons scattering of ions \citep{Condon2016}. The CORNISH 5\,GHz data (Figs. \ref{fig:overview_IRDC_G1111} - \ref{fig:overview_UCHII_G1387}) reveal that only in the HMCs and UCH{\sc ii} regions strong cm emission is present. Unfortunately, HMC\,G9.62$+$0.19 is not covered by the CORNISH survey. In order to reliably estimate the H$_{2}$ column density and mass (Sect. \ref{sec:ALMANH2}) for all protostellar sources, the free-free contribution at 3\,mm wavelengths is estimated toward the HMC and UCH{\sc ii} regions using the H(40)$\alpha$ recombination line.

	Our ALMA 3\,mm spectral setup SPR2 covers the H(40)$\alpha$ recombination line allowing us to estimate for which sources there is a significant contribution of free-free emission at 3\,mm. The integrated intensity maps of the H(40)$\alpha$ recombination line are shown in Fig. \ref{fig:Halpha_moment0}. In all IRDCs and HMPOs, there is no H(40)$\alpha$ emission detected at a line sensitivity of $\approx$0.1\,K (Table \ref{tab:ALMAlinedataproducts}) which is expected since these regions are young and is in agreement with the lack of strong cm emission (Figs. \ref{fig:overview_IRDC_G1111} - \ref{fig:overview_UCHII_G1387}). In contrast, all HMCs and UCH{\sc ii} regions show at least toward some sources H(40)$\alpha$ emission. Thus the ALMA 3\,mm continuum emission is dominated by dust for all IRDCs and HMPOs and a composite of dust and free-free emission for the HMCs and UCH{\sc ii} regions.

	All cores with compact 3\,mm continuum emission with detected H(40)$\alpha$ emission are classified as dust+ff cores (Sect. \ref{sec:ALMAfrag}, Table \ref{tab:ALMApositions}). In addition, all three cometary UCH{\sc ii} regions show extended H(40)$\alpha$ emission. Assuming local thermal equilibrium (LTE) and that the free-free emission is optically thin, we can use the H(40)$\alpha$ recombination line and the line-to-continuum ratio \citep{Condon2016} given by
	
	\begin{equation}
	\label{eq:ff}
	\frac{T_\mathrm{L}}{T_\mathrm{C}} = 7\times10^3 \biggl( \frac{\Delta \varv}{\mathrm{km\,s}^{-1}} \biggl) ^{-1} \biggl( \frac{\nu}{\mathrm{GHz}} \biggl) ^{1.1} \biggl( \frac{T_\mathrm{e}}{\mathrm{K}} \biggl) ^{-1.15} \biggl( 1+\frac{N(\mathrm{He}^+)}{N(\mathrm{H}^+)} \biggl) ^{-1}
	\end{equation}
	
	to estimate the free-free contribution at 3\,mm wavelengths. $T_\mathrm{L}$ and $\Delta \varv$ are the H(40)$\alpha$ peak intensity and line width, respectively at $\nu = 99.023$\,GHz (Table \ref{tab:ALMAline}). $T_\mathrm{e}$ and $\frac{N(\mathrm{He}^+)}{N(\mathrm{H}^+)}$ are the electron temperature and He$^{+}$/H$^{+}$ ion ratio, respectively. The corresponding free-free continuum emission is $T_\mathrm{C}$. Since the H(40)$\alpha$ emission line is covered by the SPR2 setup, we image the SPR2-only 3\,mm continuum in order to ensure that exactly the same spatial scales are traced in both data sets. The results for all HMC and UCH{\sc ii} regions are summarized in Fig. \ref{fig:freefree}. The SPR2 3\,mm continuum is shown as contours in all panels in Fig. \ref{fig:freefree}. In all panels, areas are masked where the H(40)$\alpha$ integrated intensity and SPR2 3\,mm continuum have a $S$/$N < 5$. In all bottom panels, the units are converted from units of brightness temperature $T$ (K) to intensity $I$ (mJy\,beam$^{-1}$). The H(40)$\alpha$ peak intensity $T_\mathrm{L}$ is shown in the top left panel for all regions. The line integrated intensity map is shown in the top center panel. The line width $\Delta \varv$ is estimated from the 2nd moment, corresponding to the dispersion $\sigma^2$, with $\Delta \varv = \sqrt{8 \times \mathrm{ln}(2) \times \sigma^2}$ (top right panel), assuming Gaussian-shaped line profiles. The bottom left panel shows the SPR2 3\,mm continuum emission containing both dust and free-free emission. The bottom center panel shows the 3\,mm free-free continuum emission calculated using Eq. \eqref{eq:ff} assuming an electron temperature of $T_\mathrm{e}$ of 5\,500\,K \citep[e.g.,][]{Khan2022} and $\frac{N(\mathrm{He}^+)}{N(\mathrm{H}^+)}=0.08$ \citep{Condon2016}. In the bottom right panel the pure dust emission, calculated according to $I_\mathrm{3mm,dust} = I_\mathrm{3mm,dust+ff} - I_\mathrm{3mm,ff}$, is shown. In Table \ref{tab:ffemission}, the integrated flux of the SPR2 3\,mm continuum $F_\mathrm{3mm}$, free-free $F_\mathrm{ff,3mm}$, and dust $F_\mathrm{dust,3mm}$ emission is presented for all dust+ff cores and cometary UCH{\sc ii} regions. In addition, the fraction of free-free emission at 3\,mm wavelengths, $\frac{F_\mathrm{ff,3mm}}{F_\mathrm{3mm}}$ is shown in Table \ref{tab:ffemission}.
	
	While for a few dust+ff cores (e.g., HMC\,G9.62$+$0.19 2 and HMC\,G10.47$+$0.03 1) the free-free emission contributes less than 30\% to the total 3\,mm continuum, for the remaining sources the 3\,mm emission is dominated by free-free emission ($>$60\%) at 3\,mm. For HMC\,G34.26$+$0.15 1 we might underestimate the free-free contribution at 3\,mm since \citet{Mookerjea2007} estimate that at 3\,mm the continuum is completely dominated by free-free emission, while we estimate a fraction of 69\%. Therefore, we might overestimate the mass and column density in Sect. \ref{sec:ALMANH2}.
	
\section{Temperature structure}\label{sec:ALMAtemperature}

\begin{figure*}
\includegraphics[width=0.94\textwidth]{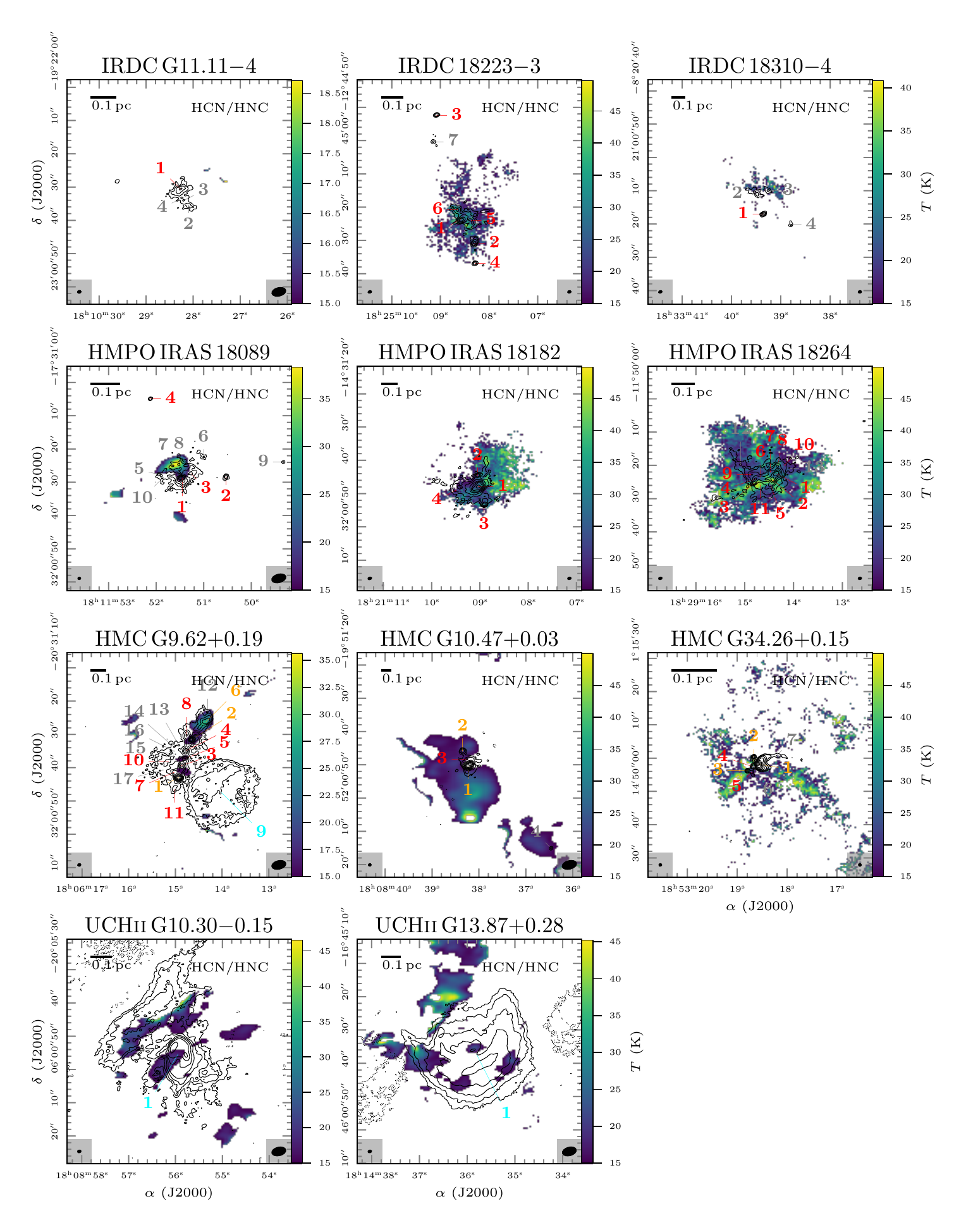}
\caption[Temperature maps derived with the HCN-to-HNC intensity ratio.]{Temperature maps derived with the HCN-to-HNC intensity ratio. In each panel, the temperature is shown in color and the 3\,mm continuum in black contours. The dotted black contour marks the $-5\sigma_\mathrm{cont}$ level. The solid black contours start at $5\sigma_\mathrm{cont}$ and contour steps increase by a factor of 2 (e.g., 5, 10, 20, $40\sigma_\mathrm{cont}$). The synthesized beam size of the continuum and line data is shown in the bottom left and right corner, respectively. The continuum fragments are classified into dust cores (red), dust+ff cores (orange), cometary UCH{\sc ii} regions (cyan), further explained in Sect. \ref{sec:ALMAfrag}. Fragments with $S$/$N < 15$ are not analyzed in this study and are labeled in grey.}
\label{fig:hcnhnctemperaturemaps}
\end{figure*}

	The temperature in the densest regions in IRDCs is typically very low, $T < 20$\,K \citep[e.g.,][]{Carey1998}, but as protostars form, they heat up the surrounding gas and dust and eventually the envelope is completely disrupted. On clump scales, the temperature can be estimated by fitting the spectral energy distribution (SED) of the dust continuum emission \citep[e.g., with \textit{Herschel} observations,][]{Ragan2012}, but the angular resolution is not sufficient to resolve individual cores embedded within their parental clumps. Molecular line emission can also be used to estimate the temperature, for example H$_{2}$CO and CH$_{3}$CN \citep[e.g.,][]{Rodon2012,Gieser2019, Gieser2021, Gieser2022, Lin2022}. Recently, \citet{Hacar2020} reported that between 15\,K and 50\,K the HCN-to-HNC intensity ratio provides a good estimate of the temperature.
	
	In this section we use both the HCN-to-HNC intensity ratio (Sect. \ref{sec:TmapHCNHNC}), tracing the cold extended gas, and CH$_{3}$CN and CH$_{3}^{13}$CN line emission (Sect. \ref{sec:ALMAch3cn}), tracing warmer gas, to create temperature maps for all regions. Azimuthal-averaged temperature profiles are computed in Sect. \ref{sec:ALMAradialTprofiles} for all protostellar sources to estimate the temperature power-law index $q$ according to Eq. \eqref{eq:temperatureprofile}.
	
	It has to be noted that these molecular temperature tracers were covered in the spectral setup SPR3 (Sect. \ref{sec:ALMAallobs}). For all regions located in Field 2 (Table \ref{tab:ALMA_regions}), the angular resolution is poorer ($\approx 3''$) compared to Field 1 and Field 3 ($\approx 1''$). Thus, the spatial scales that are traced by the ALMA observations in these molecular tracers are larger for regions in Field 2.
		
\subsection{HCN-to-HNC intensity ratio}\label{sec:TmapHCNHNC}

	\citet{Hacar2020} find an empirical relation between the gas temperature and the HCN-to-HNC line integrated intensity ratio $\frac{I(\mathrm{HCN})}{I(\mathrm{HNC})}$. The authors used IRAM 30m observations of the $1-0$ transition of HCN and HNC and compared the line integrated intensity ratios with kinetic temperature estimates derived from NH$_{3}$ observations. The empirical relation derived by \citet{Hacar2020} follows:
	\begin{equation}
	\begin{split}
	T &= 10 \times \frac{I(\mathrm{HCN})}{I(\mathrm{HNC})} &\mathrm{if} \; \frac{I(\mathrm{HCN})}{I(\mathrm{HNC})} &\leq 4\\
	T &= 3 \times \left(\frac{I(\mathrm{HCN})}{I(\mathrm{HNC})} - 4\right) + 40 &\mathrm{if} \; \frac{I(\mathrm{HCN})}{I(\mathrm{HNC})} &> 4.
	\end{split}
	\label{eq:HCNHNCratio}
	\end{equation}
	
	Our observational setup covers the $1-0$ transition of HCN and HNC (Table \ref{tab:ALMAline}) and therefore we use this approach to estimate the kinetic temperature of the extended low-density and low-temperature gas, where no CH$_{3}$CN line emission is detected. The HCN and HNC line data products are summarized in Table \ref{tab:ALMAlinedataproducts}. For both molecules, the integrated intensity is computed from $\varv_\mathrm{LSR}-30$\,km\,s$^{-1}$ to $\varv_\mathrm{LSR}+30$\,km\,s$^{-1}$. The noise in the integrated intensity maps is estimated from $\sigma_\mathrm{line}$ (Table \ref{tab:ALMAlinedataproducts}) and the number of channels $n_\mathrm{channel}$: $\sigma_\mathrm{int.intensity} = \sqrt{n_\mathrm{channel}} \times \sigma_\mathrm{line} \times \delta\varv$, with $\delta\varv = 0.8$\,km\,s$^{-1}$. All pixels with a $S$/$N < 3$ in the integrated intensity map are masked. Since the HCN line is in SPR1 and HNC in SPR3, for which all regions in Field 2 (Table \ref{tab:ALMA_regions}) have a poorer resolution of $\approx$3$''$, we imaged the HCN spectral line data using the same baselines as the HNC spectral line data (Sect. \ref{sec:ALMAobs}). The temperature is calculated according to Eq. \eqref{eq:HCNHNCratio} from the HCN-to-HNC intensity ratio. We further mask all pixels with $T_\mathrm{kin} < 15$\,K and $T_\mathrm{kin} > 50$\,K. 
	
	The temperature maps derived with the HCN-to-HNC intensity ratio are presented in Fig. \ref{fig:hcnhnctemperaturemaps}. In many cases (IRDC\,18223$-$3, HMPO\,IRAS\,18182, HMPO\,IRAS\,18264, HMC\,G10.47$+$0.03), the temperature maps are extended due to HCN and HNC emitting strong along large spatial scales. In IRDC\,G11.11$-$4, we are not able to estimate the temperature using this method since the derived temperatures are $< 15$\,K, that is the lower limit for which the HCN-to-HNC intensity ratio is valid \citep{Hacar2020}. In IRDC\,18310$-$4, a temperature can only be estimated in the outskirts of core 2 and 3. In HMPO\,IRAS\,18089, an enhanced temperature is found toward the northern part of mm 1 towards position 7 and 8. This feature can be connected to the north-south outflow \citep{Beuther2004}. Broad line wings of bipolar outflows can affect the line integrated intensities (e.g., toward HMC\,G34.26$+$0.15). In HMPO\,IRAS\,18182 a large scale E-W temperature gradient can be observed. The UCH{\sc ii} region at position HMC\,G9.62$+$0.19 6 heats up the environment with elevated temperatures in the surrounding envelope. In HMC\,G10.47$+$0.03 a clump with enhanced temperatures is found $\approx$ 15$''$ south of the continuum peak position.
	
	Since this method is only valid from $\approx$15\,K up to $\approx$50\,K \citep{Hacar2020} and the HCN and HNC lines can become optically thick in the densest regions, we use CH$_{3}$CN and CH$_{3}^{13}$CN in the next section to probe the temperature in the high-density and high-temperature regions.

\subsection{Methyl cyanide}\label{sec:ALMAch3cn}

\begin{figure*}
\includegraphics[width=0.94\textwidth]{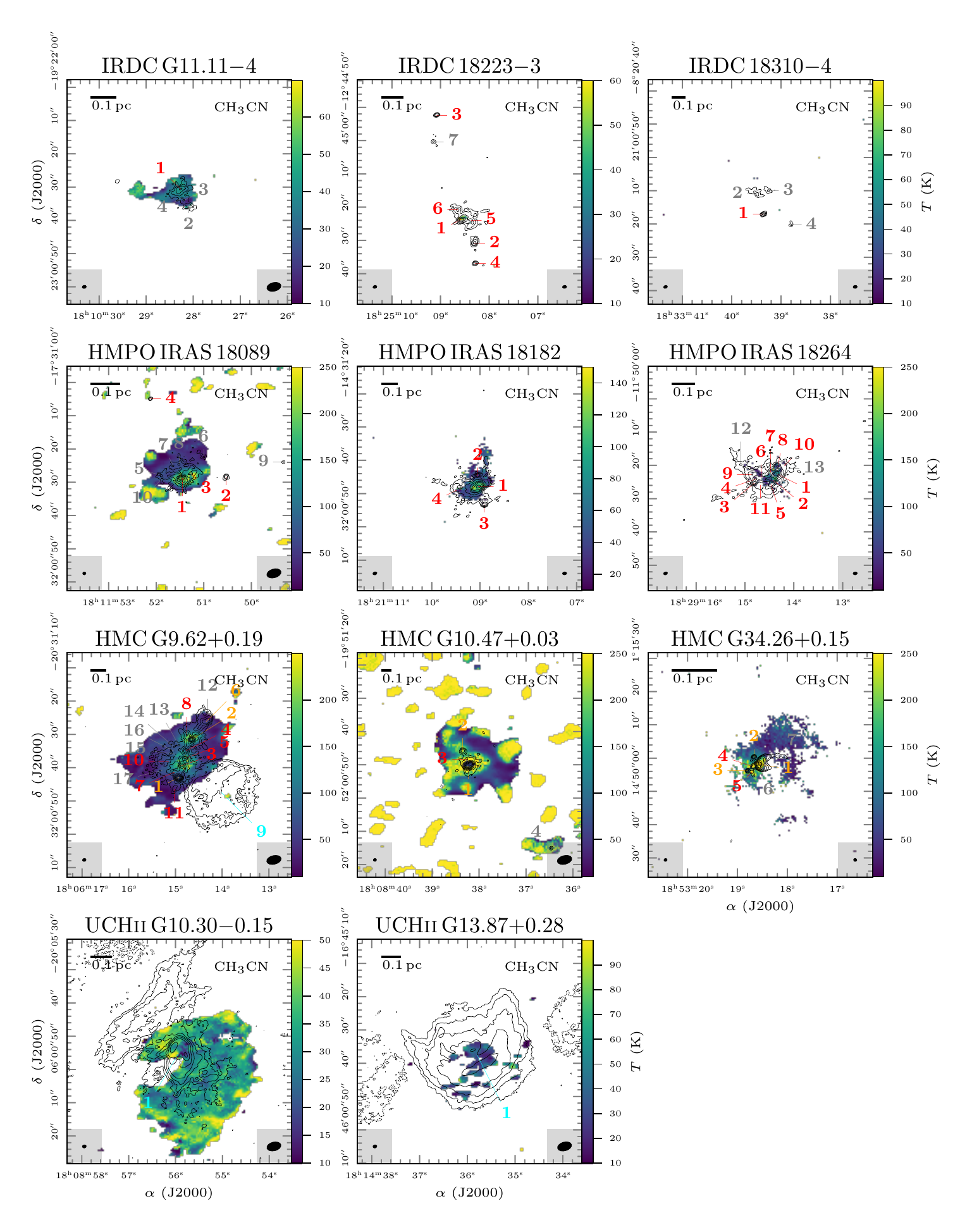}
\caption[Temperature maps derived with CH$_{3}$CN.]{Temperature maps derived with CH$_{3}$CN. In each panel, the temperature is shown in color and the 3\,mm continuum in black contours. The dotted black contour marks the $-5\sigma_\mathrm{cont}$ level. The solid black contours start at $5\sigma_\mathrm{cont}$ and contour steps increase by a factor of 2 (e.g., 5, 10, 20, $40\sigma_\mathrm{cont}$). The synthesized beam size of the continuum and spectral line data are shown in the bottom left and right corner, respectively. The continuum fragments are classified into dust cores (red), dust+ff cores (orange), cometary UCH{\sc ii} regions (cyan), further explained in Sect. \ref{sec:ALMAfrag}. Fragments with $S$/$N < 15$ are not analyzed in this study and are labeled in grey. The clumpy features surrounding HMPO\,IRAS\,18089 and HMC\,G10.47$+$0.03 are caused by side lobes.}
\label{fig:ch3cntemperaturemaps}
\end{figure*}

\begin{figure*}[!htb]
\includegraphics[width=0.99\textwidth]{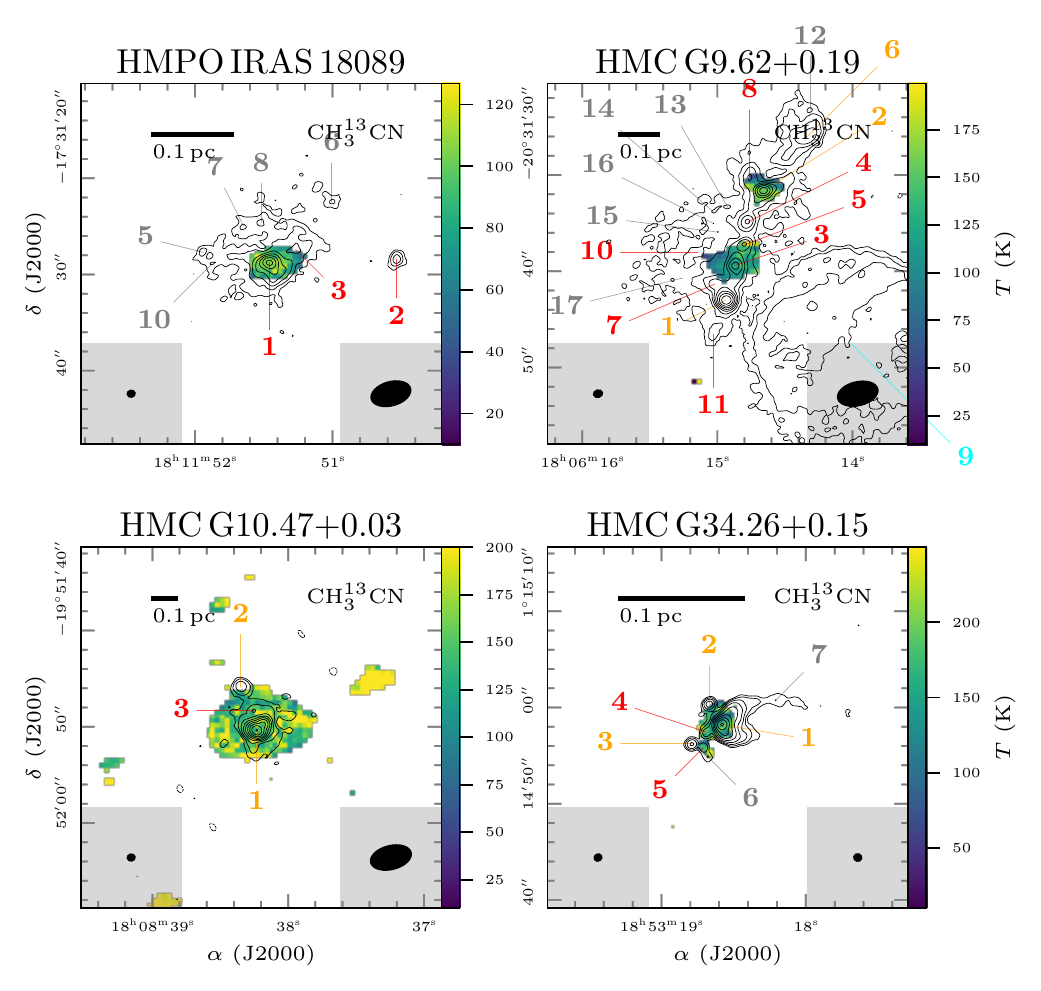}
\caption[Temperature maps derived with CH$_{3}^{13}$CN.]{Temperature maps derived with CH$_{3}^{13}$CN. In each panel, the temperature is shown in color and the 3\,mm continuum in black contours. The dotted black contour marks the $-5\sigma_\mathrm{cont}$ level. The solid black contours start at $5\sigma_\mathrm{cont}$ and contour steps increase by a factor of 2 (e.g., 5, 10, 20, $40\sigma_\mathrm{cont}$). The synthesized beam size of the continuum and spectral line data are shown in the bottom left and right corner, respectively. The continuum fragments are classified into dust cores (red), dust+ff cores (orange), cometary UCH{\sc ii} regions (cyan), further explained in Sect. \ref{sec:ALMAfrag}. Fragments with $S$/$N < 15$ are not analyzed in this study and are labeled in grey.}
\label{fig:ch313cntemperaturemaps}
\end{figure*}

	\citet{Gieser2021} model CH$_{3}$CN line emission to infer the temperature structure in the CORE sample using the radiative transfer tool \texttt{XCLASS} \citep{XCLASS}. While the CORE 1\,mm spectral setup covered the CH$_{3}$CN $J=12-11$ $K$-ladder ($E_\mathrm{u}$/$k_\mathrm{B}=69-530$\,K), in the ALMA 3\,mm setup the $J=5-4$ transitions are covered ($E_\mathrm{u}$/$k_\mathrm{B}=13-130$\,K for CH$_{3}$CN and CH$_{3}^{13}$CN, Table \ref{tab:ALMAline}). Extended CH$_{3}^{13}$CN emission is only detected toward the three HMCs and HMPO\,IRAS\,18089, and due to a high optical depth of the main isotopologue, CH$_{3}^{13}$CN can thus more reliably trace the temperature of denser inner regions.
	
	With \texttt{myXCLASSMapfit}, we fit all pixels with a peak intensity $>$5$\sigma_\mathrm{line}$ in the spectrum (Table \ref{tab:ALMAlinedataproducts}) using CH$_{3}$CN for all 11 regions and CH$_{3}^{13}$CN for the four regions in which significant emission is detected. Each parameter range (rotation temperature $T_\mathrm{rot}$, column density $N$, source size $\theta_\mathrm{Source}$, line width $\Delta\varv$, and velocity offset $\varv_\mathrm{off}$) is adjusted for each region and molecule, since the sample covers a broad range of densities and temperatures. We therefore iteratively adjust the parameter ranges that result in relatively smooth parameter maps without too many outliers due to unreliable fits. When many lines are optically thick, the algorithm converges to high temperatures \citep{Gieser2021}, we therefore set the highest possible temperature to 250\,K. To trace hotter gas layers, observations of CH$_{3}$CN and CH$_{3}^{13}$CN transitions with higher upper energy levels are required.
	
	The temperature maps of CH$_{3}$CN and CH$_{3}^{13}$CN are presented in Figs. \ref{fig:ch3cntemperaturemaps} and \ref{fig:ch313cntemperaturemaps}, respectively. Toward the likely youngest region in our sample, IRDC\,18310$-$4, CH$_{3}$CN is not detected. In IRDC\,18223$-$3, the temperature can only be estimated toward the 3\,mm continuum peak position. In IRDC\,G11.11$-$4, the CH$_{3}$CN emission is already more extended with $T \approx 50$\,K. From the IRDCs to the HMCs, the peak temperature is clearly increasing up to $\approx$250\,K (that is the upper limit set in \texttt{XCLASS}). Bright CH$_{3}$CN emission in HMPO\,IRAS\,18089 and HMC\,G10.47$+$0.03 causes strong side lobes which can be seen as artifacts toward the edges surrounding the central region. A temperature plateau is reached toward the continuum peak positions in the HMCs, and in these cases CH$_{3}^{13}$CN is a better temperature probe. CH$_{3}^{13}$CN is less extended, but clearly traces high temperatures $>$150\,K in the four regions with detected CH$_{3}^{13}$CN emission. The cometary UCH{\sc ii} region in HMC\,G9.62$+$0.19 (source 9) produces a temperature increase of $\approx$150\,K toward the edge of the filament which otherwise has a lower temperature of $\approx$50\,K in the envelope.
	
	The temperature map of UCH{\sc ii}\,G10.30$-$0.15 is very extended toward the west, and no CH$_{3}$CN is detected toward the east facing the bipolar H{\sc ii} region. In UCH{\sc ii}\,G13.87$+$0.28 the temperature map is less extended with CH$_{3}$CN being present in the cometary halo. In both UCH{\sc ii} regions, the kinetic temperature is very low, with $T < 50$\,K. There are two possible explanations for such a low temperature: either CH$_{3}$CN is present in the cold gas envelope or the line emission is produced by non-LTE effects. However, since the HCN-to-HNC intensity ratio method also infers temperatures $\lesssim$50\,K, the first explanation seems to be more likely. Therefore, the extended molecular emission in the cometary UCH{\sc ii} regions seems to stem from an envelope that either stayed cold during the evolution of the protostars or is cooling due to the expanding UCH{\sc ii} region.

\subsection{Radial temperature profiles}\label{sec:ALMAradialTprofiles}

\begin{figure}
\includegraphics[width=0.49\textwidth]{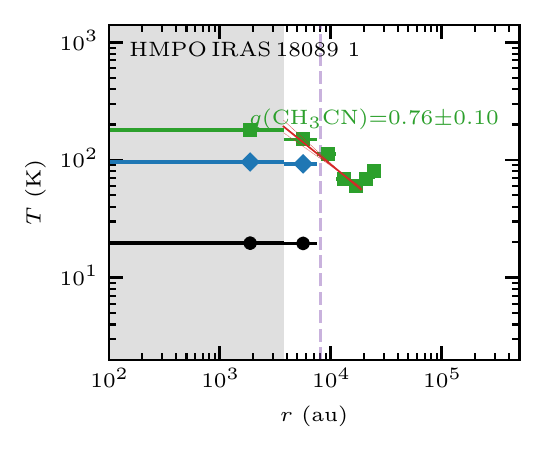}
\caption{Radial temperature profile of dust core 1 in HMPO\,IRAS\,18089. The profiles extracted from the HCN-to-HNC intensity ratio (Fig. \ref{fig:hcnhnctemperaturemaps}), CH$_{3}$CN (Fig. \ref{fig:ch3cntemperaturemaps}), and CH$_{3}^{13}$CN (Fig. \ref{fig:ch313cntemperaturemaps}) temperature maps are shown by black circles, green squares, and blue diamonds, respectively. The inner unresolved region (one beam radius) is shown as a grey-shaded area. The dashed purple vertical line indicates the outer radius $r_\mathrm{out}$ based on the continuum (Table \ref{tab:ALMApositions}). A power-law fit and its $1\sigma$ uncertainty to resolved and radially decreasing profiles is shown by the red solid and dashed lines, respectively (Sect. \ref{sec:ALMAradialTprofiles}). The figures for the remaining sources are shown in Fig. \ref{fig:ALMATradapp}.}
\label{fig:ALMATrad}
\end{figure}

	The temperature maps created with the HCN-to-HNC intensity ratio, CH$_{3}$CN, and CH$_{3}^{13}$CN allow us to derive radial temperature profiles of the protostellar sources. The radial temperature profiles are approximated by a power-law profile with temperature power-law index $q$, $T \sim r^{-q}$ (Eq. \ref{eq:temperatureprofile}). For each of the protostellar sources and temperature tracers, the azimuthal-averaged temperature profile is calculated along seven beams, that is sufficient to fully resolve the radial profiles, starting at the continuum peak position of the source (Table \ref{tab:ALMApositions}). The temperature profiles are binned in steps of half the synthesized beam size.
	
	In order to derive the radial temperature profile, we fit a power-law profile (Eq. \ref{eq:temperatureprofile}) to continuous profiles that are resolved along at least two beams and that are radially decreasing. In the fitting, we exclude the innermost data point that is smeared out by the limiting beam size (highlighted as a gray shaded area in Fig. \ref{fig:ALMATrad}) if more than three radial steps are otherwise available for a fit. The profiles are fitted as long as continuous data points are available and as long as the profiles are not increasing with increasing radius along radial steps. The fit results including the inner radius $r_\mathrm{in}$, temperature $T_\mathrm{in} = T(r_\mathrm{in})$, and temperature power-law index $q$ are summarized in Table \ref{tab:ALMAradialtemp}. The radial temperature profile of the dust core HMPO\,IRAS\,18089 1 is shown in Fig. \ref{fig:ALMATrad} and in Fig. \ref{fig:ALMATradapp} for all remaining protostellar sources that could be fitted by a power-law profile. 
	
	The temperature profiles derived from the HCN-to-HNC intensity ratio tracing the colder envelope are flat $q \approx 0.1-0.3$, while the CH$_{3}$CN profiles tracing the hotter gas can be steeper with values ranging between $q \approx 0.1-0.8$. The temperature profiles are steep $q > 0.5$ toward sources that are in the HMC regions, but also already toward dust cores in HMPO regions (dust cores 1 and 3 in HMPO\,IRAS\,18089, dust core 1 in HMPO\,IRAS\,18182). The observed CH$_{3}^{13}$CN radial profiles (five sources in total) are either not resolved or flat and are therefore not fitted. With HCN and HNC we therefore trace the temperature of the colder larger scale envelope where the cores are embedded in, while high-density tracers such as CH$_{3}$CN can be used to infer the temperature of the protostellar core.
	
	The beam-averaged temperature $\overline T$ for all three tracers is computed for all sources from the temperature maps (Figs. \ref{fig:hcnhnctemperaturemaps}, \ref{fig:ch3cntemperaturemaps}, and \ref{fig:ch313cntemperaturemaps}) in order to estimate the H$_{2}$ column density and mass (Sect. \ref{sec:ALMANH2}). The results of the beam-averaged temperatures are summarized in Table \ref{tab:ALMAbeamavgtempNM}. By constraint, the HCN-to-HNC intensity ratio only traces temperatures up to 50\,K, in addition, the low upper energy level ($E_\mathrm{u}$/$k_\mathrm{B} =4$\,K) of the HCN and HNC emission line is only sensitive to the colder envelope and might become optically thick toward the denser regions. Therefore, the temperatures might be considerably lower compared to the temperatures derived with CH$_{3}$CN and CH$_{3}^{13}$CN. In most cases, the CH$_{3}^{13}$CN beam-averaged temperature is lower than the CH$_{3}$CN beam-averaged temperature. This can be attributed to the fact that if most CH$_{3}$CN transitions become optically thick, the fitting algorithm in \texttt{XCLASS} converges toward the upper limit of the rotation temperature that we set to 250\,K. A similar effect was found for the CORE sample, where in regions with a high H$_{2}$CO line optical depth, the derived H$_{2}$CO rotation temperature is higher than the CH$_{3}$CN rotation temperature \citep{Gieser2021}. In Sect. \ref{sec:ALMAdiscussion} we discuss evolutionary trends of the temperature profiles and compare the results with the CORE and CORE-extension regions \citep{Gieser2021,Gieser2022}.
	
	A high line optical depth of the CH$_{3}$CN lines causes \texttt{XCLASS} to converge to 250\,K, that is the set upper limit (Sect. \ref{sec:ALMAch3cn}). This regime is reached toward the inner regions of dust+ff cores 1 in HMC\,G10.47$+$0.03 and HMC\,G34.26$+$0.15 (Fig. \ref{fig:ch3cntemperaturemaps}). However, since the CH$_{3}$CN emission is extended, more data points in the optically thin regime are available for a reliable radial temperature profile fit (Fig. \ref{fig:ALMATradapp}).
	
\section{Density profiles}\label{sec:ALMAdensity}

		In this section, we derive radial density profiles of the protostellar sources using the 3\,mm continuum data. Interferometric observations filter out extended emission and missing flux can therefore be an issue. The radial density profile with power-law index $p$ (Eq. \ref{eq:densityprofile}) can be best estimated from the continuum data in the $uv$ plane \citep[e.g.,][]{Adams1991,Looney2003,Zhang2009,Beuther2007B} considering the visibility and temperature power-law indices, $\alpha$ and $q$, with 
	\begin{equation}
	\label{eq:uvanalysis}
	\alpha = p + q - 3.
	\end{equation}
Since bright nearby sources can affect the visibility profiles of individual fragmented sources, we first subtract the emission of other dust and/or dust+ff cores (Sect. \ref{sec:ALMAsourcesub}) and then compute and fit the complex-averaged visibility profiles $\overline V$ (Sect. \ref{sec:ALMAvisibprofile}).

\subsection{Source subtraction}\label{sec:ALMAsourcesub}

	For each dust and/or dust+ff core, we subtract the emission of the remaining dust and/or dust+ff cores within a region by modeling their emission with a circular Gaussian profile using the task \texttt{uvmodelfit} in \texttt{CASA}.
	
	 The source model is Fourier-transformed to the $uv$ plane with the \texttt{ft} task in \texttt{CASA} and subtracted from the data using the \texttt{uvsub} task in \texttt{CASA}. Source subtraction is only necessary when multiple dust and/or dust+ff cores are present within one region, therefore this step is not necessary for IRDC\,G11.11$-$4, IRDC\,18310$-$4, UCH{\sc ii}\,G10.30$-$0.15, and UCH{\sc ii}\,G13.87$+$0.28 (Fig. \ref{fig:ALMAcontinuum} and Table \ref{app:fragprops}). Since HMC\,G9.62$+$0.19 9 is an extended cometary UCH{\sc ii} region (Fig. \ref{fig:ALMAcontinuum}) with a complex morphology and cannot be described by a simple model, we refrain from modeling and subtracting the emission of this source. For the remaining regions, we carefully check by imaging the source-subtracted data that the emission of each source is approximately removed without over-subtracting the emission and does not affect the emission of nearby sources. Since most of the sources are embedded in fainter extended envelopes, it is not possible to completely model and remove the emission of those extended envelopes by a simple Gaussian model. However, with imaging the core-subtracted data we ensure that the bright source emission is subtracted well.
	
	As an example, before computing the visibility profile of dust core 2 in HMPO\,IRAS\,18182, we subtract the emission of the remaining dust cores 1, 3, and 4. One caveat is that the continuum data of the dust+ff cores and cometary UCH{\sc ii} regions contain not only dust, but also extended free-free emission (Sect. \ref{sec:ALMAffemission}) that might have an impact on the derived density profile.
	
\subsection{Visibility profiles}\label{sec:ALMAvisibprofile}

\begin{figure}[!htb]
\includegraphics[width=0.49\textwidth]{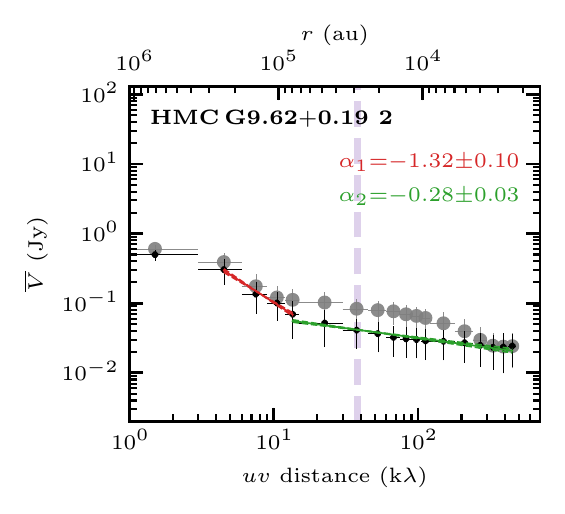}
\caption{Visibility profile of dust+ff core 2 in HMC\,G9.62$+$0.19. The profile of the non-core-subtracted and core-subtracted data is shown in grey and black, respectively (further explained in Sect. \ref{sec:ALMAsourcesub}). Two power-law profiles, tracing roughly the clump and core scales, are fitted to the core-subtracted data shown in red and green, respectively. The bottom axis shows the $uv$ distance in k$\lambda$ and the top axis is the corresponding spatial scale. The purple dashed line indicates the source diameter (Table \ref{tab:ALMApositions}). The figures for the remaining sources are shown in Fig. \ref{fig:ALMAvisibilityprofileapp}.}
\label{fig:ALMAvisibilityprofile}
\end{figure}

	The complex-averaged visibility profiles $\overline V$ as a function of $uv$ distance are computed using the \texttt{plotms} task in \texttt{CASA}. The phase center is shifted to the corresponding source position (Table \ref{tab:ALMAdens}). For a comparison of how the subtraction of the remaining cores in the region is impacting the visibility profiles, we also compute the complex-averaged visibility profiles $\overline V$ using the original 3\,mm continuum data, i.e. with no source subtraction. 
	
	Since the number of long baselines is much smaller than the number of short baselines, the visibility profiles are binned with bin sizes of 3, 15, and 60\,k$\lambda$ in the ranges of $0-15$, $15-120$, and $120-470$\,k$\lambda$, respectively. Since the shortest baseline corresponds to $\approx$2.3\,k$\lambda$, the first binned data point ($0-3$\,k$\lambda$) is expected to suffer from missing flux. The visibility profile of dust+ff core 2 in HMC\,G9.62$+$0.19 is shown in Fig. \ref{fig:ALMAvisibilityprofile} as an example and in Fig. \ref{fig:ALMAvisibilityprofileapp} for all remaining protostellar sources.
	
	In contrast to the CORE and CORE-extension cores \citep{Gieser2021,Gieser2022} which all have visibility profiles that can be described by a single power-law profile, $\overline V \sim s^{\alpha}$, most of the sources in this ALMA sample are better described by two power-law profiles with with slope $\alpha_1$, at short baselines (large spatial scales), and with slope $\alpha_2$, at longer baselines (small spatial scales). We therefore fit two power-law profiles with $\alpha_1$ and $\alpha_2$ to the observed binned visibility profiles. The breakpoint between the two slopes is determined using the \texttt{pwlf} python package that allows a continuous piecewise linear function to be fitted to the logarithmic visibility profiles. In order to avoid a breakpoint at the shortest and longest $uv$-distances where only a few data points would be available, the breakpoint is allowed to be between 10\,k$\lambda$ and 100\,k$\lambda$. The slopes $\alpha_1$ and $\alpha_2$ roughly correspond to clump ($>$0.1\,pc) and core ($<$0.1\,pc) scales, respectively. The first binned data point, suffering from missing flux, is not included in the fit. The corresponding density profiles are calculated according to Eq. \eqref{eq:uvanalysis} using the temperature power-law index $q$ measured toward each source (Table \ref{tab:ALMAradialtemp}). The results for $\alpha_1$, $\alpha_2$, $p_1$, and $p_2$ are summarized in Table \ref{tab:ALMAdens}. 
	
	Potential reasons for the presence of two power-law profiles in the ALMA data, compared to one power-law profile in the NOEMA data are discussed in more detail in Sect. \ref{sec:ALMAdiscussioncore}. The most likely reason is that at the longer $uv$-distances, ALMA has a higher sensitivity due to the array consisting of more antennas with longer baselines compared to NOEMA. We therefore can more easily identify several power-law profiles in the ALMA data.
	
	In Figs. \ref{fig:ALMAvisibilityprofile} and \ref{fig:ALMAvisibilityprofileapp} it is clearly visible that the source subtraction can have a big impact on the visibility profiles when bright sources are located within a single region (e.g., for sources in the HMC\,G9.62$+$0.19 and HMC\,G34.26$+$0.15 regions). Toward sources in bright regions (HMC\,G10.47$+$0.03 and HMC\,G34.26$+$0.15), the visibility profiles flatten at short $uv$ distances. A possible explanation could be the presence of free-free emission. However, this would imply extended free-free emission, while we find that the free-free contribution is concentrated toward the sources and is not extended (Fig. \ref{fig:freefree}). A physical explanation could be that the envelope is disrupted by the protostars themselves or by neighboring sources resulting in a steep density profile and thus a flat visibility profile at short $uv$-distances.
	
	In a few regions, a significant contribution of free-free emission is present at 3\,mm wavelengths as evaluated in Sect. \ref{sec:ALMAffemission} (Table \ref{tab:ffemission}). As an example, for UCH{\sc ii}\,G10.30$-$0.15 and UCH{\sc ii}\,G13.87$+$0.28 the free-free contribution is greater than 80\%. In these cases, the visibility profiles are a composite of dust and free-free emission, while in the calculation of the density power-law index optically thin dust emission is assumed (Sect. \ref{sec:ALMANH2}).

	When multiple sources are present within one region (e.g., IRDC\,18223$-$3 and HMPO\,IRAS\,18182), the visibility slope at clump scales ($\alpha_1$) are similar (Table \ref{tab:ALMAdens}). This can be explained by the fact that at small baselines, the cores are all covered by such short baselines (and corresponding larger angular resolution) and even though the phase center is shifted to the source position, it does not impact the clump scale in which the cores are embedded in. Except for the sources with a flattening at small $uv$ distances, there is a trend such that the visibility profiles $\alpha_1$, corresponding to the large scale structures, steepen from the IRDC to UCH{\sc ii} regions implying a flattening of the density profile $p_1$. 
	
	The visibility slope at the smaller core scales ($\alpha_2$) are nearly flat, $\alpha_2 \approx 0$,  implying unresolved sources in the IRDC stages and become then also steeper, with a slope up to $\alpha_2 \approx -1.0$. A detailed discussion of the physical structure and evolutionary trends is presented in Sect. \ref{sec:ALMAdiscussion}.

\section{Molecular hydrogen column density and mass estimates}\label{sec:ALMANH2}

	The H$_{2}$ column density $N$(H$_{2}$) and mass $M$ of the sources can be estimated from the 3\,mm continuum emission according to 
	\begin{equation}
	\label{eq:H2calc}
	N(\mathrm{H}_2) = \int n(\mathrm{H}_2) \mathrm{d}s = \frac{F_{\nu}^{\mathrm{peak}} \eta }{\mu m_{\mathrm{H}} \Omega_\mathrm{A} \kappa_{\nu} B_{\nu}(T)},
	\end{equation}
and
	\begin{equation}
	\label{eq:Mcalc}
	M = \frac{F_{\nu} \eta d^2}{\kappa_{\nu} B_{\nu}(T)},
	\end{equation} 
with the assumption that the emission stems from dust and is optically thin. Therefore, for the dust+ff cores and cometary UCH{\sc ii} regions that all have free-free emission at 3\,mm, we correct the peak intensity $I_\mathrm{3mm}$ and integrated flux $F_\mathrm{3mm}$ by the fraction of estimated free-free emission listed in Table \ref{tab:ffemission}. For the temperature, we assume either the beam-averaged temperature $\overline T$ based on the CH$_{3}^{13}$CN temperature maps if detected, CH$_{3}$CN if otherwise detected, and HCN-to-HNC intensity ratio otherwise (Table \ref{tab:ALMAbeamavgtempNM}). We use the same values for the following parameters as for the CORE and CORE-extension regions \citep{Gieser2021,Gieser2022} with gas-to-dust mass ratio $\eta = 150$ \citep{Draine2011} and mean molecular weight $\mu = 2.8$ \citep{Kauffmann2008}. The dust opacity is extrapolated from 1.3\,mm \citep[$\kappa_0 = 0.9$\,g\,cm$^{-2}$,][]{Ossenkopf1994} to 3\,mm wavelengths with $\kappa_\nu = \kappa_0 \left(\frac{\nu}{\nu_0}\right)^\beta$, resulting in $\kappa_\nu = 0.12$\,g\,cm$^{-2}$ assuming $\beta=2$.
	
	The results for $N$(H$_{2}$) and $M$ are summarized in Table \ref{tab:ALMAbeamavgtempNM}. The optical depth $\tau^\mathrm{cont}_\nu$ calculated according to 	
	\begin{equation}
	\label{eq:opticaldepth}
	\tau_{\nu}^{\mathrm{cont}} = -\mathrm{ln}\bigg( 1 - \frac{I_{\nu}}{B_{\nu}(T)} \bigg).
	\end{equation}
is also listed in Table \ref{tab:ALMAbeamavgtempNM}. Since the properties are derived from interferometric observations, the results are lower limits due to potential missing flux (Sect. \ref{sec:ALMAobs}). For HMC\,G34.26$+$0.15 1, the optical depth cannot be estimated since the numerator is larger than the denominator in Eq. \eqref{eq:opticaldepth}, most likely because we overestimate the 3\,mm dust emission as discussed in Sect. \ref{sec:ALMAffemission}, and therefore the optical depth is estimated to be high. In addition, for dust+ff core 1 in HMC\,G10.47$+$0.03, the optical depth becomes high as well, $\tau^\mathrm{cont}_\nu = 0.5$. For the remaining sources, the optical depth $\tau^\mathrm{cont}_\nu$ is much smaller than 1. Indeed, the brightness temperatures are high toward the continuum peak position in HMC\,G10.47$+$0.03 and HMC\,G34.26$+$0.15 with $T_\mathrm{B} \approx 40$\,K and $T_\mathrm{B} \approx 300$\,K, respectively, while for the remaining regions $T_\mathrm{B} < 10$\,K holds. 

	The H$_{2}$ column densities vary between $1\times10^{23}$\,cm$^{-2}$ and $2\times10^{26}$\,cm$^{-2}$ and the masses range between 0.1\,$M_\odot$ and 150\,$M_\odot$. While the estimated core masses are below 8\,$M_\odot$ for most cores, all regions are embedded within massive clumps of at least 1\,000\,$M_\odot$ (Table \ref{tab:ALMA_regions}). The estimated core masses, based on the interferometric observations, are lower limits since the extended emission is filtered out. Thus, considering mass estimates on the core and clump scales, and taking into account the high bolometric luminosities (Table \ref{tab:ALMA_regions}), suggests that the regions are forming massive stars. However, not all of the cores will be massive enough to host a high-mass star ($M_\star > 8 M_\odot$). The highest mass is found toward HMC\,G10.47$+$0.03 1 with $M \approx 150$\,$M_\odot$. \citet{Cesaroni2010} detect three sources toward this position at 1.6\,cm with the VLA at a higher angular resolution of 0\as1. This indicates that this region is currently forming a compact cluster of massive stars.

\section{Discussion}\label{sec:ALMAdiscussion}

	In this study, we characterized the physical properties toward a sample of 11 HMSFRs observed with ALMA based on the the 3\,mm continuum and spectral line emission. The 3\,mm continuum data reveal a high degree of fragmentation (Fig. \ref{fig:ALMAcontinuum}) and fragmented objects, identified using \texttt{cumpfind} \citep{Williams2004}, are classified into dust cores (compact 3\,mm emission), dust+ff cores (with compact 3\,mm emission, H(40)$\alpha$, and cm emission), and cometary UCH{\sc ii} regions (with extended 3\,mm emission, H(40)$\alpha$, and cm emission), summarized in Table \ref{tab:ALMApositions}. Cores with $S$/$N < 15$ were not analyzed in this study due to insufficient sensitivity and angular resolution.
	
	In Sect. \ref{sec:ALMAdiscussionclump}, the region properties (peak intensity $I_\mathrm{3mm}^\mathrm{region}$ and flux density $F_\mathrm{3mm}^\mathrm{region}$) derived with the ALMA observations are compared to the large scale clump properties derived from the ATLASGAL survey \citep{Urquhart2018,Urquhart2022}. In Sect. \ref{sec:ALMAdiscussioncore} the evolutionary trends of the physical properties on core scales are discussed and compared with the results of the CORE \citep{Gieser2021} and CORE-extension \citep{Gieser2022} studies.
	
\subsection{Evolutionary trends of the clump properties}\label{sec:ALMAdiscussionclump}

\begin{figure*}[!htb]
\centering
\includegraphics[width=0.24\textwidth]{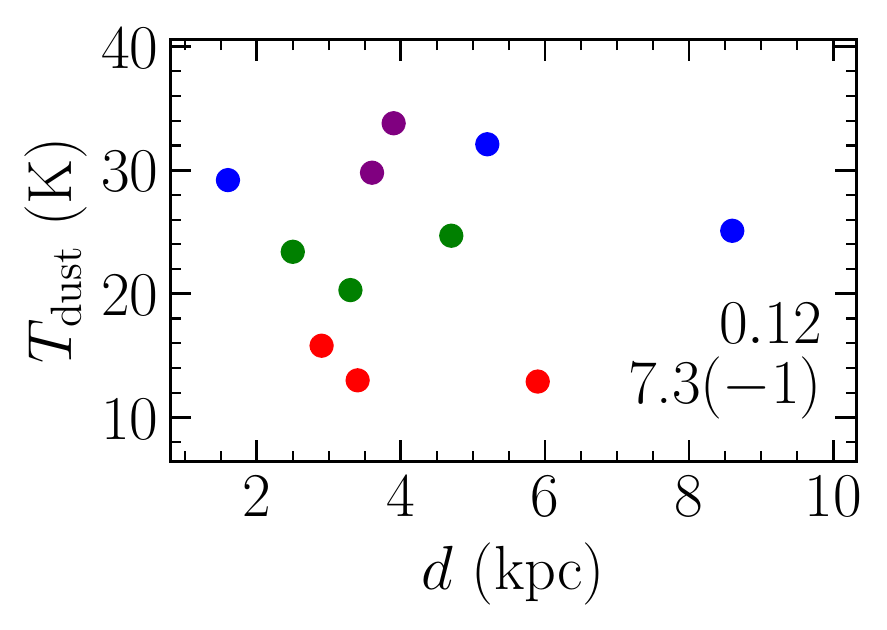}
\includegraphics[width=0.24\textwidth]{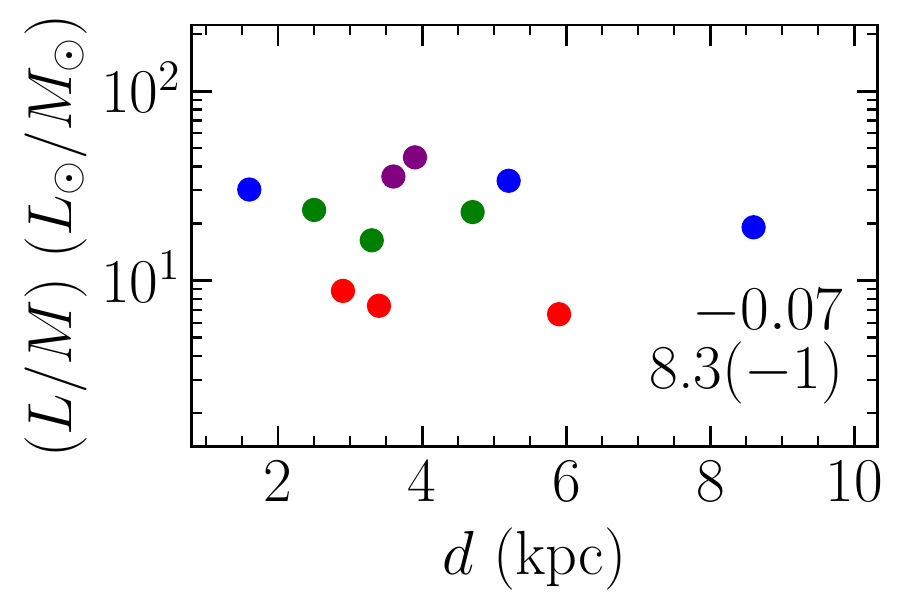}
\includegraphics[width=0.24\textwidth]{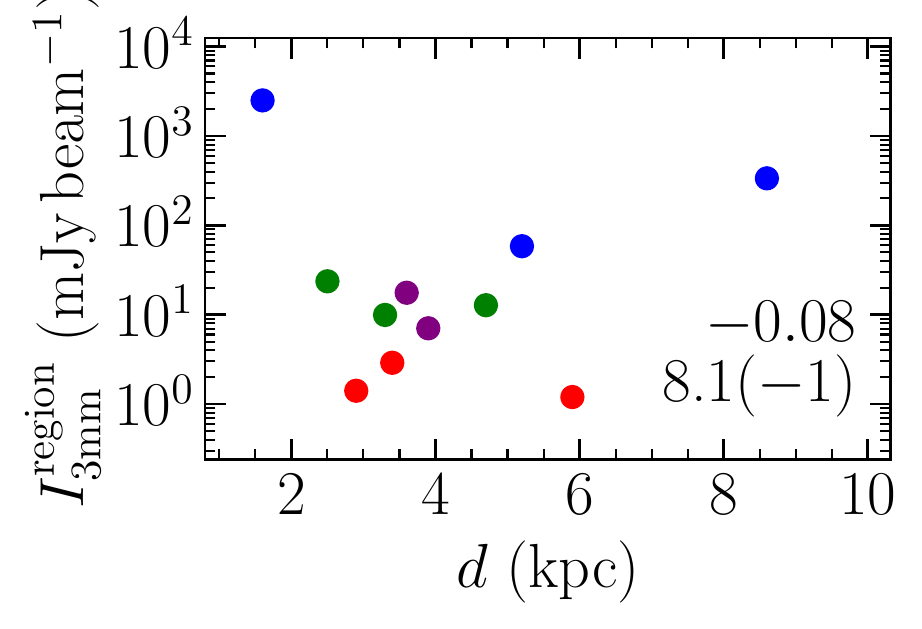}
\includegraphics[width=0.24\textwidth]{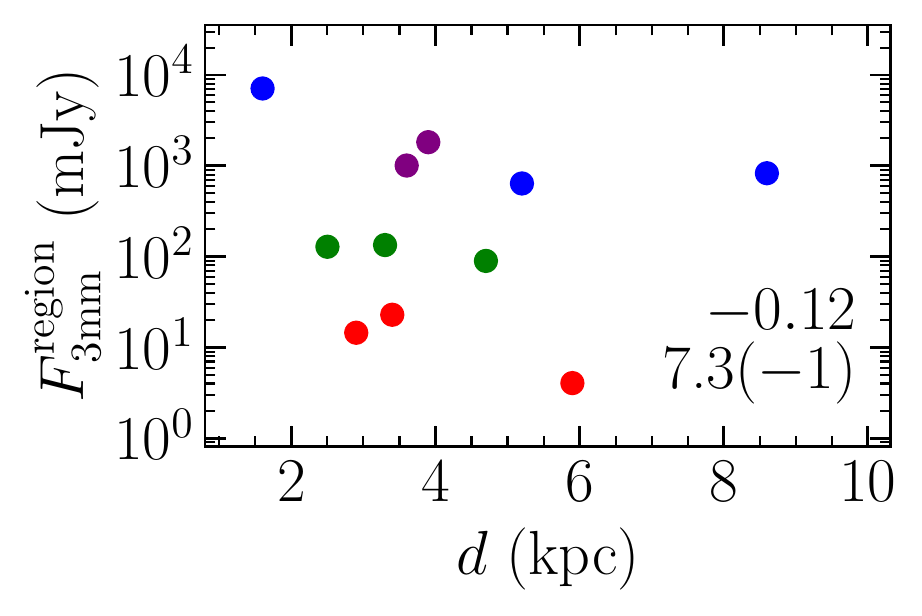}\\
\includegraphics[width=0.24\textwidth]{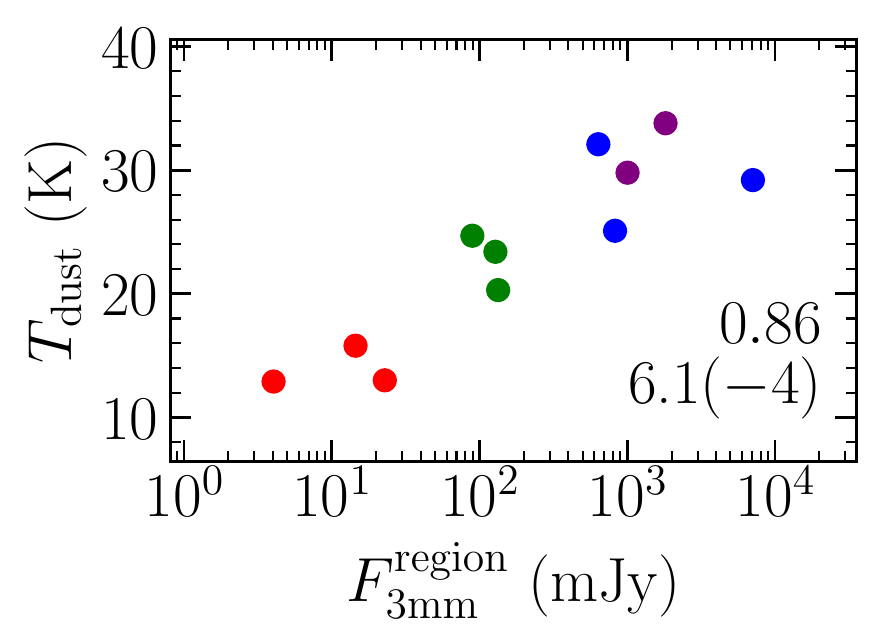}
\includegraphics[width=0.24\textwidth]{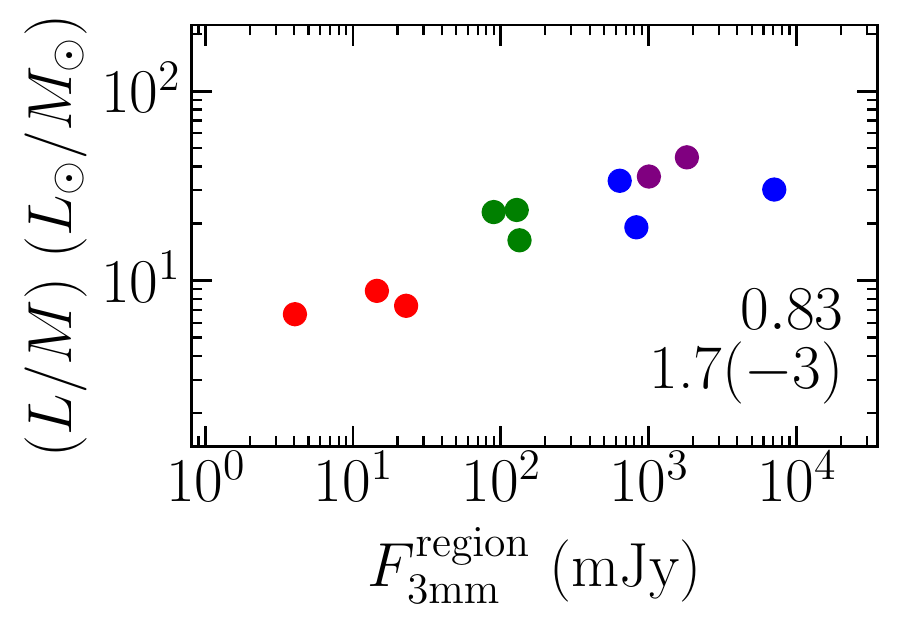}
\includegraphics[width=0.24\textwidth]{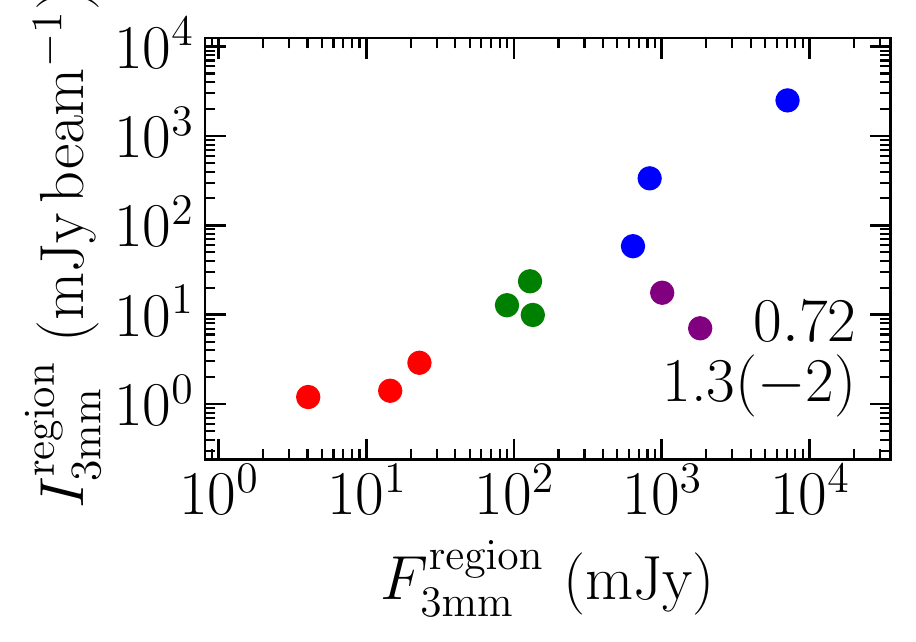}
\includegraphics[width=0.24\textwidth]{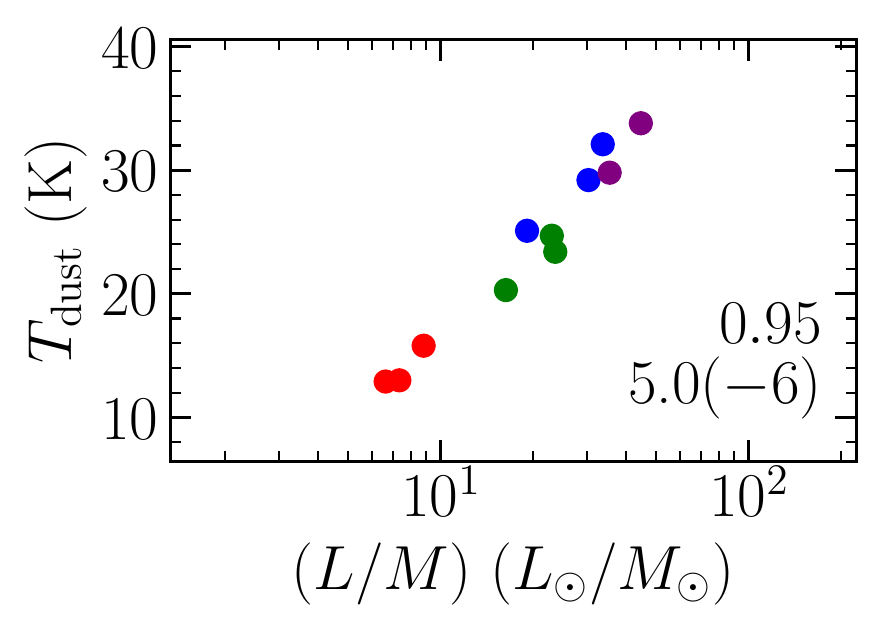}\\
\includegraphics[width=0.48\textwidth]{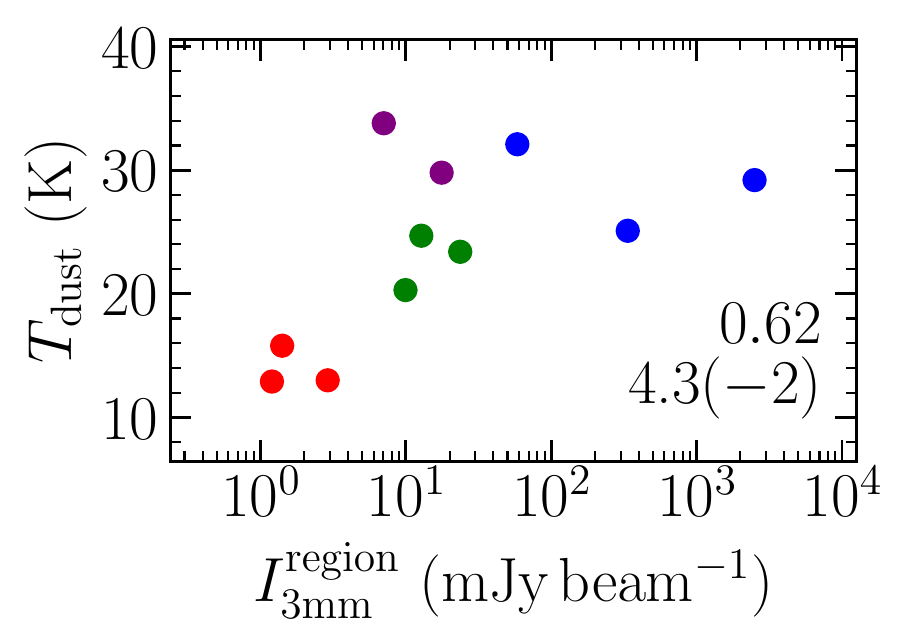}
\includegraphics[width=0.48\textwidth]{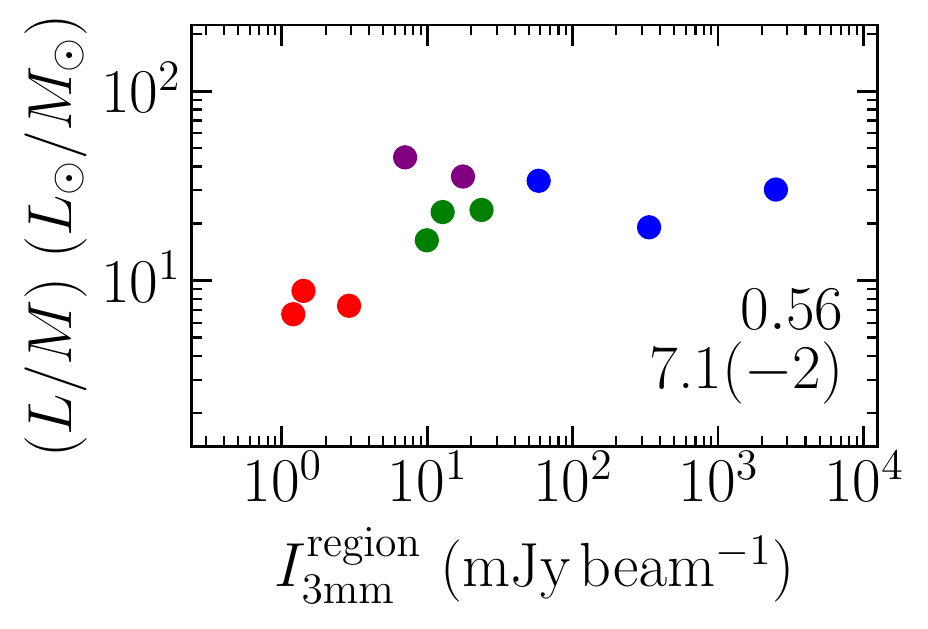}
\caption[Comparison of the clump parameters of the ALMA sample.]{Comparison of the clump parameters of the regions. The temperature $T_\mathrm{dust}$, luminosity $L$, mass $M$, and distance $d$ are listed in Table \ref{tab:ALMA_regions} and are taken from the ATLASGAL survey \citep{Schuller2009,Urquhart2018}. The region peak intensity $I^\mathrm{region}_\mathrm{3mm}$ and flux density $F^\mathrm{region}_\mathrm{3mm}$ of the ALMA 3\,mm continuum data is listed in Table \ref{tab:ALMAcontinuumdataproducts}. The colors of the data points highlight the four evolutionary stages (red: IRDC, green: HMPO, blue: HMC, and purple: UCH{\sc ii} region). The Spearman correlation coefficient $r_\mathrm{S}$ and $p$-value, with a(b) meaning a$\times10^{\mathrm{b}}$, are shown in the bottom right corner.}
\label{fig:ALMAregioncomparison}
\end{figure*}

	All regions in our ALMA sample are also covered by the ATLASGAL survey \citep{Schuller2009} targeting dust emission on clump scales in the Galactic plane. The 870\,$\upmu$m images are presented in the top right panel in Figs. \ref{fig:overview_IRDC_G1111} $-$ \ref{fig:overview_UCHII_G1387}. The clump sizes cover roughly the FOV of the ALMA 3\,mm observations, we therefore compare in this section the ATLASGAL clump properties with the ALMA 3\,mm peak intensity $I_\mathrm{3mm}^\mathrm{region}$ and flux density $F_\mathrm{3mm}^\mathrm{region}$ of each region.
	
	The distance $d$, luminosity $L$, mass $M$, and dust temperature $T_\mathrm{dust}$ are taken from the ATLASGAL study by \citet{Urquhart2018} and are listed in Table \ref{tab:ALMA_regions}. The clump distances $d$ are estimated using different methods, including maser parallaxes and kinematic distances based on radial velocities. The luminosity, mass, and dust temperature are derived by fitting the SED using additional MIR and FIR data from, for example, the Galactic Legacy Infrared Mid-Plane Survey Extraordinaire \citep[GLIMPSE,][]{Benjamin2003}, the MIPS GALactic plane survey \citep[MIPSGAL,][]{Carey2009}, and the \textit{Herschel} infrared GALactic plane survey \citep[Hi-GAL][]{Molinari2010}. The CORE and CORE-extension regions are not covered by ATLASGAL and therefore a consistent method that could derive these parameters is not available. We therefore do not include the clump properties of CORE and CORE-extension in this analysis.
	
	The ALMA 3\,mm peak intensity $I^\mathrm{region}_\mathrm{3mm}$ and flux density $F^\mathrm{region}_\mathrm{3mm}$ are listed in Table \ref{tab:ALMAcontinuumdataproducts}. The flux density is integrated considering the area with emission $>$5$\sigma_\mathrm{cont}$. In contrast to the single-dish observations with \textit{Spitzer}, \textit{Herschel}, and APEX telescopes used for the SED fit in \citet{Urquhart2018}, the interferometric ALMA observations suffer from spatial filtering (on scales larger than $>$60$''$, Sect. \ref{sec:ALMAobs}) that can be an issue especially for the extended UCH{\sc ii} regions that have widespread emission within the full ALMA primary beam (Fig. \ref{fig:ALMAcontinuum}). In addition to dust emission, at 3\,mm wavelengths there is a contribution from free-free emission toward evolved protostars in the HMC and UCH{\sc ii} regions (Sect. \ref{sec:ALMAffemission}).
	
	In Fig. \ref{fig:ALMAregioncomparison} relations between the following clump parameters are presented: $d$, $L$/$M$, $T_\mathrm{dust}$, $I^\mathrm{region}_\mathrm{3mm}$, and $F^\mathrm{region}_\mathrm{3mm}$. The $L$/$M$ ratio is an indicator of evolutionary stage increasing from IRDC to UCH{\sc ii} regions \citep[e.g.,][]{Sridharan2002,Molinari2008,Molinari2010,Maud2015,Molinari2016, Urquhart2018,Molinari2019}. In color, the different evolutionary stages (IRDC - HMPO - HMC - UCH{\sc ii}) based on the classification by \citet{Gerner2014,Gerner2015} are highlighted. In order to quantify potential correlations, we use the Spearman correlation coefficient $r_\mathrm{S}$, that is zero for uncorrelated data and $\pm$1 for correlated and anticorrelated data, respectively \citep{Cohen1988}. The $p$-value indicates the probability that an uncorrelated data set would have with this Spearman correlation coefficient value. The Spearman correlation coefficient $r_\mathrm{S}$ and $p$-value are shown in Fig. \ref{fig:ALMAregioncomparison} for each parameter pair.
	
	The top row in Fig. \ref{fig:ALMAregioncomparison} shows the relation of the properties as a function of distance $d$. With $|r_\mathrm{S}| < 0.12$ for all parameters, $T_\mathrm{dust}$, $L$/$M$, $I^\mathrm{region}_\mathrm{3mm}$, and $F^\mathrm{region}_\mathrm{3mm}$ are not correlated with distance $d$. This implies that our sample is not biased by distance such that more luminous regions would be at larger distances. However, there are clear evolutionary trends, highlighted in different colors, for $T_\mathrm{dust}$, $L$/$M$, and $F^\mathrm{region}_\mathrm{3mm}$, with these values increasing with evolutionary stage, but independent of distance. For example, the dust temperature $T_\mathrm{dust}$ increases from $\approx$15\,K (IRDCs), $\approx$20\,K (HMPOs), $\approx$30\,K (HMCs and UCH{\sc ii} regions). For the peak intensity $I^\mathrm{region}_\mathrm{3mm}$, this trend is seen from the IRDC to the HMC stages, while for the UCH{\sc ii} regions, the peak intensity $I^\mathrm{region}_\mathrm{3mm}$ decreases compared to the HMC stage. Since both UCH{\sc ii} regions have a cometary morphology with extended emission (Fig. \ref{fig:ALMAcontinuum}), the decrease of $I^\mathrm{region}_\mathrm{3mm}$ compared to previous evolutionary stages can be explained by the expanding envelope driven by the central protostar resulting in a lower surface brightness. In addition, the material is continuously fed onto the central protostar, while at the same time matter is expelled through the outflow.
	
	For the remaining parameter pairs (excluding the distance $d$), we find positive correlations with $r_\mathrm{S} \geq 0.56$ (all panels in the second and third row in Fig. \ref{fig:ALMAregioncomparison}), with $T_\mathrm{dust}$, $L$/$M$, and $F^\mathrm{region}_\mathrm{3mm}$ increasing with evolutionary stage.
	
	\citet{Urquhart2022} investigated the evolutionary trends of $\approx$5\,000 ATLASGAL clumps using a slightly different evolutionary scheme (quiescent - protostellar - YSO - H{\sc ii} region). The clumps were visually classified based on additional MIR and FIR data from GLIMPSE, MIPSGAL, and Hi-GAL. For example, quiescent clumps have no counterpart at wavelengths of 70\,$\upmu$m and below, while YSOs have 70\,$\upmu$m, 24\,$\upmu$m and $4-8\upmu$m counterparts, but show no cm emission. Based on the cumulative distribution of $T_\mathrm{dust}$ and $L$/$M$, \citet{Urquhart2022} also find that these parameters increase with evolutionary stage.
	
	Despite the fact that the sample analyzed in this work covers only 11 HMSFRs compared to, for example, the $\approx$10\,000 clumps studied in the ATLASGAL survey, we still cover a broad range in clump properties as shown in Fig. \ref{fig:ALMAregioncomparison}. The four evolutionary stages form relatively distinct regions in these correlation plots, and in the next section we will investigate if and how the physical properties in HMSFRs correlate on smaller core scales, such as the temperature and density profiles, and mass. Since a high degree of fragmentation is observed in this sample (Fig. \ref{fig:ALMAcontinuum} and Table \ref{tab:ALMApositions}), it is expected that the evolutionary trends become less prominent due to multiple protostars, potentially at different evolutionary stages, present within a single clump.
	
\subsection{Evolutionary trends on core scales}\label{sec:ALMAdiscussioncore}

\begin{figure}[!htb]
\centering
\includegraphics[width=0.49\textwidth]{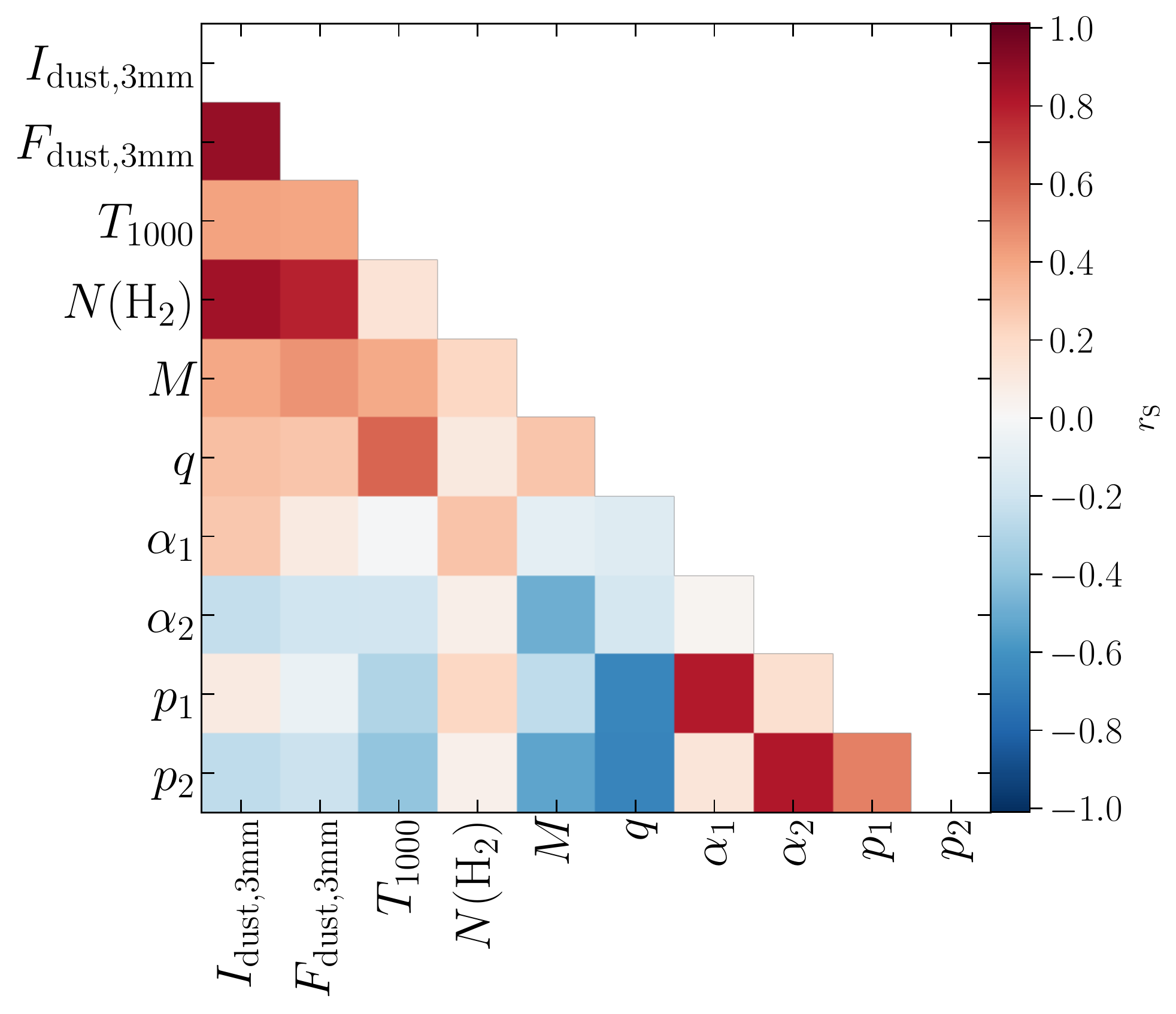}
\caption[Correlations of the physical parameters of the ALMA, CORE, and CORE-extension samples.]{Correlations of the physical parameters of the ALMA, CORE, and CORE-extension samples.}
\label{fig:corecorr}
\end{figure}

	In this section, we analyze relations between physical properties on the smaller core scales, i.e., the physical structure of the dust cores, dust+ff cores, and cometary UCH{\sc ii} regions of the sample analyzed in this study and the cores of the CORE and CORE-extension sample \citep{Gieser2021,Gieser2022}. Following the nomenclature of this study, the cores in the CORE and CORE-extension studies can be classified as dust cores. We focus on the following physical properties that were derived in all three studies.
	
	 The 3\,mm peak intensity $I_\mathrm{dust,3mm}$ and integrated flux density $F_\mathrm{dust,3mm}$ of the dust emission are estimated from the continuum data. For the regions observed with ALMA at 3\,mm, we use the free-free corrected 3\,mm continuum data (Sect. \ref{sec:ALMAffemission}). For the CORE and CORE-extension sample, we extrapolate the 1\,mm peak intensity and flux density to 3\,mm wavelengths assuming black body emission with $I_\nu \sim \nu^2$ and $F_\nu \sim \nu^2$. The masses $M$ and H$_{2}$ column density $N$(H$_{2}$) were calculated in all studies according to Eq. \ref{eq:Mcalc} and \ref{eq:H2calc}, respectively.
	 
	 In order to account for the fact that the regions are at various distances and were observed at slightly different angular resolutions, we extrapolate the radial temperature profiles to a common radius of 1\,000\,au, $T_{1000} = T$($r$=1\,000\,au). The radial temperature profiles were taken from the H$_{2}$CO, CH$_{3}$CN, and CH$_{3}^{13}$CN temperature maps derived with \texttt{XCLASS} or the temperature map derived using the HCN-to-HNC intensity ratio \citep{Hacar2020}.
	 
	The observed temperature power-law index $q$ is derived from the radial temperature profiles and the visibility power-law index $\alpha$ is estimated from the continuum visibility profiles. Since we find that most of the ALMA sources are best fitted by two power-law profiles, $\alpha_1$ and $\alpha_2$, we assume $\alpha_1=\alpha_2=\alpha$ for the CORE and CORE-extension dust cores and check for correlations for both profiles separately. The indices $\alpha_1$ and $\alpha_2$ roughly correspond to the clump and core scales, respectively (Sect. \ref{sec:ALMAdensity}).

	The density profiles $p_1$ and $p_2$ are calculated according to Eq. \eqref{eq:uvanalysis} using $\alpha_1$ and $\alpha_2$, respectively. We use for each source the observed temperature power-law index $q$. Therefore, we have two density power-law indices, $p_1$ and $p_2$ tracing the clump and core scales, respectively. For the dust cores in the CORE and CORE extension sample $p_1 = p_2$ holds.

\begin{figure*}[!htb]
\centering
\includegraphics[width=0.32\textwidth]{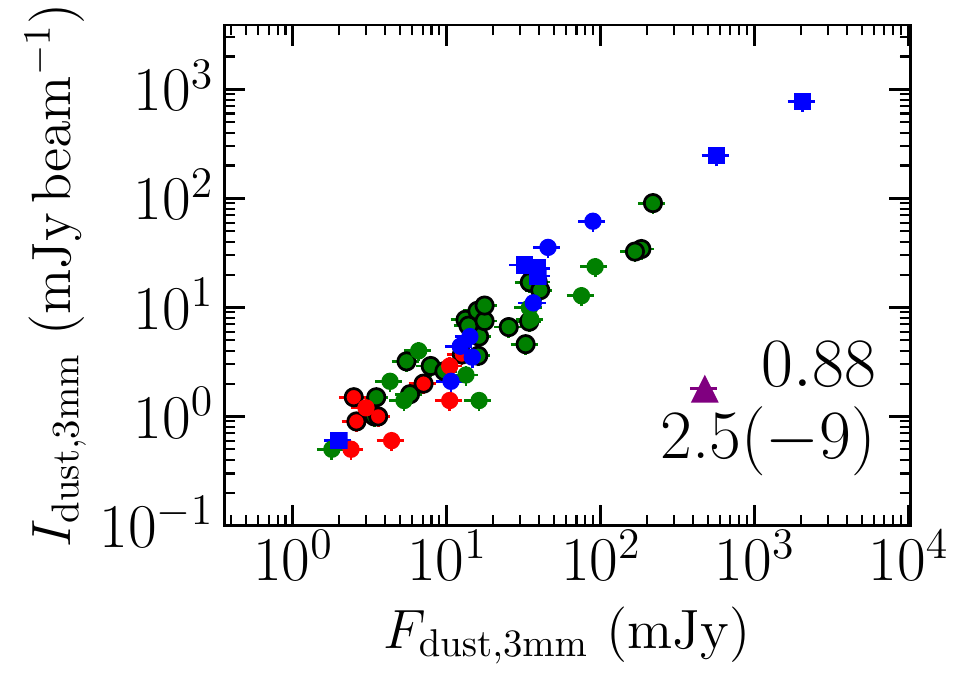}
\includegraphics[width=0.32\textwidth]{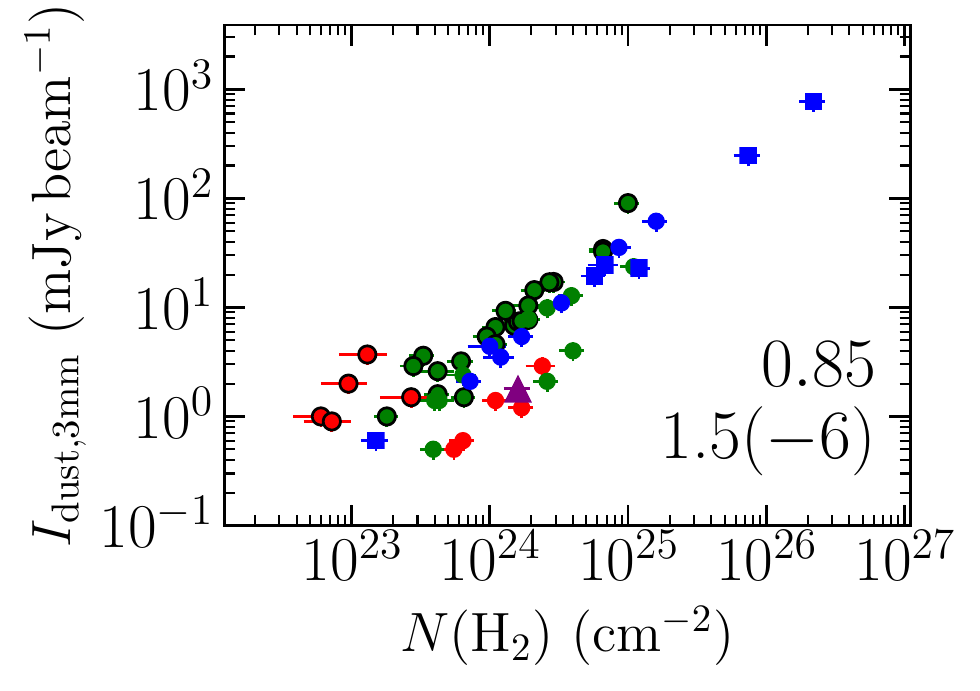}
\includegraphics[width=0.32\textwidth]{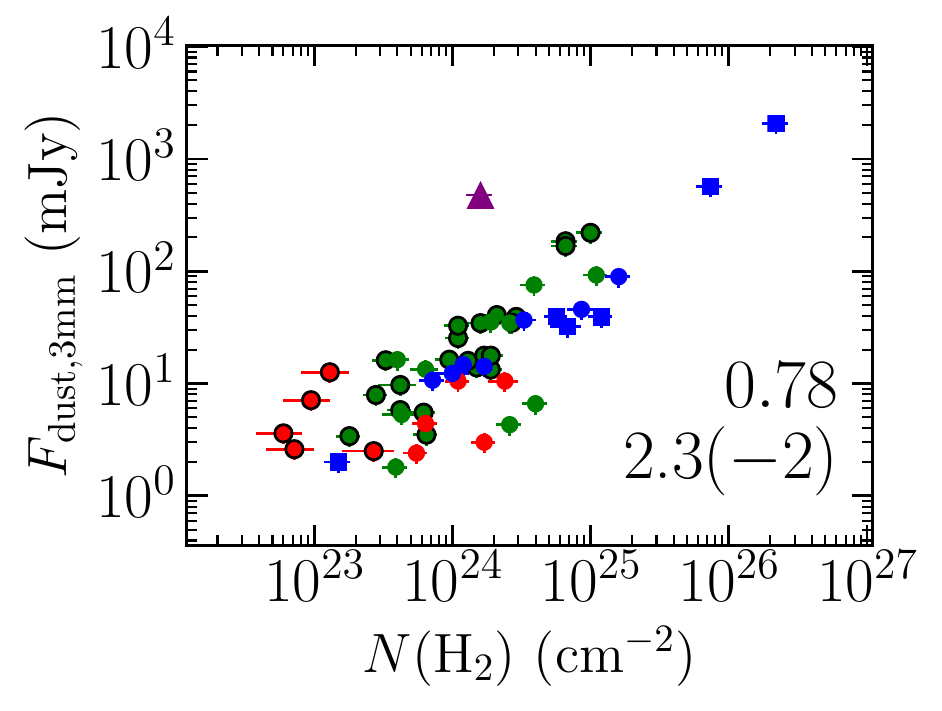}\\
\includegraphics[width=0.32\textwidth]{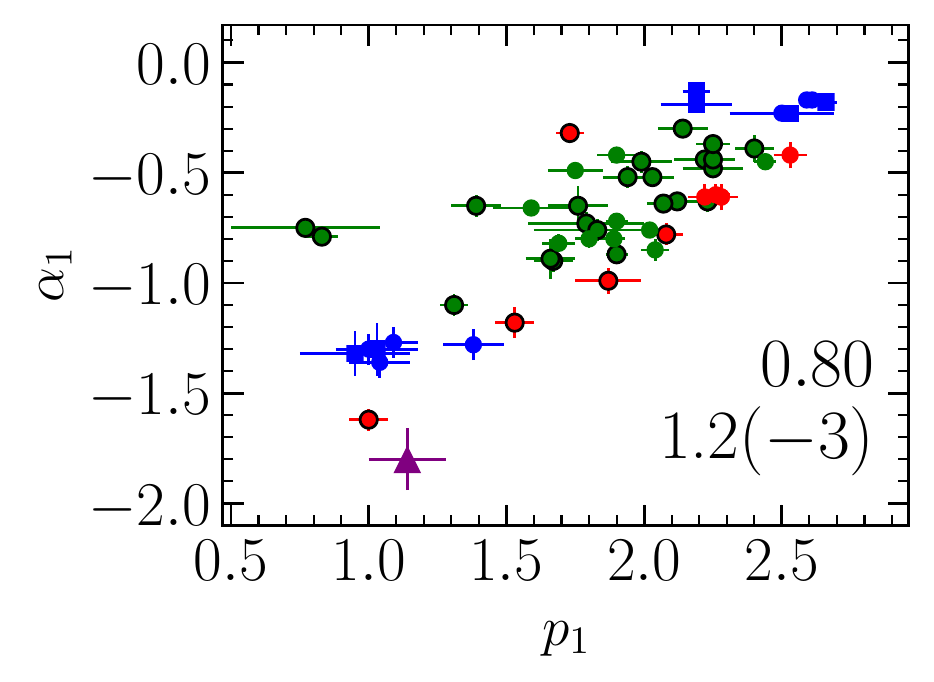}
\includegraphics[width=0.32\textwidth]{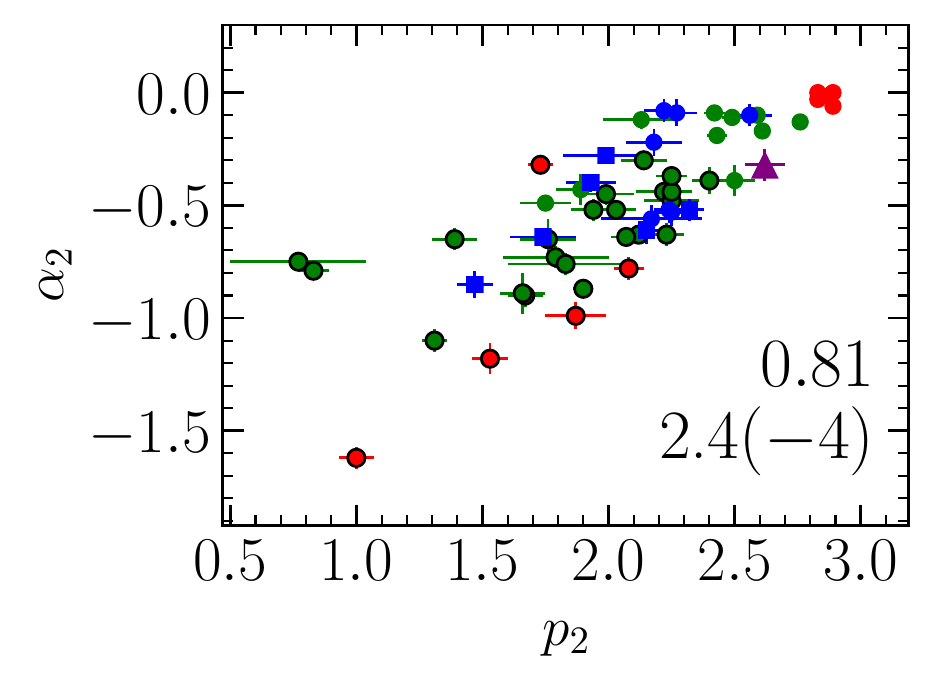}
\includegraphics[width=0.32\textwidth]{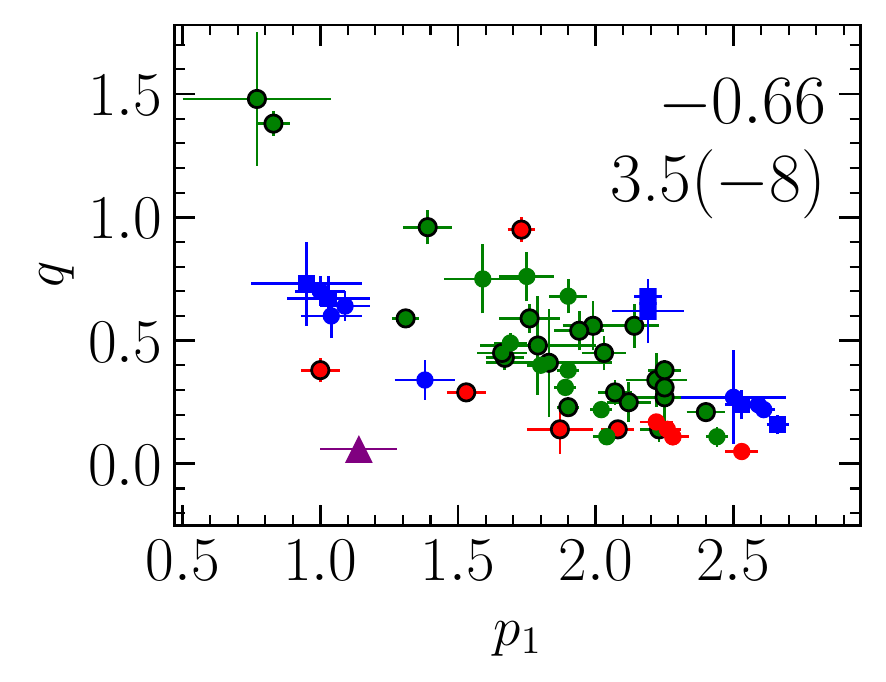}\\
\includegraphics[width=0.32\textwidth]{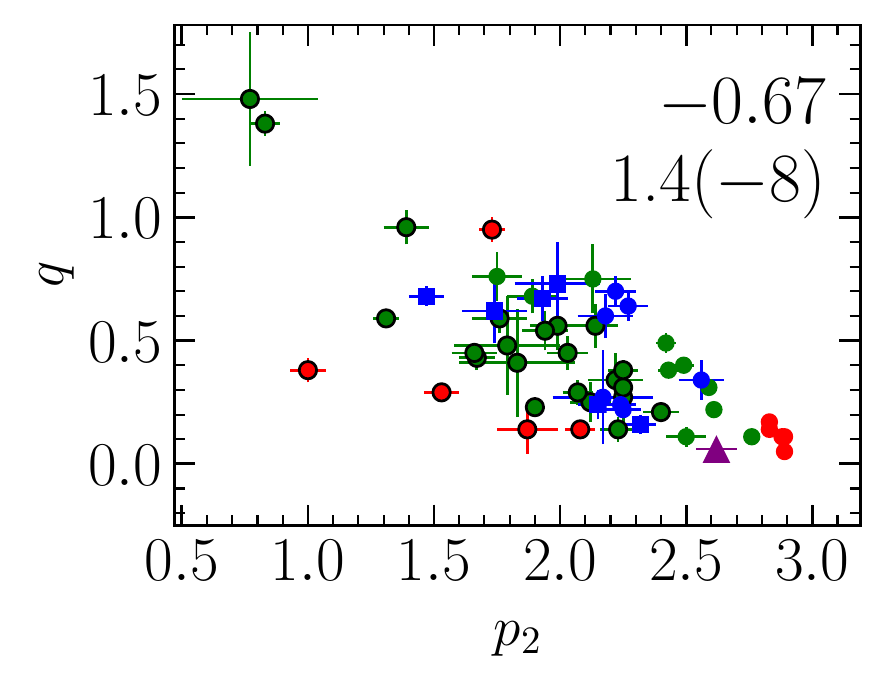}
\includegraphics[width=0.32\textwidth]{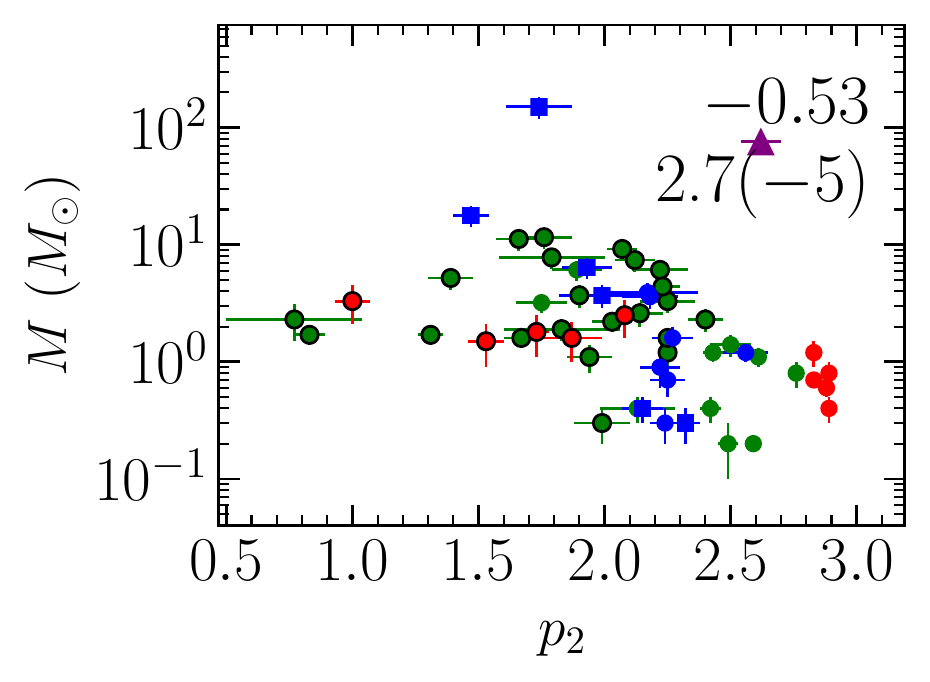}
\includegraphics[width=0.32\textwidth]{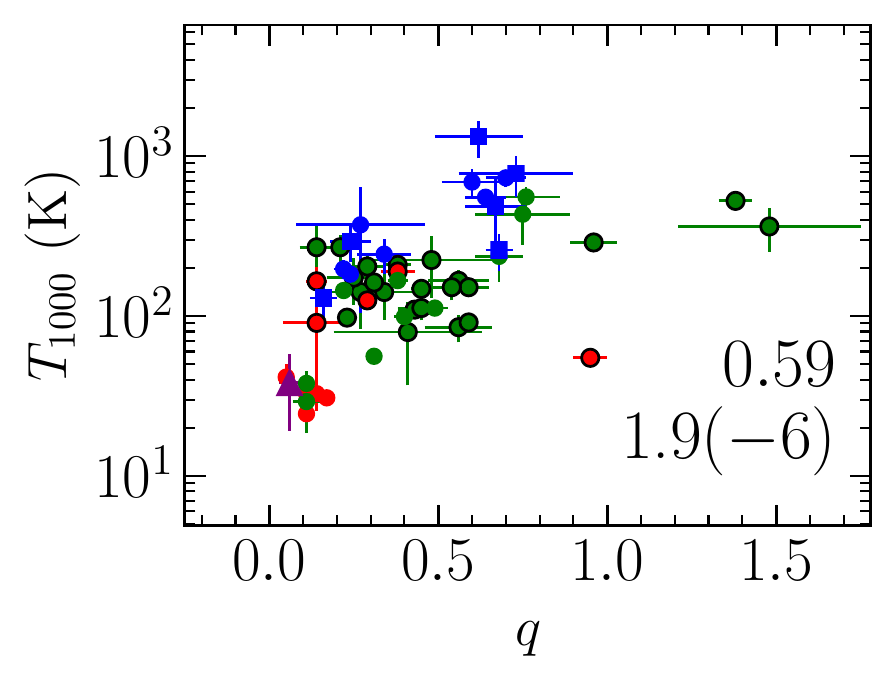}
\caption[Correlations of the physical parameters of the ALMA, CORE, and CORE-extension sample.]{Correlations with $|r_\mathrm{S}| > 0.5$ of the physical parameters of the ALMA, CORE, and CORE-extension sample. The color of the data points highlight the four evolutionary stages (red: IRDC, green: HMPO, blue: HMC, and purple: UCH{\sc ii} region). The data points of the cores of the CORE (green) and CORE-extension (red) sample are enclosed in black rings. The dust cores, dust+ff cores and cometary UCH{\sc ii} regions are shown by circles, squares and triangles, respectively. The Spearman correlation coefficient $r_\mathrm{S}$ and $p$-value, with a(b) meaning a$\times10^{\mathrm{b}}$, are shown in each panel.}
\label{fig:corestatisticplots}
\end{figure*}

	In total, a complete set of these physical parameters can be derived for 56 protostellar sources: 22 dust cores in the CORE sample \citep{Gieser2021}; 5 dust cores in the CORE-extension sample \citep{Gieser2022}; and 22 dust cores, 6 dust+ff cores and 1 cometary UCH{\sc ii} region in this study. To investigate potential correlations between pairs of these parameters, we compute the Spearman correlation coefficient. Figure \ref{fig:corecorr} summarizes the Spearman correlation coefficient $r_\mathrm{S}$ for all parameter pairs. While for some parameters, e.g., $\alpha_1$ and $M$, we do not find hints for potential correlations, other parameters show clear correlations, e.g., $q$ and $T_{1000}$. Thus in Fig. \ref{fig:corestatisticplots} the parameter pairs with $|r_\mathrm{S}| > 0.5$ are shown, indicating weak ($|r_\mathrm{S}| \approx 0.5$) to high (anti)correlations ($|r_\mathrm{S}| = 1.0$), and in the following we focus on the discussion of these parameter pairs. Different colors in Fig. \ref{fig:corestatisticplots} highlight the different evolutionary stages based on the larger clump scales (IRDC - HMPO - HMC - UCH{\sc ii}). We group the dust cores CORE-extension and CORE regions into the IRDC and HMPO stage, respectively \citep{Gieser2021,Gieser2022}.
	
	We find positive correlations of $I_\mathrm{dust,3mm}$ with $F_\mathrm{dust,3mm}$ and $N$(H$_{2}$). The integrated flux density $F_\mathrm{dust,3mm}$ is positively correlated with $N$(H$_{2}$). Evolutionary trends for these parameters are clearly seen from dust cores in IRDCs to the dust+ff cores in HMCs. The peak intensity is lower for the cometary UCH{\sc ii} region compared to the other sources. From dust+ff cores to cometary UCH{\sc ii} regions, the dust peak intensity decreases significantly. This can be explained by the fact that the protostellar radiation disrupts the surrounding gas and dust envelope. The surface brightness (corresponding to the intensity) is therefore decreasing as well and since the dust is slowly expelled into the more diffuse ISM, the flux density within our limited sensitivity decreases as well. Regarding the estimated temperatures, the central protostar in the cometary UCH{\sc ii} region is expected to be much hotter than the $\approx 36$\,K estimated with the extended CH$_{3}$CN line emission (Table \ref{tab:ALMAradialtemp}).

	With the observed CH$_{3}$CN and CH$_{3}^{13}$CN emission lines and angular resolution, gas layers up to $200-250$\,K can be traced, however, higher J-transitions are required to constrain the temperature profiles in higher density regions. In addition, the angular resolution and line sensitivity only allows for the radial temperature profiles to be only marginally resolved, typically only along $2-3$ beams, that are challenging to reliably fit (Figs. \ref{fig:ALMATrad} and \ref{fig:ALMATradapp}).

	The positive correlation between the pair of $\alpha_1$ and $p_1$ and of $\alpha_2$ and $p_2$ is due to the fact that $p$ is calculated taking into account $\alpha$ in Eq. \eqref{eq:uvanalysis}. If these flat visibility slopes, with $\alpha_1 > -0.3$, are not considered, there is a clear trend such that the density profile on clump scales, $p_1$, flattens from the dust cores in IRDCs ($p_1 \approx 2.2$) to dust+ff cores in HMCs ($p_1 \approx 1.2$). This can be explained by the fact that steep density profiles are expected for collapsing and accreting clumps, while feedback of the massive stars disrupts the surrounding envelope, e.g. due to powerful outflows or an expanding UCH{\sc ii} region. This flattening of the density power-law index on clump scales has also been observed by \citet{Beuther2002} where their ``strong molecular sources'' have $p \approx 1.9$ and their more evolved ``cm sources'' have $p \approx 1.5$. We do not find strong evidence for evolutionary trends of the density profile on the core scales, $p_2$. Most of the objects seem to be collapsing and accreting toward the gravitational center and thus toward $p_2 \approx 2$ a high density of data points of cores of dust and dust+ff cores is found, but overall there is a large scatter from $1.0$ to $2.5$ (excluding the unresolved dust cores in the IRDCs with $\alpha_2\approx 0$). The presence of compact free-free emission at 3\,mm wavelengths can impact the visibility profiles and thus also the derived density profile. It is therefore necessary to analyze the density profile on cores scales in an even larger sample, preferentially at shorter wavelengths, where the contribution of free-free emission is less severe, e.g. with the ALMAGAL project that targets $\approx$1\,000 HMSFRs in a similar spectral setup as CORE.

	The observed temperature power-law index $q$ is anticorrelated with the density power-law indices $p_1$ and $p_2$. There is a clear evolutionary trend such that the temperature profiles steepens from the IRDC ($q \approx 0.1$) to the UCH{\sc ii} regions ($q \approx 0.7$), while for the cometary UCH{\sc ii} region the profile becomes flat. This suggests that an embedded source is steadily heating up the envelope. This is also confirmed by the fact that the slope $q$ becomes steeper with $T_{1000}$ increasing from 20\,K up to 1\,000\,K. The cometary UCH{\sc ii} region has a low value of $T_{1000}$ ($\approx 30$\,K). An increase of the temperature power-law index $q$ compared to the typical value of 0.3$-$0.4 has been observed toward several HMCs in the literature \citep{Beltran2018,Mottram2020}. We also find in the analysis of the 22 cores of the CORE sample, that the distribution of $q$ has a spread \citep{Gieser2021}. The four data points with $q > 0.9$ are most likely outliers due to unreliable fits. In the HMC models by \citet{Nomura2004} and \citet{Osorio2009}, in which the temperature profile is calculated self-consistently, the temperature profiles also steepen on core scales with $q > 0.4$.

	An unexpected result of the density profile analysis is that for most of the sources, we find that the visibility profiles follow at least two power-law profiles (Sect. \ref{sec:ALMAdensity}), while for the CORE and CORE-extension samples, a single power-law profile was sufficient \citep{Gieser2021,Gieser2022}. The traced spatial scales of NOEMA and ALMA observations are not too different ($2-500$\,k$\lambda$ and $10-600$\,k$\lambda$, respectively), but NOEMA has less long baselines resulting in higher noise at long baselines in the visibility profiles \citep{Gieser2021,Gieser2022} compared to ALMA (Fig. \ref{fig:ALMAvisibilityprofile}). To test if the two power-law slopes are an effect of observed wavelength (1\,mm versus 3\,mm), we aim to obtain additional 3\,mm observations of the CORE sample. A pilot study toward the IRAS\,23385, IRAS\,23033 and NGC7538S CORE regions will investigate the differences between the 1\,mm and 3\,mm data (Gieser et al. in prep.). The fact that the sources in the ALMA regions show two power-law slopes can be explained by the young cores in the IRDCs containing unresolved point sources. In an inside-out collapse picture, the cores grow in mass and size, and the density profile of the cores in the HMPOs aligns with the underlying clump density profile. In these cases, a single visibility power-law profile is observed, in our cases for the CORE sample \citep{Gieser2021} and for dust core 1 in HMPO\,IRAS\,18089 (Fig. \ref{fig:ALMAvisibilityprofileapp}) for which we also find $\alpha_1 = \alpha_2$.

	We find evolutionary trends on core scales, but there are clear overlaps in contrast to the larger scale clump properties discussed in Sect. \ref{sec:ALMAdiscussionclump}. In our analysis of the physical properties, we assume that the envelopes are spherically symmetric, however, the presence of outflows and the inclination might also impact, for example, the temperature and density profiles. High angular resolution data of HMSFRs clearly show that a single classification of the evolutionary stage based on the clump properties is not sufficient to describe the evolutionary phases of protostellar sources embedded within a single clump. We therefore aim to use in a future study the physical-chemical model \texttt{MUSCLE} \citep{Gerner2014,Gerner2015} to derive chemical timescales, $\tau_\mathrm{chem}$, in combination with MIR, FIR, and cm data in order to better characterize and classify individual protostellar sources on core scales in clustered HMSFRs.

\section{Summary and conclusions}\label{sec:ALMAconclusions}

	In this study we analyzed ALMA 3\,mm continuum and line observations of 11 high-mass star-forming regions at evolutionary stages from protostars in young infrared dark clouds to evolved UCH{\sc ii} regions. In addition, we made use of archival data at MIR, FIR, and cm wavelengths in order to characterize the regions along different wavelengths. In particular, we made use of the H(40)$\alpha$ recombination line to distinguish between dust and free-free emission in the ALMA 3\,mm continuum data.
	
	At an angular resolution of $\approx$1$''$, we observe a high degree of fragmentation in the regions and the fragmentation was quantified using the \texttt{clumpfind} algorithm. We classify the fragments into protostellar sources with $S$/$N > 15$, while cores with $S$/$N < 15$ are not further analyzed. The protostellar sources are divided further into dust cores (compact mm emission originating from dust), dust+ff cores (compact mm and additional H(40)$\alpha$ and cm emission) and cometary UCH{\sc ii} regions (extended mm and additional H(40)$\alpha$ and cm emission) for which we detect 37, 8, and 3 sources, respectively. In our analysis, we exclude 24 cores due to insufficient sensitivity and/or angular resolution. Our findings are summarized as follows:

\begin{enumerate}

\item We create temperature maps using the HCN-to-HNC intensity ratio to trace the low-temperature regime and with \texttt{XCLASS} we model and fit CH$_{3}$CN and CH$_{3}^{13}$CN line emission to trace the high-temperature regime. Radial temperature profiles with power-law index $q$ (Eq. \ref{eq:temperatureprofile}) are computed for all protostellar sources. We find that there is a spread in $q$ between $\approx 0.1 - 0.7$.

\item The density profiles (Eq. \ref{eq:densityprofile}) are estimated from the 3\,mm continuum visibility profiles. In contrast to the CORE and CORE-extension regions \citep{Gieser2021,Gieser2022}, most of the visibility profiles are best explained by two profiles with varying slope $\alpha_1$ and $\alpha_2$ instead of a single profile. The visibility slopes $\alpha_1$ and $\alpha_2$ approximately trace the clump and core scales, respectively. The estimated density power-law index $p_1$ and $p_2$ (Eq. \ref{eq:uvanalysis}) varies between 1 and 2.6, and 1.6 and 3, respectively.

\item Using the peak intensity $I_\mathrm{dust,3mm}$ and integrated flux density $F_\mathrm{dust,3mm}$ of the 3\,mm dust, we estimate the mass and H$_{2}$ column density of the protostellar sources. We find a large spread in H$_{2}$ column density, $N($H$_{2}) \approx 10^{23} - 10^{26}$\,cm$^{-2}$, and mass, $M \approx 0.1 - 150$\,$M_\odot$, within the protostellar sources.

\item Comparing the 3\,mm peak intensity and region-integrated flux density ($I^\mathrm{region}_\mathrm{3\,mm}$ and $F^\mathrm{region}_\mathrm{3\,mm}$) with clump properties ($T_\mathrm{dust}$, $L$/$M$) derived by the ATLASGAL survey \citep{Urquhart2018}, we find that there are clear evolutionary trends of $T_\mathrm{dust}$, $L$/$M$, and $F^\mathrm{region}_\mathrm{3\,mm}$ increasing from IRDC to UCH{\sc ii} regions. The peak intensity $I^\mathrm{region}_\mathrm{3\,mm}$ is increasing from the IRDC to HMC stage and then significantly decreases in the UCH{\sc ii} regions. This can be explained by the fact that the expanding extended envelope has a low surface brightness compared to compact cores in the IRDC to HMC stages. Even though our sample consists of only 11 regions, we still cover a broad range in physical properties on clump scales.

\item We combined the results of the physical structure on core scales derived in this study with the results of the CORE and CORE extension sample \citep{Gieser2021,Gieser2022} and analyzed a sample of a total of 56 protostellar sources for correlations and evolutionary trends. We find that the temperature at a characteristic radius of 1\,000\,au, $T_{1000}$, is increasing with evolutionary stage and at the same time, the temperature power-law index $q$ steepens from $q \approx 0.1$ to $q \approx 0.7$ from dust cores in IRDCs to dust+ff cores in HMCs. The density profile $p_1$ on clump scales, flattens from $p_1 \approx 2.2$ in the IRDC stage to $p_1 \approx 1.2$ in UCH{\sc ii} regions, while for the density profile $p_2$ on core scales, we do not find evidence for evolutionary trends, with a mean of $p \approx 2$ considering the full sample, but with an overall scatter from 1.0 to 2.5. These results provide invaluable observational constraints for physical models describing the formation of high-mass stars.

\end{enumerate}

	In this study, we focused on the characterization of the physical properties 11 regions observed with ALMA at high angular resolution. We now aim to further investigate the physical and chemical properties using the rich molecular line data covered by our ALMA observations (Table \ref{tab:almaspwsummary}) and apply the physical-chemical model \texttt{MUSCLE} \citep{Gerner2014,Gerner2015} to these regions in order to estimate chemical timescales $\tau_\mathrm{chem}$ of the protostellar sources \citep{Gieser2021,Gieser2022}. With additional MIR, FIR, and cm data at high angular resolution, we aim to better characterize and classify the evolutionary stages on core scales, compared to larger clump scales, considering that multiple fragmented protostellar sources in different evolutionary phases are typically embedded within a single clump.

\begin{acknowledgements}
	The authors would like to thank the anonymous referee whose comments helped improve the clarity of this paper. C.G. and H.B. acknowledge support from the European Research Council under the Horizon 2020 Framework Programme via the ERC Consolidator Grant CSF-648505. This paper makes use of the following ALMA data: ADS/JAO.ALMA\#2018.1.00424.S ALMA is a partnership of ESO (representing its member states), NSF (USA) and NINS (Japan), together with NRC (Canada), MOST and ASIAA (Taiwan), and KASI (Republic of Korea), in cooperation with the Republic of Chile. The Joint ALMA Observatory is operated by ESO, AUI/NRAO and NAOJ. This research made use of Astropy\footnote{\url{http://www.astropy.org}}, a community-developed core Python package for Astronomy \citep{Astropy2013, Astropy2018}.
\end{acknowledgements}

\bibliographystyle{aa} 
\bibliography{bibliography} 

\begin{appendix}

\section{Summary of the ALMA observations}\label{app:almaoverview}

	Tables \ref{tab:almaobssummary} and \ref{tab:almaspwsummary} give an detailed overview of the ALMA 3\,mm observations and spectral setup, respectively. The data calibration and imaging of this data set are explained in detail in Sect. \ref{sec:ALMAobs}.

\begin{table*}[!htb]
\caption[Overview of the ALMA 3\,mm observation blocks.]{Overview of the ALMA 3\,mm observation blocks (Cycle 6, project code 2018.1.00424.S). The target regions were grouped into three fields in order to minimize off-source time (Table \ref{tab:ALMAcontinuumdataproducts}).}
\label{tab:almaobssummary}
\centering
\begin{tabular}{llrrrrrrr}
\hline \hline
Field & SPR & Array & Antennae & Date & On-source & PWV & shortest & longest\\
 & & configuration & & & integration & & \multicolumn{2}{c}{baseline}\\
 & & & && (min) & (mm) & (m) & (m)\\
\hline
Field 1 & SPR1 & C43-4 & 47 & 2018-10-28 21:06 & 20.4 & 1.3 & 15.1 & 1397.8\\ 
Field 1 & SPR1 & ACA & 11 & 2018-12-28 18:08 & 48.6 & 6.0 & 8.9 & 48.9\\ 
Field 1 & SPR2 & C43-4 & 41 & 2018-12-15 19:34 & 26.6 & 4.1 & 15.1 & 740.4\\ 
Field 1 & SPR2 & ACA & 10 & 2018-12-30 15:27 & 51.5 & 5.0 & 8.9 & 48.9\\ 
Field 1 & SPR3 & C43-4 & 48 & 2018-10-27 21:54 & 20.6 & 1.3 & 15.1 & 1397.8\\ 
Field 1 & SPR3 & ACA & 11 & 2018-12-28 17:24 & 44.2 & 6.0 & 8.9 & 48.9\\ 
\hline
Field 2 & SPR1 & C43-4 & 46 & 2018-11-20 17:54 & 47.7 & 1.4 & 15.1 & 1397.8\\ 
Field 2 & SPR1 & C43-4 & 46 & 2018-11-20 18:42 & 48.6 & 1.5 & 15.1 & 1397.8\\ 
Field 2 & SPR1 & ACA & 11 & 2019-05-10 09:30 & 84.2 & 5.0 & 8.9 & 48.0\\ 
Field 2 & SPR1 & ACA & 11 & 2019-05-20 07:12 & 89.0 & 2.0 & 8.9 & 48.0\\ 
Field 2 & SPR1 & ACA & 11 & 2019-05-23 07:36 & 83.8 & 1.0 & 8.9 & 48.0\\ 
Field 2 & SPR2 & C43-4 & 41 & 2019-05-03 11:22 & 55.8 & 1.0 & 15.1 & 740.4\\ 
Field 2 & SPR2 & ACA & 11 & 2019-01-26 15:04 & 76.7 & 4.0 & 8.9 & 48.0\\ 
Field 2 & SPR2 & ACA & 11 & 2019-03-29 08:16 & 83.2 & 5.0 & 8.9 & 48.0\\ 
Field 2 & SPR2 & ACA & 11 & 2019-03-29 09:39 & 84.0 & 5.0 & 8.9 & 48.0\\ 
Field 2 & SPR2 & ACA & 11 & 2019-03-29 11:06 & 54.6 & 5.0 & 8.9 & 48.0\\ 
Field 2 & SPR3 & C43-1 & 46 & 2019-01-10 18:20 & 41.3 & 2.4 & 15.1 & 313.7\\ 
Field 2 & SPR3 & ACA & 11 & 2019-05-20 08:41 & 83.7 & 2.0 & 8.9 & 48.0\\ 
Field 2 & SPR3 & ACA & 11 & 2019-05-23 08:59 & 83.4 & 1.0 & 8.9 & 48.0\\ 
\hline
Field 3 & SPR1 & C43-4 & 46 & 2018-11-16 21:38 & 43.6 & 2.0 & 15.1 & 1397.8\\ 
Field 3 & SPR1 & C43-4 & 46 & 2018-11-19 20:31 & 44.3 & 1.9 & 15.1 & 1397.8\\ 
Field 3 & SPR1 & ACA & 11 & 2019-01-16 13:15 & 74.8 & 5.0 & 8.9 & 48.9\\ 
Field 3 & SPR1 & ACA & 12 & 2019-01-17 12:35 & 78.6 & 5.0 & 8.9 & 48.9\\ 
Field 3 & SPR2 & C43-4 & 41 & 2019-04-17 11:20 & 61.8 & 3.3 & 15.0 & 783.5\\ 
Field 3 & SPR2 & ACA & 10 & 2019-04-01 12:04 & 87.0 & 4.0 & 8.9 & 48.9\\ 
Field 3 & SPR2 & ACA & 10 & 2019-04-17 12:17 & 86.9 & 3.0 & 8.9 & 44.7\\ 
Field 3 & SPR3 & C43-4 & 47 & 2018-10-30 22:39 & 50.2 & 1.4 & 15.1 & 1397.8\\ 
Field 3 & SPR3 & ACA & 11 & 2019-01-16 14:43 & 74.9 & 5.0 & 8.9 & 48.9\\ 
Field 3 & SPR3 & ACA & 11 & 2019-01-19 14:25 & 74.9 & 5.0 & 8.9 & 48.9\\ 
\hline
\end{tabular}
\end{table*}

\begin{table*}[!htb]
\caption[Overview of the spectral windows observed with ALMA.]{Overview of the spectral windows observed with ALMA.}
\label{tab:almaspwsummary}
\centering
\begin{tabular}{llrrrrl}
\hline \hline
SPR & spw & Central & Bandwidth & number of & channel & Targeted\\
& & frequency & & channels & spacing & species\\
 & & $\nu_0$ & & $n_\mathrm{channel}$ & $\delta \nu$ &\\
 & & (GHz) & (GHz) & & (MHz) &\\
\hline
SPR1 & 0 & 86.676 & 0.117 & 480 & 0.244 & HCO\\ 
SPR1 & 1 & 86.759 & 0.117 & 480 & 0.244 & H$^{13}$CO$^{+}$\\ 
SPR1 & 2 & 86.852 & 0.117 & 480 & 0.244 & SiO, CH$_{3}$CH$_{2}$CN\\ 
SPR1 & 3 & 86.345 & 0.117 & 480 & 0.244 & H$^{13}$CN\\ 
SPR1 & 4 & 87.44 & 0.117 & 480 & 0.244 & CCH\\ 
SPR1 & 5 & 88.637 & 0.117 & 480 & 0.244 & HCN, CH$_{3}$OH\\ 
SPR1 & 6 & 87.93 & 0.117 & 480 & 0.244 & HNCO\\ 
SPR1 & 7 & 87.322 & 0.117 & 480 & 0.244 & CCH\\ 
SPR1 & 8 & 97.721 & 0.117 & 480 & 0.244 & $^{34}$SO\\ 
SPR1 & 9 & 97.987 & 0.117 & 480 & 0.244 & CS\\ 
SPR1 & 10 & 96.926 & 0.117 & 480 & 0.244 & CH$_{3}$CH$_{2}$CN\\ 
SPR1 & 11 & 96.994 & 0.117 & 480 & 0.244 & O$^{13}$CS, H$^{13}$CCCN\\ 
SPR1 & 12 & 99.331 & 0.117 & 480 & 0.244 & CH$_{3}$OCH$_{3}$\\ 
SPR1 & 13 & 100.736 & 0.117 & 480 & 0.244 & HC$_{3}$N, CH$_{3}$OCHO\\ 
SPR1 & 14 & 99.305 & 0.117 & 480 & 0.244 & SO\\ 
SPR1 & 15 & 99.872 & 0.117 & 480 & 0.244 & CCS\\ 
\hline
SPR2 & 0 & 108.786 & 0.117 & 480 & 0.244 & $^{13}$CN\\ 
SPR2 & 1 & 110.207 & 0.117 & 480 & 0.244 & $^{13}$CO\\ 
SPR2 & 2 & 109.469 & 0.117 & 480 & 0.244 & OCS, HNCO, HC$_{3}$N\\ 
SPR2 & 3 & 109.788 & 0.117 & 480 & 0.244 & C$^{18}$O, NH$_{2}$CHO\\ 
SPR2 & 4 & 109.506 & 1.875 & 1920 & 0.977 & C$^{15}$N, SiS, HC$_{3}$N, HC$_{5}$N, CH$_{3}$OCH$_{3}$\\ 
SPR2 & 5 & 97.006 & 1.875 & 1920 & 0.977 & C$^{33}$S, H$_{2}^{13}$CS, C$_{2}$H$_{5}$OH, CH$_{3}$NCO\\ 
SPR2 & 6 & 98.806 & 1.875 & 1920 & 0.977 & H$\alpha$, CH$_{3}$CHO, CH$_{3}$OCHO, CH$_{3}$COCH$_{3}$\\ 
\hline
SPR3 & 0 & 104.035 & 0.117 & 480 & 0.244 & CH$_{3}$CH$_{2}$CN, SO$_{2}$, CH$_{3}$OH\\ 
SPR3 & 1 & 102.224 & 0.117 & 480 & 0.244 & HSCN, CH$_{3}$SH\\ 
SPR3 & 2 & 103.047 & 0.117 & 480 & 0.244 & H$_{2}$CS\\ 
SPR3 & 3 & 103.223 & 0.117 & 480 & 0.244 & NH$_{2}$CHO\\ 
SPR3 & 4 & 104.207 & 0.117 & 480 & 0.244 & CH$_{2}$CHCN, SO$_{2}$, CH$_{3}$OCH$_{3}$\\ 
SPR3 & 5 & 104.88 & 0.117 & 480 & 0.244 & HOCN\\ 
SPR3 & 6 & 104.717 & 0.117 & 480 & 0.244 & $^{13}$C$^{18}$O, CH$_{3}$OCH$_{3}$\\ 
SPR3 & 7 & 105.564 & 0.117 & 480 & 0.244 & CH$_{3}$OH, CH$_{3}$OCH$_{3}$\\ 
SPR3 & 8 & 90.269 & 0.117 & 480 & 0.244 & $^{15}$NNH$^{+}$, CH$_{3}$OCHO\\ 
SPR3 & 9 & 90.669 & 0.117 & 480 & 0.244 & HNC, CCS\\ 
SPR3 & 10 & 90.985 & 0.117 & 480 & 0.244 & $^{13}$C$^{34}$S, HC$_{3}$N, CH$_{3}$OCH$_{3}$\\ 
SPR3 & 11 & 91.945 & 0.117 & 480 & 0.244 & CH$_{3}^{13}$CN\\ 
SPR3 & 12 & 91.64 & 0.117 & 480 & 0.244 & CH$_{3}$COCH$_{3}$\\ 
SPR3 & 13 & 91.985 & 0.117 & 480 & 0.244 & CH$_{3}$CN\\ 
SPR3 & 14 & 93.179 & 0.117 & 480 & 0.244 & N$_{2}$H$^{+}$\\ 
SPR3 & 15 & 92.415 & 0.117 & 480 & 0.244 & CH$_{2}$CHCN, CH$_{3}$OH\\ 
\hline
\end{tabular}
\end{table*}

\setlength{\tabcolsep}{5pt}

\section{Overview of the sample}\label{app:sampleoverview}

	A multi-wavelength and multi-scale overview of the target regions are presented in Figs. \ref{fig:overview_IRDC_G1111} - \ref{fig:overview_UCHII_G1387} using archival VLA cm, ATLASGAL submm, \textit{Herschel} FIR and \textit{Spitzer} MIR data in comparison with the ALMA 3\,mm observations. We note that mass and luminosity estimates in those descriptions might differ from the ATLASGAL results shown in Table \ref{tab:ALMA_regions}, for example, due to a different assumed source distance.

\subsection{IRDC\,G11.11$-$4 (Fig. \ref{fig:overview_IRDC_G1111})}

	The IRDC\,G11.11$-$4, also referred to as the ``Snake'', is a filamentary cloud \citep{Carey1998} that shows a spatial correlation between dust extinction maps and submm emission \citep[][see also Fig. \ref{fig:overview_IRDC_G1111}]{Johnstone2003, Henning2010}. \citet{Henning2010} detected 18 cores along the filament, two of them with masses $>$50\,$M_\odot$. \citet{Wang2014} studied the hierarchical fragmentation from scales of 1\,pc down to 0.01\,pc in the filament and found chemical differentiation between fragmented cores indicating that the cores are at different evolutionary phases.
	
	 We observed with ALMA the region referred to as ``P1'' in the literature which already shows signatures of active star formation with 8\,$\upmu$m emission, a molecular outflow \citep{Johnstone2003}, and the presence of CH$_{3}$OH and H$_{2}$O masers \citep{Pillai2006a}. By modeling the SED of the dust continuum, \citet{Pillai2006a} infer a luminosity of 1\,200\,$L_\odot$ that is powered by a Zero Age Main Sequence (ZAMS) star of 8\,$M_\odot$. The authors infer an envelope mass and temperature of 500\,$M_\odot$ and 19\,K, respectively, and elevated temperatures around the protostar ($\approx$60\,K). \citet{Rosero2014} detected radio continuum emission at 1.3\,cm and 6\,cm with the VLA which they attributed to a thermal ionized jet, while at 3\,mm wavelengths the emission is dominated by dust.

\subsection{IRDC\,18223$-$3 (Fig. \ref{fig:overview_IRDC_IRAS18223})}

	IRDC\,18223$-$3 \citep[``peak 3'' studied by][]{Beuther2002} is located in the same filament as IRAS\,18223$-$1243, an H{\sc ii} region \citep{Bronfman1996} with bright radio continuum emission \citep{Dewangan2018}. \citet{Beuther2005b} studied the region with the Plateau de Bure Interferometer (PdBI) at 3\,mm wavelengths, while no emission is detected at MIR wavelengths except for weak emission at 4.5\,$\upmu$m around the source tracing shock-excited gas by an outflow. The fact that there is no MIR source detected in the \textit{Spitzer} IRAC bands and only a small outflow mass of $M_\mathrm{out} = 3.4$\,$M_\odot$ is estimated, the authors suspected that this source is a very young protostar before the HMPO phase.
	
	 Weak protostellar MIR emission became detectable with more sensitive \textit{Spitzer} MIPS observations at 24 and 70\,$\upmu$m \citep{Beuther2007C}. The authors fit the SED and find that most of the mass ($M \approx 580$\,$M_\odot$) is enclosed at low temperatures of $T \approx 15$\,K, while a small fraction ($M \approx 0.01$\,$M_\odot$) has elevated temperatures of $T \geq 51$\,K with a total luminosity of 180\,$L_\odot$ and thus the source is expected to evolve into a high-mass star. The outflow is directed from the northwest-southeast with a disk that is likely to be edge-on \citep{Fallscheer2009}.
	 
	 \citet{Beuther2015} studied the larger scale filament with PdBI mosaic observations and the authors detect 12 cores with a separation of 0.4\,pc which is consistent with thermal Jeans fragmentation. However, a high mass-to-length ratio of $\approx$1\,000\,$M_\odot$\,pc$^{-1}$ in the filament requires additional support against further collapse. The filament and its environment were studied by \citet{Dewangan2018} with $^{13}$CO line emission. The authors find two clouds at slightly different velocities at 45\,km\,s$^{-1}$ and 51\,km\,s$^{-1}$ that seem to be connected at the intermediate velocities suggesting that about 1\,Myr ago a cloud-cloud collision could have occurred which could then have triggered the formation of high-mass stars.
 
\subsection{IRDC\,18310$-$4 (Fig. \ref{fig:overview_IRDC_IRAS18310})}

	IRDC\,18310$-$4 \citep[``peak 4'' in ][]{Beuther2002} is located in the same filament as IRAS\,18310$-$0825. The region, being dark up to 100\,$\upmu$m, was a prestellar clump candidate with a mass reservoir of $\approx$800\,$M_\odot$ without any signposts of active star formation \citep{Beuther2013}. The line width of N$_{2}$H$^{+}$ decreases toward the center of the clump \citep{Tackenberg2014}. 
	
	Weak 70\,$\upmu$m emission was detected toward one of the fragments \citep{Beuther2015B} and the magnetic field is relatively weak ($\approx$2.6\,mG) compared to more evolved HMPOs and HMCs \citep{Beuther2018B}. Recent ALMA observations revealed that among 11 cores with masses ranging from 1.1 to 19\,$M_\odot$ at least four cores show outflows traced by CO and SiO emission such that star formation has been taking place for at least $10^{4}$\,yr \citep{Morii2021}.

\subsection{HMPO\,IRAS\,18089 (Fig. \ref{fig:overview_HMPO_IRAS18089})}

	IRAS\,18089$-$1732 is a massive HMPO with $M \approx 2\,000$\,$M_\odot$ with H$_{2}$O and CH$_{3}$OH maser emission \citep{Beuther2004}. The authors found a collimated north-south outflow revealed by SiO emission, while CH$_{3}$OCHO emission is confined to the central core region with a velocity gradient almost perpendicular to the outflow tracing the rotating envelope/disk structure. Sub-arcsecond resolution observations show that the core does not further fragment \citep{Beuther2005C}, however, the variability of CH$_{3}$OH maser emission with a period of $\approx$30\,d could be explained by a binary system \citep{Goedhart2009}. 
	
	The rich molecular content and a temperature of 350\,K derived with CH$_{3}$CN indicates that the central region has already formed a hot core in the inner region \citep{Beuther2005C}. \citet{Sanhueza2021} studied the region at 700\,au scales with ALMA and find that the dust and magnetic field have spiral-like filamentary features and the H$^{13}$CO$^{+}$ emission shows a complex velocity gradient suggesting that the filaments are rotating and infalling. An analysis of the energy budget shows that gravity is dominant over rotation, turbulence, and the magnetic field \citep{Sanhueza2021}. The authors estimate an infall rate of $0.9-2.5\times10^{-4}$\,$M_\odot$\,yr$^{-1}$.
 
\subsection{HMPO\,IRAS\,18182 (Fig. \ref{fig:overview_HMPO_IRAS18182})}

	HMPO\,IRAS\,18182 is a massive isolated core \citep[$M \approx 1500$\,$M_\odot$ and $L \approx 10^{4.3}$\,$L_\odot$,][]{Beuther2002} with detected OH, H$_{2}$O, and CH$_{3}$OH maser emission \citep{Sanna2010}. \citet{Beuther2006} resolve no further fragmentation down to their angular resolution of $3''$ and derive a temperature of 150\,K in the central region using CH$_{3}$CN line emission. The authors find that the outflow has a quadrupolar morphology and that the molecular emission peaks $1-2''$ off from the continuum peak position. With multi-epoch CH$_{3}$OH maser observations, \citet{Sanna2010} estimate that the central object has a mass of 35\,$M_\odot$ assuming that the emission stems from a rotating disk.

\subsection{HMPO\,IRAS\,18264 (Fig. \ref{fig:overview_HMPO_IRAS18264})}

	IRAS\,18264 is an extended green object (EGO) with widespread emission in 4.5\,$\upmu$m \textit{Spitzer} data that rises from shocked H$_{2}$ gas in the outflow \citep{Cyganowski2008} and has associated H$_{2}$O and CH$_{3}$OH masers \citep{Beuther2002C,Chen2011b}. Sensitive VLA observations at 1.3\,cm and 6\,cm reveal seven cm sources in the region with spectral indices in the range from $-$0.2 to $+$1.1 \citep{Rosero2016} at these wavelengths. \citet{Issac2020} find that IRAS\,18264 is a protocluster with multiple radio as well as mm components detected. The C$^{17}$O line has an inverse P-Cygni profile tracing infalling material and multiple bipolar outflows are traced by C$^{18}$O \citep{Issac2020}.

\subsection{HMC\,G9.62$+$0.19 (Fig. \ref{fig:overview_HMC_G0962})}

	The G9.62$+$0.19 star-forming complex consists of protostars at different evolutionary stages, from an embedded HMC to evolved UCH{\sc ii} regions, revealed by multi-wavelength observations from NIR to radio wavelengths \citep{Cesaroni1994,Testi1998,Linz2005}. The NIR emission is not only dominated by the SED of the YSO, but there are also contributions of the circumstellar matter and outflows, and the non-detection of the UCH{\sc ii} regions is consistent with a high extinction \citep{Linz2005}.
	
	High angular resolution observations at 1\,mm carried out by \citet{Liu2017} reveal that the filament contains at least 12 dense cores at different evolutionary stages from starless cores to the known UCH{\sc ii} regions, with three cores harboring molecular outflows. The authors propose that the expanding UCH{\sc ii} regions compress the gas of the clump in a filament in which sequential star formation is then triggered. The feedback of the UCH{\sc ii} regions results in a lack of low-mass star formation \citep{Liu2017}.
	
	\citet{Liu2020} studied the spatial correlation of the molecular emission using ALMA 3\,mm observations using the histogram of oriented gradients (HOG) method \citep{Soler2019} that was also applied to the molecular lines of the CORE-extension regions \citep{Gieser2022}. The authors find that the molecules can be grouped to species that either trace extended emission (CS, HCN, HCO$^{+}$), shocks (SiO), or the dense cores (SO, H$^{13}$CN, HC$_{3}$N, CH$_{3}$OH). Widespread SiO emission with narrow line widths seems to stem from either colliding flows or the H{\sc ii} regions \citep{Liu2020}. A sketch of the regions with the dense cores at evolutionary stages under the influence of the H{\sc ii} regions is given in Fig. 10 in \citet{Liu2017} and Fig. 20 in \citet{Liu2020}.

\subsection{HMC\,G10.47$+$0.03 (Fig. \ref{fig:overview_HMC_G1047})}

	The G10.47$+$0.03 region contains several MIR sources including a HMC, as well as multiple UCH{\sc ii} regions and strong maser activity \citep[][and references within]{Pascucci2004}. \citet{Cesaroni2010} found that three UCH{\sc ii} regions are located within the HMC itself. Based on CH$_{3}$CN line emission the temperature in the innermost 1\as5 is estimated to be 160\,K \citep{Olmi1996}. Molecular emission lines reveals that there is outflowing gas as well as infalling material \citep{Rolffs2011}. The authors find that the density toward the central 10\,000\,au flattens due to pressure (e.g., thermal, turbulent, wind-driven) driving expansion. \citet{Gorai2020} detect precursors of prebiotic molecules toward the HMC, including HNCO, NH$_{2}$CHO, and CH$_{3}$NCO.

\subsection{HMC\,G34.26$+$0.15 (Fig. \ref{fig:overview_HMC_G3426})}

	G34.26$+$0.15 contains a cometary UCH{\sc ii} region and two unresolved UCH{\sc ii} regions \citep{Reid1985}. The cometary UCH{\sc ii} region harbors a line-rich HMC \citep{Macdonald1996}. \citet{Mookerjea2007} find that the HMC is externally heated by the UCH{\sc ii} region with no internal heating. Toward the region, many COMs such as CH$_{3}$OCHO, CH$_{3}$OCH$_{3}$, and NH$_{2}$CHO are detected \citep{Mookerjea2007}. Ammonia (NH$_{3}$) lines have inverse P-Cygni profiles indicating infalling material \citep{Hajigholi2016}.

\subsection{UCH{\sc ii}\,G10.30$-$0.15 (Fig. \ref{fig:overview_UCHII_G1030})}

	The UCH{\sc ii} region G10.30$-$0.15 is part of the W31 star-forming complex containing a cluster of stars and YSOs at different evolutionary stages \citep{Beuther2011} and a large scale bipolar H{\sc ii} region launched by a massive O8$-$O9 star \citep{Deharveng2015}. The UCH{\sc ii} region G10.30$-$0.15 is associated with CH$_{3}$OH maser emission and the central object is a B0 star \citep{Deharveng2015}. The authors studied the clustered nature of the W31 region and find that five clumps, including the clump containing G10.30$-$0.15 at the waist of the bipolar H{\sc ii} region, are a result of triggered star formation.

\subsection{UCH{\sc ii}\,G13.87$+$0.28 (Fig. \ref{fig:overview_UCHII_G1387})}

	G13.87$+$0.28 is a cometary UCH{\sc ii} region, with a bright head and an extended low brightness tail. The morphology and kinematics of the ionized and molecular gas a consistent with a bow shock launched by a stellar wind of a massive O6.5 ZAMS star \citep{Garay1994}. High velocity components of 22\,GHz water masers are likely associated with an outflow \citep{Xi2015}.

\begin{figure*}
\centering
\includegraphics[width=0.36\textwidth]{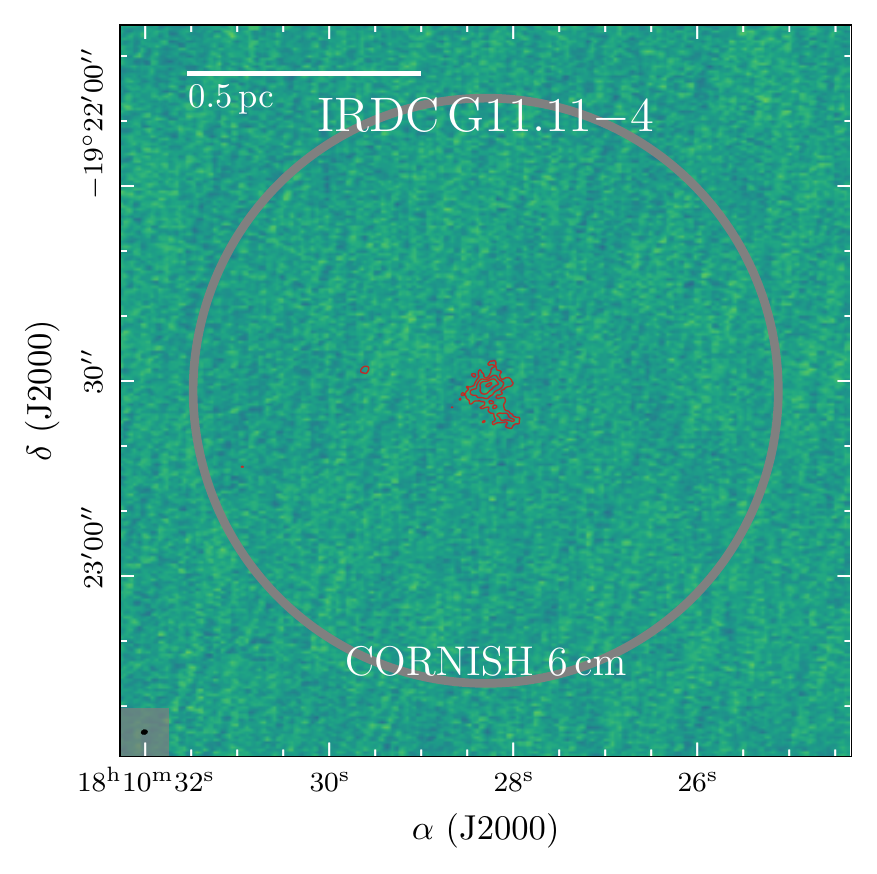}
\includegraphics[width=0.36\textwidth]{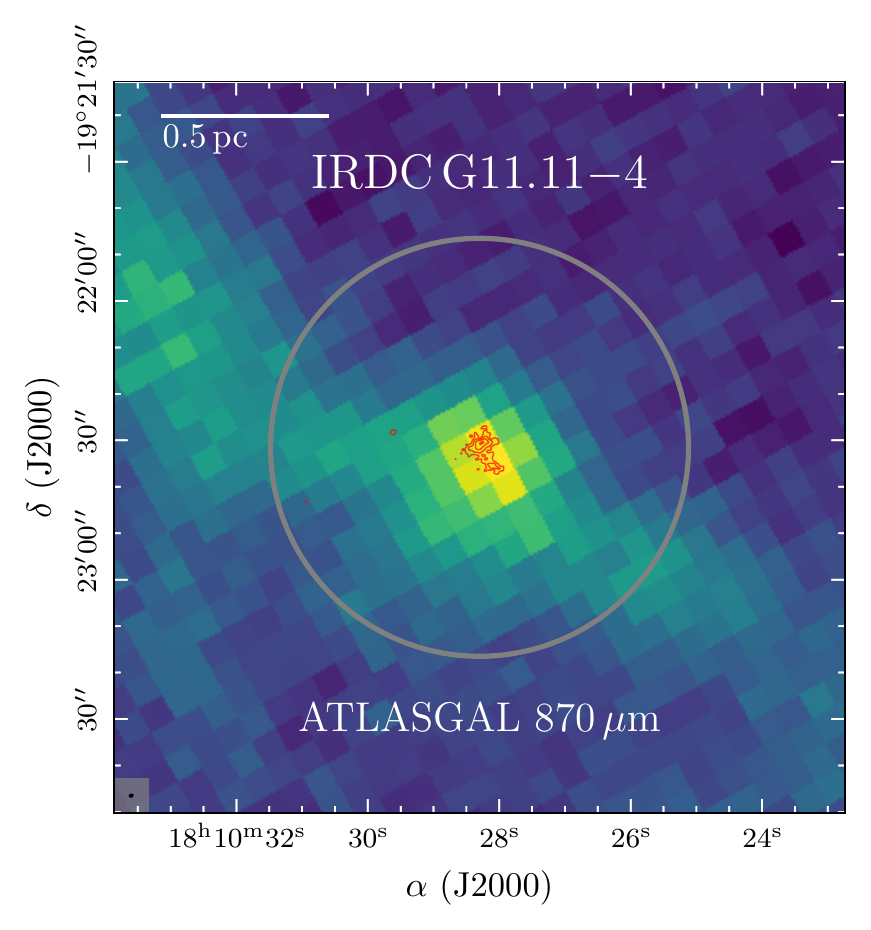}\\
\includegraphics[width=0.36\textwidth]{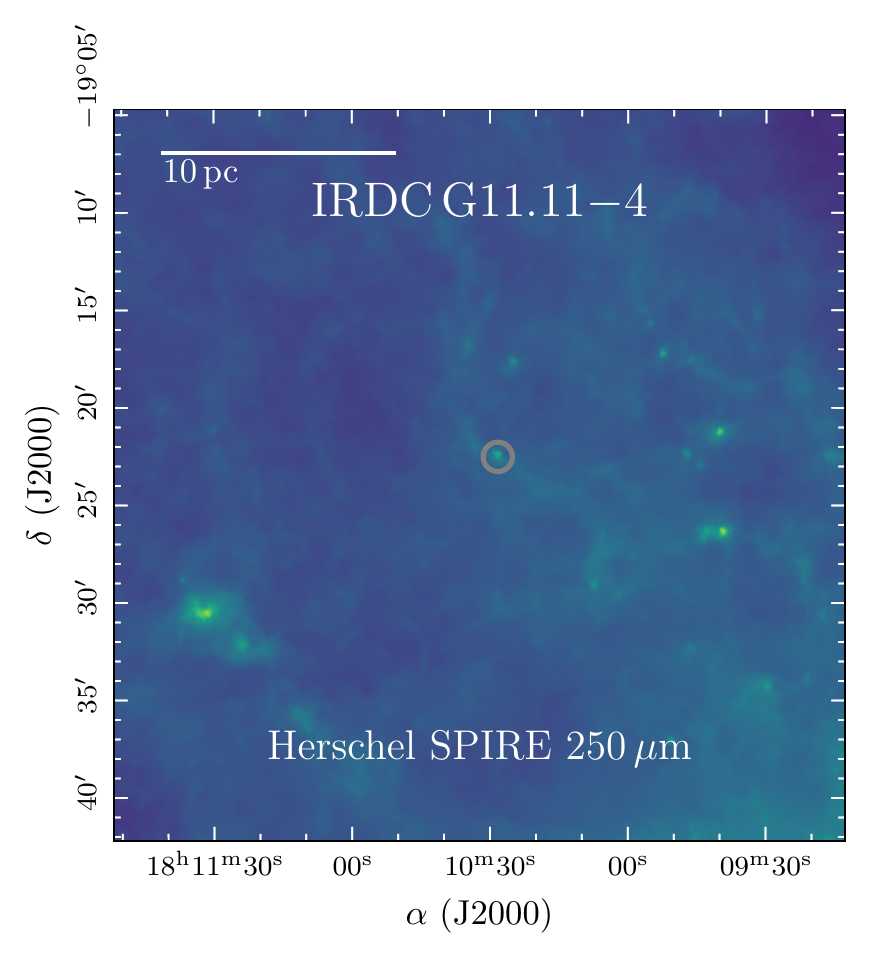}
\includegraphics[width=0.36\textwidth]{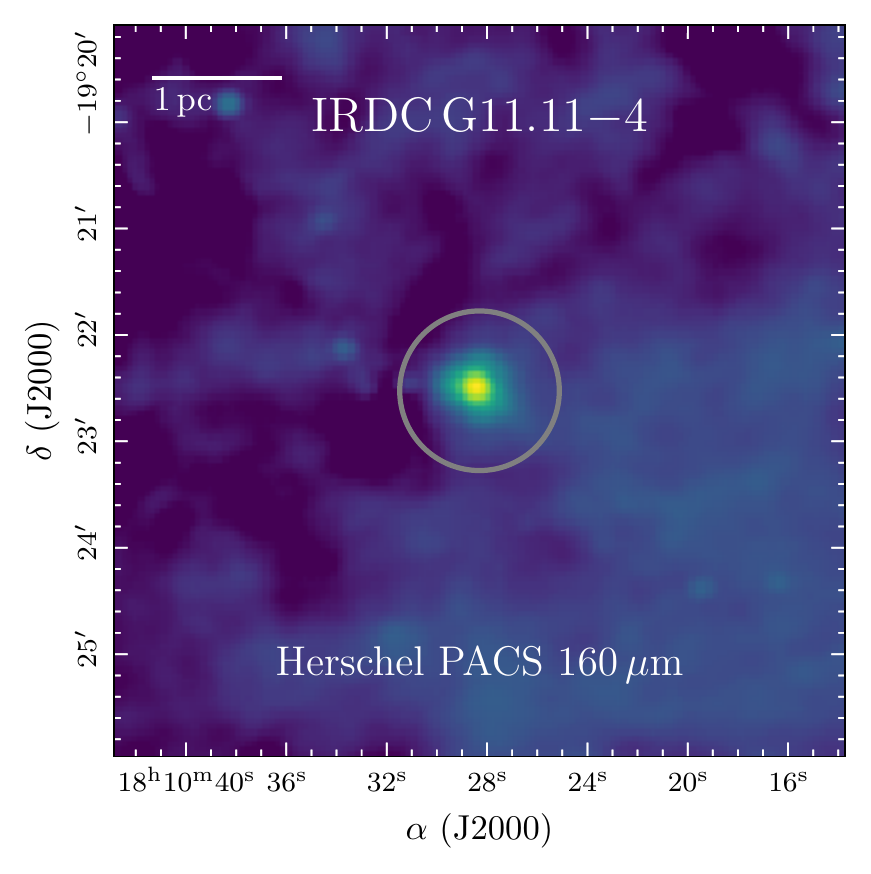}\\
\includegraphics[width=0.36\textwidth]{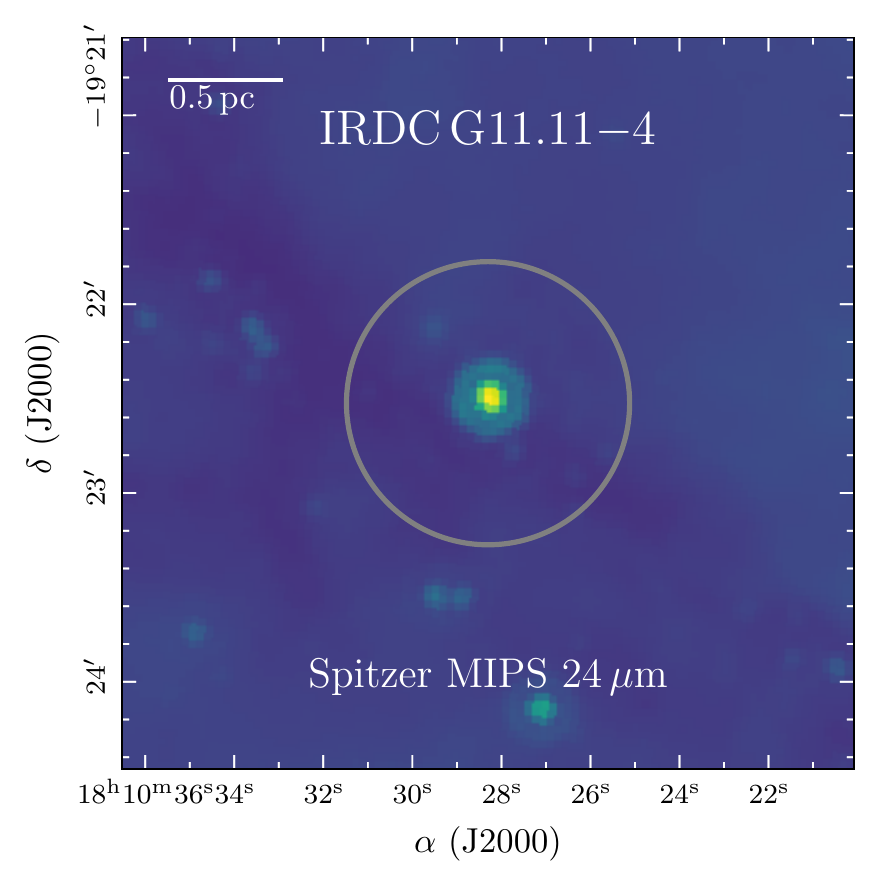}
\includegraphics[width=0.36\textwidth]{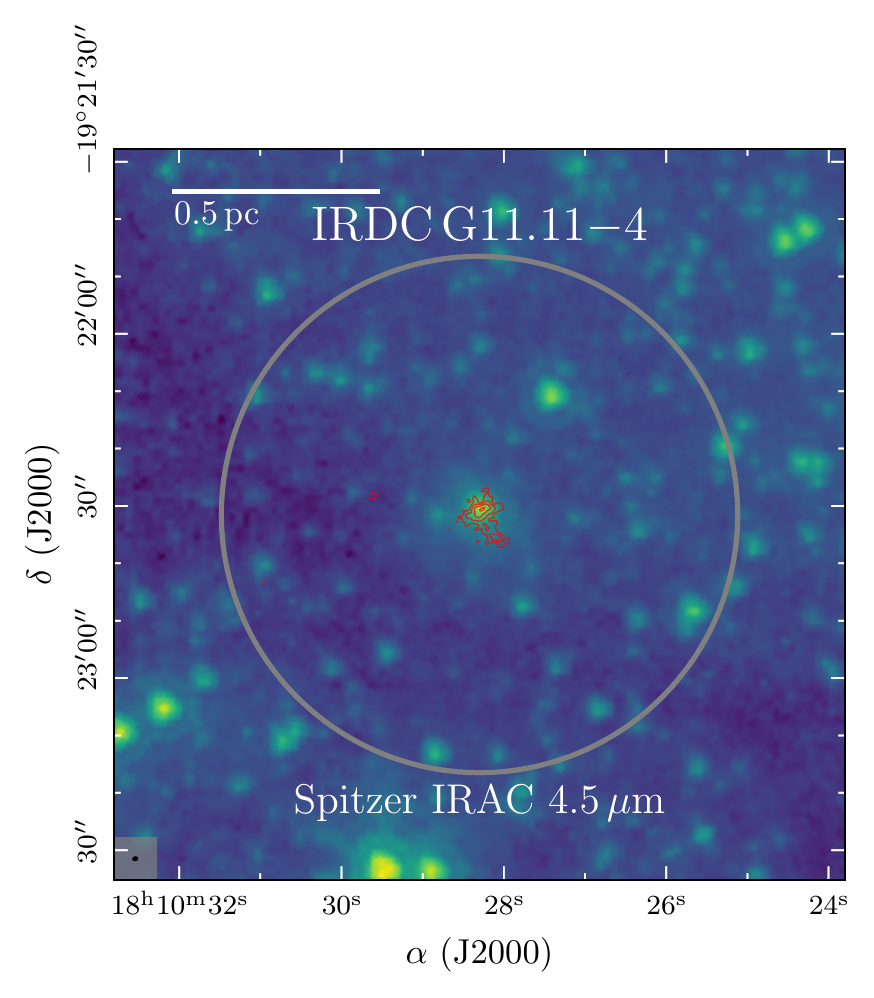}
\caption[Overview of IRDC\,G11.11$-$4.]{Multi wavelength overview of IRDC\,G11.11$-$4. In color, CORNISH 6\,cm, ATLASGAL 870$\upmu$m, \textit{Herschel} SPIRE 250\,$\upmu$m, \textit{Herschel} PACS 70\,$\upmu$m, \textit{Spitzer} MIPS 24\,$\upmu$m, and \textit{Spitzer} IRAC 4.5\,$\upmu$m data are presented as labeled. In all panels, the ALMA primary beam size is indicated by a grey circle. In the top right and left and bottom right panel, the ALMA 3\,mm continuum data are shown by red contours. The dotted red contour marks the $-5\sigma_\mathrm{cont}$ level. The solid red contours start at $5\sigma_\mathrm{cont}$ and contour steps increase by a factor of 2 (e.g., 5, 10, 20, $40\sigma_\mathrm{cont}$). The ALMA synthesized beam size is shown in the bottom left corner.}
\label{fig:overview_IRDC_G1111}
\end{figure*}

\begin{figure*}
\centering
\includegraphics[width=0.38\textwidth]{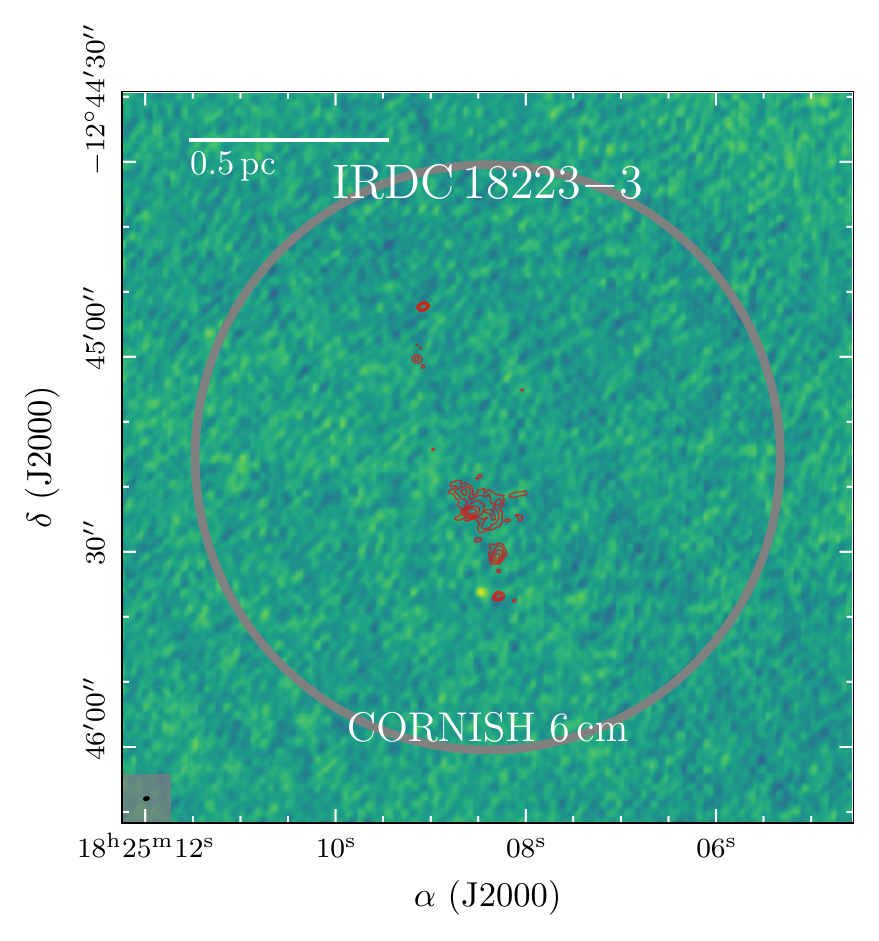}
\includegraphics[width=0.38\textwidth]{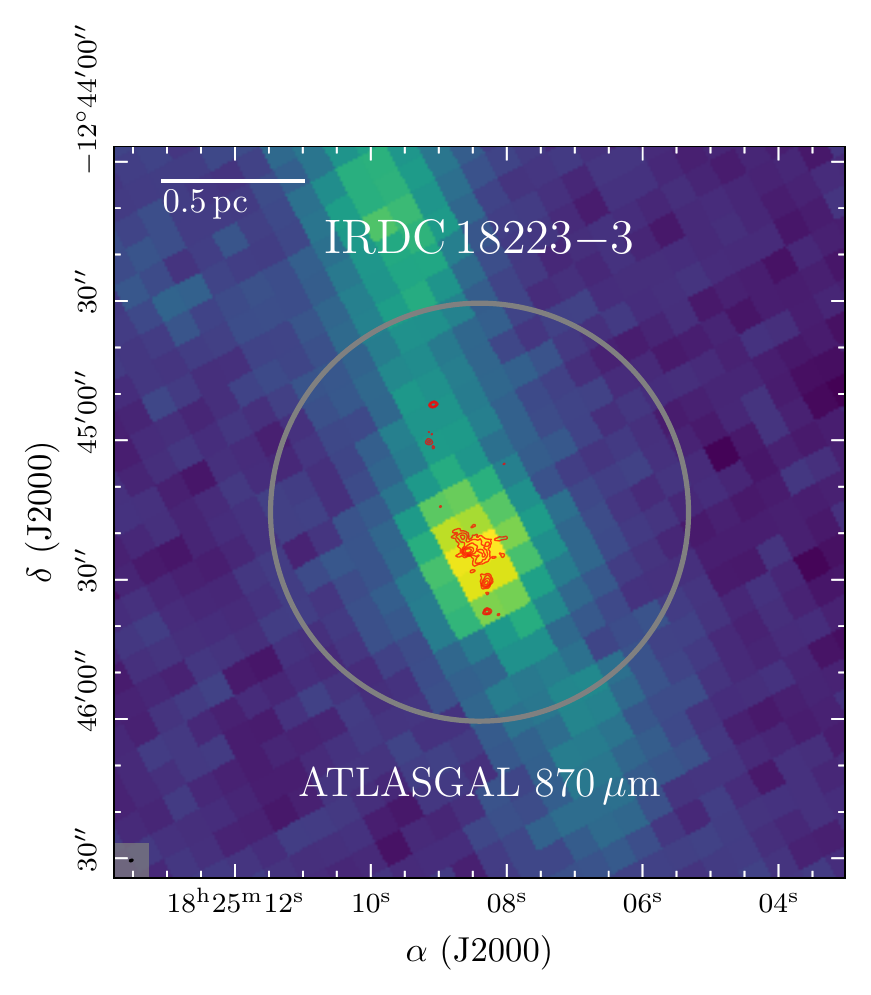}\\
\includegraphics[width=0.38\textwidth]{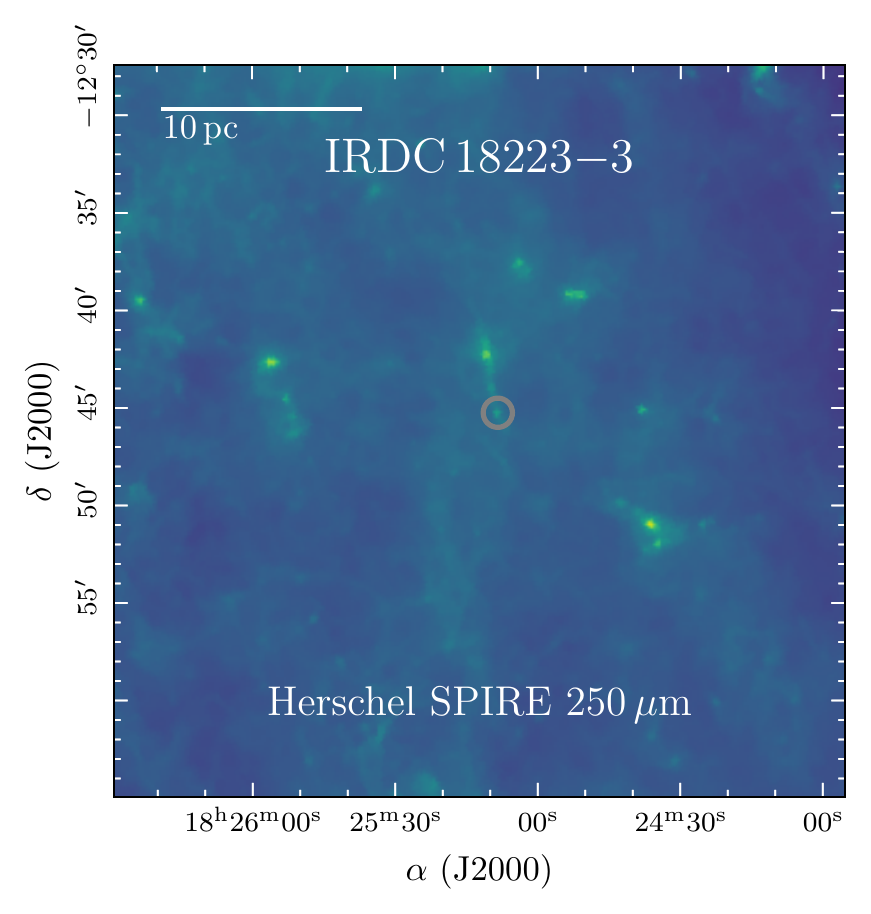}
\includegraphics[width=0.38\textwidth]{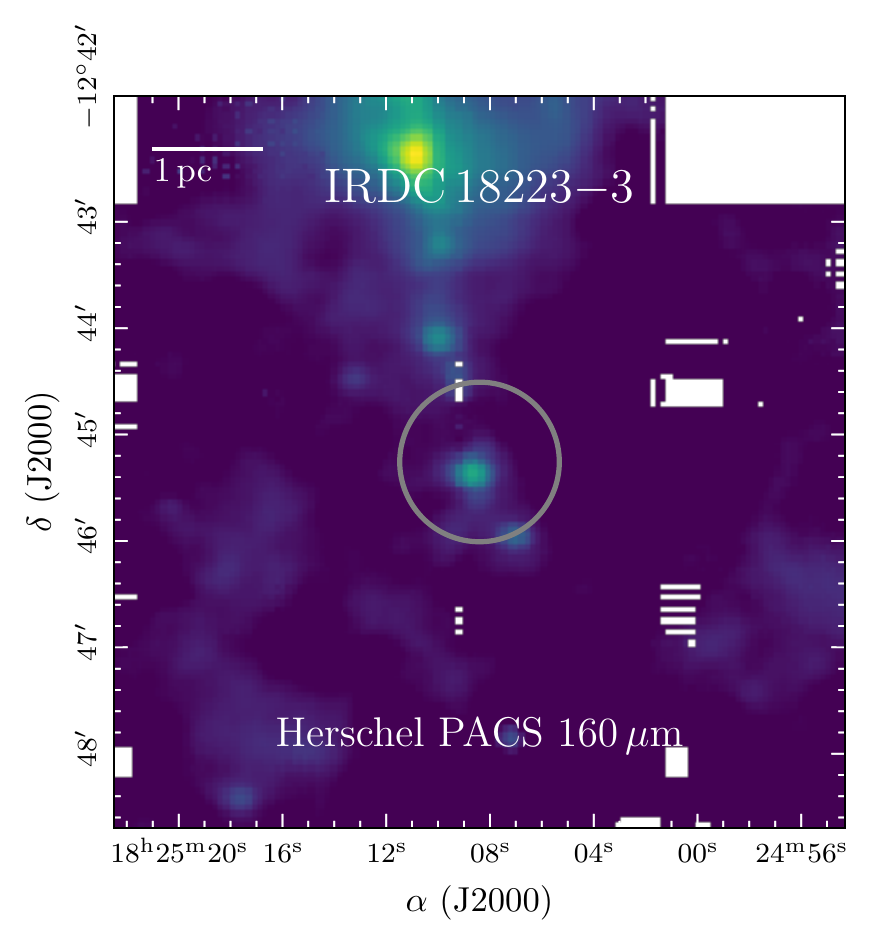}\\
\includegraphics[width=0.38\textwidth]{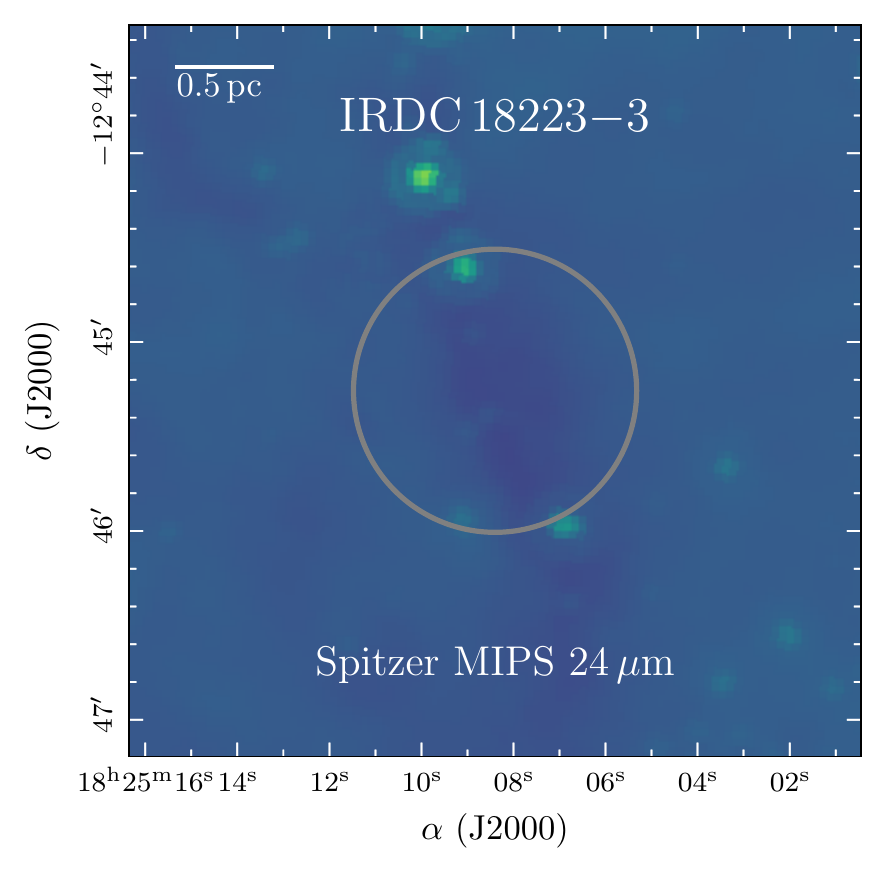}
\includegraphics[width=0.38\textwidth]{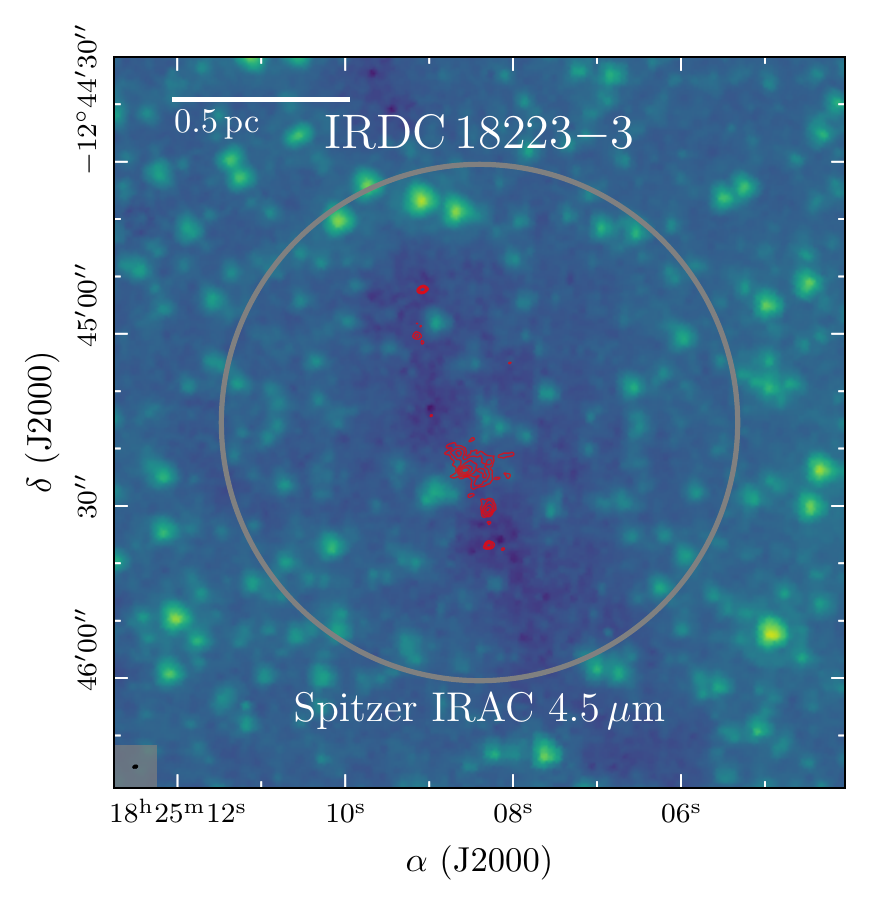}
\caption[Overview of IRDC\,18223$-$3.]{The same as Fig. \ref{fig:overview_IRDC_G1111}, but for IRDC\,18223$-$3.}
\label{fig:overview_IRDC_IRAS18223}
\end{figure*}

\begin{figure*}[!htb]
\centering\includegraphics[width=0.39\textwidth]{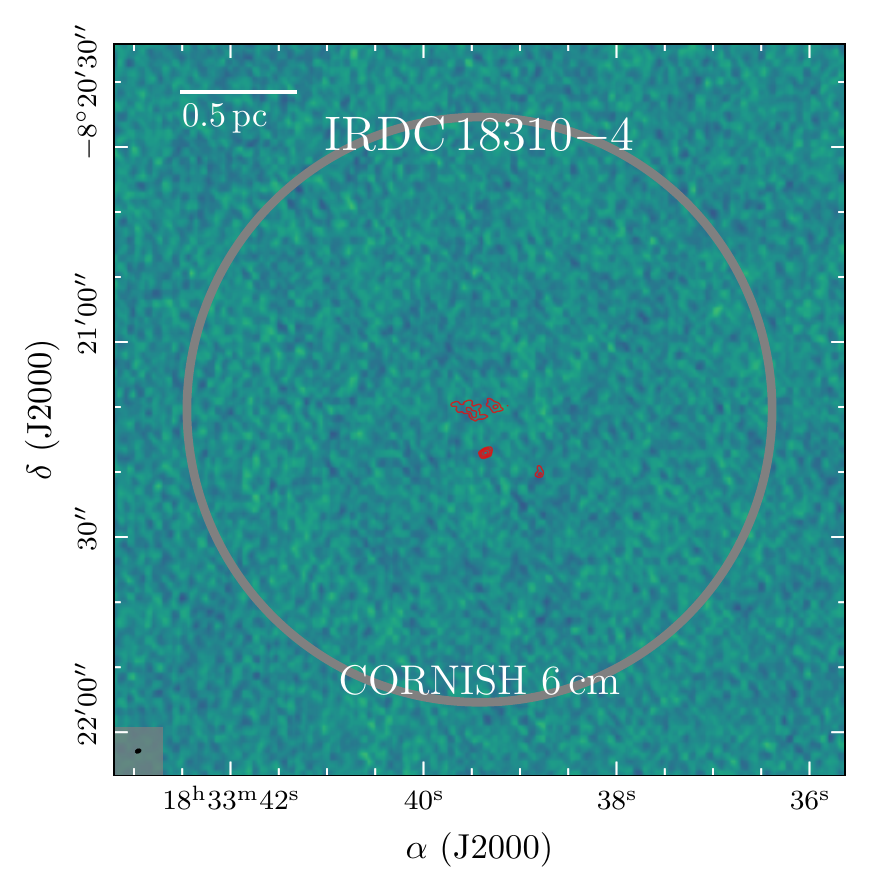}
\includegraphics[width=0.39\textwidth]{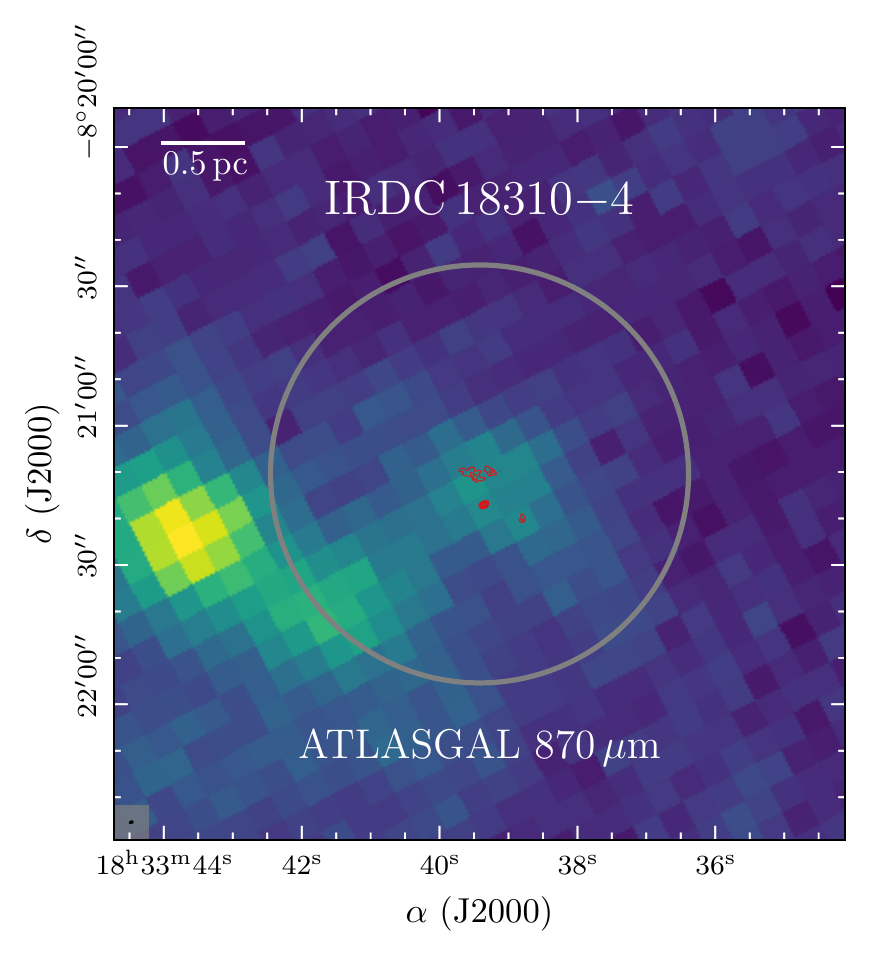}\\
\includegraphics[width=0.39\textwidth]{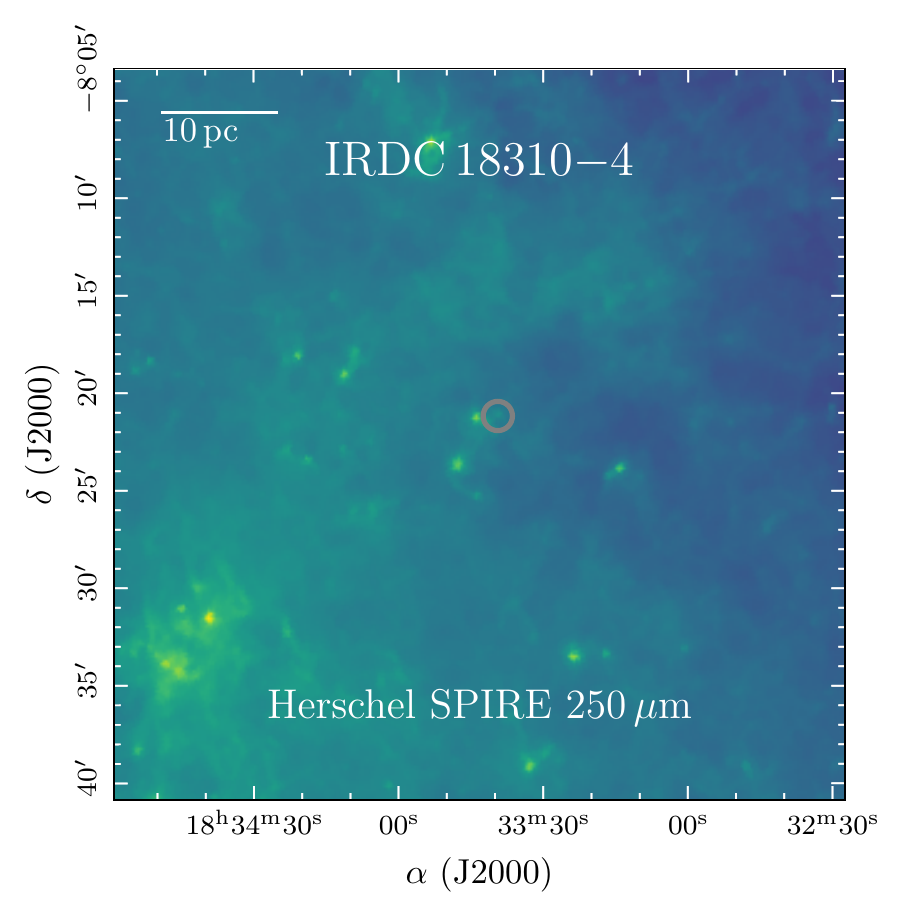}
\includegraphics[width=0.39\textwidth]{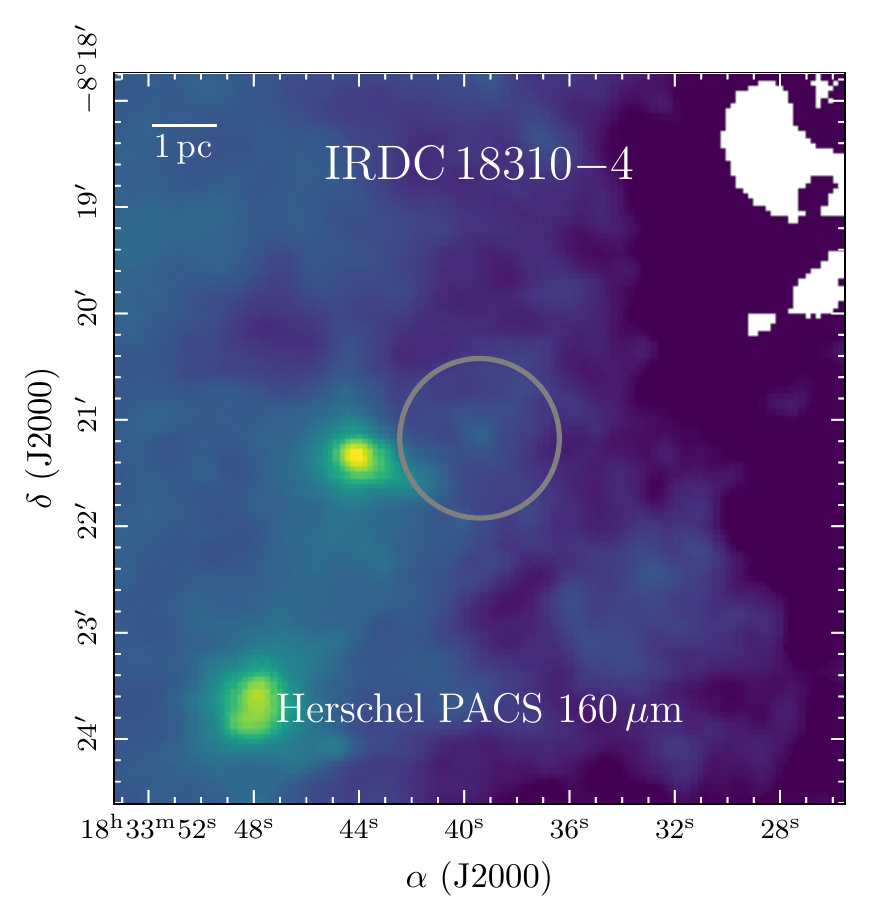}\\
\includegraphics[width=0.39\textwidth]{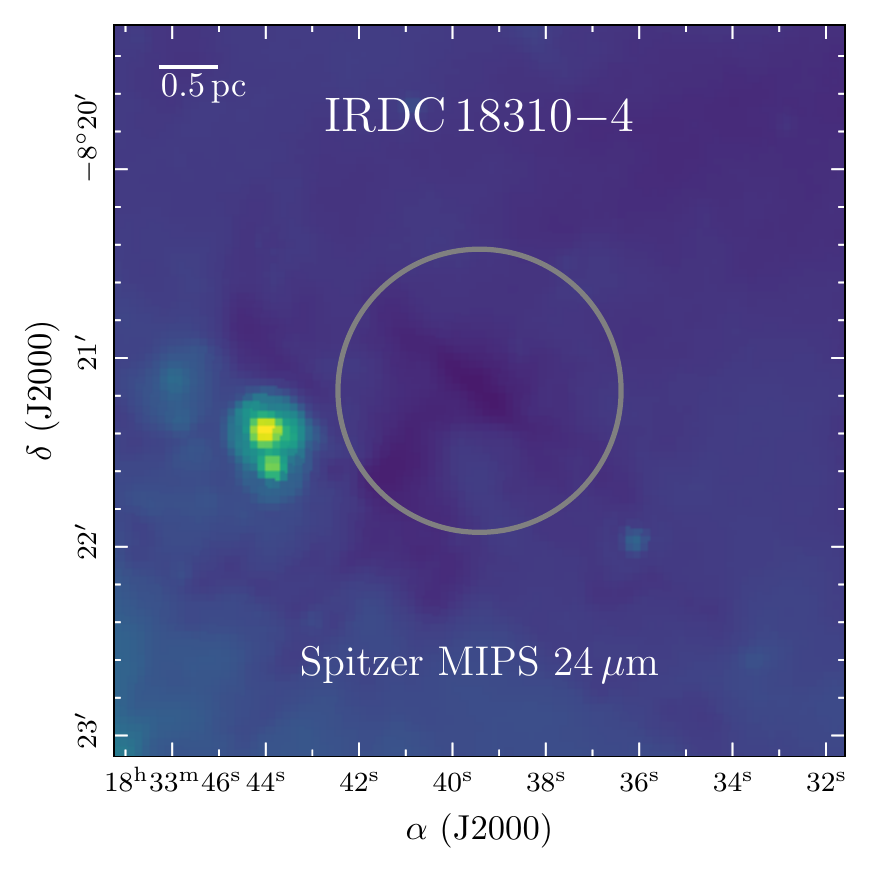}
\includegraphics[width=0.39\textwidth]{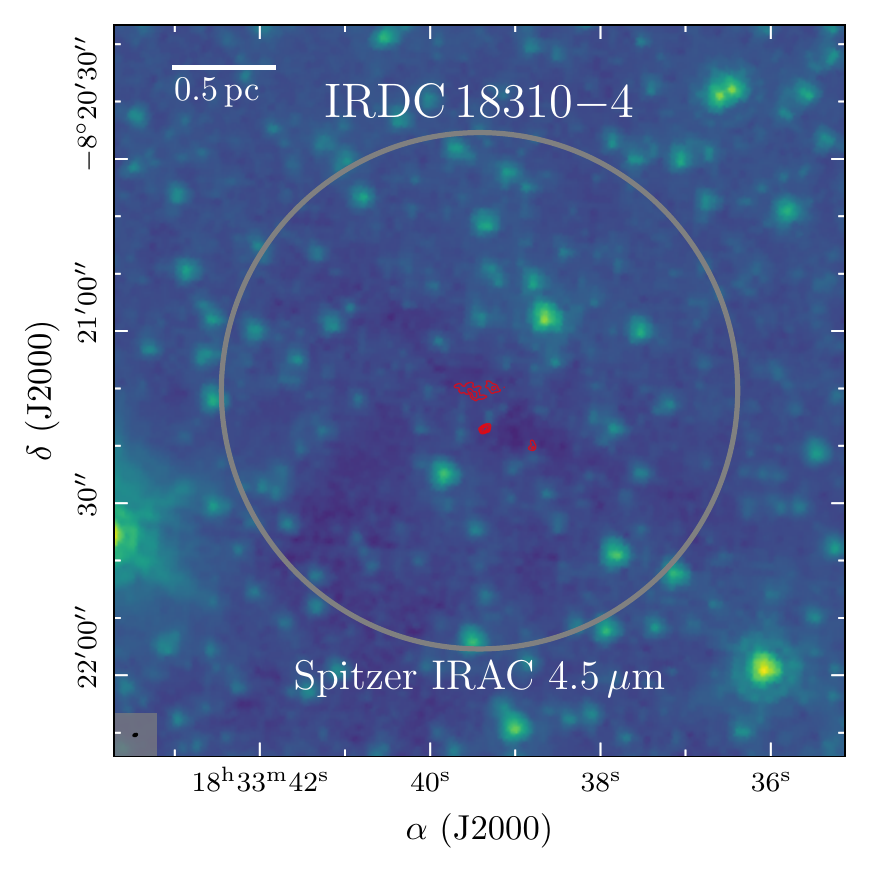}
\caption[Overview of IRDC\,18310$-$4.]{The same as Fig. \ref{fig:overview_IRDC_G1111}, but for IRDC\,18310$-$4.}
\label{fig:overview_IRDC_IRAS18310}
\end{figure*}

\begin{figure*}
\centering
\includegraphics[width=0.39\textwidth]{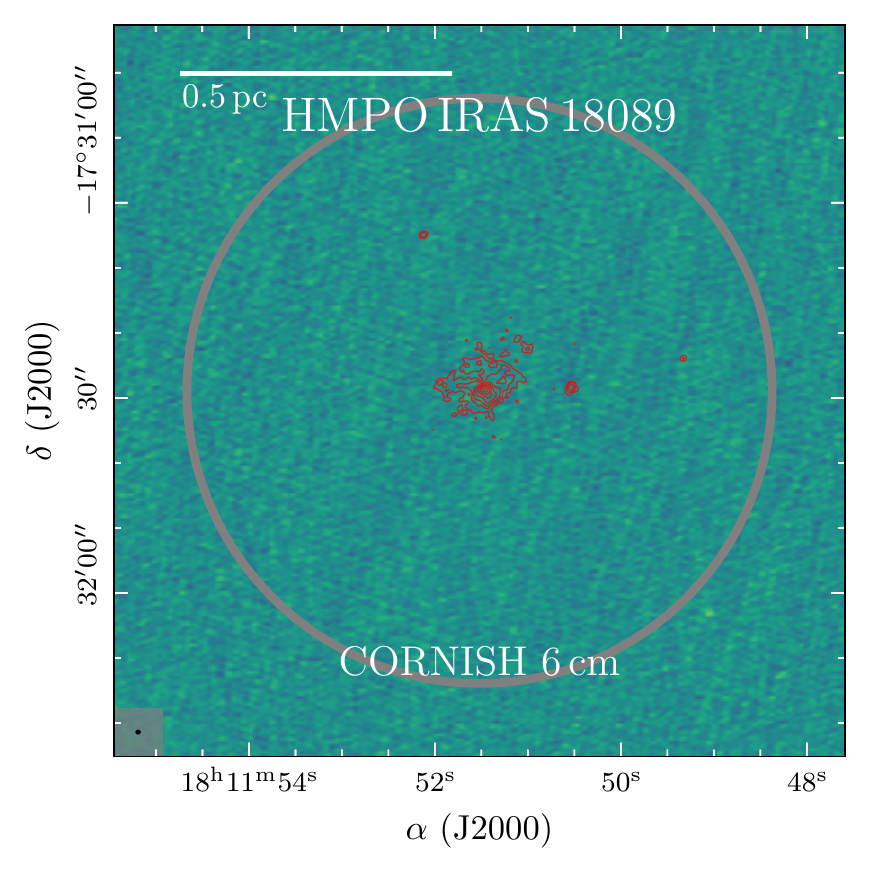}
\includegraphics[width=0.39\textwidth]{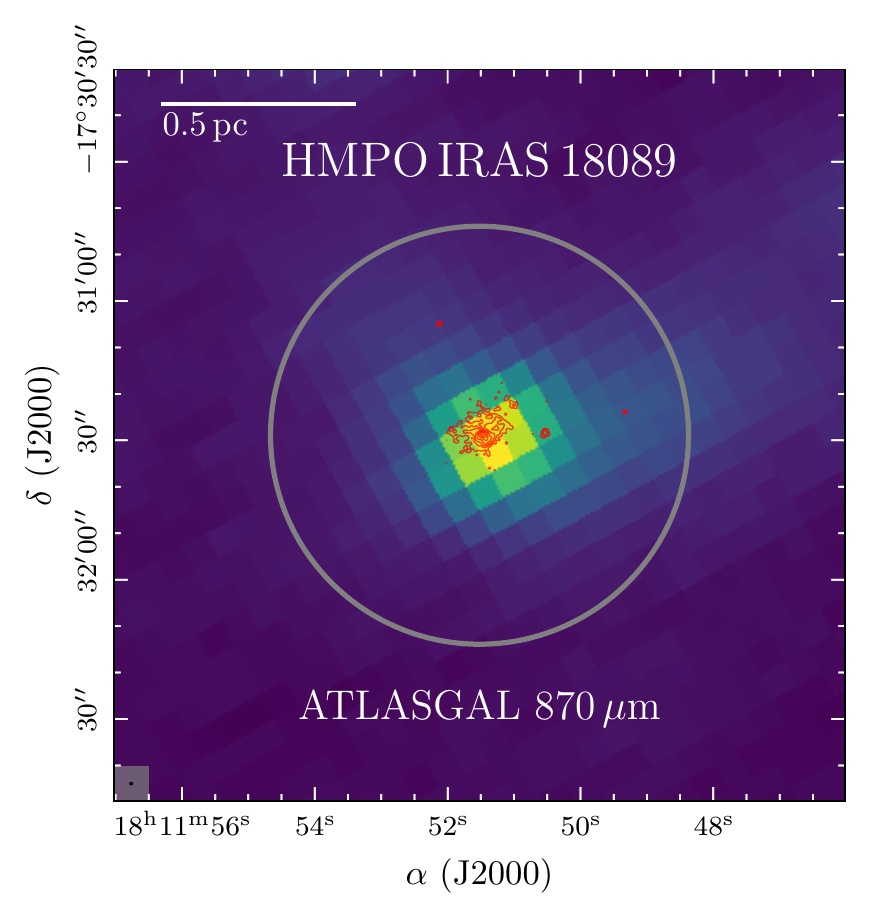}\\
\includegraphics[width=0.39\textwidth]{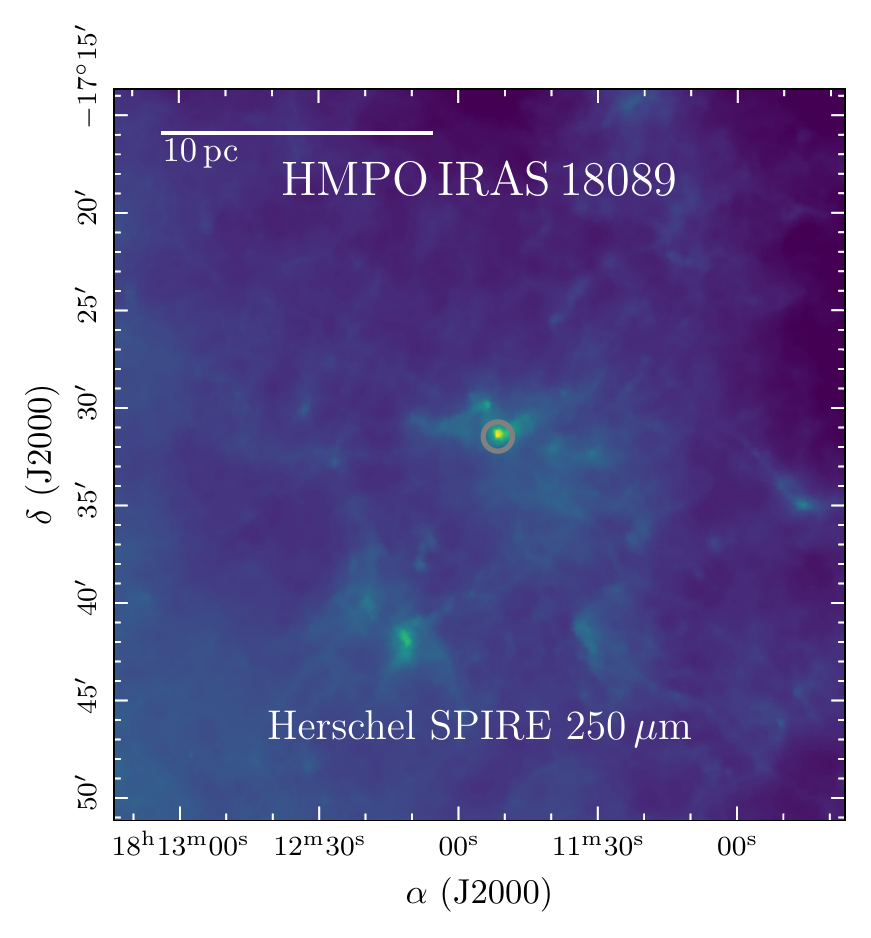}
\includegraphics[width=0.39\textwidth]{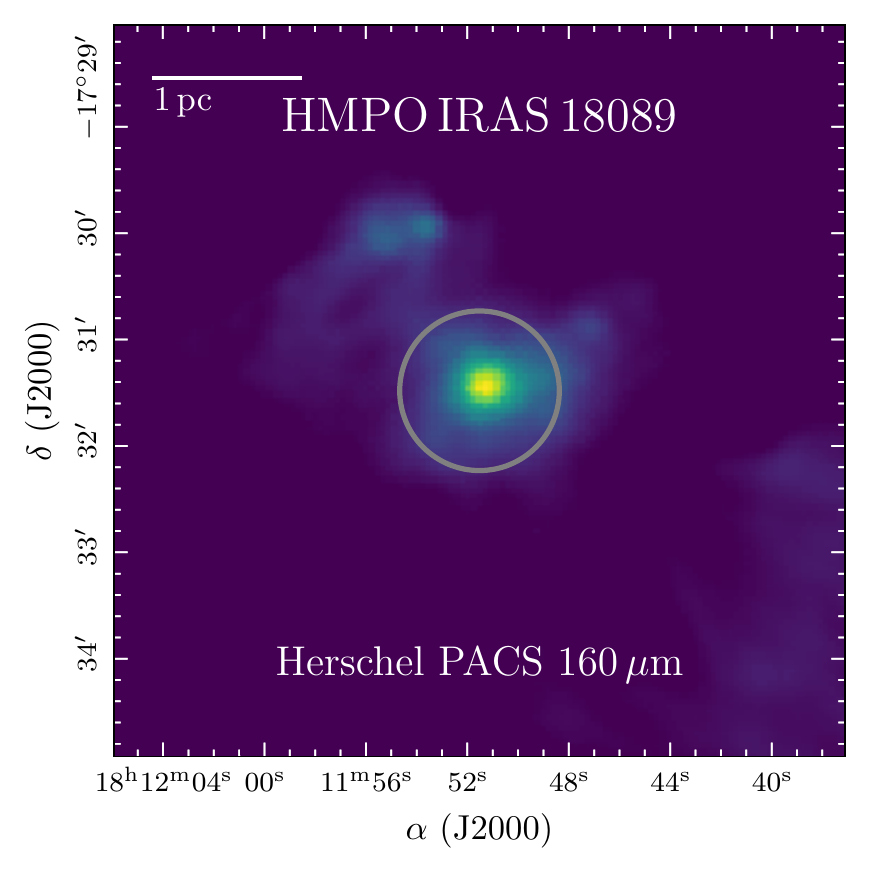}\\
\includegraphics[width=0.39\textwidth]{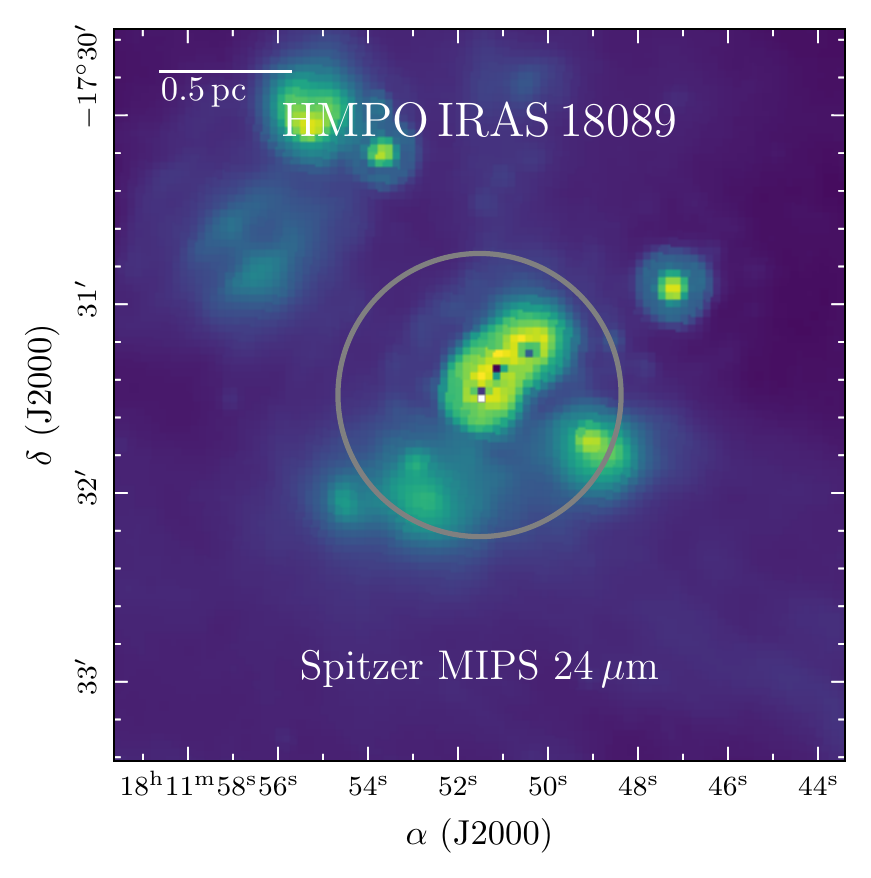}
\includegraphics[width=0.39\textwidth]{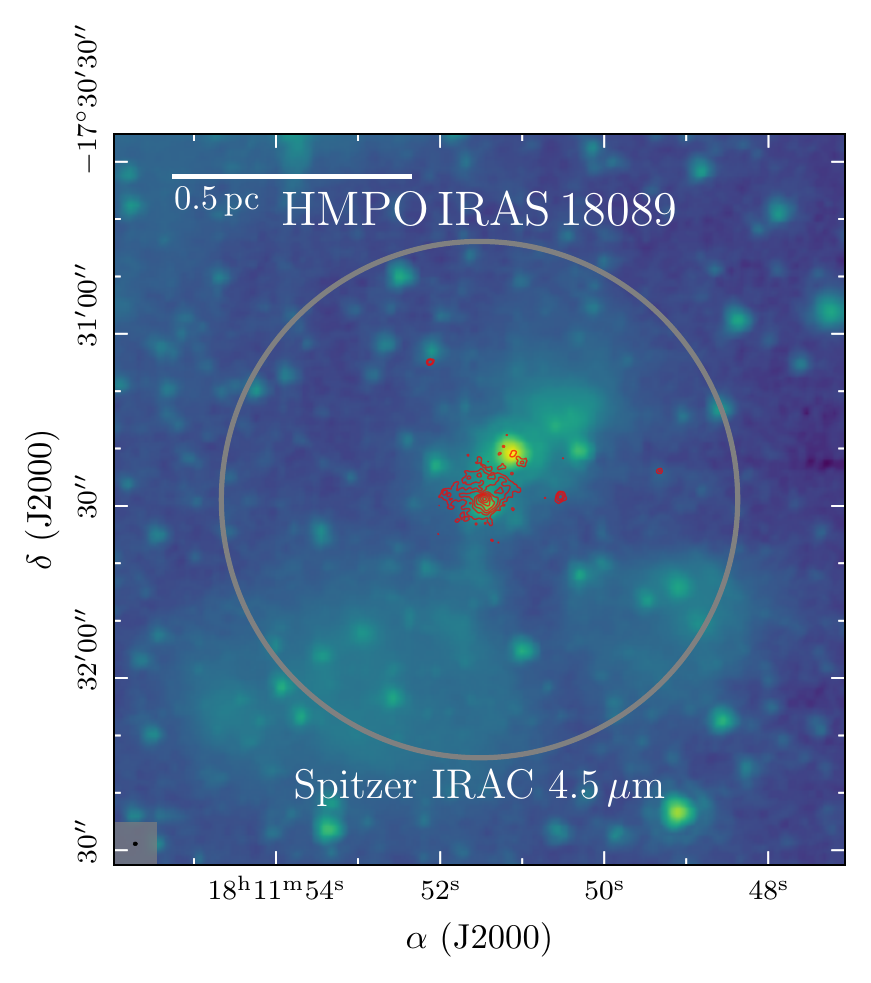}
\caption[Overview of HMPO\,IRAS\,18089.]{The same as Fig. \ref{fig:overview_IRDC_G1111}, but for HMPO\,IRAS\,18089.}
\label{fig:overview_HMPO_IRAS18089}
\end{figure*}

\begin{figure*}
\centering
\includegraphics[width=0.39\textwidth]{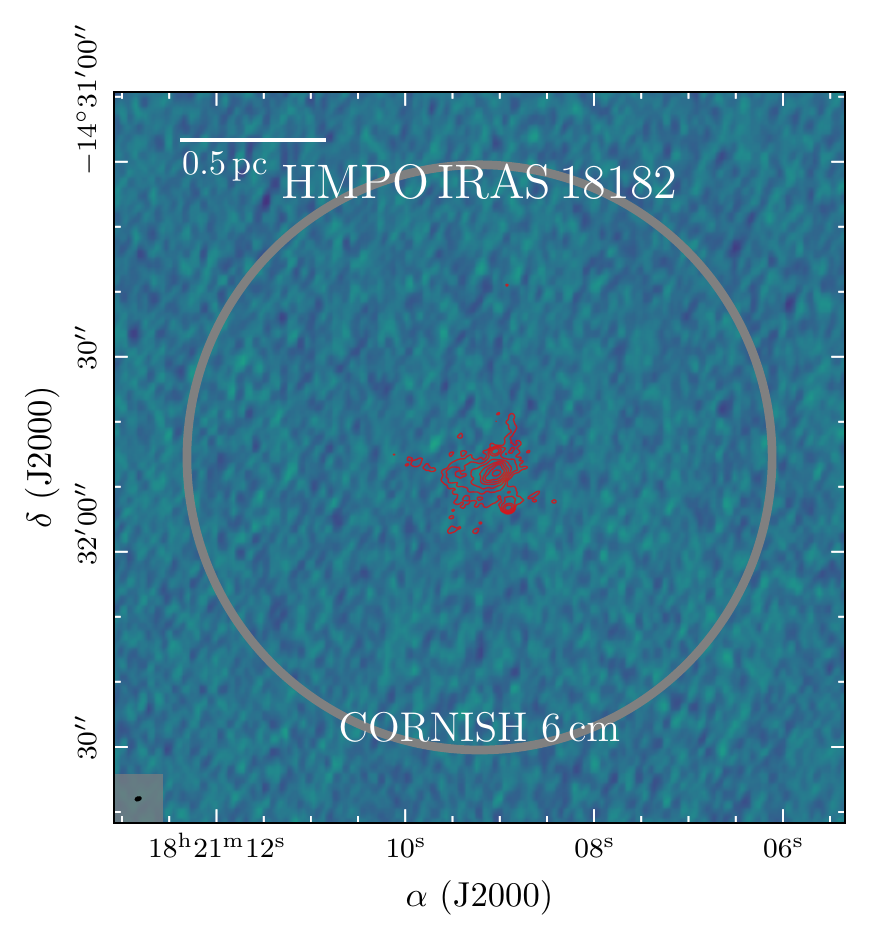}
\includegraphics[width=0.39\textwidth]{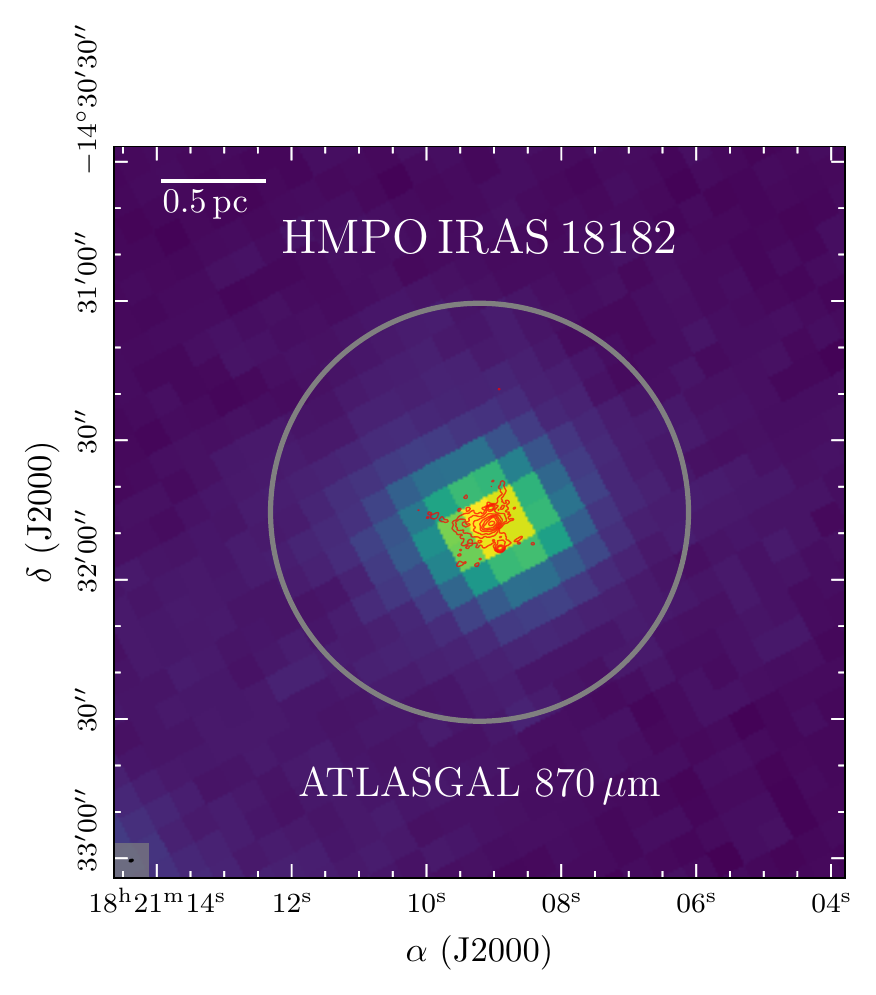}\\
\includegraphics[width=0.39\textwidth]{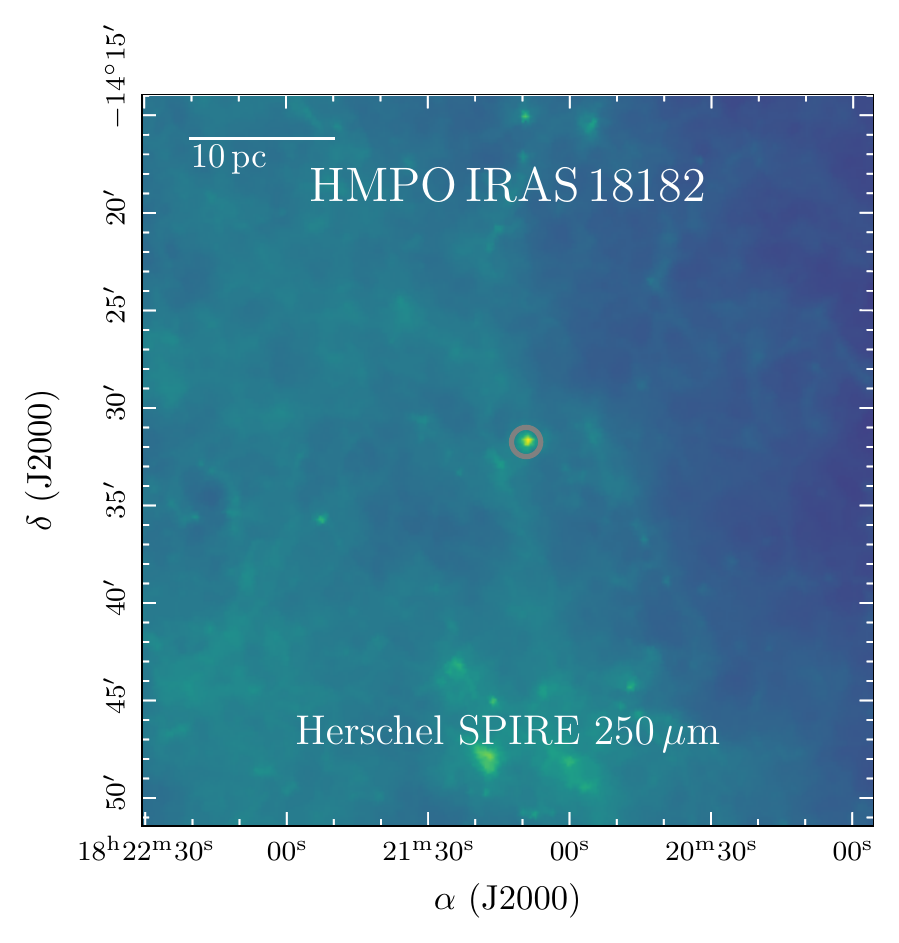}
\includegraphics[width=0.39\textwidth]{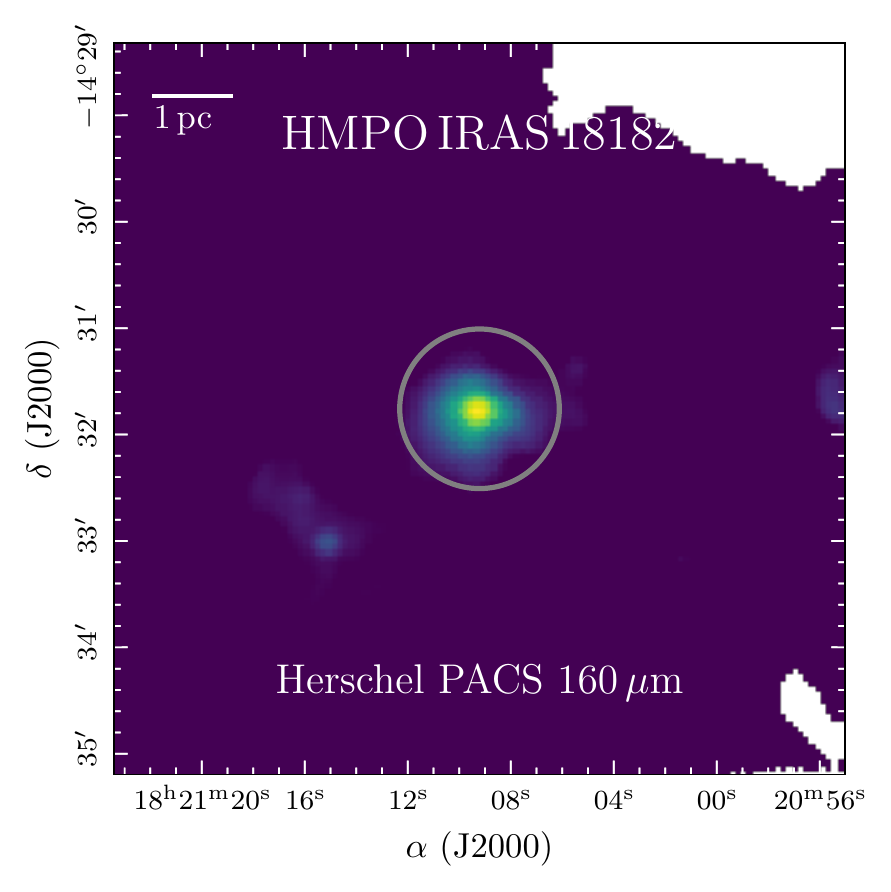}\\
\includegraphics[width=0.39\textwidth]{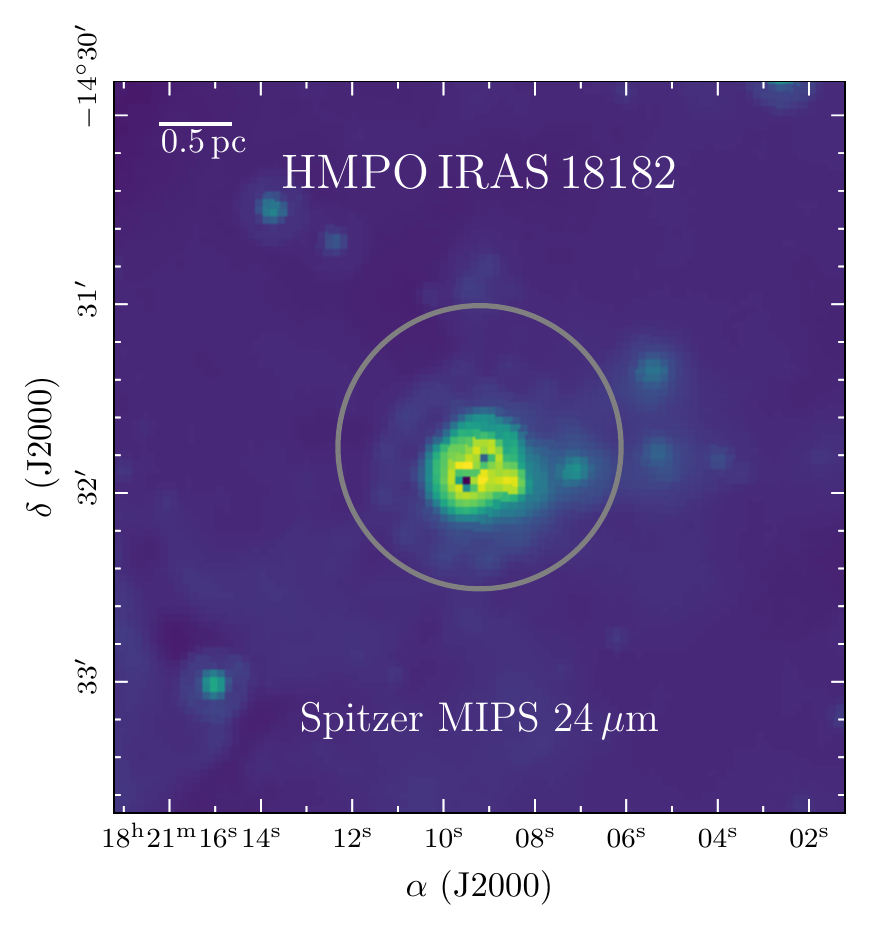}
\includegraphics[width=0.39\textwidth]{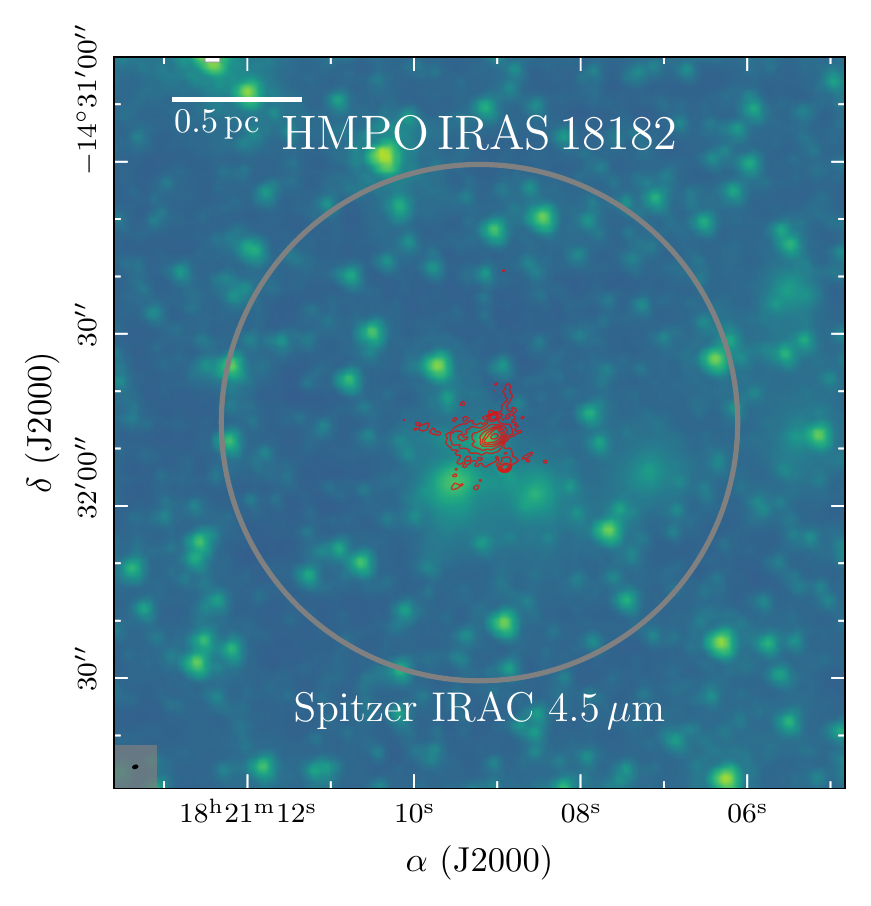}
\caption[Overview of HMPO\,IRAS\,18182.]{The same as Fig. \ref{fig:overview_IRDC_G1111}, but for HMPO\,IRAS\,18182.}
\label{fig:overview_HMPO_IRAS18182}
\end{figure*}

\begin{figure*}
\centering
\includegraphics[width=0.39\textwidth]{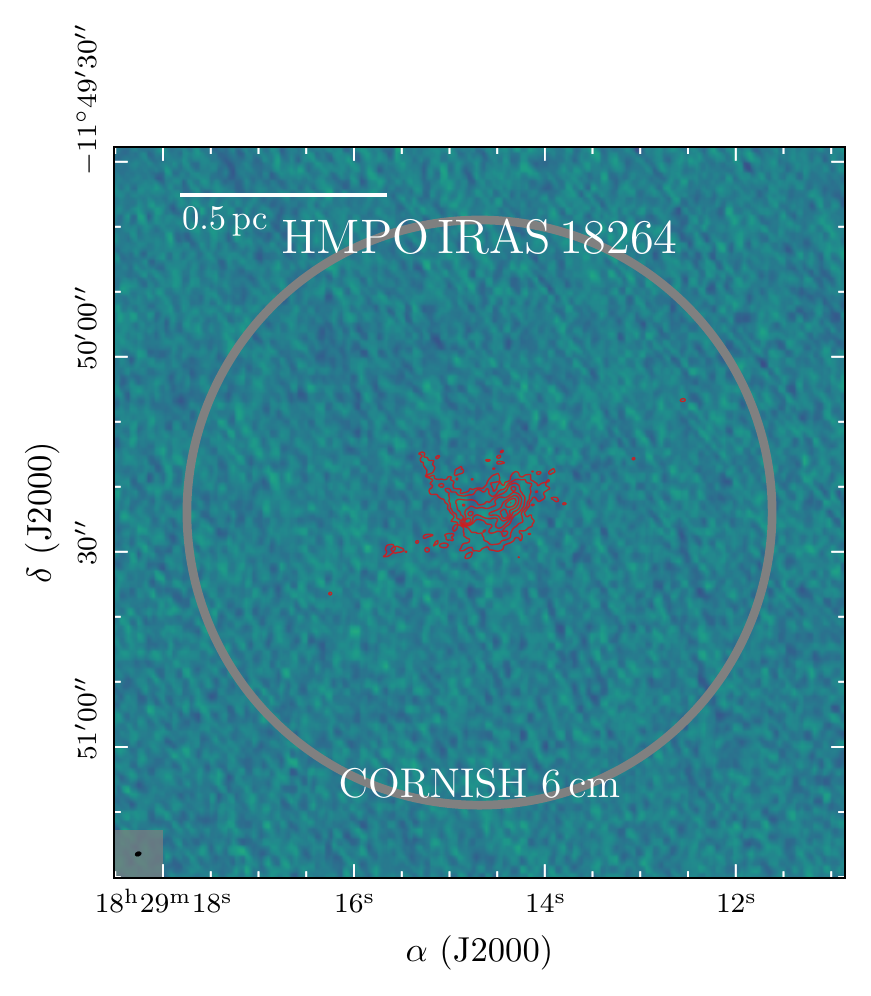}
\includegraphics[width=0.39\textwidth]{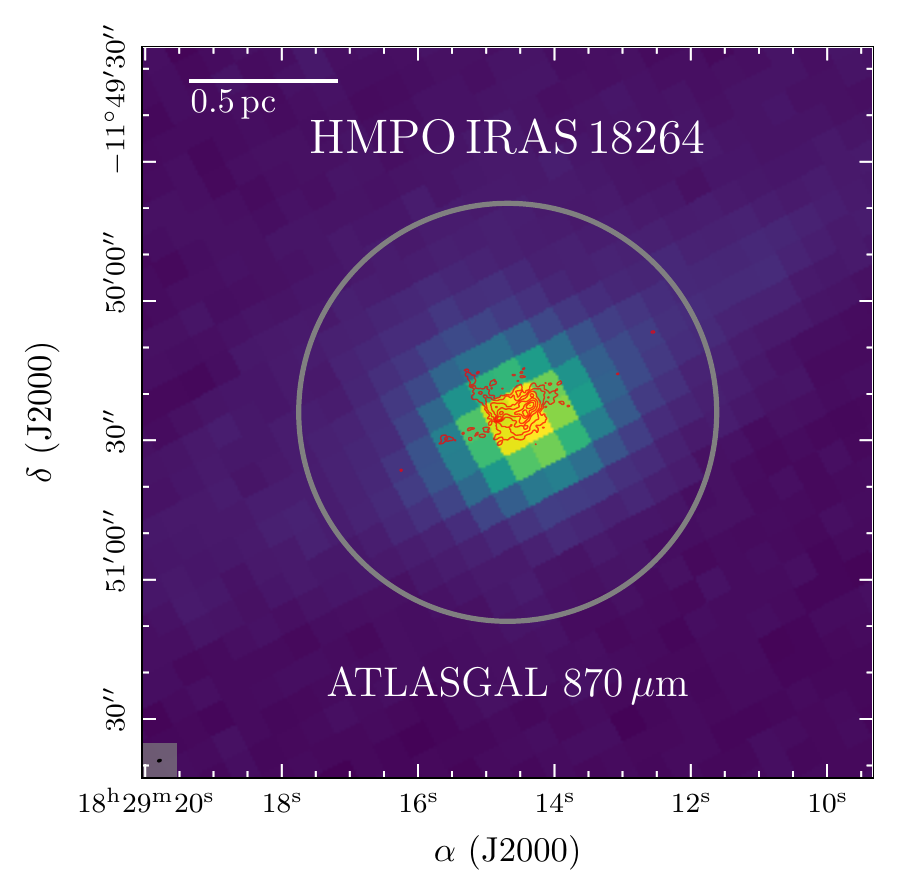}\\
\includegraphics[width=0.39\textwidth]{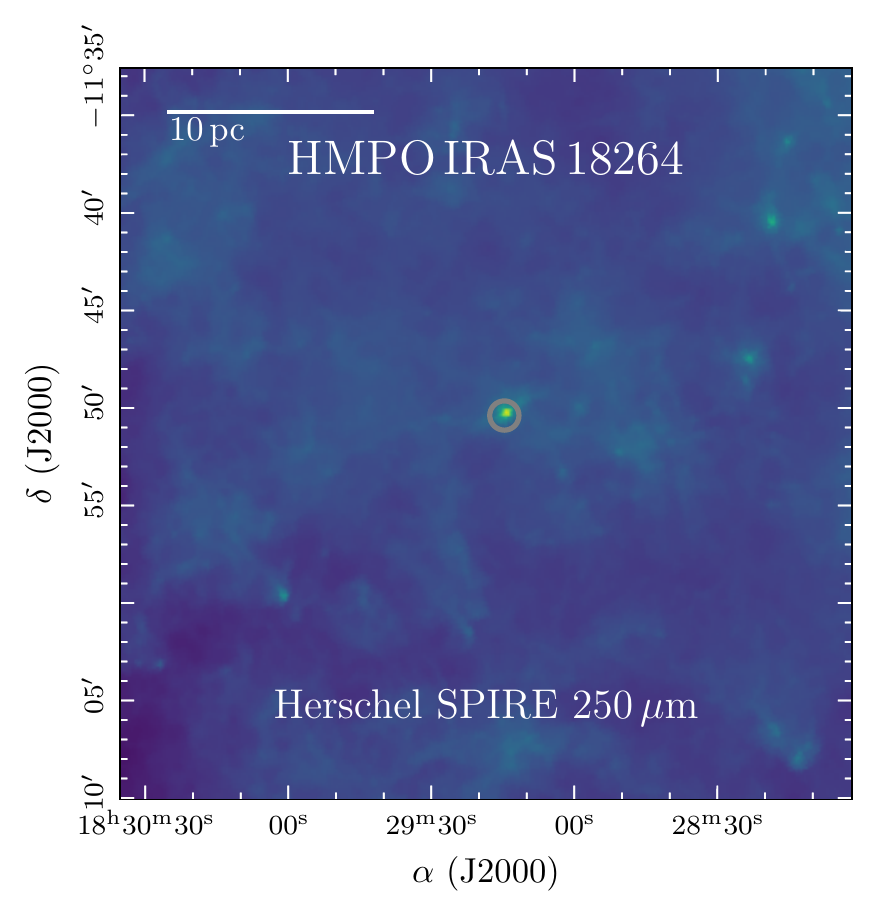}
\includegraphics[width=0.39\textwidth]{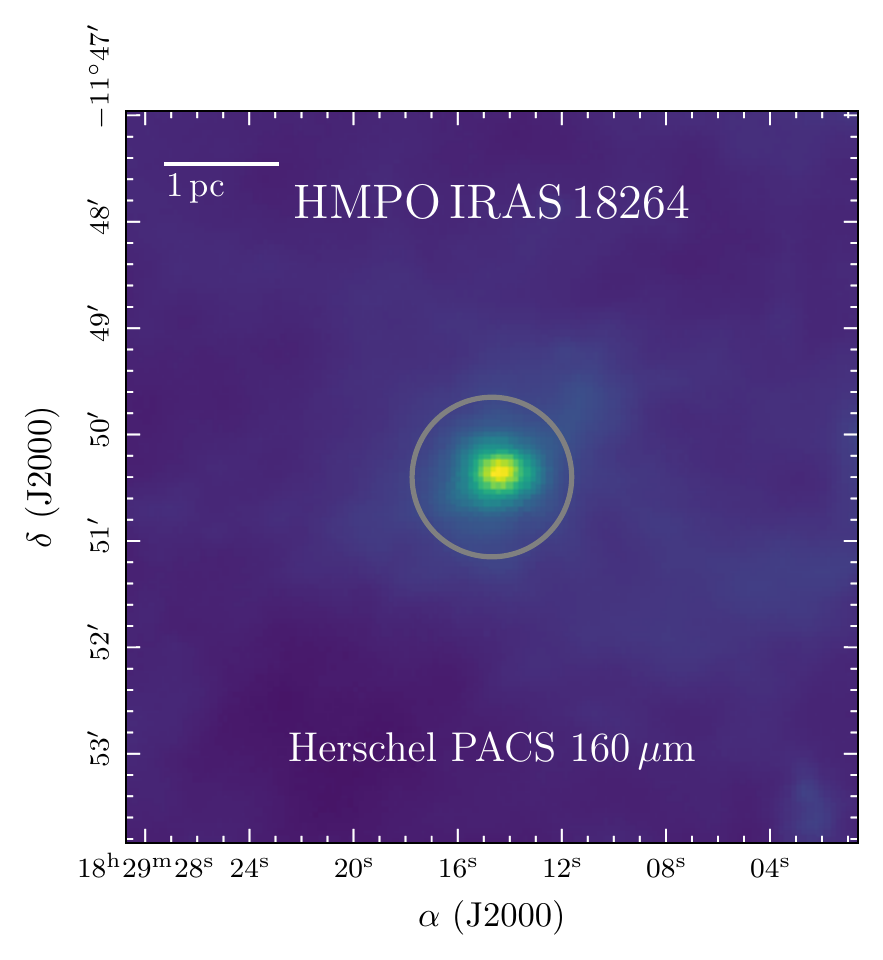}\\
\includegraphics[width=0.39\textwidth]{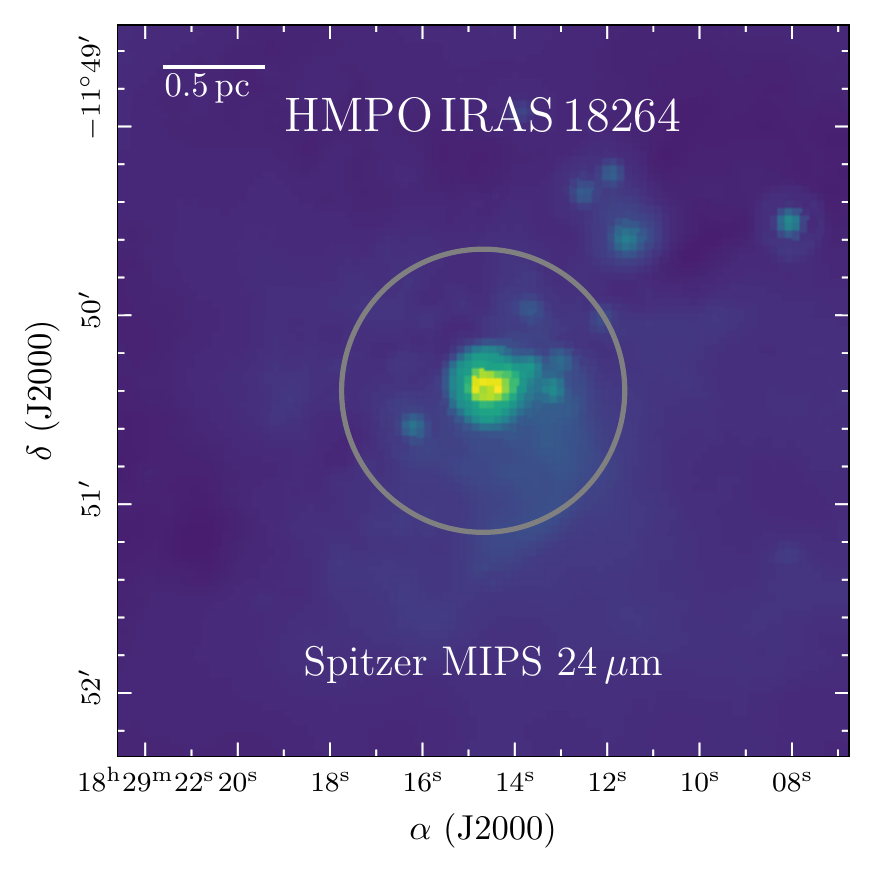}
\includegraphics[width=0.39\textwidth]{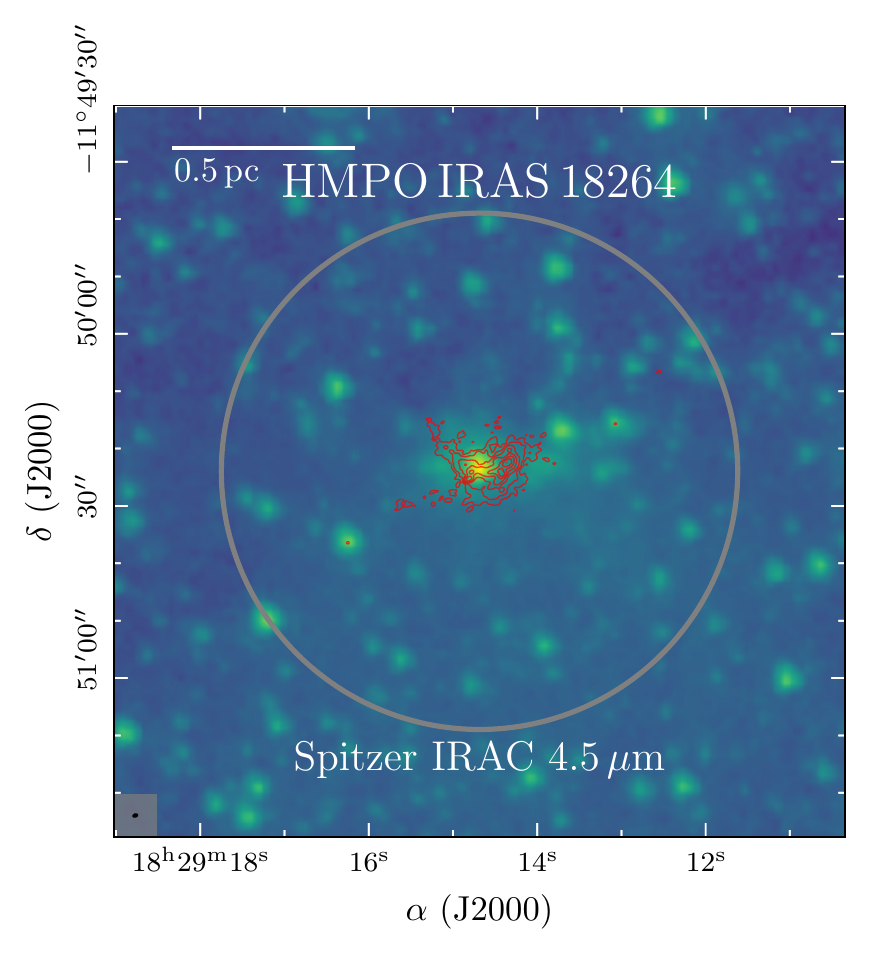}
\caption[Overview of HMPO\,IRAS\,18264.]{The same as Fig. \ref{fig:overview_IRDC_G1111}, but for HMPO\,IRAS\,18264.}
\label{fig:overview_HMPO_IRAS18264}
\end{figure*}

\begin{figure*}
\centering
\includegraphics[width=0.39\textwidth]{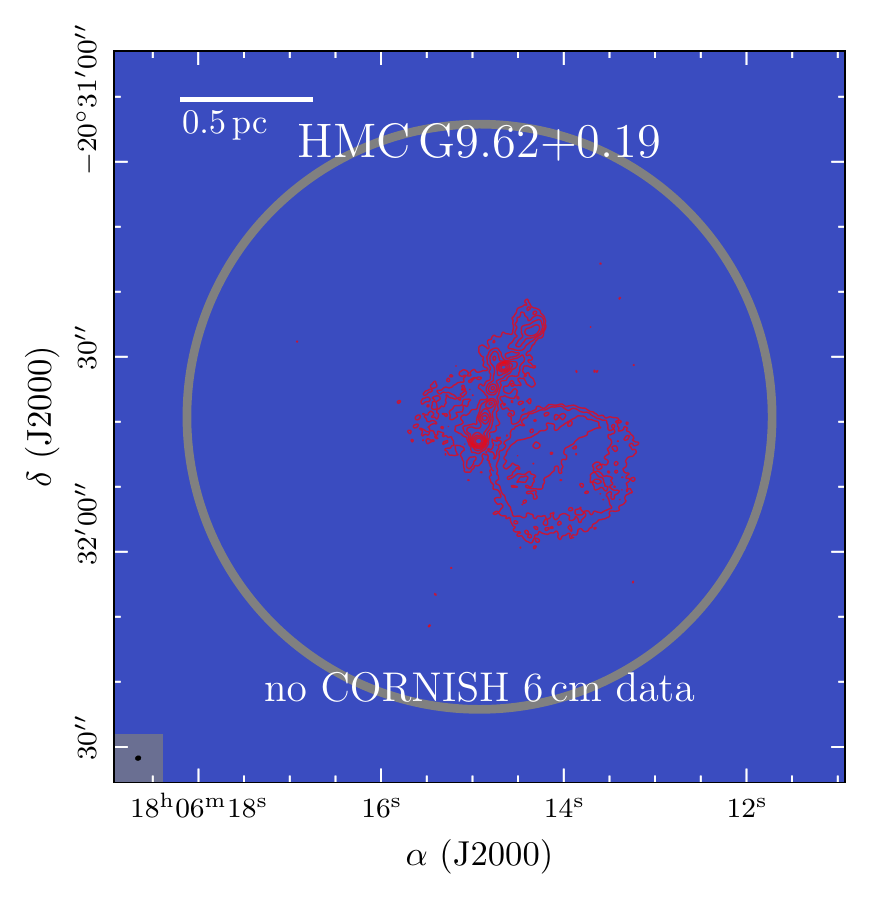}
\includegraphics[width=0.39\textwidth]{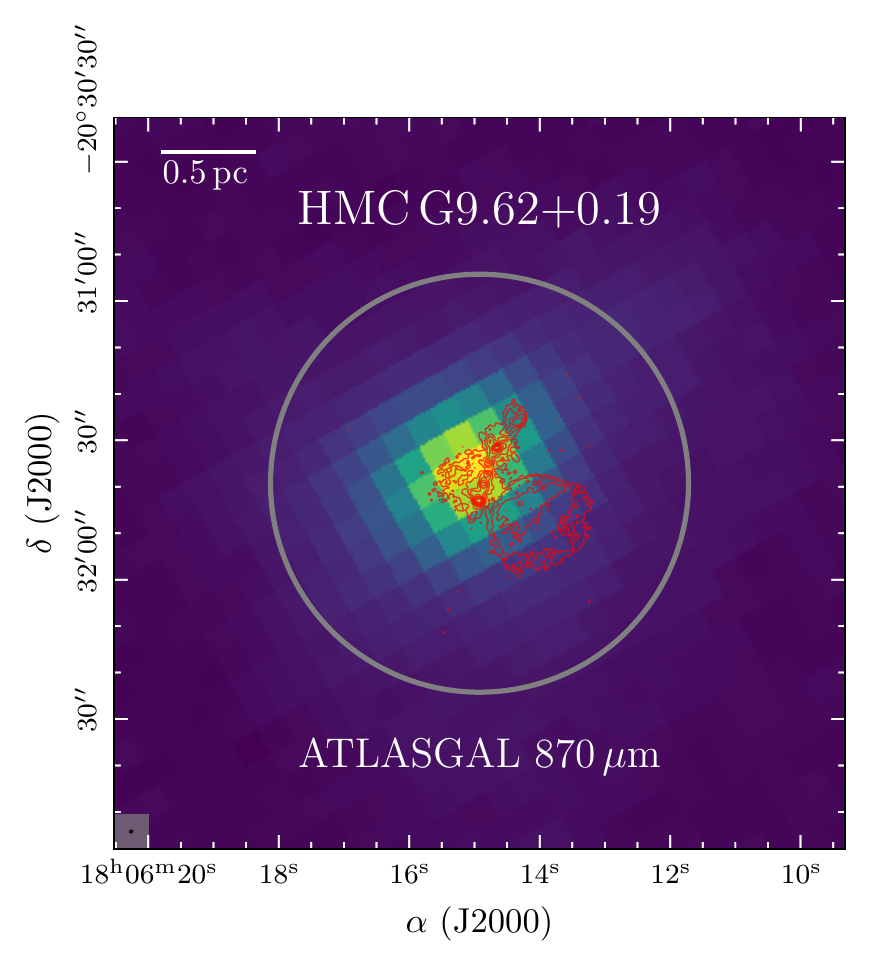}\\
\includegraphics[width=0.39\textwidth]{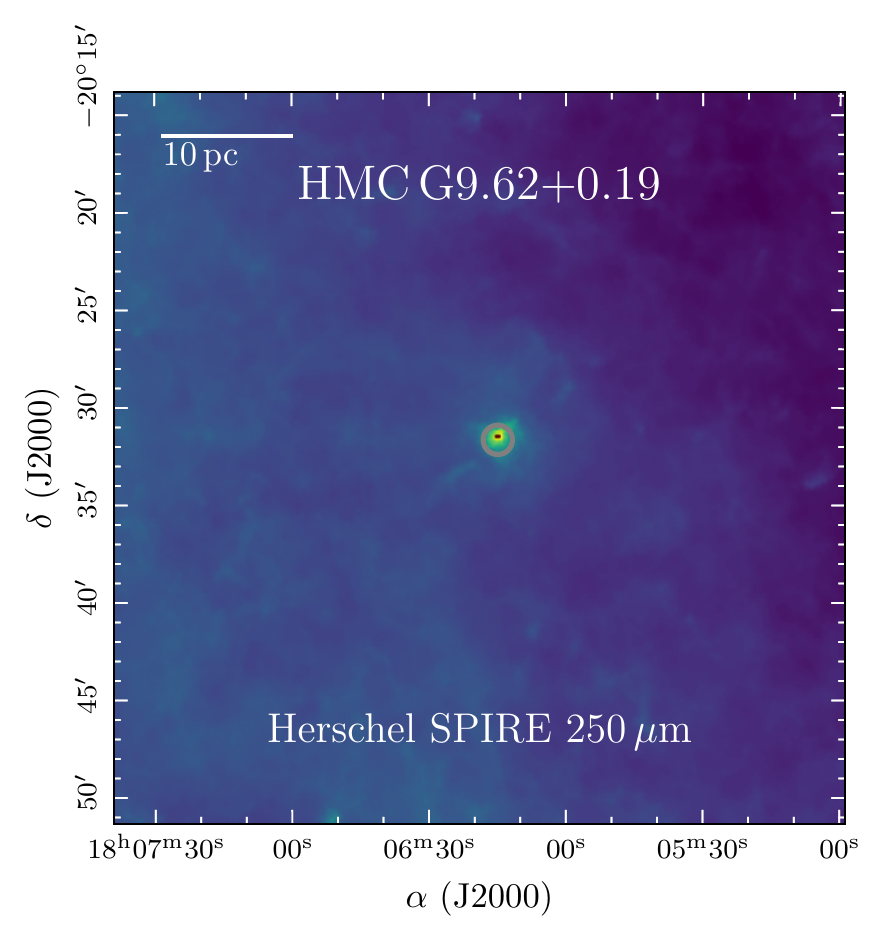}
\includegraphics[width=0.39\textwidth]{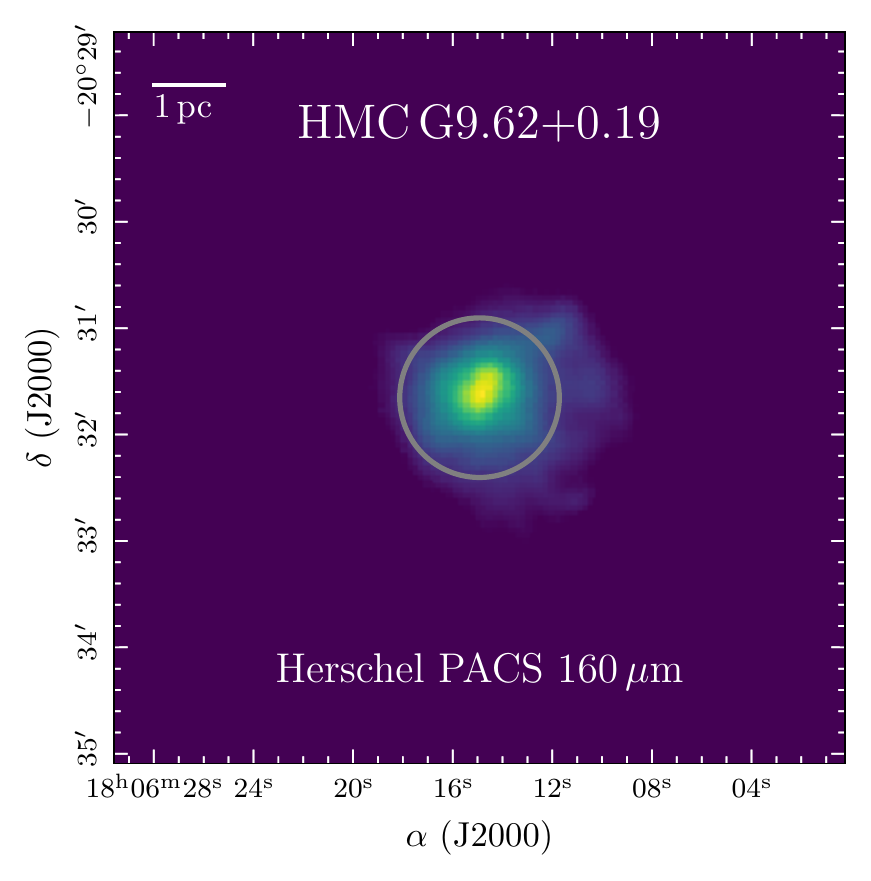}\\
\includegraphics[width=0.39\textwidth]{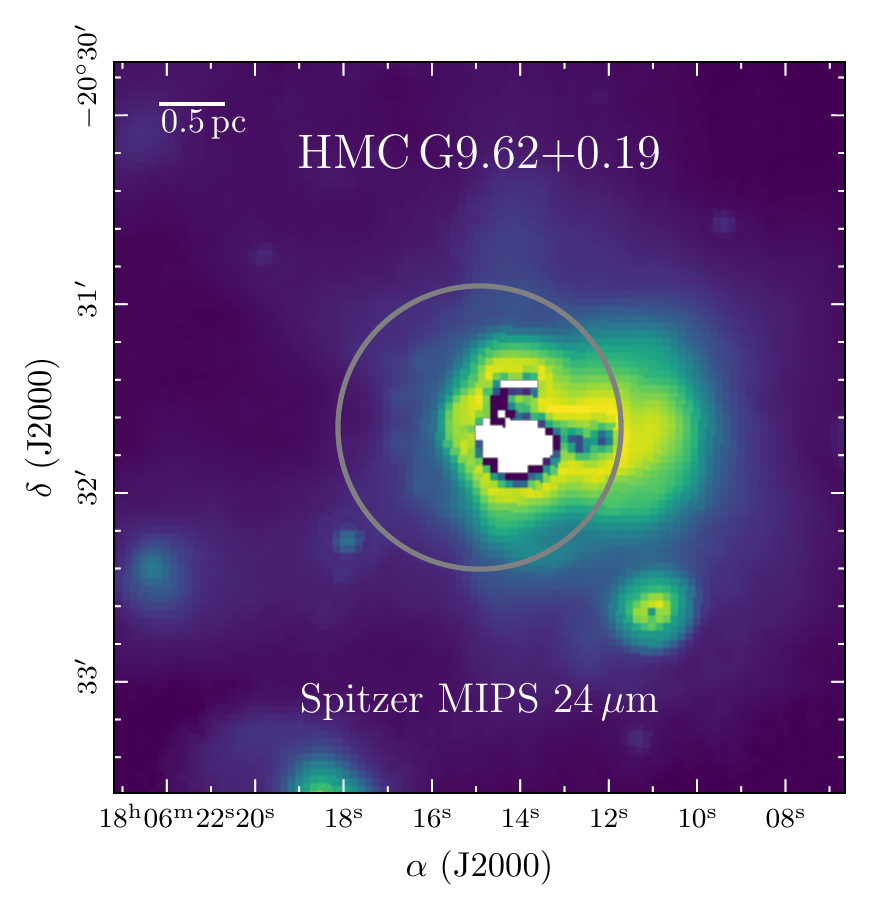}
\includegraphics[width=0.39\textwidth]{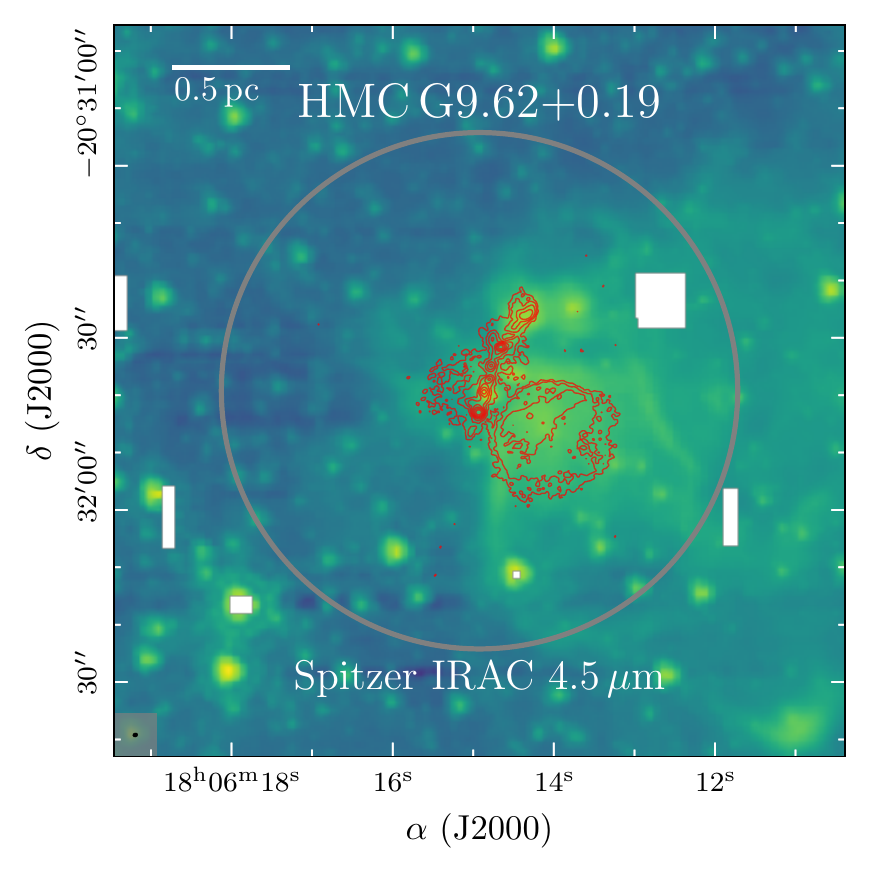}
\caption[Overview of HMC\,G9.62$+$0.19.]{The same as Fig. \ref{fig:overview_IRDC_G1111}, but for HMC\,G9.62$+$0.19. The region was not covered by the CORNISH survey.}
\label{fig:overview_HMC_G0962}
\end{figure*}

\begin{figure*}
\centering
\includegraphics[width=0.39\textwidth]{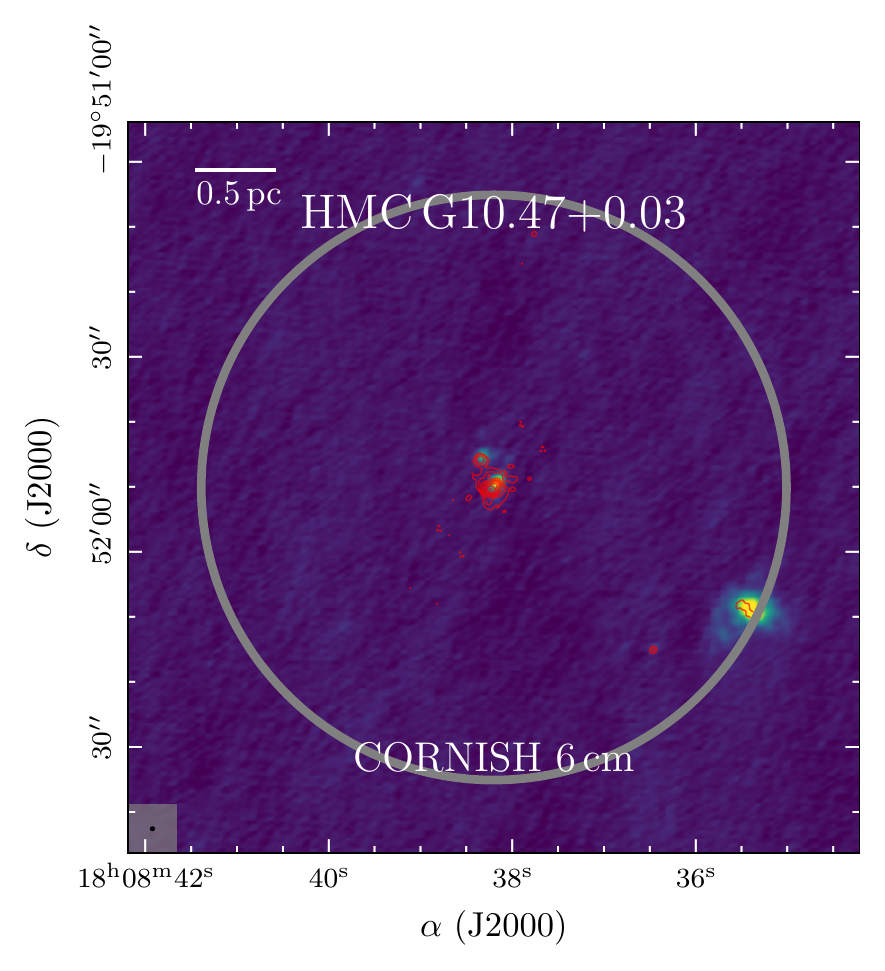}
\includegraphics[width=0.39\textwidth]{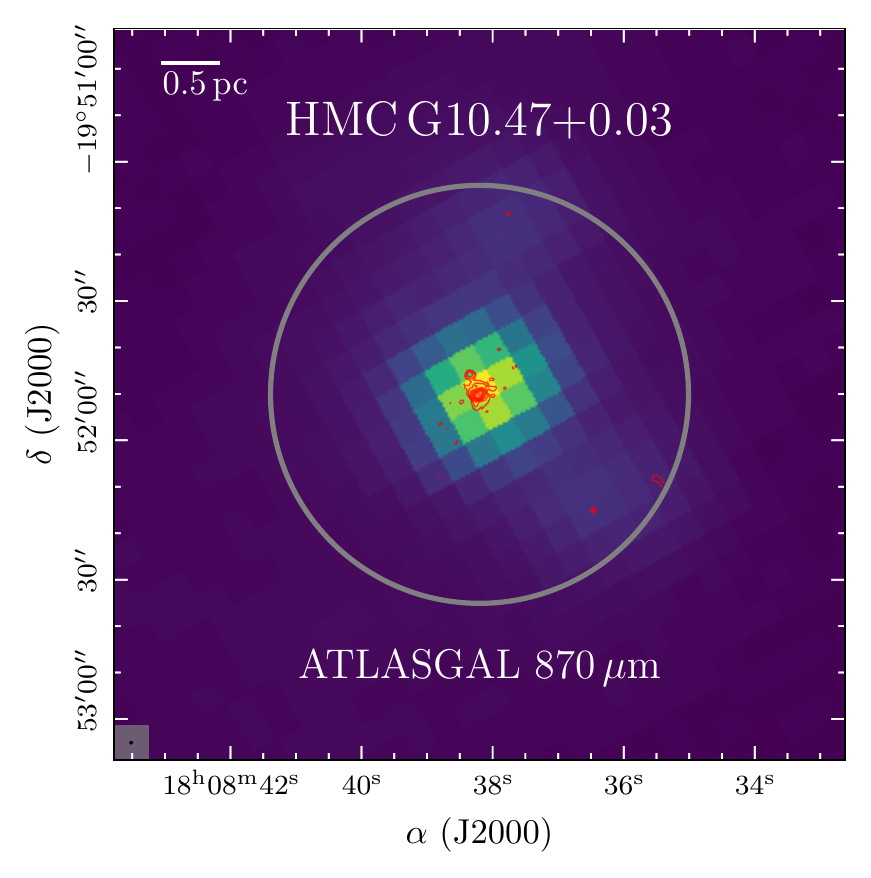}\\
\includegraphics[width=0.39\textwidth]{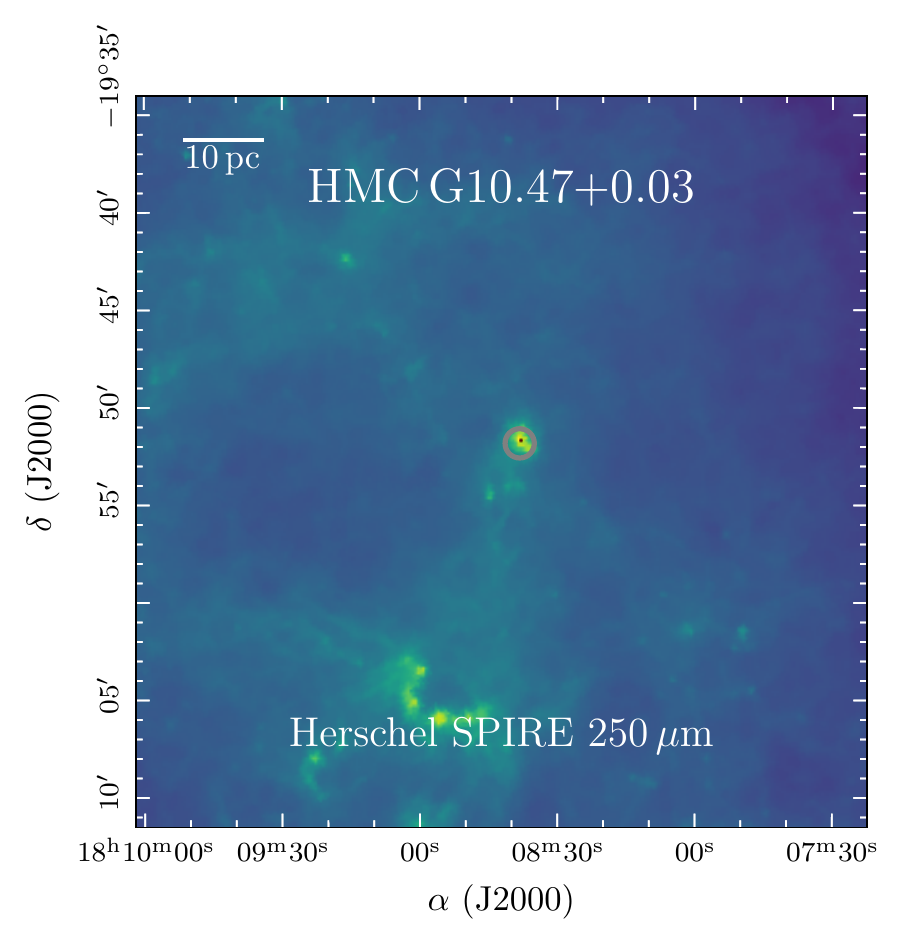}
\includegraphics[width=0.39\textwidth]{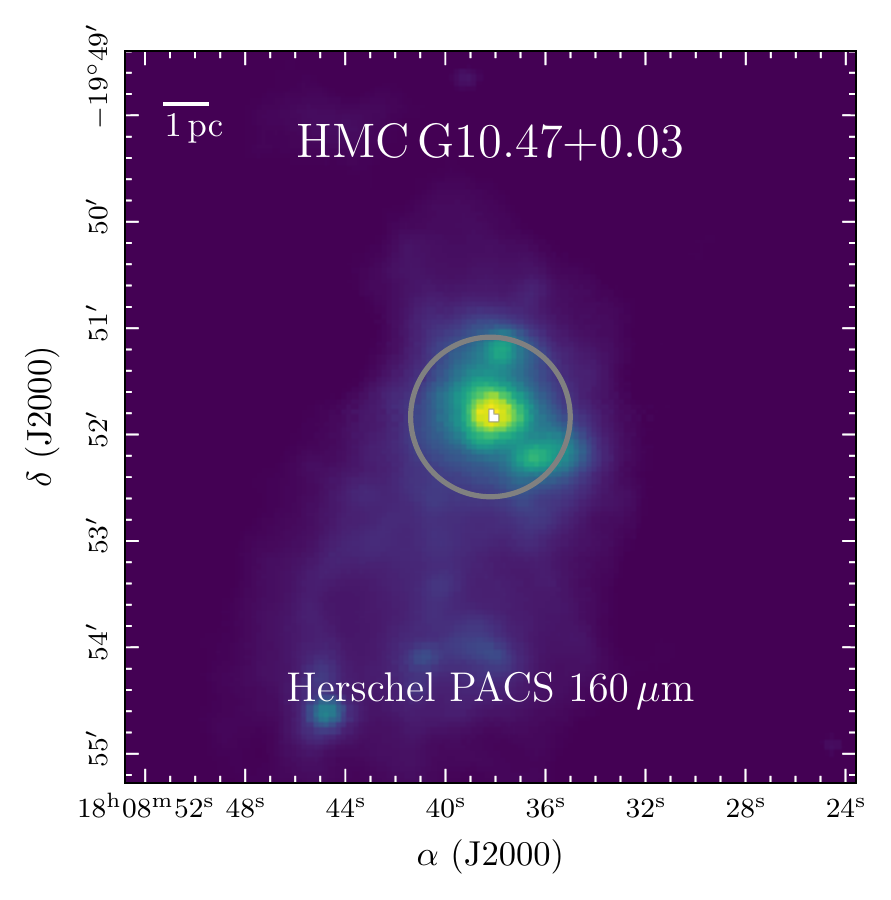}\\
\includegraphics[width=0.39\textwidth]{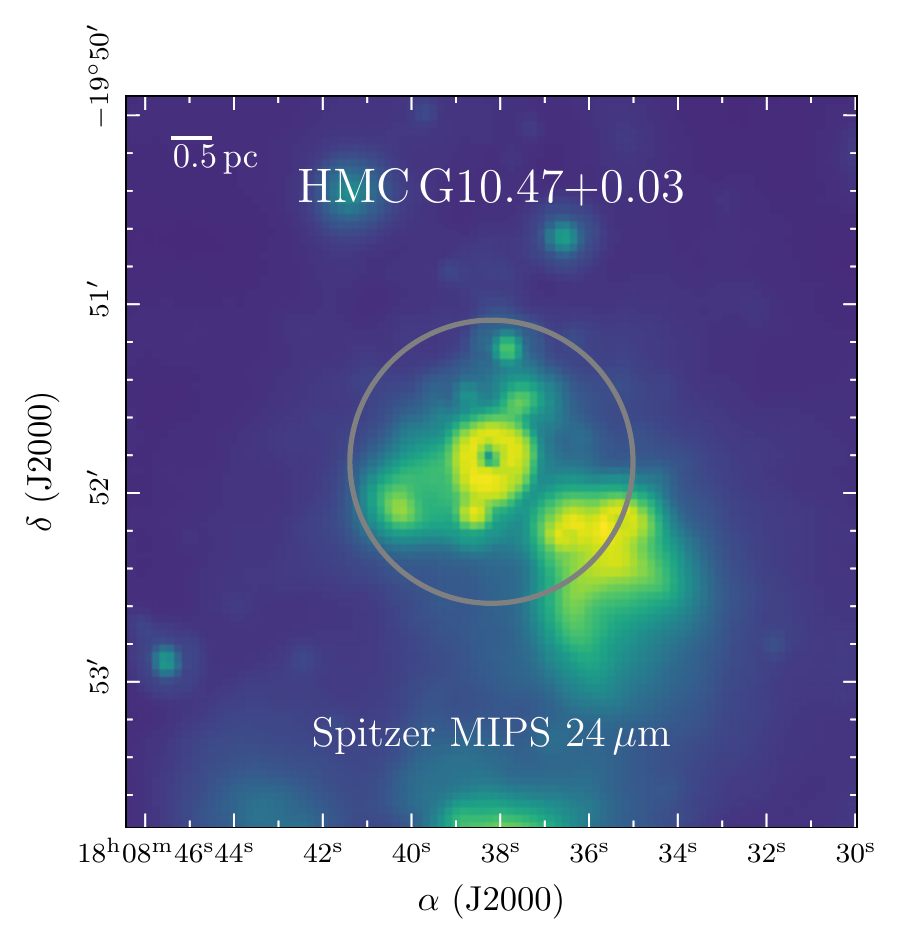}
\includegraphics[width=0.39\textwidth]{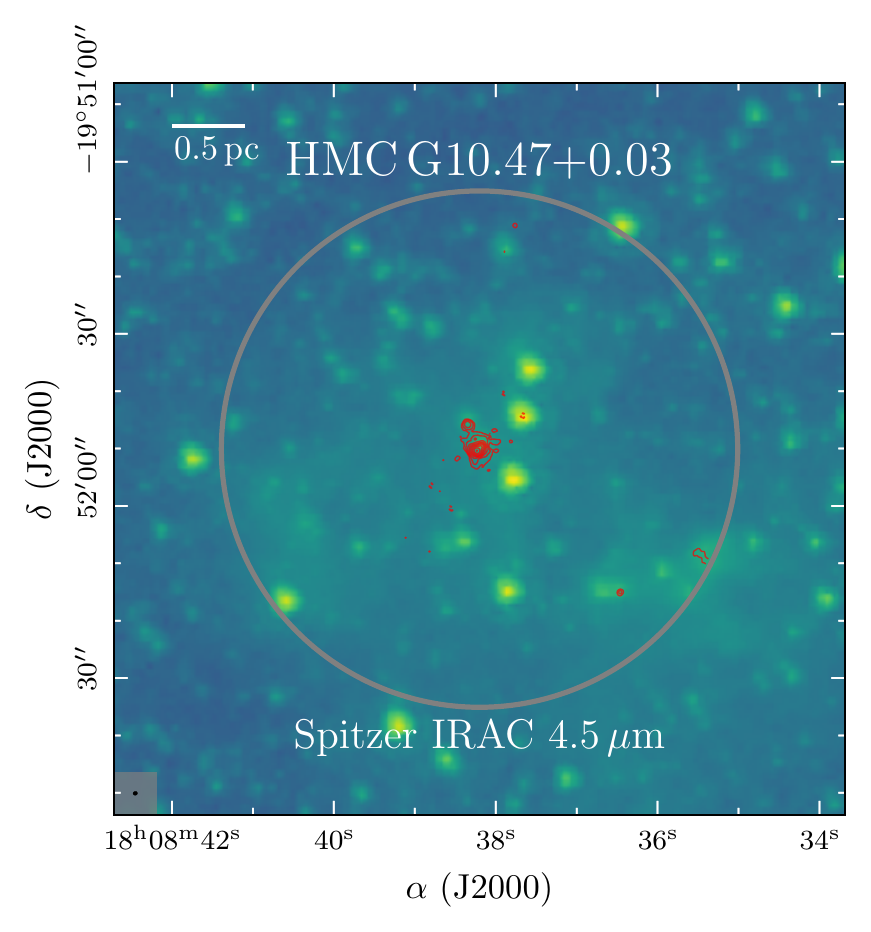}
\caption[Overview of HMC\,G10.47$+$0.03.]{The same as Fig. \ref{fig:overview_IRDC_G1111}, but for HMC\,G10.47$+$0.03.}
\label{fig:overview_HMC_G1047}
\end{figure*}

\begin{figure*}
\centering
\includegraphics[width=0.39\textwidth]{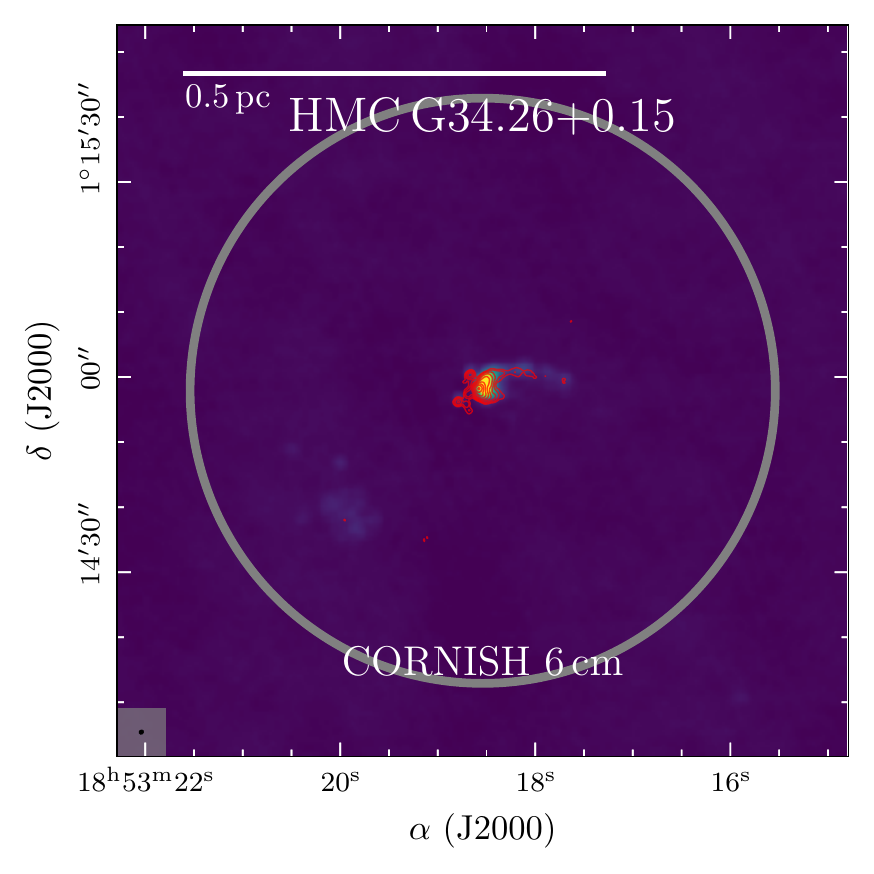}
\includegraphics[width=0.39\textwidth]{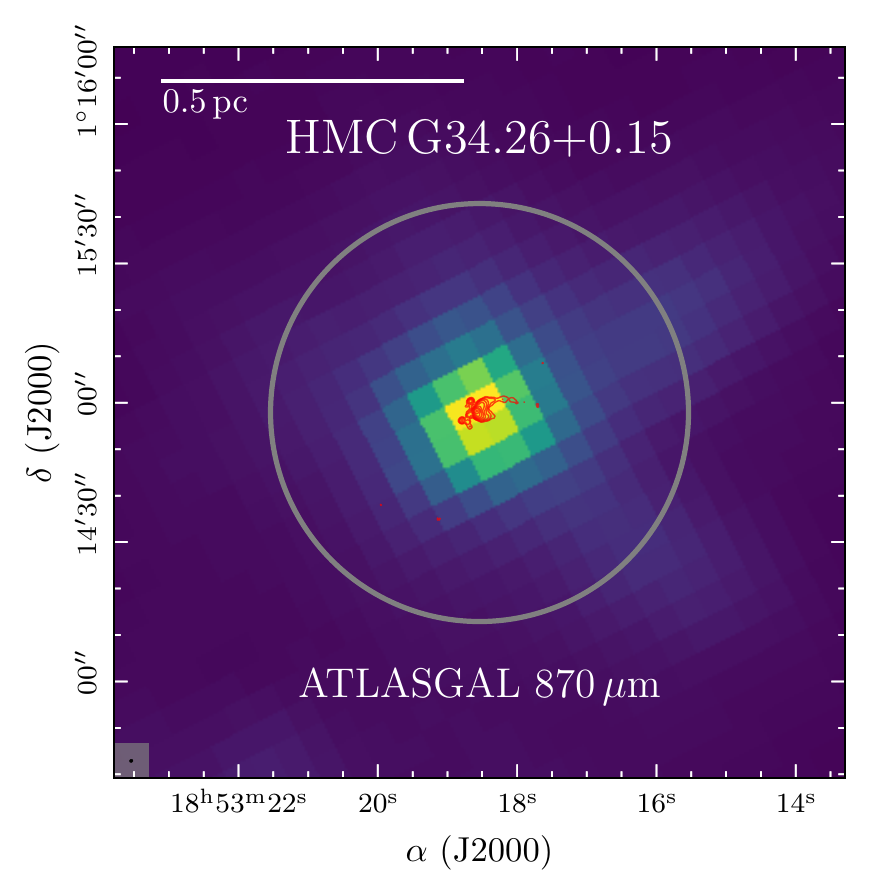}\\
\includegraphics[width=0.39\textwidth]{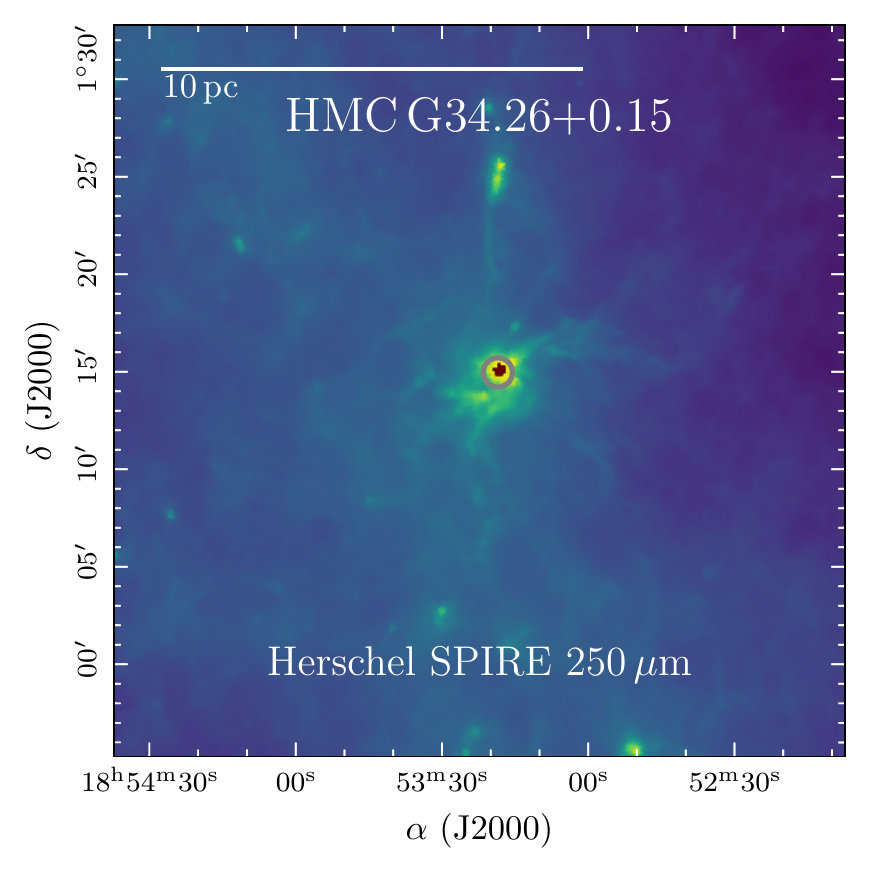}
\includegraphics[width=0.39\textwidth]{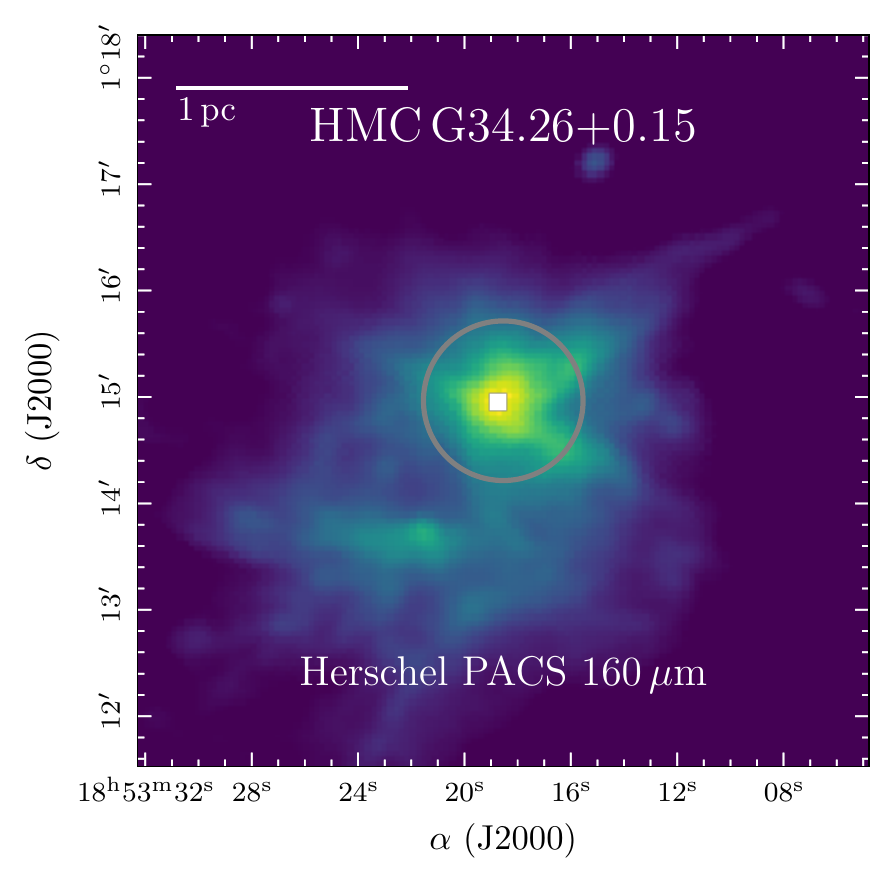}\\
\includegraphics[width=0.39\textwidth]{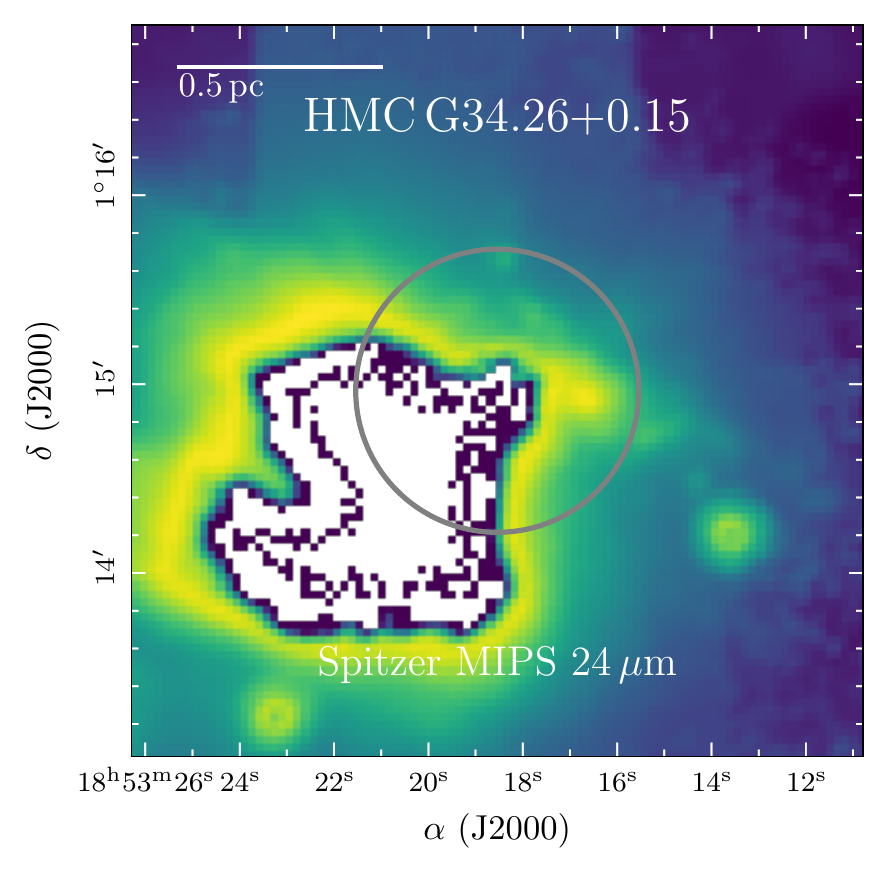}
\includegraphics[width=0.39\textwidth]{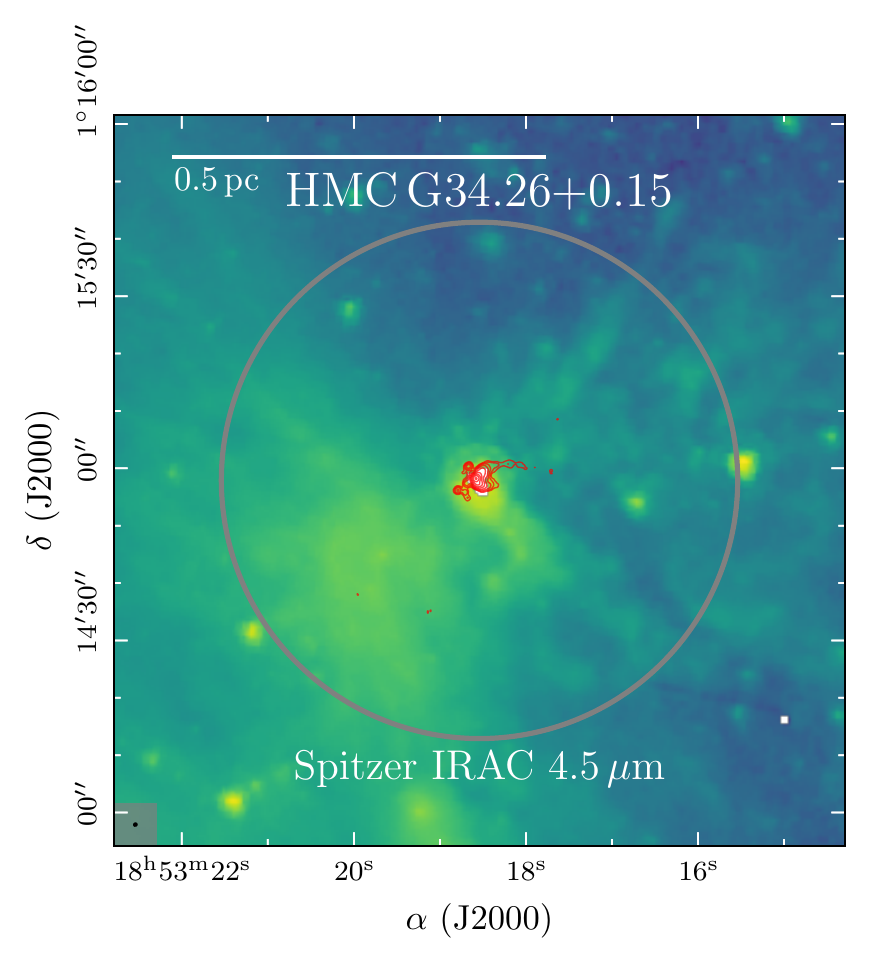}
\caption[Overview of HMC\,G34.26$+$0.15.]{The same as Fig. \ref{fig:overview_IRDC_G1111}, but for HMC\,G34.26$+$0.15.}
\label{fig:overview_HMC_G3426}
\end{figure*}

\begin{figure*}
\centering
\includegraphics[width=0.39\textwidth]{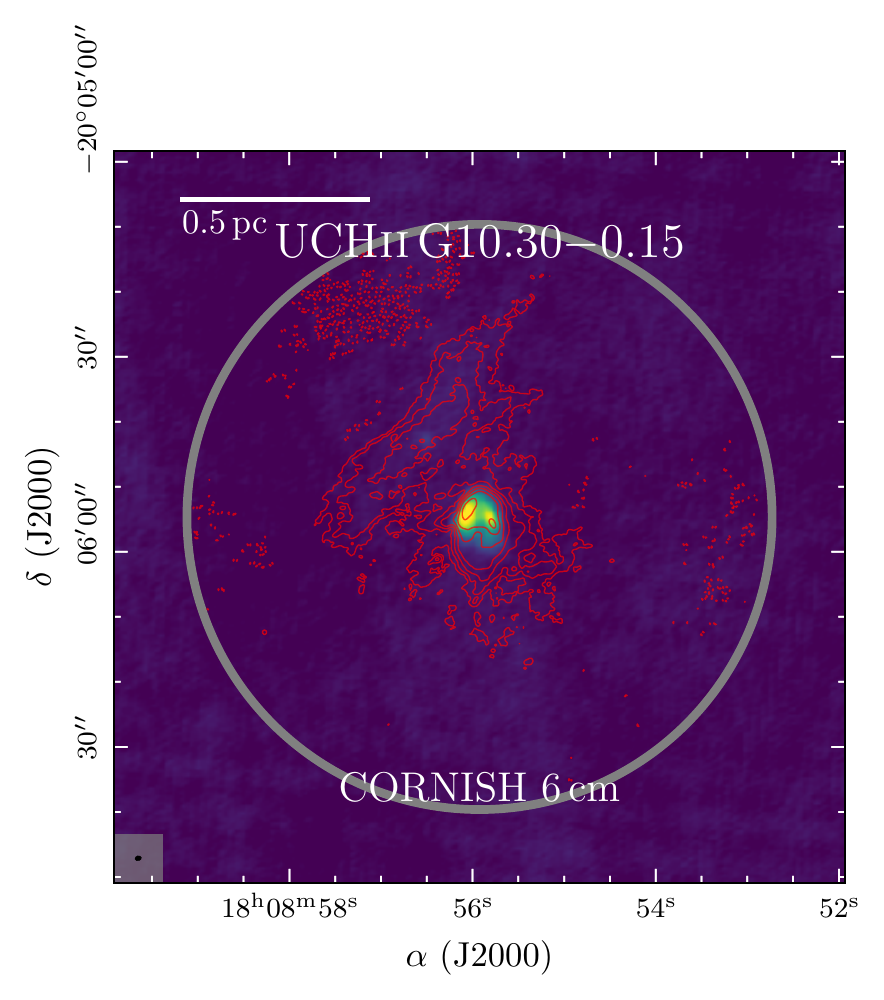}
\includegraphics[width=0.39\textwidth]{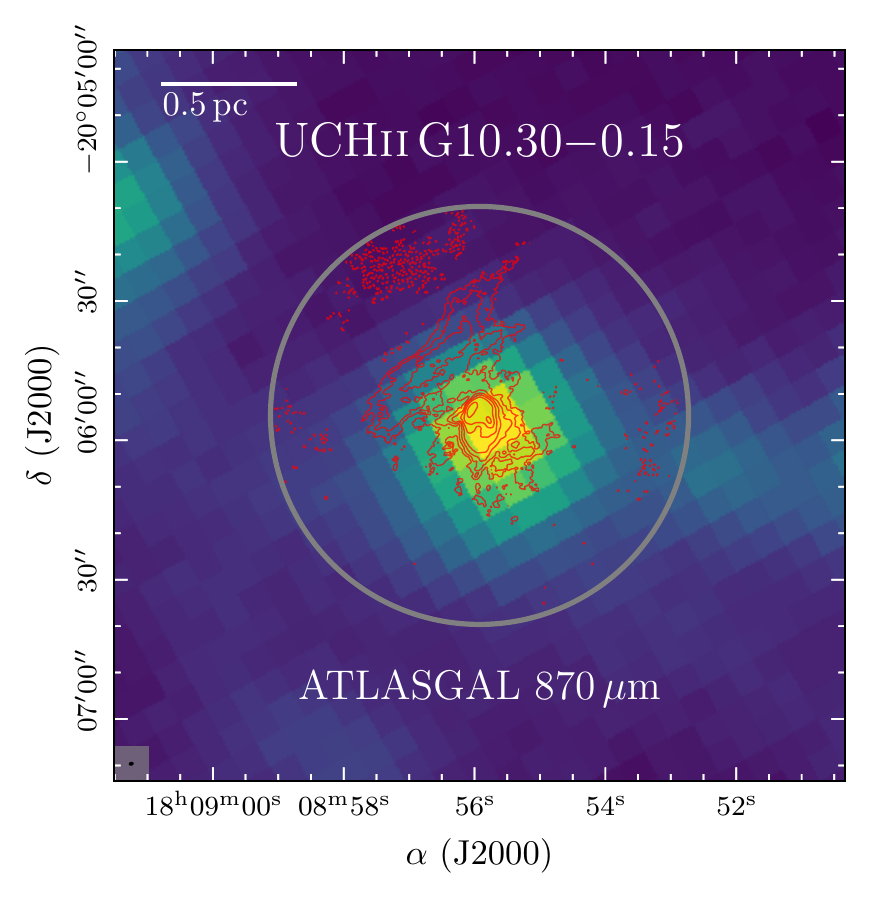}\\
\includegraphics[width=0.39\textwidth]{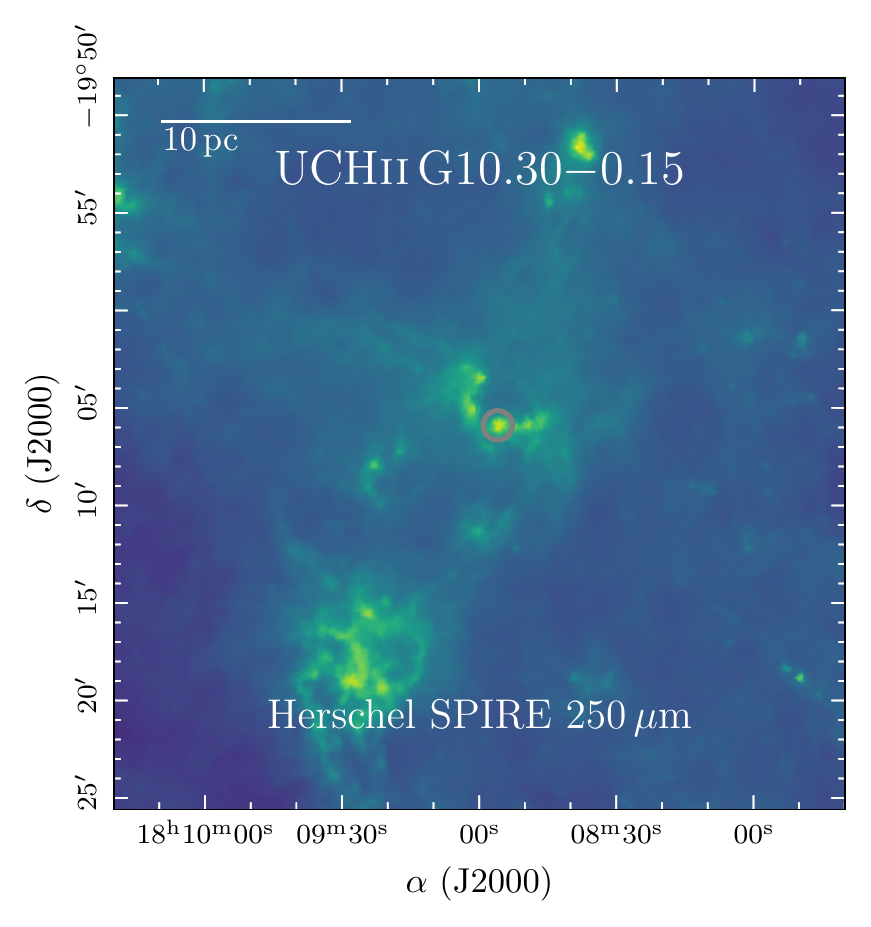}
\includegraphics[width=0.39\textwidth]{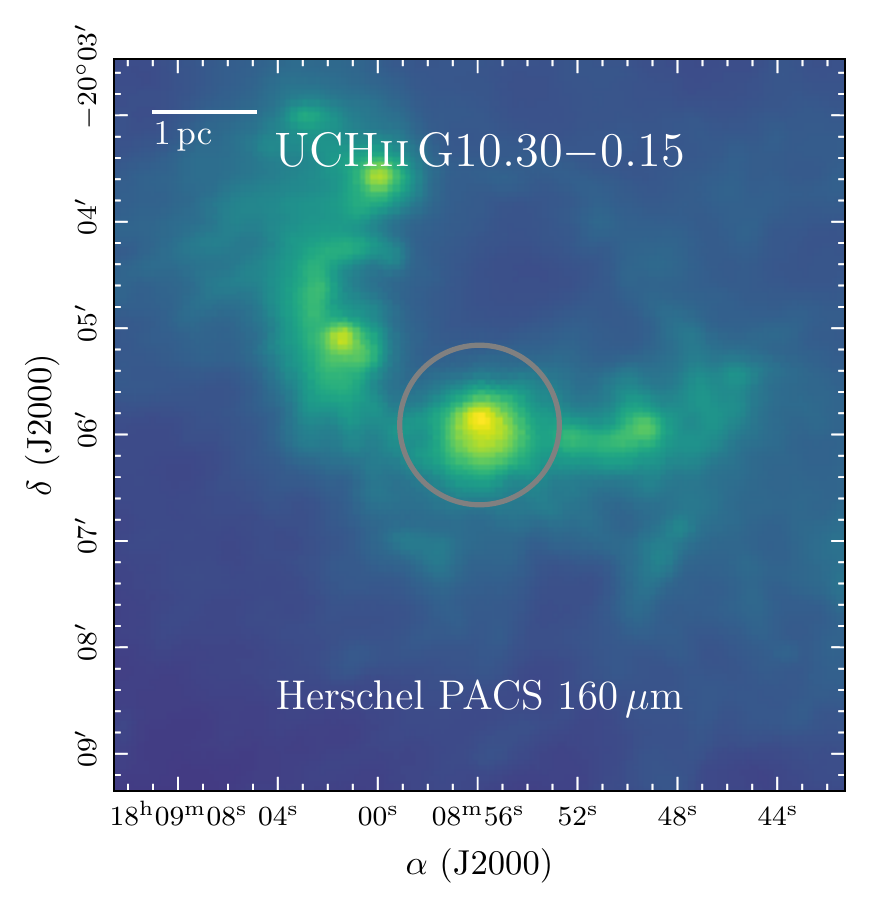}\\
\includegraphics[width=0.39\textwidth]{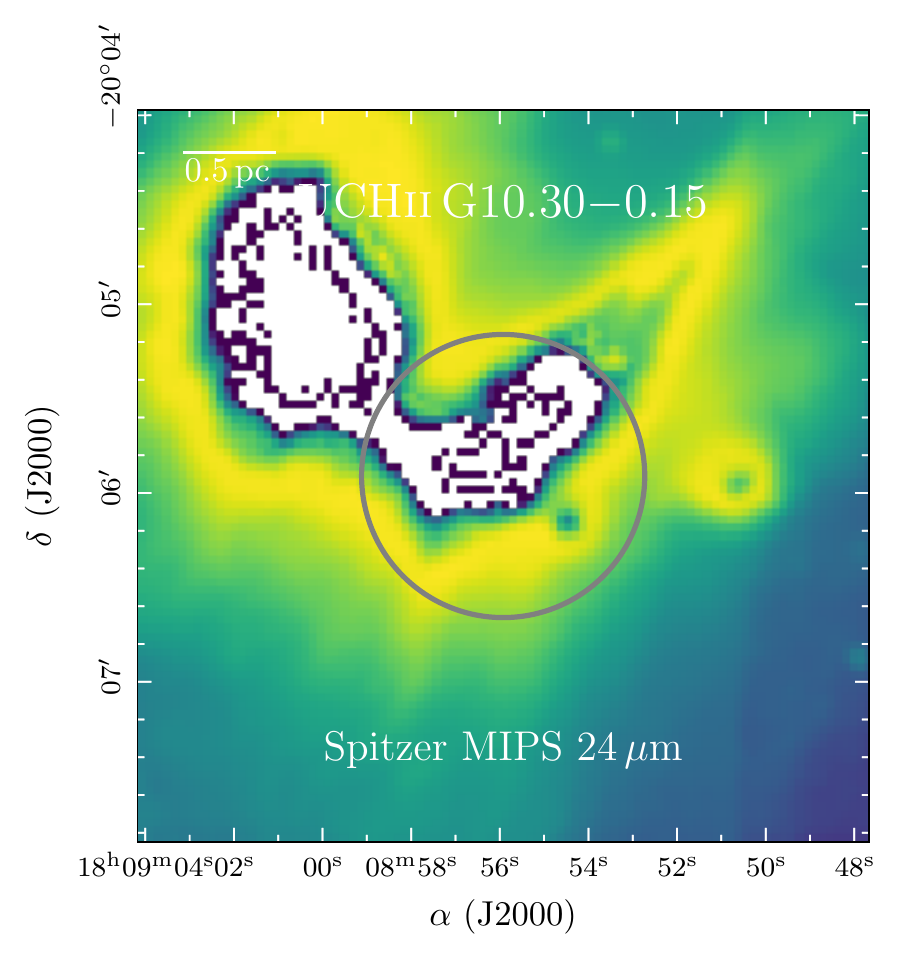}
\includegraphics[width=0.39\textwidth]{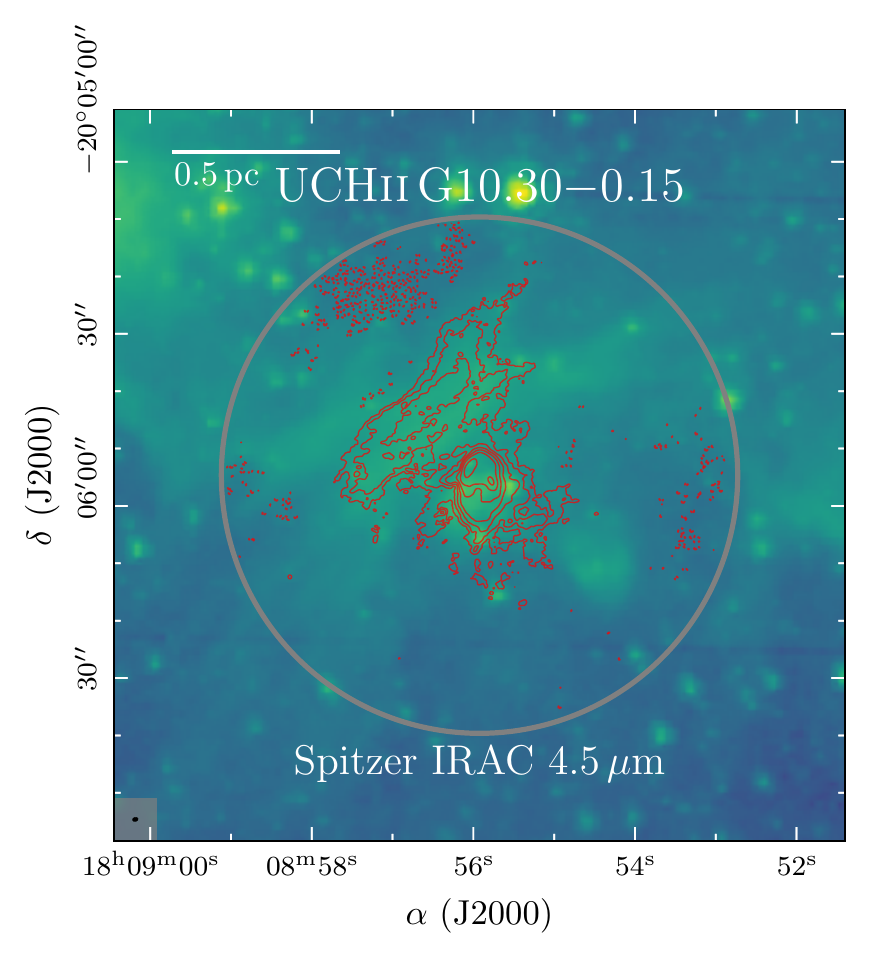}
\caption[Overview of UCH{\sc ii}\,G10.30$-$0.15.]{The same as Fig. \ref{fig:overview_IRDC_G1111}, but for UCH{\sc ii}\,G10.30$-$0.15.}
\label{fig:overview_UCHII_G1030}
\end{figure*}

\begin{figure*}
\centering
\includegraphics[width=0.39\textwidth]{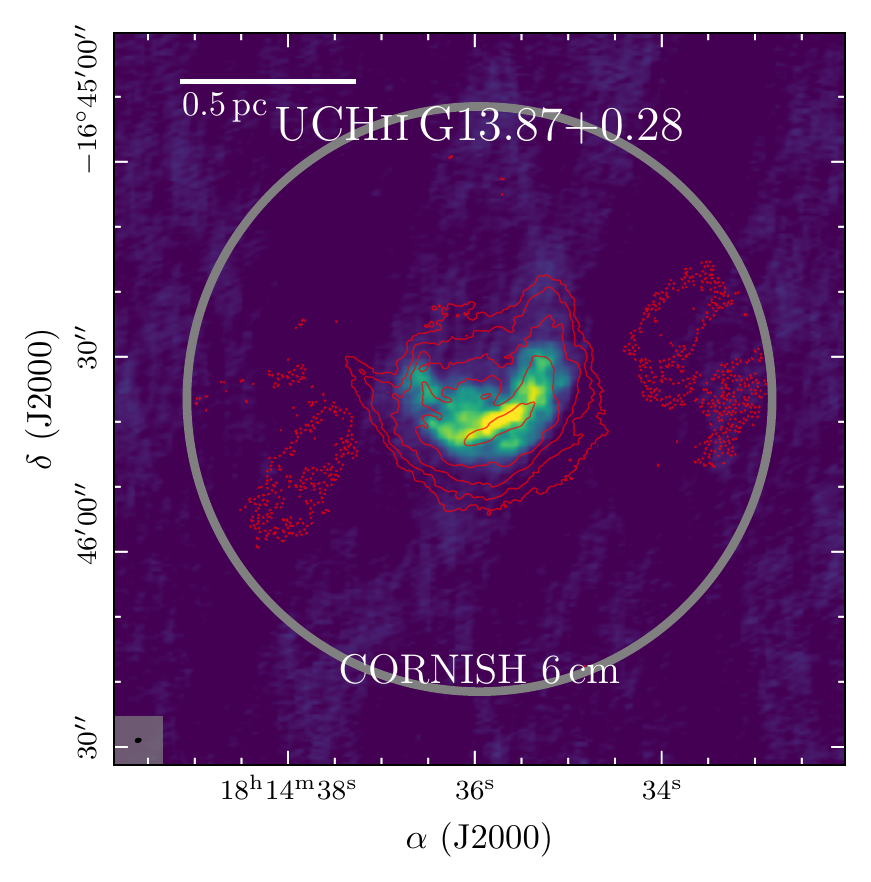}
\includegraphics[width=0.39\textwidth]{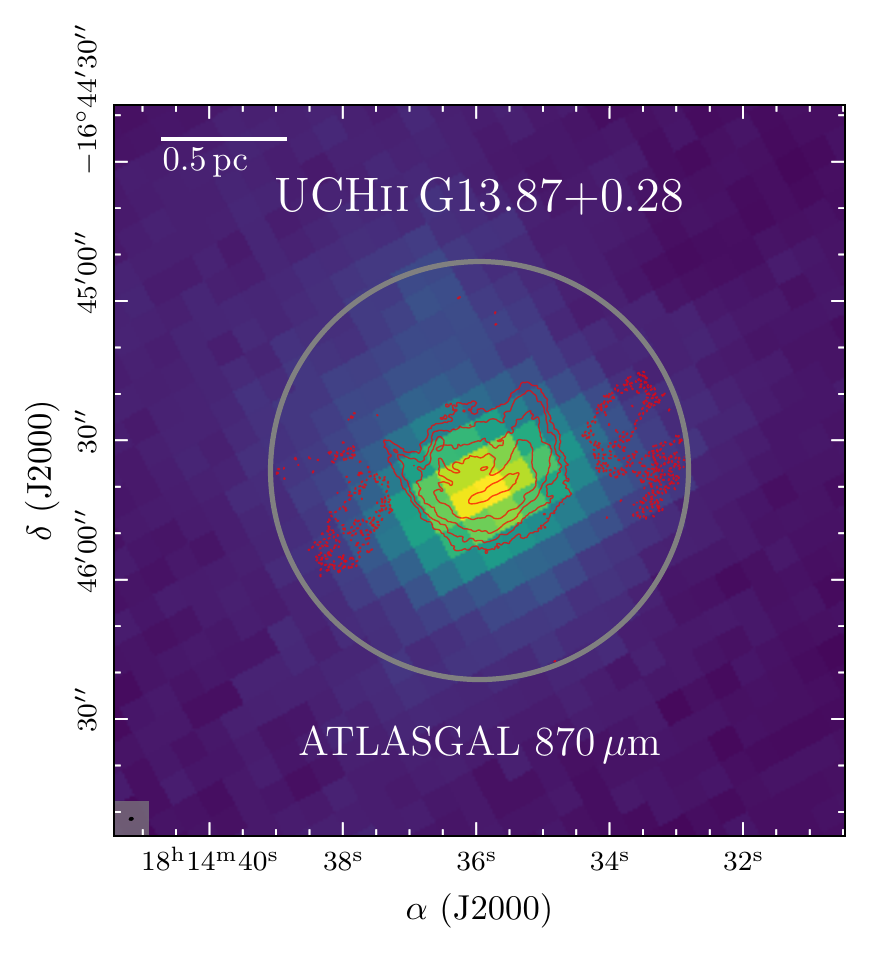}\\
\includegraphics[width=0.39\textwidth]{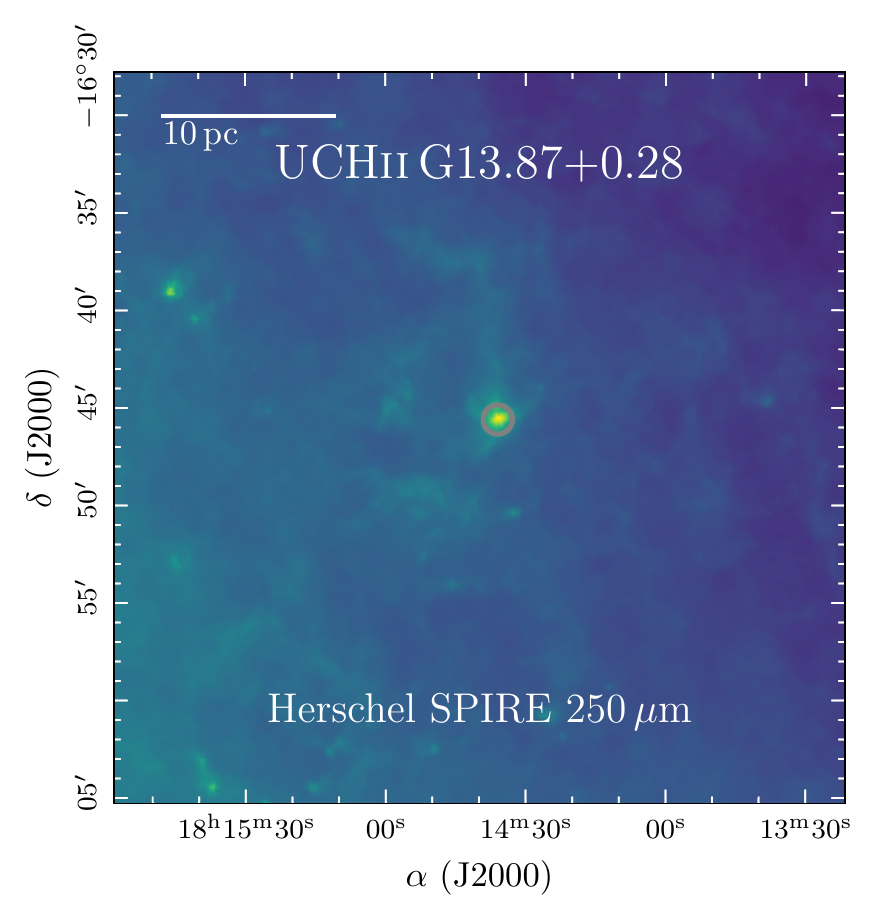}
\includegraphics[width=0.39\textwidth]{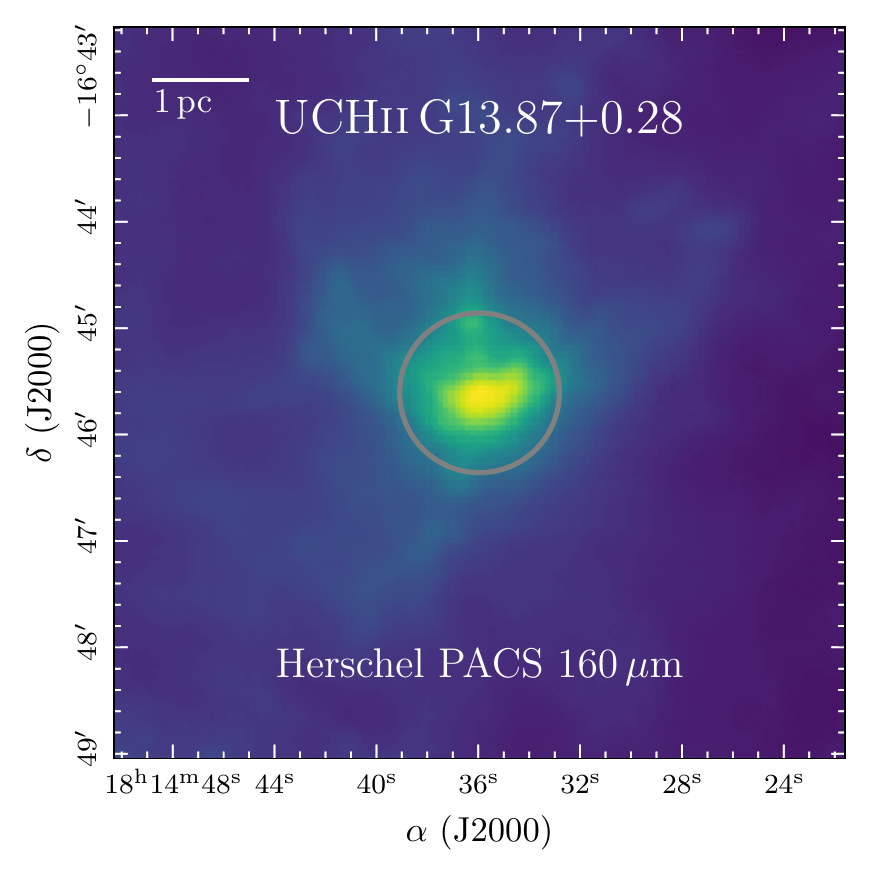}\\
\includegraphics[width=0.39\textwidth]{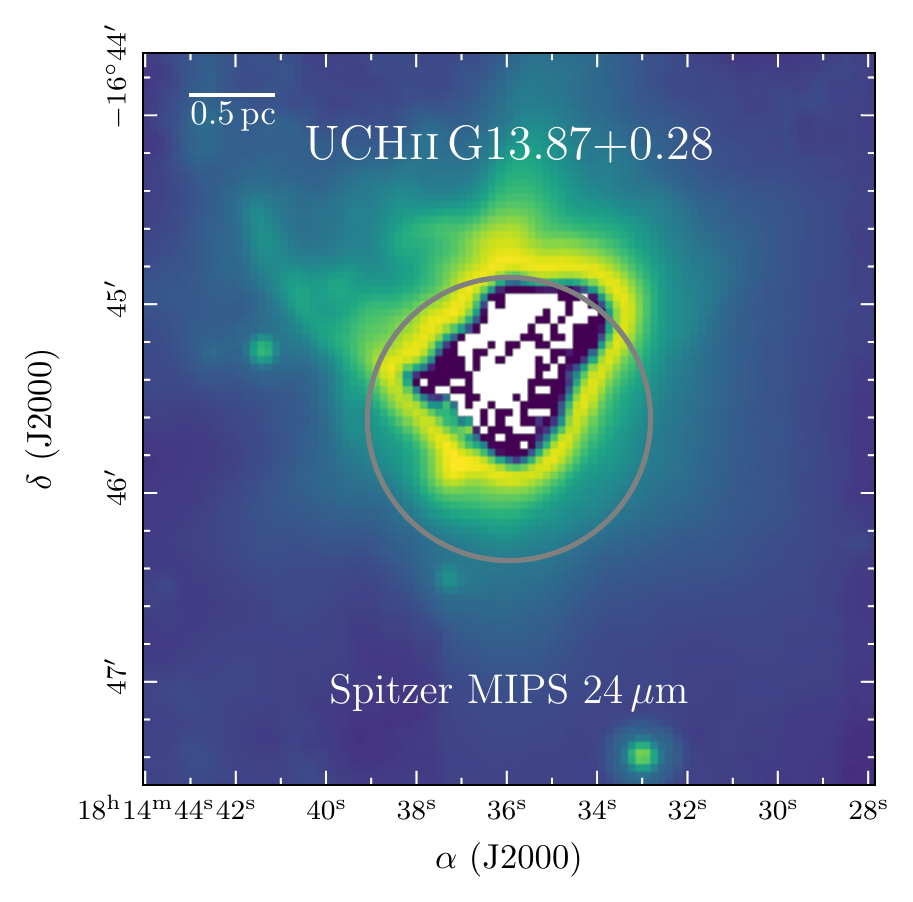}
\includegraphics[width=0.39\textwidth]{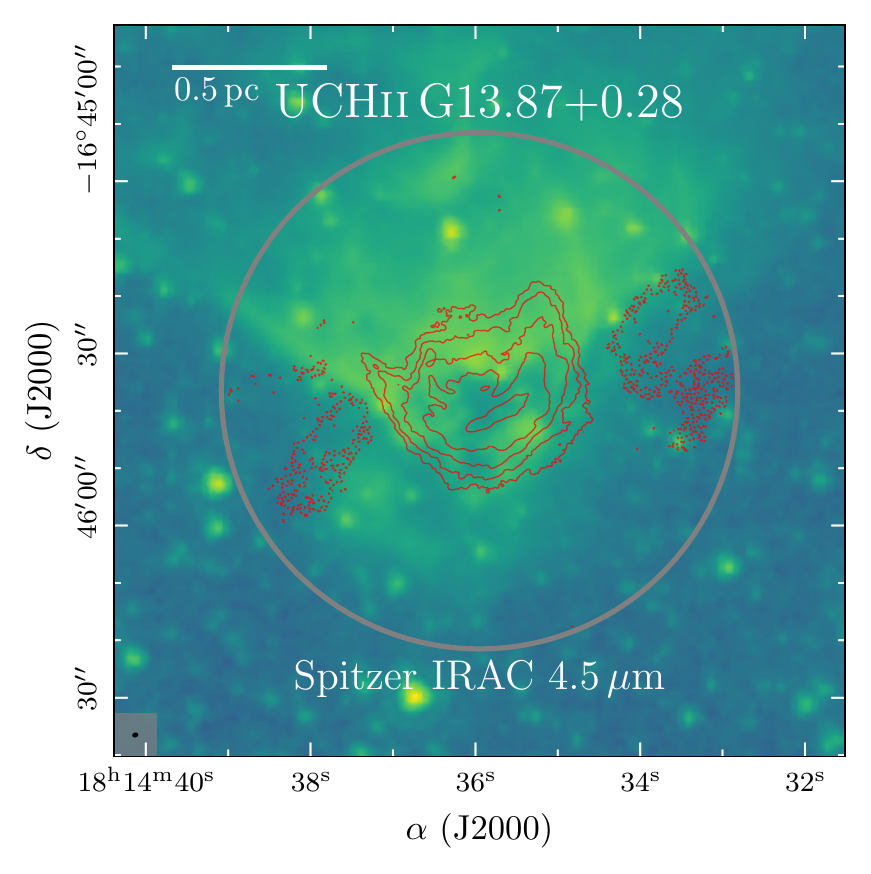}
\caption[Overview of UCH{\sc ii}\,G13.87$+$0.28.]{The same as Fig. \ref{fig:overview_IRDC_G1111}, but for UCH{\sc ii}\,G13.87$+$0.28.}
\label{fig:overview_UCHII_G1387}
\end{figure*}

\section{Source properties}\label{app:fragprops}

	The fragmentation properties of the regions based on the 3\,mm continuum emission are analyzed in Sect. \ref{sec:ALMAfrag}. Table \ref{tab:ALMApositions} summarizes the properties derived with \texttt{clumpfind}, such as position and peak intensity of all fragments. We classify the fragments into protostellar sources (including dust cores, dust+ff cores and cometary UCH{\sc ii} regions), while cores with $S$/$N < 15$ are not further analyzed in this study due to insufficient sensitivity and angular resolution. The dust cores have compact mm emission, while dust+ff cores have in addition H(40)$\alpha$ recombination line (Fig. \ref{fig:Halpha_moment0}) and CORNISH 6\,cm emission (Figs. \ref{fig:overview_IRDC_G1111} - \ref{fig:overview_UCHII_G1387}). Our sample also covers three cometary UCH{\sc ii} regions with extended, cometary-shaped, mm emission as well as H(40)$\alpha$ recombination line and 6\,cm emission.
	
	Table \ref{tab:ALMAradialtemp} summarizes the radial temperature fit results derived in Sect. \ref{sec:ALMAradialTprofiles} for all protostellar sources. Fig. \ref{fig:ALMATrad} shows the radial temperature profile and fit for dust core 1 in HMPO\,IRAS\,18089 and for the remaining sources the profiles are shown in Fig. \ref{fig:ALMATradapp}. The beam-averaged temperature (Sect. \ref{sec:ALMAtemperature}), and H$_{2}$ column density and mass (Sect. \ref{sec:ALMANH2}) are listed in Table \ref{tab:ALMAbeamavgtempNM} for all protostellar sources. 
	
	Table \ref{tab:ALMAdens} summarizes the fit results of the visibility profiles and corresponding density profiles for all protostellar sources (Sect. \ref{sec:ALMAvisibprofile}). Fig. \ref{fig:ALMAvisibilityprofile} shows the visibility profile and fit for HMC\,G9.62$+$0.19\,2 and for the remaining sources the profiles are shown in Fig. \ref{fig:ALMAvisibilityprofileapp}.
	
\begin{figure*}
\includegraphics[width=0.32\textwidth]{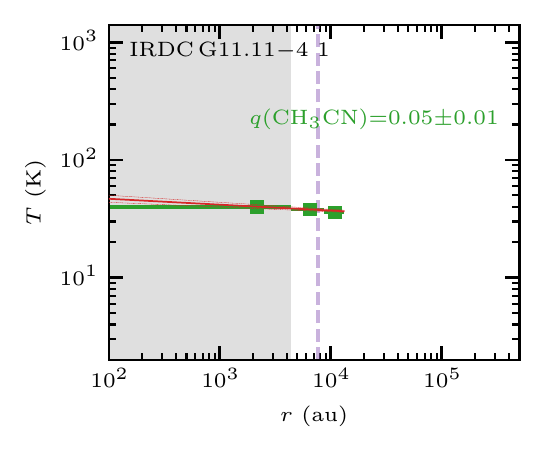}
\includegraphics[width=0.32\textwidth]{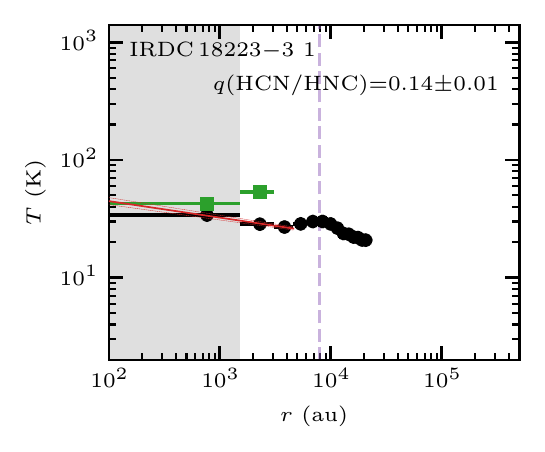}
\includegraphics[width=0.32\textwidth]{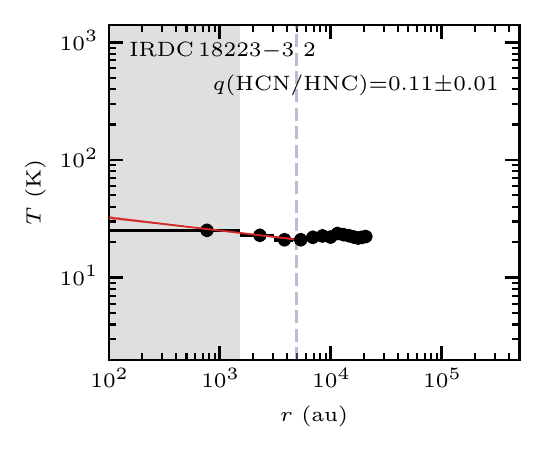}\\
\includegraphics[width=0.32\textwidth]{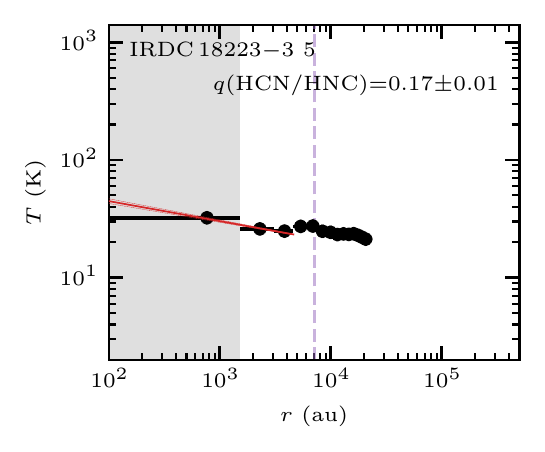}
\includegraphics[width=0.32\textwidth]{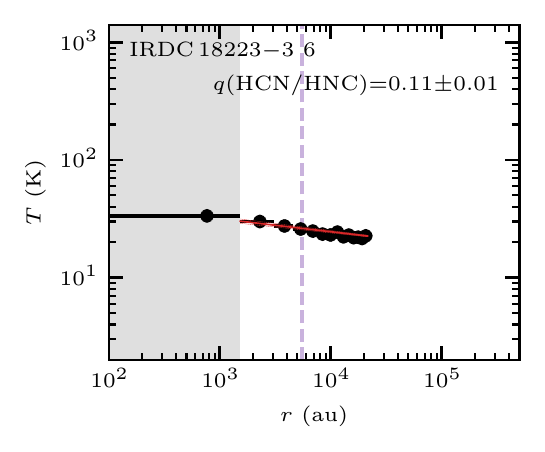}
\includegraphics[width=0.32\textwidth]{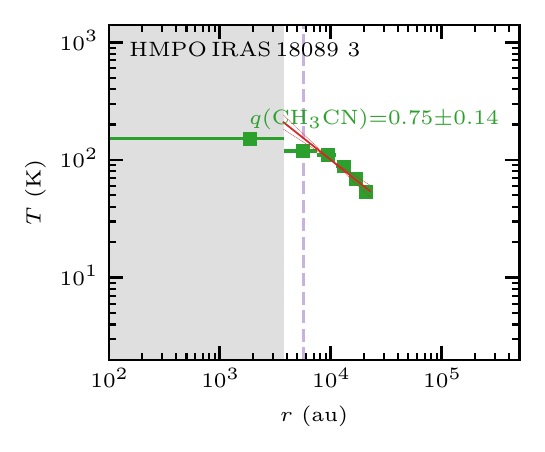}\\
\includegraphics[width=0.32\textwidth]{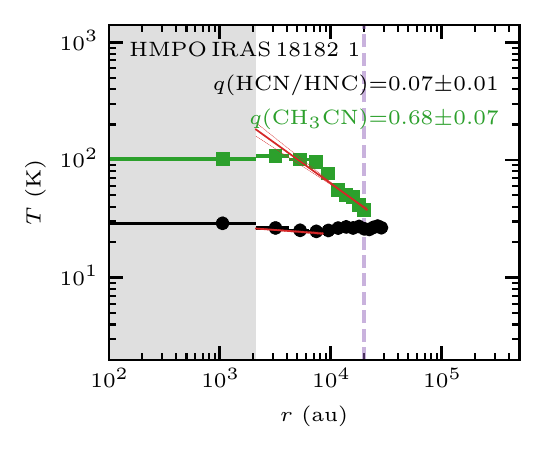}
\includegraphics[width=0.32\textwidth]{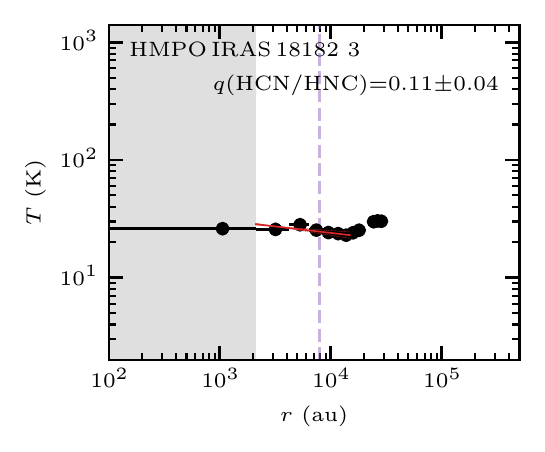}
\includegraphics[width=0.32\textwidth]{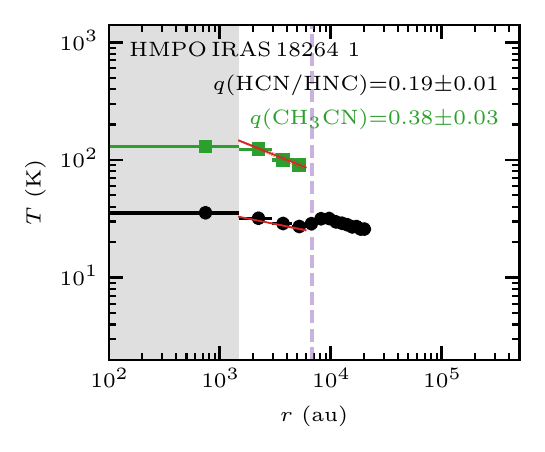}\\
\includegraphics[width=0.32\textwidth]{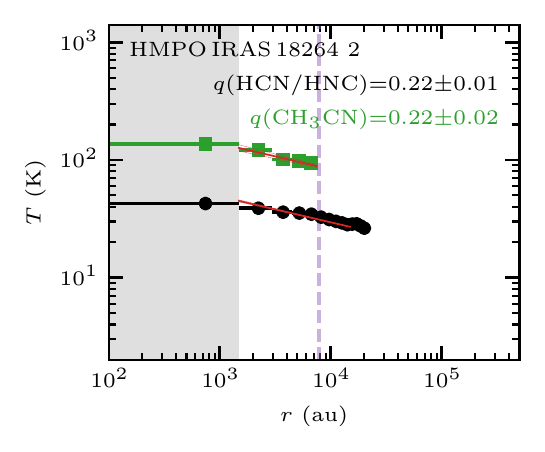}
\includegraphics[width=0.32\textwidth]{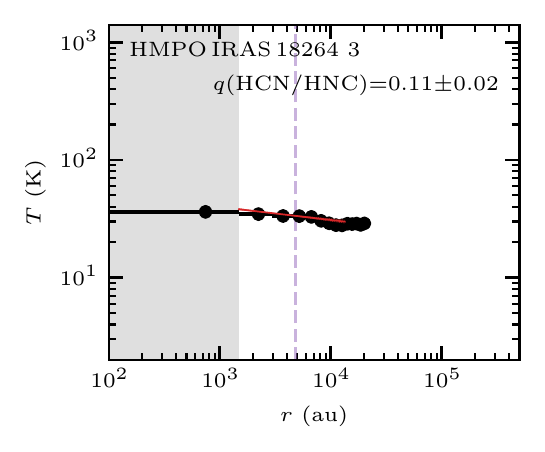}
\includegraphics[width=0.32\textwidth]{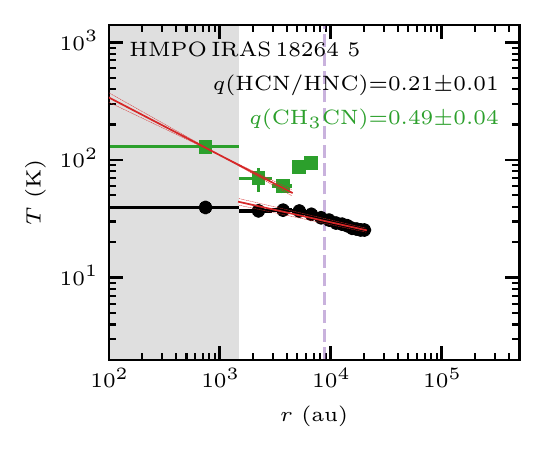}
\caption{Radial temperature profiles. The profiles extracted from the HCN-to-HNC intensity ratio (Fig. \ref{fig:hcnhnctemperaturemaps}), CH$_{3}$CN (Fig. \ref{fig:ch3cntemperaturemaps}), and CH$_{3}^{13}$CN (Fig. \ref{fig:ch313cntemperaturemaps}) temperature maps are shown by black circles, green squares, and blue diamonds, respectively. The inner unresolved region (one beam radius) is shown as a grey-shaded area. The dashed purple vertical line indicates the outer radius $r_\mathrm{out}$ of the continuum (Table \ref{tab:ALMApositions}). A power-law fit and its $1\sigma$ uncertainty to resolved and radially decreasing profiles is shown by the red solid and dashed lines, respectively (Sect. \ref{sec:ALMAradialTprofiles}). The radial temperature profile of dust core 1 in HMPO\,IRAS\,18089 is shown in Fig. \ref{fig:ALMATrad}.}
\label{fig:ALMATradapp}
\end{figure*}

\begin{figure*}
\ContinuedFloat
\captionsetup{list=off,format=cont}
\includegraphics[width=0.32\textwidth]{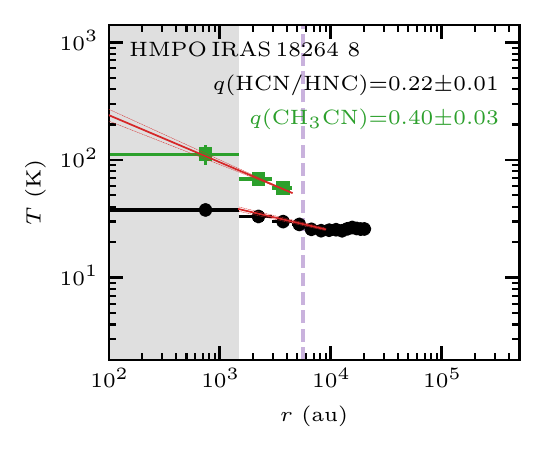}
\includegraphics[width=0.32\textwidth]{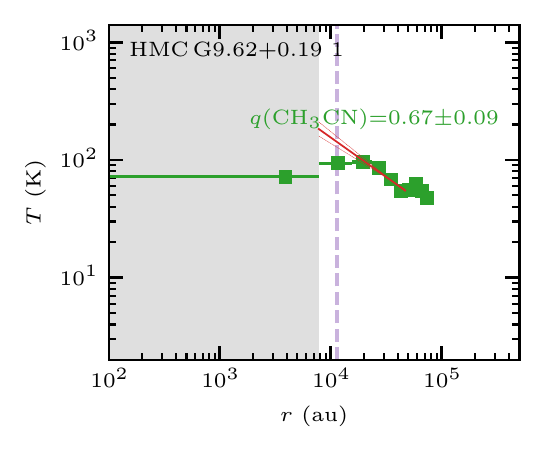}
\includegraphics[width=0.32\textwidth]{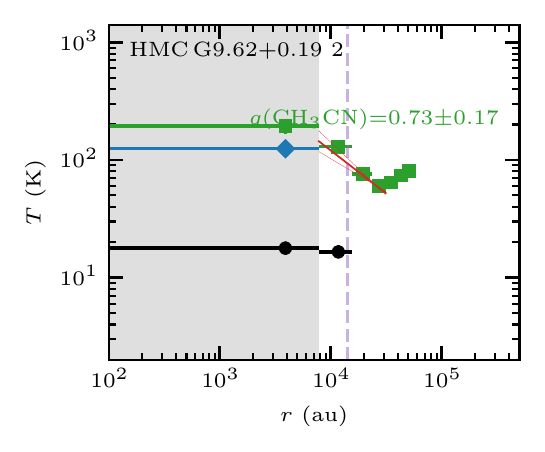}\\
\includegraphics[width=0.32\textwidth]{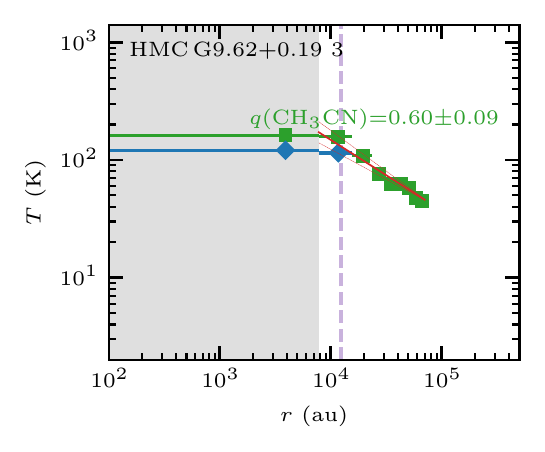}
\includegraphics[width=0.32\textwidth]{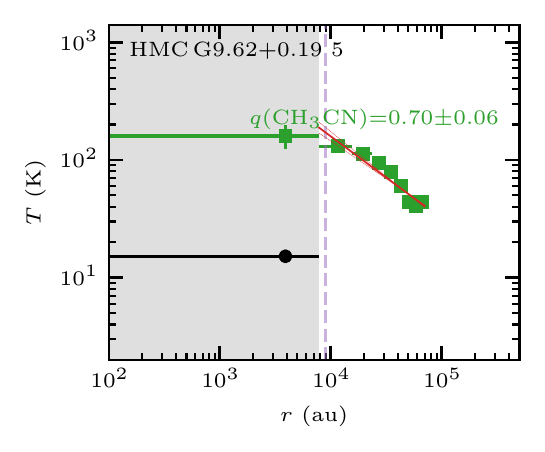}
\includegraphics[width=0.32\textwidth]{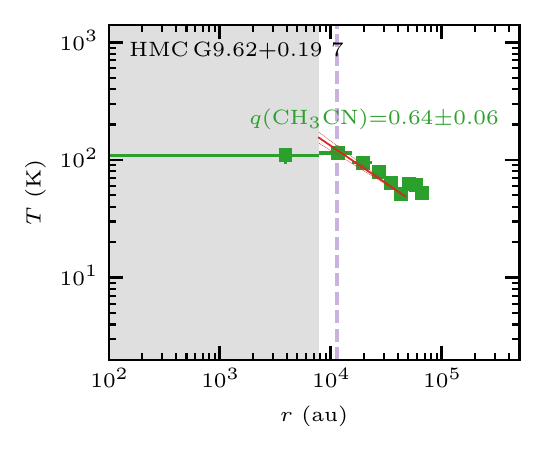}\\
\includegraphics[width=0.32\textwidth]{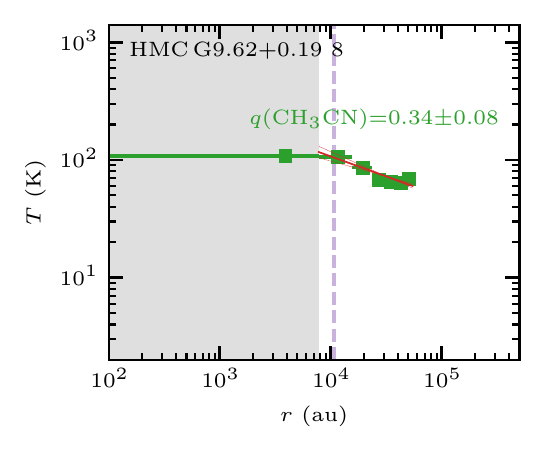}
\includegraphics[width=0.32\textwidth]{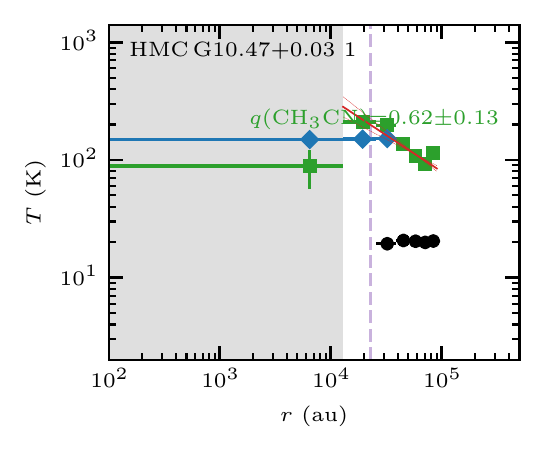}
\includegraphics[width=0.32\textwidth]{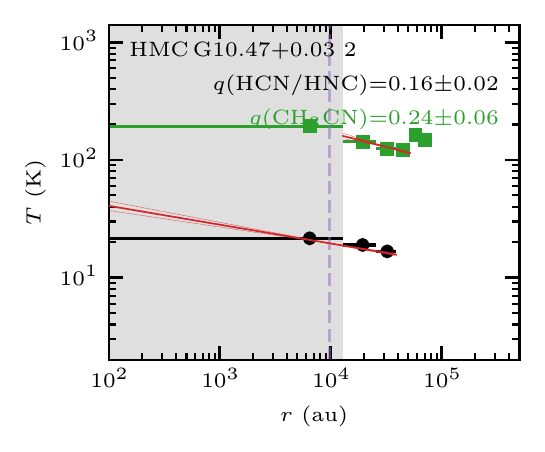}\\
\includegraphics[width=0.32\textwidth]{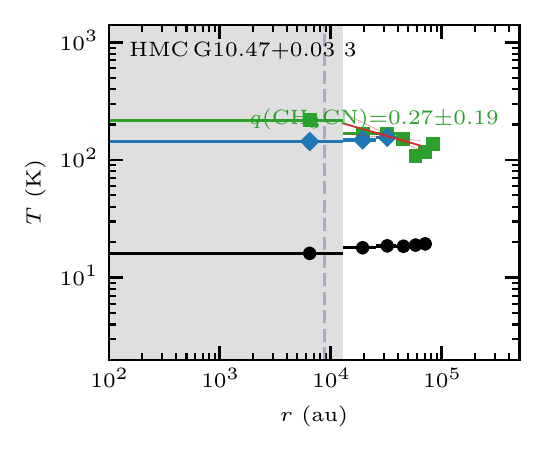}
\includegraphics[width=0.32\textwidth]{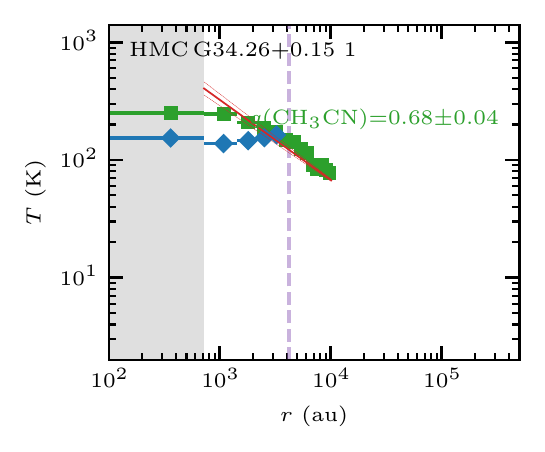}
\includegraphics[width=0.32\textwidth]{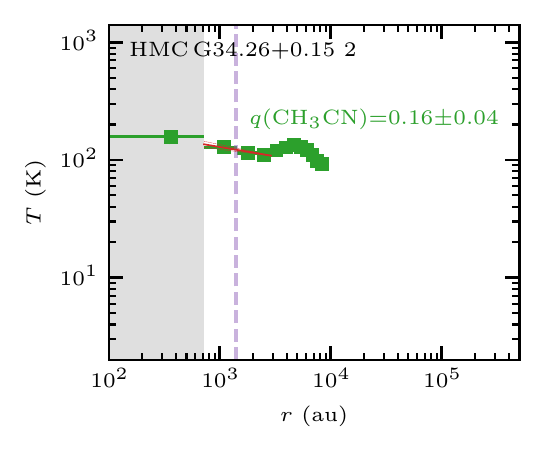}
\caption{Radial temperature profiles. The profiles extracted from the HCN-to-HNC intensity ratio (Fig. \ref{fig:hcnhnctemperaturemaps}), CH$_{3}$CN (Fig. \ref{fig:ch3cntemperaturemaps}), and CH$_{3}^{13}$CN (Fig. \ref{fig:ch313cntemperaturemaps}) temperature maps are shown by black circles, green squares, and blue diamonds, respectively. The inner unresolved region (one beam radius) is shown as a grey-shaded area. The dashed purple vertical line indicates the outer radius $r_\mathrm{out}$ of the continuum (Table \ref{tab:ALMApositions}). A power-law fit and its $1\sigma$ uncertainty to resolved and radially decreasing profiles is shown by the red solid and dashed lines, respectively (Sect. \ref{sec:ALMAradialTprofiles}). The radial temperature profile of dust core 1 in HMPO\,IRAS\,18089 is shown in Fig. \ref{fig:ALMATrad}.}
\end{figure*}

\begin{figure*}
\ContinuedFloat
\captionsetup{list=off,format=cont}
\includegraphics[width=0.32\textwidth]{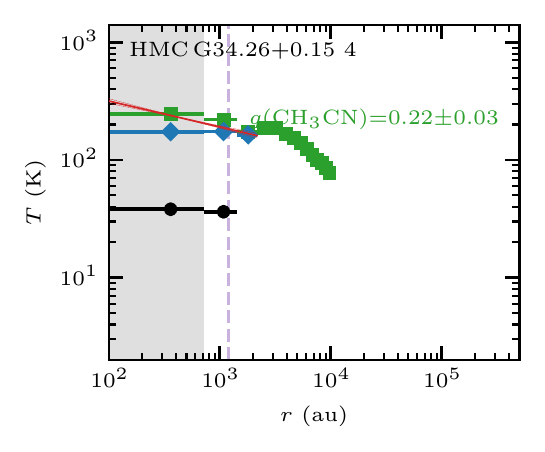}
\includegraphics[width=0.32\textwidth]{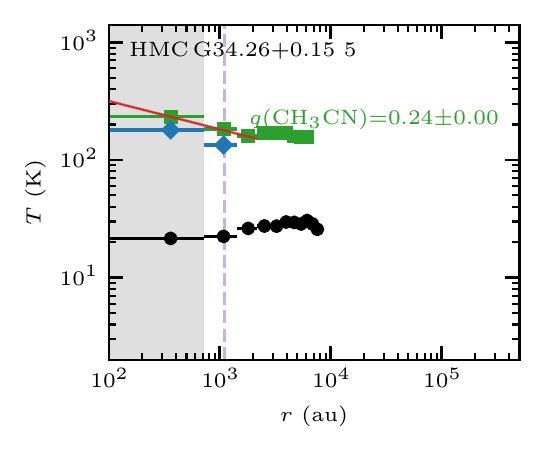}
\includegraphics[width=0.32\textwidth]{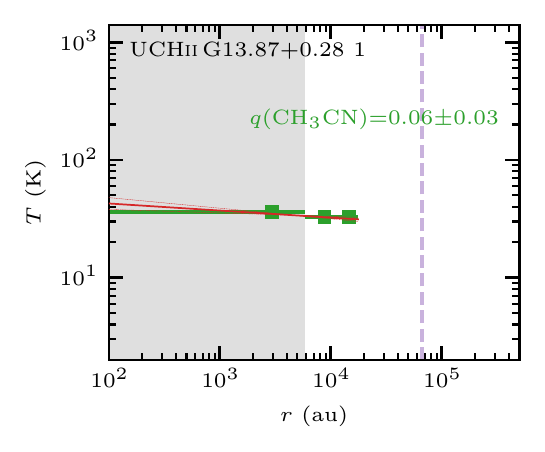}
\caption{Radial temperature profiles. The profiles extracted from the HCN-to-HNC intensity ratio (Fig. \ref{fig:hcnhnctemperaturemaps}), CH$_{3}$CN (Fig. \ref{fig:ch3cntemperaturemaps}), and CH$_{3}^{13}$CN (Fig. \ref{fig:ch313cntemperaturemaps}) temperature maps are shown by black circles, green squares, and blue diamonds, respectively. The inner unresolved region (one beam radius) is shown as a grey-shaded area. The dashed purple vertical line indicates the outer radius $r_\mathrm{out}$ of the continuum (Table \ref{tab:ALMApositions}). A power-law fit and its $1\sigma$ uncertainty to resolved and radially decreasing profiles is shown by the red solid and dashed lines, respectively (Sect. \ref{sec:ALMAradialTprofiles}). The radial temperature profile of dust core 1 in HMPO\,IRAS\,18089 is shown in Fig. \ref{fig:ALMATrad}.}
\end{figure*}

\begin{figure*}
\includegraphics[width=0.32\textwidth]{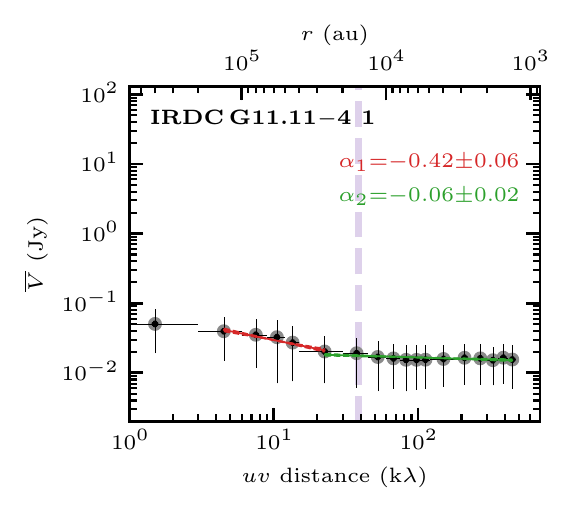}
\includegraphics[width=0.32\textwidth]{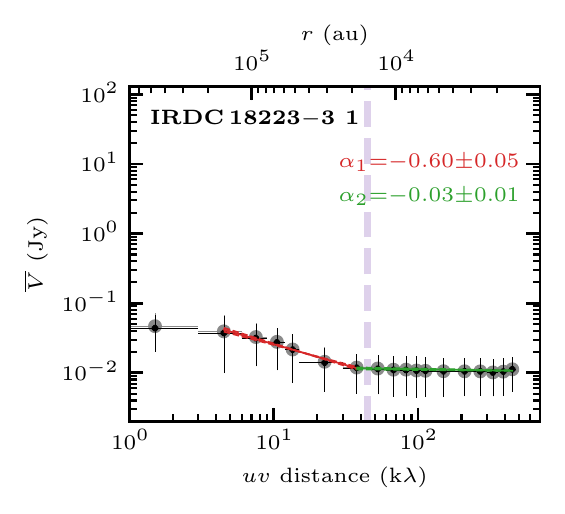}
\includegraphics[width=0.32\textwidth]{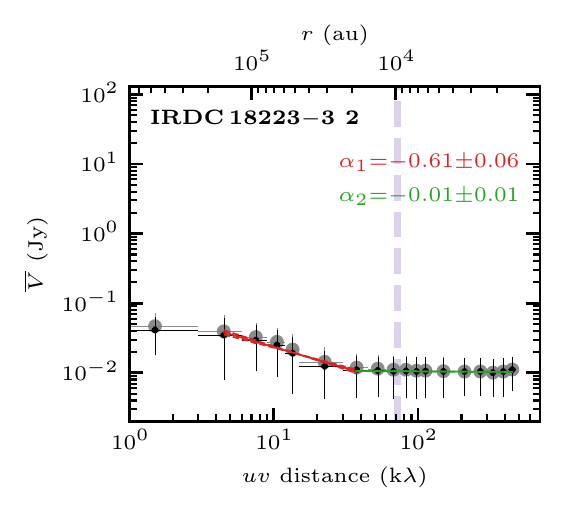}\\
\includegraphics[width=0.32\textwidth]{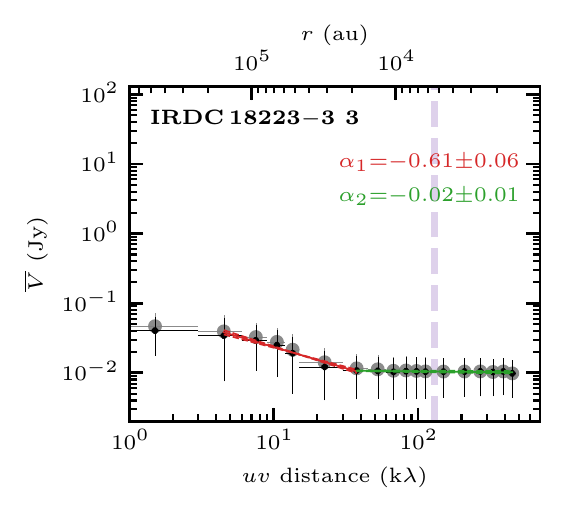}
\includegraphics[width=0.32\textwidth]{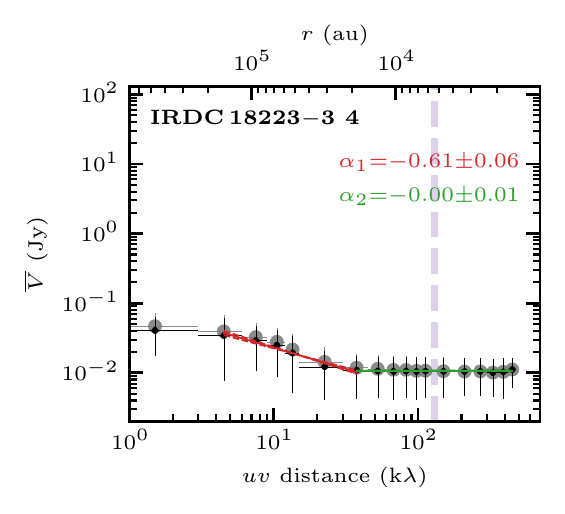}
\includegraphics[width=0.32\textwidth]{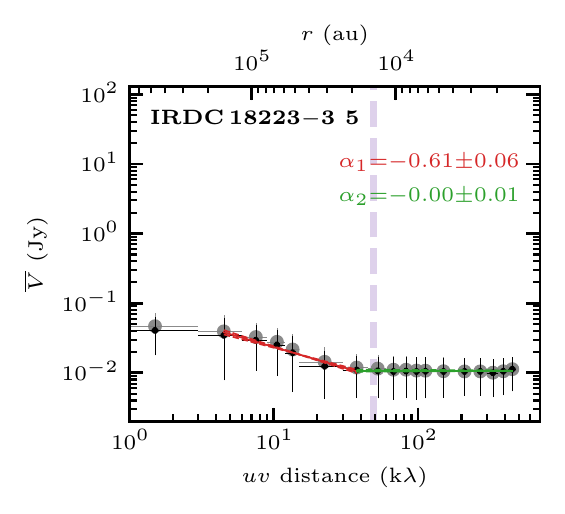}\\
\includegraphics[width=0.32\textwidth]{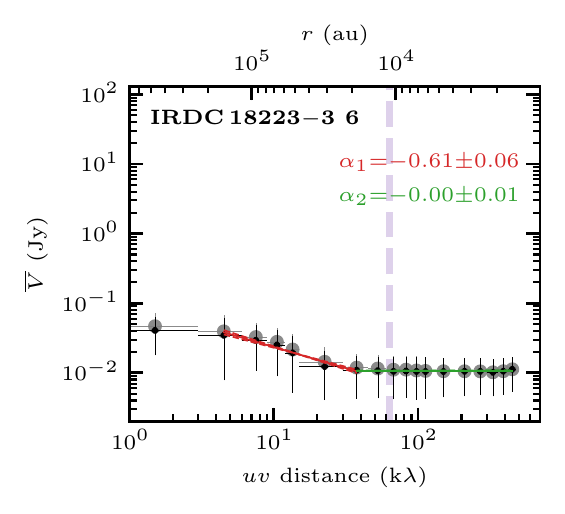}
\includegraphics[width=0.32\textwidth]{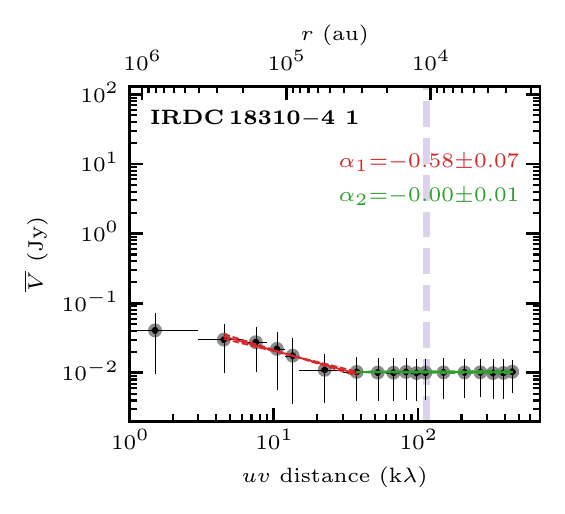}
\includegraphics[width=0.32\textwidth]{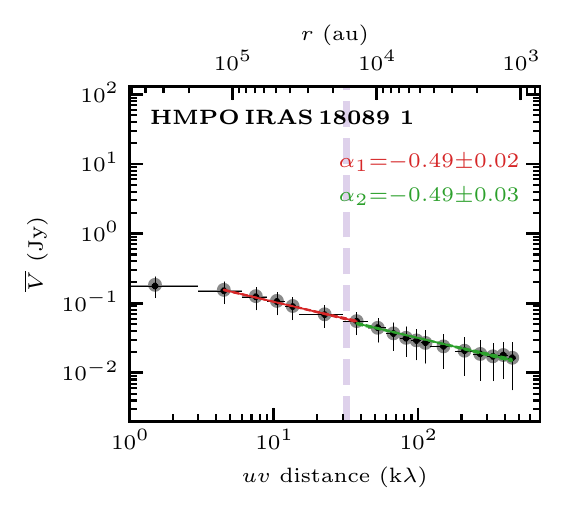}\\
\includegraphics[width=0.32\textwidth]{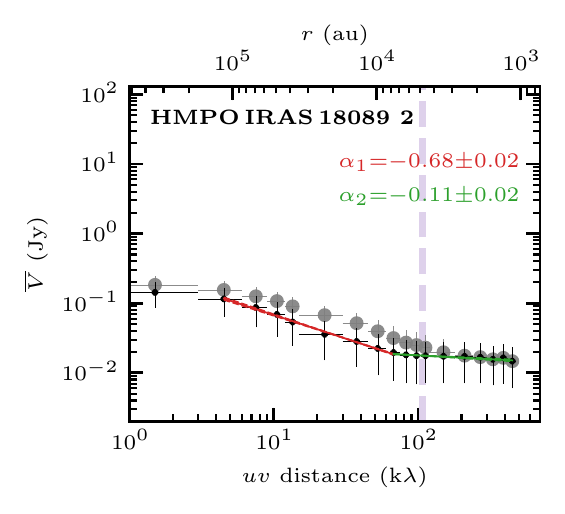}
\includegraphics[width=0.32\textwidth]{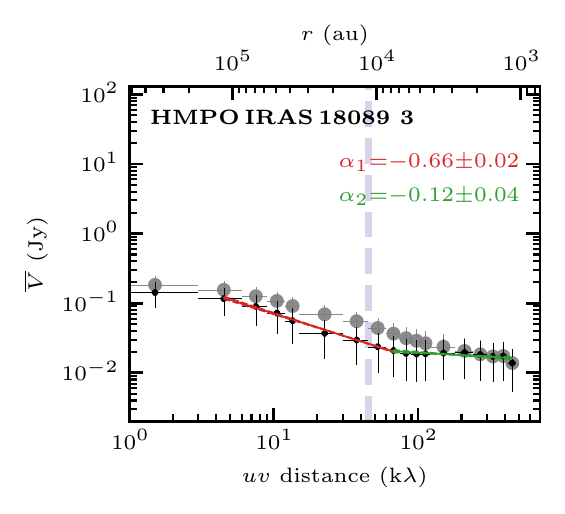}
\includegraphics[width=0.32\textwidth]{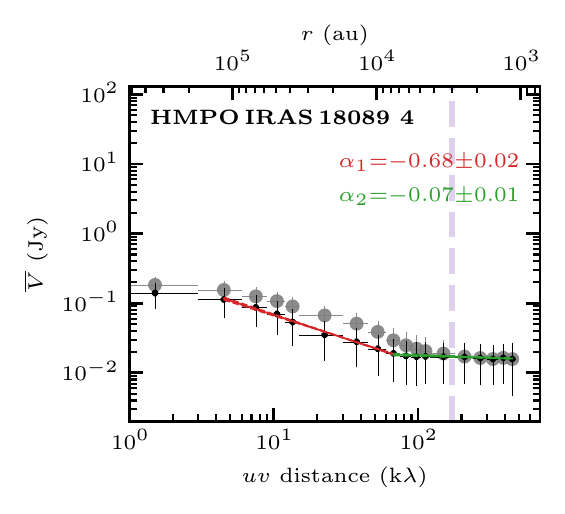}
\caption{Visibility profiles. The profile of the non-core-subtracted and core-subtracted data is shown in grey and black, respectively (further explained in Sect. \ref{sec:ALMAsourcesub}). Two power-law profiles, tracing the clump and core scales, are fitted to the core-subtracted data shown in red and green, respectively. The bottom axis shows the $uv$ distance in k$\lambda$ and the top axis is the corresponding spatial scale. The purple dashed line indicates the diameter of the sources (Table \ref{tab:ALMApositions}). The visibility profile of dust+ff core 2 in HMC\,G9.62$+$.19 is shown in Fig. \ref{fig:ALMAvisibilityprofile}.}
\label{fig:ALMAvisibilityprofileapp}
\end{figure*}

\begin{figure*}
\ContinuedFloat
\captionsetup{list=off,format=cont}
\includegraphics[width=0.32\textwidth]{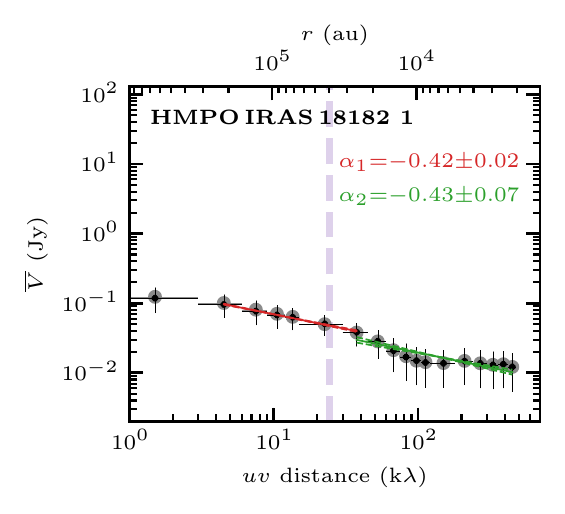}
\includegraphics[width=0.32\textwidth]{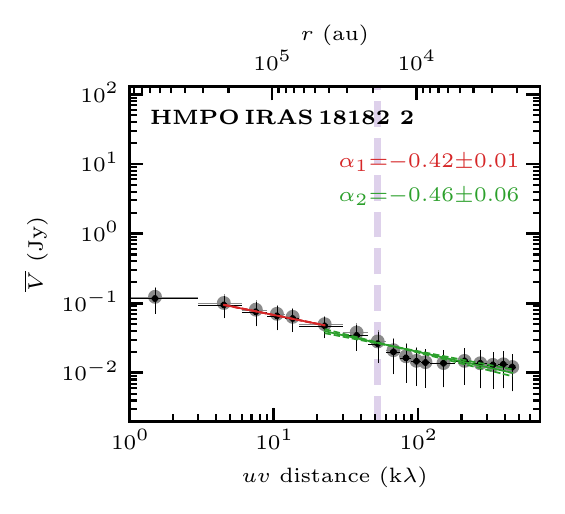}
\includegraphics[width=0.32\textwidth]{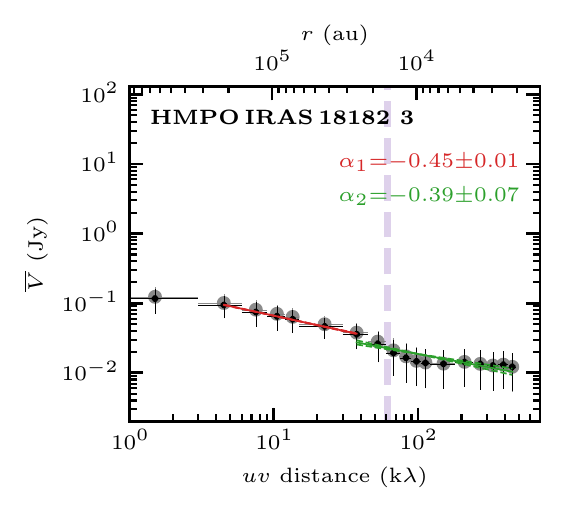}\\
\includegraphics[width=0.32\textwidth]{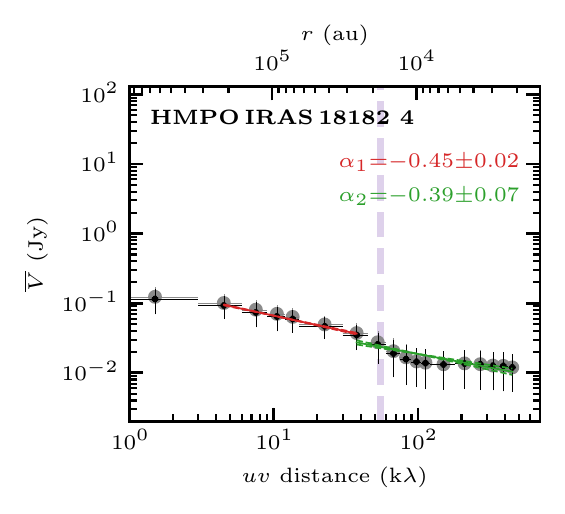}
\includegraphics[width=0.32\textwidth]{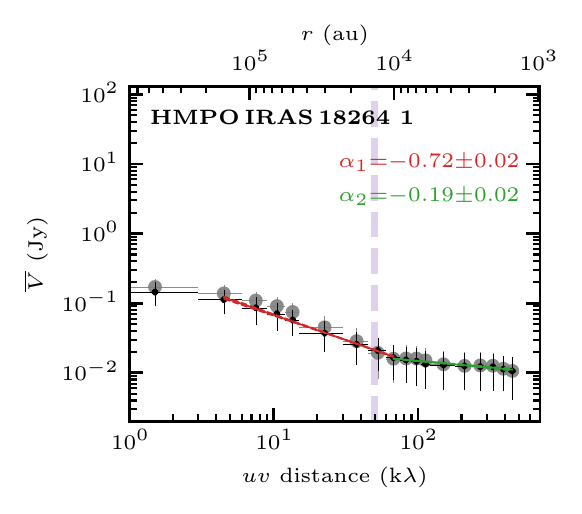}
\includegraphics[width=0.32\textwidth]{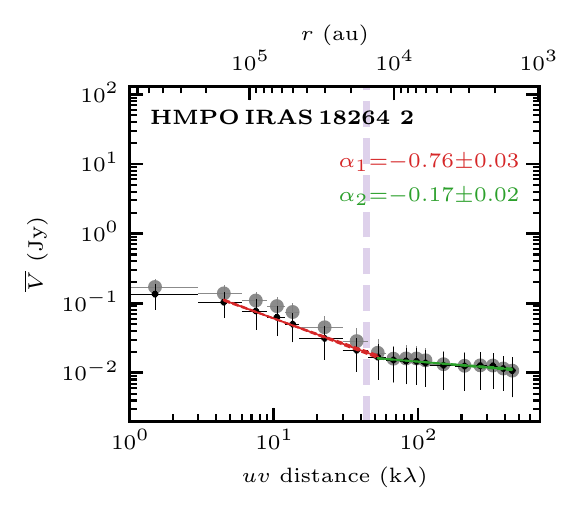}\\
\includegraphics[width=0.32\textwidth]{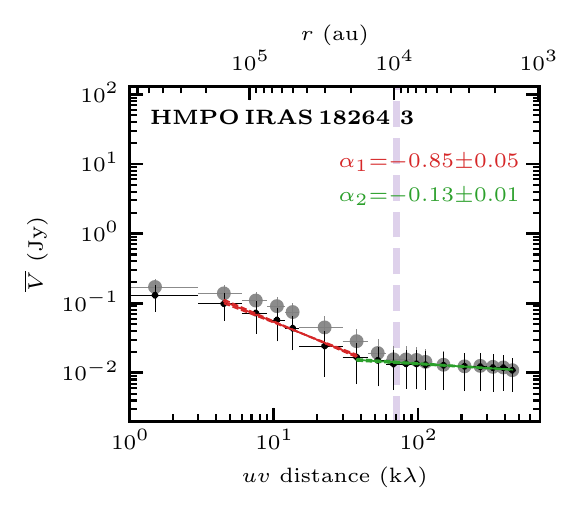}
\includegraphics[width=0.32\textwidth]{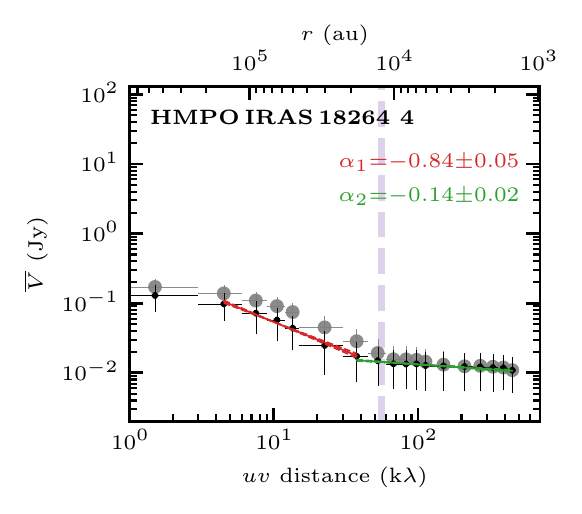}
\includegraphics[width=0.32\textwidth]{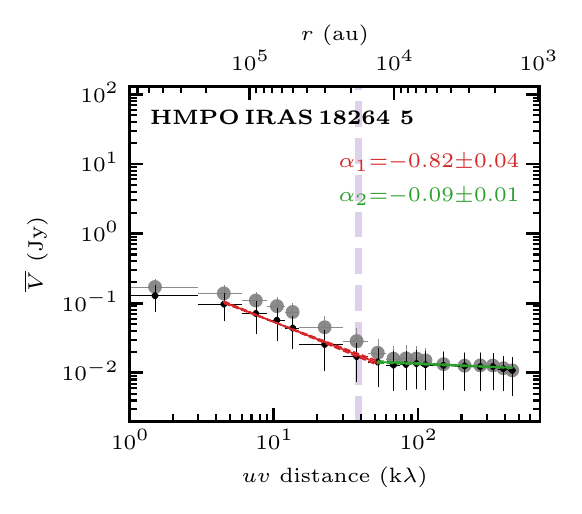}\\
\includegraphics[width=0.32\textwidth]{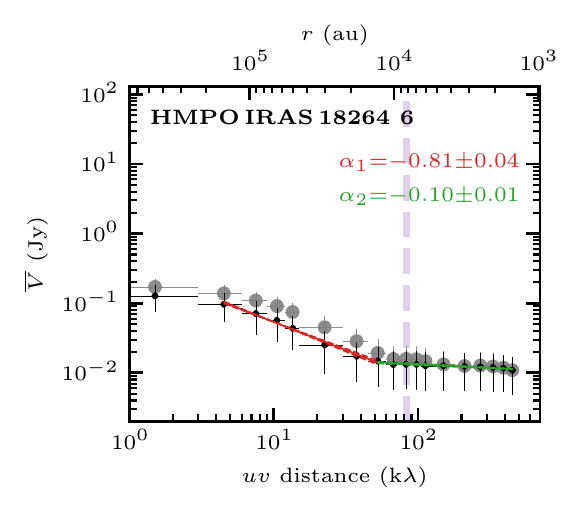}
\includegraphics[width=0.32\textwidth]{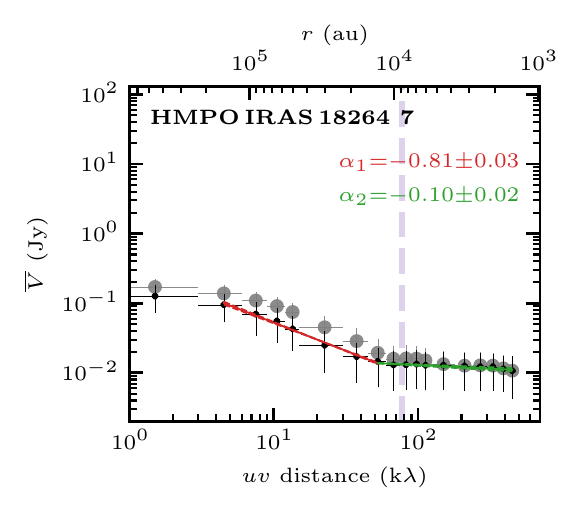}
\includegraphics[width=0.32\textwidth]{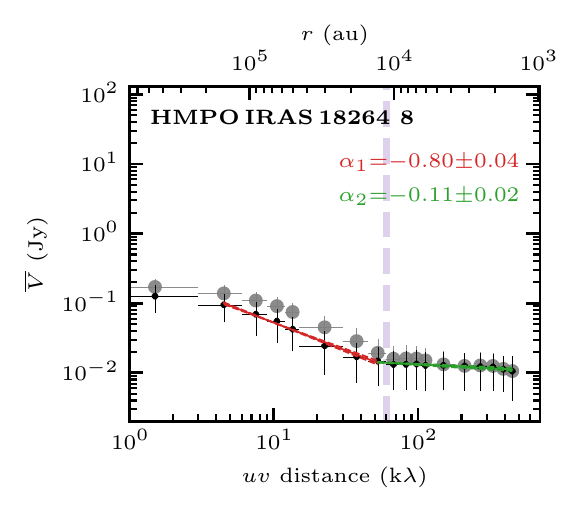}
\caption{Visibility profiles. The profile of the non-core-subtracted and core-subtracted data is shown in grey and black, respectively (further explained in Sect. \ref{sec:ALMAsourcesub}). Two power-law profiles, tracing the clump and core scales, are fitted to the core-subtracted data shown in red and green, respectively. The bottom axis shows the $uv$ distance in k$\lambda$ and the top axis is the corresponding spatial scale. The purple dashed line indicates the diameter of the sources (Table \ref{tab:ALMApositions}). The visibility profile of dust+ff core 2 in HMC\,G9.62$+$.19 is shown in Fig. \ref{fig:ALMAvisibilityprofile}.}
\end{figure*}

\begin{figure*}
\ContinuedFloat
\captionsetup{list=off,format=cont}
\includegraphics[width=0.32\textwidth]{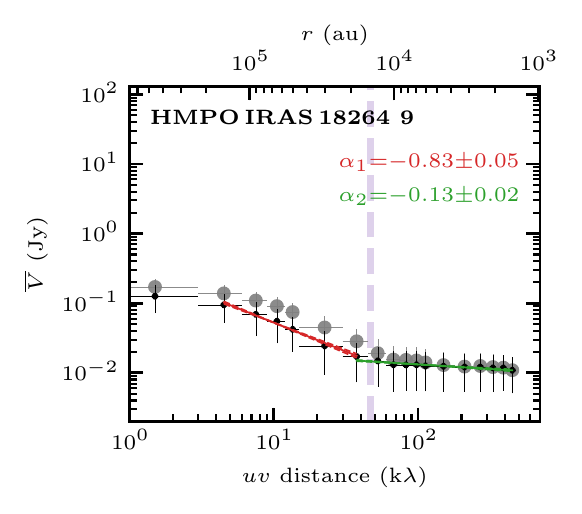}
\includegraphics[width=0.32\textwidth]{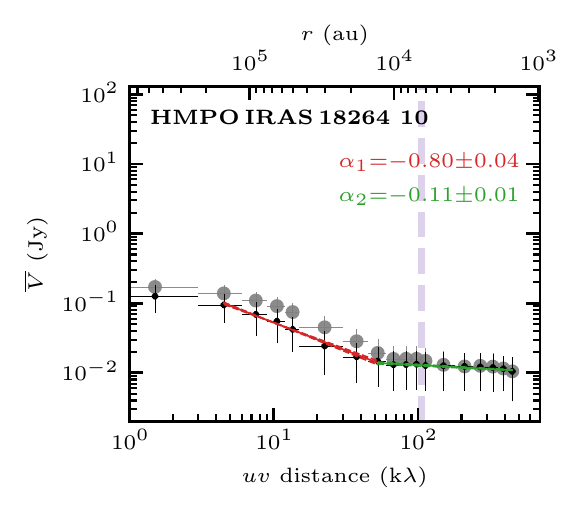}
\includegraphics[width=0.32\textwidth]{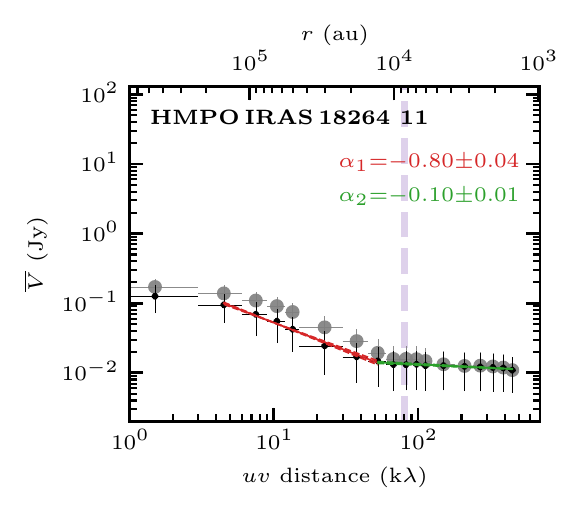}\\
\includegraphics[width=0.32\textwidth]{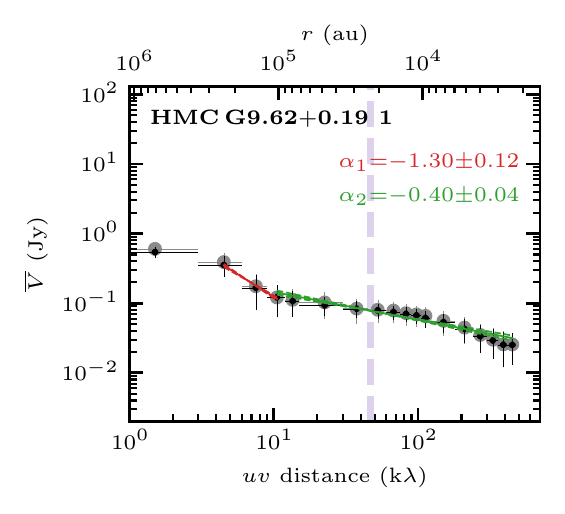}
\includegraphics[width=0.32\textwidth]{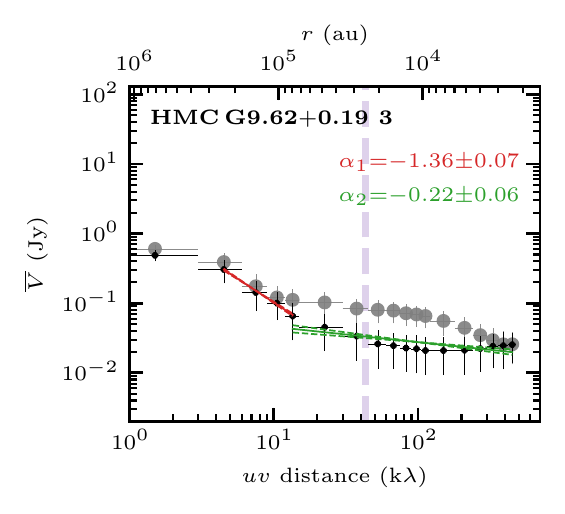}
\includegraphics[width=0.32\textwidth]{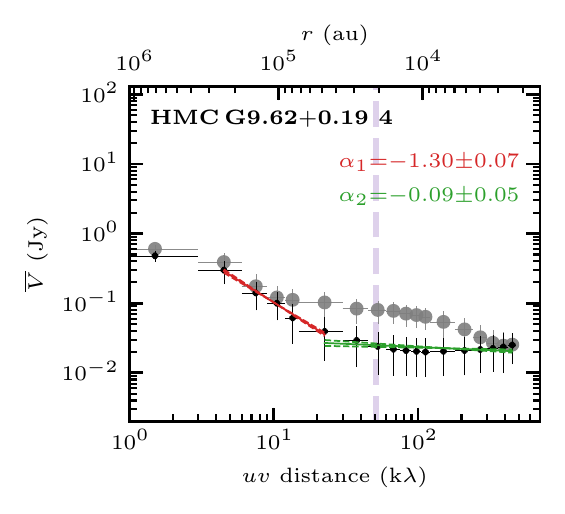}\\
\includegraphics[width=0.32\textwidth]{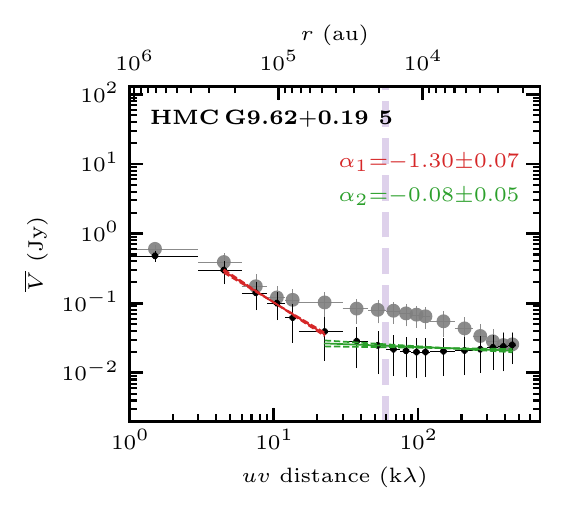}
\includegraphics[width=0.32\textwidth]{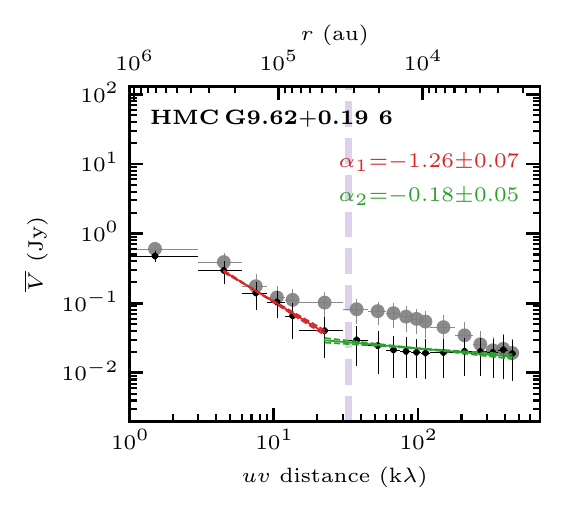}
\includegraphics[width=0.32\textwidth]{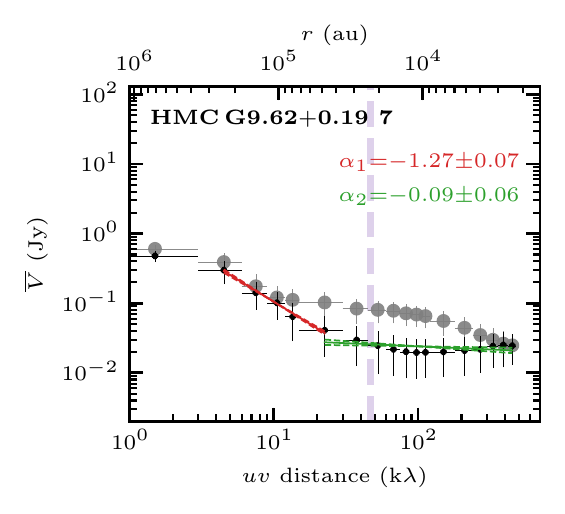}\\
\includegraphics[width=0.32\textwidth]{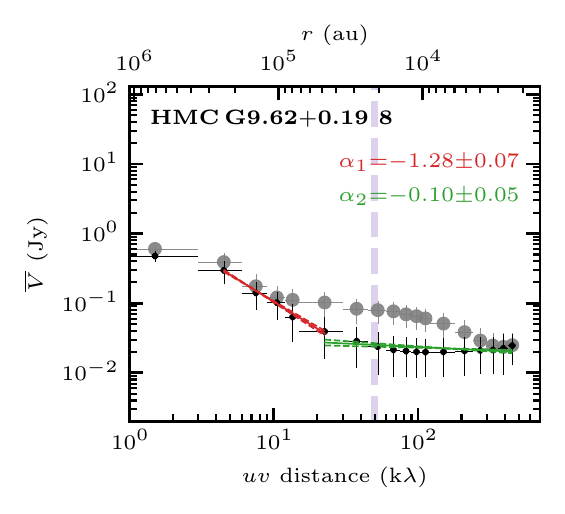}
\includegraphics[width=0.32\textwidth]{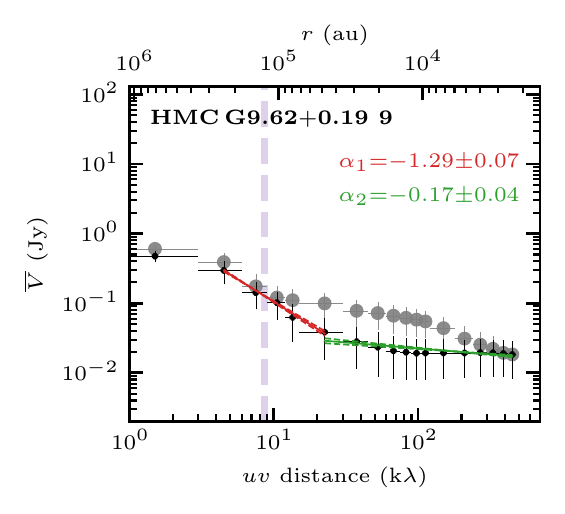}
\includegraphics[width=0.32\textwidth]{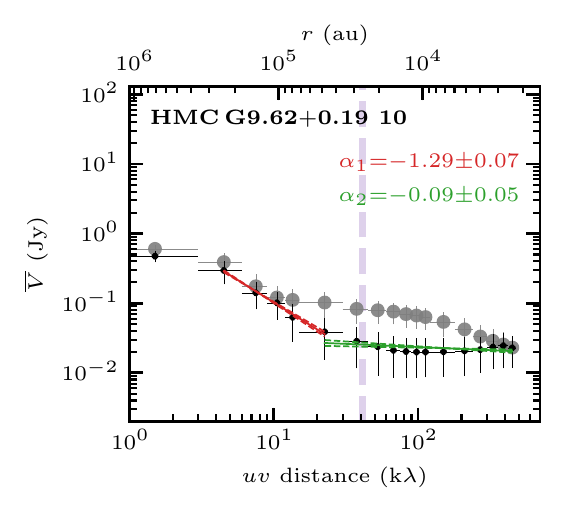}
\caption{Visibility profiles. The profile of the non-core-subtracted and core-subtracted data is shown in grey and black, respectively (further explained in Sect. \ref{sec:ALMAsourcesub}). Two power-law profiles, tracing the clump and core scales, are fitted to the core-subtracted data shown in red and green, respectively. The bottom axis shows the $uv$ distance in k$\lambda$ and the top axis is the corresponding spatial scale. The purple dashed line indicates the diameter of the sources (Table \ref{tab:ALMApositions}). The visibility profile of dust+ff core 2 in HMC\,G9.62$+$.19 is shown in Fig. \ref{fig:ALMAvisibilityprofile}.}
\end{figure*}

\begin{figure*}
\ContinuedFloat
\captionsetup{list=off,format=cont}
\includegraphics[width=0.32\textwidth]{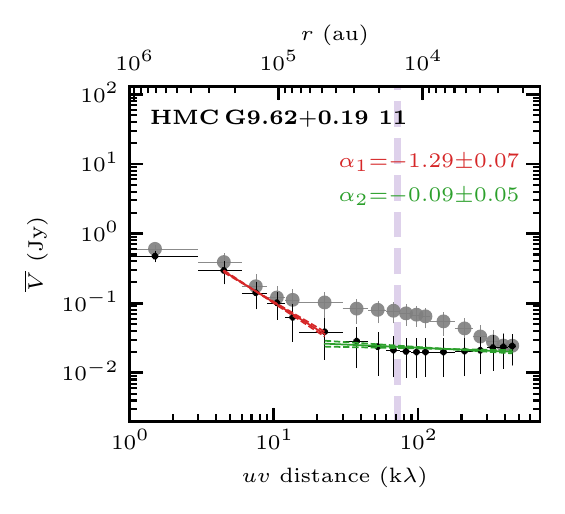}
\includegraphics[width=0.32\textwidth]{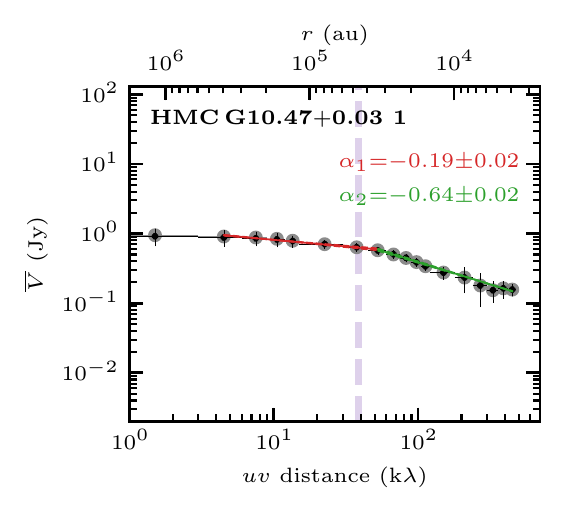}
\includegraphics[width=0.32\textwidth]{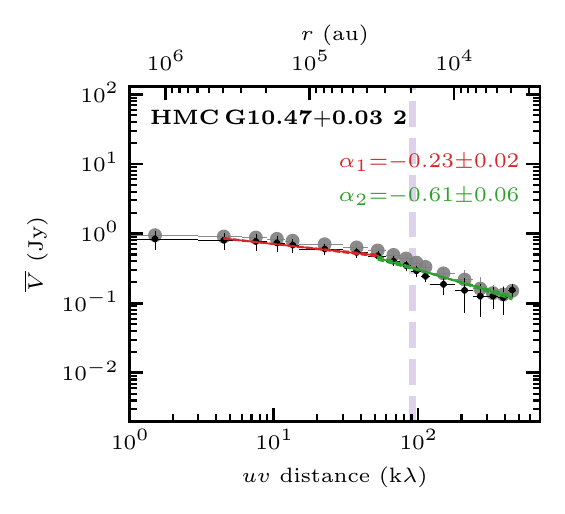}\\
\includegraphics[width=0.32\textwidth]{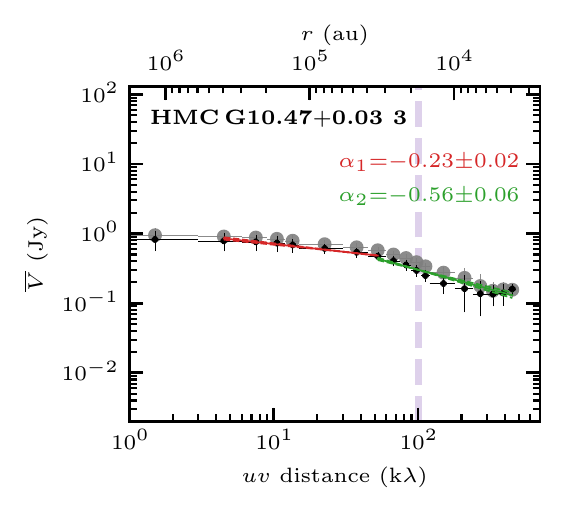}
\includegraphics[width=0.32\textwidth]{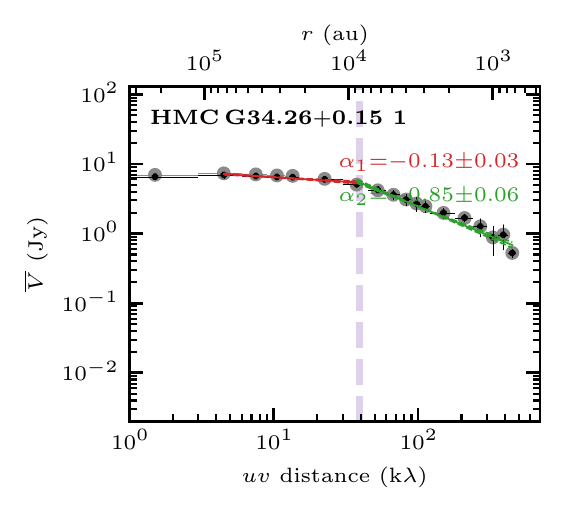}
\includegraphics[width=0.32\textwidth]{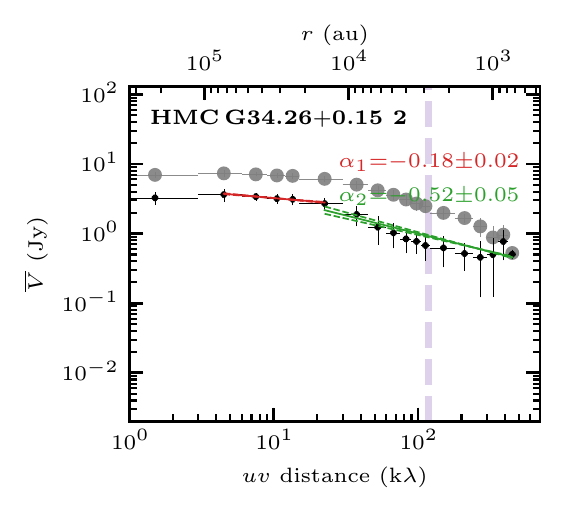}\\
\includegraphics[width=0.32\textwidth]{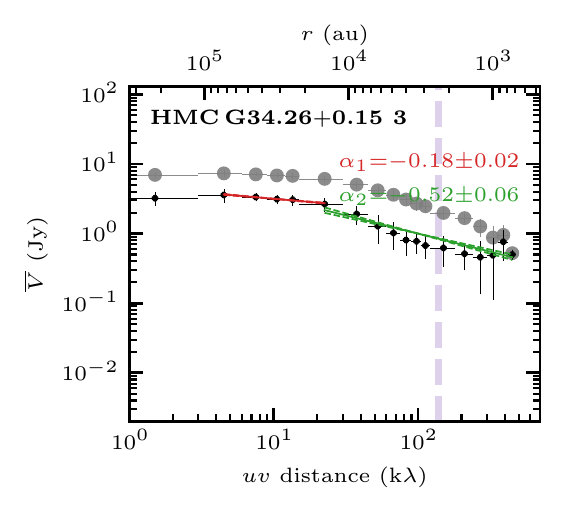}
\includegraphics[width=0.32\textwidth]{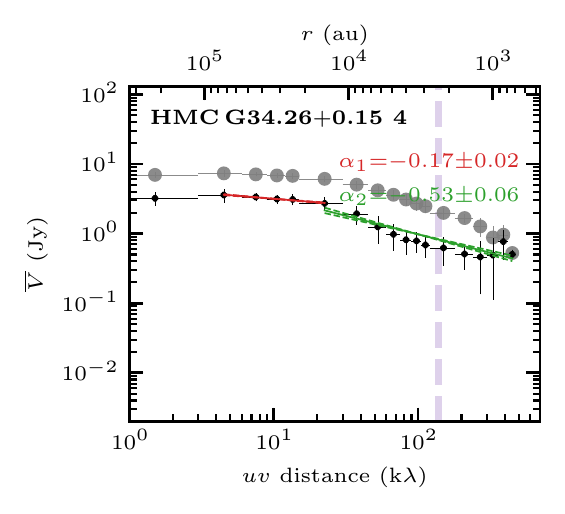}
\includegraphics[width=0.32\textwidth]{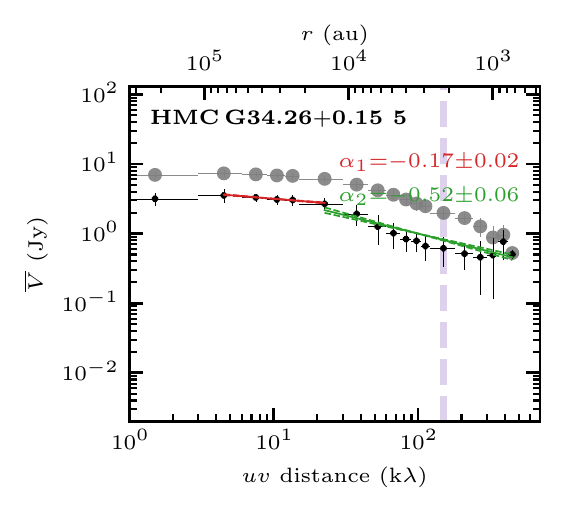}\\
\includegraphics[width=0.32\textwidth]{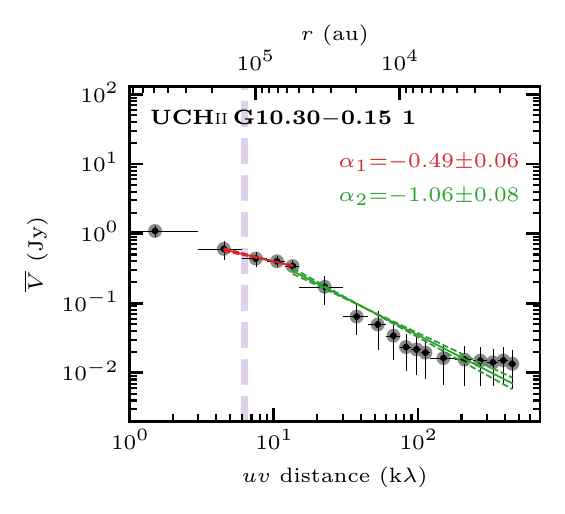}
\includegraphics[width=0.32\textwidth]{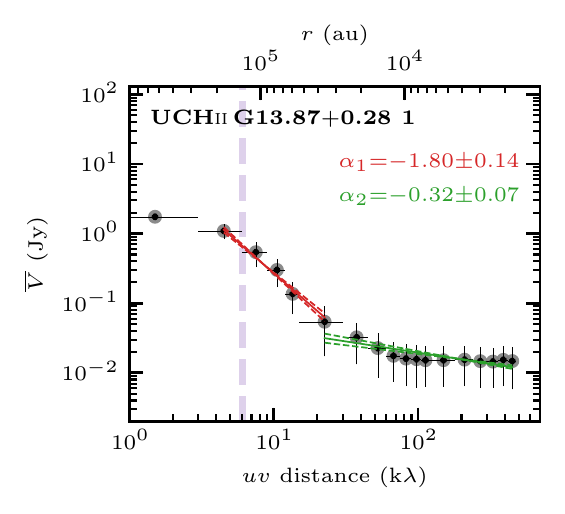}
\caption{Visibility profiles. The profile of the non-core-subtracted and core-subtracted data is shown in grey and black, respectively (further explained in Sect. \ref{sec:ALMAsourcesub}). Two power-law profiles, tracing the clump and core scales, are fitted to the core-subtracted data shown in red and green, respectively. The bottom axis shows the $uv$ distance in k$\lambda$ and the top axis is the corresponding spatial scale. The purple dashed line indicates the diameter of the sources (Table \ref{tab:ALMApositions}). The visibility profile of dust+ff core 2 in HMC\,G9.62$+$.19 is shown in Fig. \ref{fig:ALMAvisibilityprofile}.}
\end{figure*}

\onecolumn
\begin{longtable}{lrrrrrrrl}
\caption{Overview of the fragmented continuum sources in the ALMA sample. The properties are determined using the \texttt{clumpfind} algorithm on the ALMA 3\,mm continuum (Fig. \ref{fig:ALMAcontinuum}), except for the extended cometary UCH{\sc ii} regions (HMC\,G9.62$+$0.19 9, UCH{\sc ii}\,G10.30$-$0.15 1, and UCH{\sc ii}\,G13.87$+$0.28 1), for which the properties were estimated considering the full extent of the structure. The fragments are classified into dust cores, dust+ff cores and cometary UCH{\sc ii} regions, further explained in Sect. \ref{sec:ALMAfrag}. Cores with $S$/$N < 15$ are not further analyzed in this study and are not classified. The relative positions, $\Delta\alpha$ and $\Delta\delta$, are given with respect to the phase center of the region (Table \ref{tab:ALMA_regions}).}
\label{tab:ALMApositions}
\setlength{\tabcolsep}{0pt}
\\
\hline\hline
Position & $\alpha$ & $\delta$ & $\Delta\alpha$ & $\Delta\delta$ & $I_{3\mathrm{mm}}$ & $F_{3\mathrm{mm}}$ & $r_\mathrm{out}$ & classification\\
 & J(2000) & J(2000) & ($''$) & ($''$) & (mJy & (mJy) & (au) &\\
 & & & & & beam$^{-1}$) & & &\\
\hline
\endfirsthead
\caption[]{continued.}\\
\hline\hline
Position & $\alpha$ & $\delta$ & $\Delta\alpha$ & $\Delta\delta$ & $I_{3\mathrm{mm}}$ & $F_{3\mathrm{mm}}$ & $r_\mathrm{out}$ & classification\\
 & J(2000) & J(2000) & ($''$) & ($''$) & (mJy & (mJy) & (au) &\\
 & & & & & beam$^{-1}$) & & &\\
\hline
\endhead
\hline
\endfoot
IRDC\,G11.11$-$4 1 & 18:10:28.28 & $-$19:22:30.7 & $-0.45$ & $+0.90$ & $1.4$ & $10$ & $7\,700$ & dust core\\ 
IRDC\,G11.11$-$4 2 & 18:10:28.11 & $-$19:22:35.5 & $-2.85$ & $-3.90$ & $0.48$ & $2.1$ & $4\,400$ & \\ 
IRDC\,G11.11$-$4 3 & 18:10:28.22 & $-$19:22:34.2 & $-1.35$ & $-2.55$ & $0.37$ & $0.87$ & $2\,900$ & \\ 
IRDC\,G11.11$-$4 4 & 18:10:28.25 & $-$19:22:33.4 & $-0.90$ & $-1.80$ & $0.36$ & $0.64$ & $2\,400$ & \\ 
\hline 
IRDC\,18223$-$3 1 & 18:25:08.58 & $-$12:45:23.9 & $+2.55$ & $-8.25$ & $2.9$ & $11$ & $7\,900$ & dust core\\ 
IRDC\,18223$-$3 2 & 18:25:08.31 & $-$12:45:30.9 & $-1.50$ & $-15.30$ & $1.1$ & $3.0$ & $4\,900$ & dust core\\ 
IRDC\,18223$-$3 3 & 18:25:09.09 & $-$12:44:52.5 & $+9.90$ & $+23.10$ & $0.81$ & $0.89$ & $2\,700$ & dust core\\ 
IRDC\,18223$-$3 4 & 18:25:08.30 & $-$12:45:37.1 & $-1.65$ & $-21.45$ & $0.67$ & $0.75$ & $2\,700$ & dust core\\ 
IRDC\,18223$-$3 5 & 18:25:08.38 & $-$12:45:24.0 & $-0.45$ & $-8.40$ & $0.57$ & $4.4$ & $7\,100$ & dust core\\ 
IRDC\,18223$-$3 6 & 18:25:08.66 & $-$12:45:20.9 & $+3.60$ & $-5.25$ & $0.51$ & $2.4$ & $5\,500$ & dust core\\ 
IRDC\,18223$-$3 7 & 18:25:09.16 & $-$12:45:00.5 & $+10.95$ & $+15.15$ & $0.26$ & $0.30$ & $2\,200$ & \\ 
\hline 
IRDC\,18310$-$4 1 & 18:33:39.37 & $-$08:21:17.1 & $-0.90$ & $-6.60$ & $1.2$ & $1.4$ & $5\,300$ & dust core\\ 
IRDC\,18310$-$4 2 & 18:33:39.49 & $-$08:21:11.1 & $+0.90$ & $-0.60$ & $0.31$ & $1.7$ & $9\,400$ & \\ 
IRDC\,18310$-$4 3 & 18:33:39.27 & $-$08:21:10.1 & $-2.40$ & $+0.45$ & $0.27$ & $0.62$ & $5\,800$ & \\ 
IRDC\,18310$-$4 4 & 18:33:38.80 & $-$08:21:20.4 & $-9.30$ & $-9.90$ & $0.25$ & $0.29$ & $4\,000$ & \\ 
\hline 
HMPO\,IRAS\,18089 1 & 18:11:51.47 & $-$17:31:28.9 & $-0.90$ & $+0.15$ & $24$ & $92$ & $8\,100$ & dust core\\ 
HMPO\,IRAS\,18089 2 & 18:11:50.54 & $-$17:31:28.6 & $-14.10$ & $+0.45$ & $1.9$ & $3.4$ & $2\,400$ & dust core\\ 
HMPO\,IRAS\,18089 3 & 18:11:51.28 & $-$17:31:27.5 & $-3.60$ & $+1.50$ & $1.4$ & $16$ & $5\,700$ & dust core\\ 
HMPO\,IRAS\,18089 4 & 18:11:52.13 & $-$17:31:05.0 & $+8.55$ & $+24.00$ & $1.1$ & $1.1$ & $1\,500$ & dust core\\ 
HMPO\,IRAS\,18089 5 & 18:11:51.95 & $-$17:31:27.8 & $+6.00$ & $+1.20$ & $0.84$ & $2.2$ & $2\,500$ & \\ 
HMPO\,IRAS\,18089 6 & 18:11:51.02 & $-$17:31:22.6 & $-7.35$ & $+6.45$ & $0.67$ & $1.5$ & $2\,000$ & \\ 
HMPO\,IRAS\,18089 7 & 18:11:51.66 & $-$17:31:25.1 & $+1.80$ & $+3.90$ & $0.66$ & $2.6$ & $2\,700$ & \\ 
HMPO\,IRAS\,18089 8 & 18:11:51.53 & $-$17:31:24.7 & $+0.00$ & $+4.35$ & $0.62$ & $3.2$ & $2\,900$ & \\ 
HMPO\,IRAS\,18089 9 & 18:11:49.34 & $-$17:31:24.1 & $-31.35$ & $+4.95$ & $0.61$ & $0.50$ & $1\,200$ & \\ 
HMPO\,IRAS\,18089 10 & 18:11:51.89 & $-$17:31:29.0 & $+5.10$ & $-0.00$ & $0.60$ & $1.9$ & $2\,400$ & \\ 
\hline 
HMPO\,IRAS\,18182 1 & 18:21:09.05 & $-$14:31:48.0 & $-2.55$ & $-2.40$ & $13$ & $75$ & $20\,000$ & dust core\\ 
HMPO\,IRAS\,18182 2 & 18:21:09.06 & $-$14:31:44.7 & $-2.40$ & $+0.90$ & $2.1$ & $4.9$ & $9\,200$ & dust core\\ 
HMPO\,IRAS\,18182 3 & 18:21:08.92 & $-$14:31:53.4 & $-4.35$ & $-7.80$ & $2.1$ & $4.3$ & $7\,900$ & dust core\\ 
HMPO\,IRAS\,18182 4 & 18:21:09.49 & $-$14:31:49.1 & $+3.90$ & $-3.45$ & $0.44$ & $3.4$ & $8\,800$ & dust core\\ 
\hline 
HMPO\,IRAS\,18264 1 & 18:29:14.37 & $-$11:50:22.6 & $-4.80$ & $+1.50$ & $9.9$ & $35$ & $6\,800$ & dust core\\ 
HMPO\,IRAS\,18264 2 & 18:29:14.43 & $-$11:50:24.6 & $-3.90$ & $-0.45$ & $7.8$ & $35$ & $7\,800$ & dust core\\ 
HMPO\,IRAS\,18264 3 & 18:29:14.82 & $-$11:50:25.5 & $+1.80$ & $-1.35$ & $4.0$ & $6.6$ & $4\,800$ & dust core\\ 
HMPO\,IRAS\,18264 4 & 18:29:14.79 & $-$11:50:24.3 & $+1.35$ & $-0.15$ & $3.1$ & $11$ & $6\,100$ & dust core\\ 
HMPO\,IRAS\,18264 5 & 18:29:14.48 & $-$11:50:25.9 & $-3.15$ & $-1.80$ & $2.4$ & $13$ & $8\,800$ & dust core\\ 
HMPO\,IRAS\,18264 6 & 18:29:14.68 & $-$11:50:23.8 & $-0.15$ & $+0.30$ & $1.9$ & $6.2$ & $4\,100$ & dust core\\ 
HMPO\,IRAS\,18264 7 & 18:29:14.49 & $-$11:50:22.2 & $-3.00$ & $+1.95$ & $1.7$ & $4.5$ & $4\,400$ & dust core\\ 
HMPO\,IRAS\,18264 8 & 18:29:14.34 & $-$11:50:20.4 & $-5.25$ & $+3.75$ & $1.4$ & $5.3$ & $5\,600$ & dust core\\ 
HMPO\,IRAS\,18264 9 & 18:29:14.86 & $-$11:50:22.9 & $+2.40$ & $+1.20$ & $1.3$ & $8.3$ & $7\,300$ & dust core\\ 
HMPO\,IRAS\,18264 10 & 18:29:14.20 & $-$11:50:20.2 & $-7.20$ & $+3.90$ & $0.49$ & $1.1$ & $3\,200$ & dust core\\ 
HMPO\,IRAS\,18264 11 & 18:29:14.69 & $-$11:50:26.1 & $+0.00$ & $-1.95$ & $0.48$ & $1.8$ & $4\,200$ & dust core\\ 
HMPO\,IRAS\,18264 12 & 18:29:15.09 & $-$11:50:19.9 & $+5.85$ & $+4.20$ & $0.36$ & $2.6$ & $5\,800$ & \\ 
HMPO\,IRAS\,18264 13 & 18:29:14.10 & $-$11:50:21.0 & $-8.70$ & $+3.15$ & $0.32$ & $1.2$ & $4\,000$ & \\ 
\hline 
HMC\,G9.62$+$0.19 1 & 18:06:14.94 & $-$20:31:43.1 & $+0.15$ & $-3.75$ & $58$ & $100$ & $11\,400$ & dust+ff core\\ 
HMC\,G9.62$+$0.19 2 & 18:06:14.67 & $-$20:31:31.7 & $-3.75$ & $+7.65$ & $24$ & $49$ & $14\,100$ & dust+ff core\\ 
HMC\,G9.62$+$0.19 3 & 18:06:14.88 & $-$20:31:39.5 & $-0.75$ & $-0.15$ & $11$ & $37$ & $12\,400$ & dust core\\ 
HMC\,G9.62$+$0.19 4 & 18:06:14.78 & $-$20:31:35.0 & $-2.10$ & $+4.35$ & $4.5$ & $14$ & $10\,500$ & dust core\\ 
HMC\,G9.62$+$0.19 5 & 18:06:14.82 & $-$20:31:37.4 & $-1.65$ & $+1.95$ & $4.4$ & $12$ & $9\,000$ & dust core\\ 
HMC\,G9.62$+$0.19 6 & 18:06:14.36 & $-$20:31:26.3 & $-8.10$ & $+13.05$ & $3.5$ & $34$ & $16\,300$ & dust+ff core\\ 
HMC\,G9.62$+$0.19 7 & 18:06:15.03 & $-$20:31:41.5 & $+1.35$ & $-2.10$ & $3.5$ & $15$ & $11\,400$ & dust core\\ 
HMC\,G9.62$+$0.19 8 & 18:06:14.77 & $-$20:31:30.4 & $-2.25$ & $+9.00$ & $2.1$ & $11$ & $10\,700$ & dust core\\ 
HMC\,G9.62$+$0.19 9 & 18:06:14.03 & $-$20:31:47.6 & $-12.75$ & $-8.25$ & $2.2$ & $343$ & $62\,400$ & cometary UCH{\sc ii} region\\ 
HMC\,G9.62$+$0.19 10 & 18:06:15.16 & $-$20:31:38.2 & $+3.15$ & $+1.20$ & $0.85$ & $10$ & $13\,000$ & dust core\\ 
HMC\,G9.62$+$0.19 11 & 18:06:15.04 & $-$20:31:46.7 & $+1.50$ & $-7.35$ & $0.75$ & $3.7$ & $7\,500$ & dust core\\ 
HMC\,G9.62$+$0.19 12 & 18:06:14.32 & $-$20:31:23.3 & $-8.55$ & $+16.05$ & $0.54$ & $1.1$ & $4\,500$ & \\ 
HMC\,G9.62$+$0.19 13 & 18:06:14.93 & $-$20:31:33.5 & $+0.00$ & $+5.85$ & $0.54$ & $1.3$ & $4\,600$ & \\ 
HMC\,G9.62$+$0.19 14 & 18:06:15.04 & $-$20:31:34.0 & $+1.50$ & $+5.40$ & $0.52$ & $2.5$ & $6\,700$ & \\ 
HMC\,G9.62$+$0.19 15 & 18:06:15.01 & $-$20:31:36.1 & $+1.05$ & $+3.30$ & $0.50$ & $0.73$ & $3\,400$ & \\ 
HMC\,G9.62$+$0.19 16 & 18:06:15.04 & $-$20:31:35.2 & $+1.50$ & $+4.20$ & $0.49$ & $0.71$ & $3\,500$ & \\ 
HMC\,G9.62$+$0.19 17 & 18:06:15.27 & $-$20:31:40.9 & $+4.80$ & $-1.50$ & $0.49$ & $3.4$ & $8\,300$ & \\ 
\hline 
HMC\,G10.47$+$0.03 1 & 18:08:38.24 & $-$19:51:50.5 & $+0.45$ & $-0.30$ & $336$ & $776$ & $22\,800$ & dust+ff core\\ 
HMC\,G10.47$+$0.03 2 & 18:08:38.36 & $-$19:51:46.0 & $+2.10$ & $+4.20$ & $7.2$ & $23$ & $9\,700$ & dust+ff core\\ 
HMC\,G10.47$+$0.03 3 & 18:08:38.26 & $-$19:51:48.4 & $+0.75$ & $+1.80$ & $5.4$ & $14$ & $8\,800$ & dust core\\ 
HMC\,G10.47$+$0.03 4 & 18:08:36.48 & $-$19:52:15.1 & $-24.45$ & $-24.90$ & $3.2$ & $3.4$ & $4\,900$ & \\ 
\hline 
HMC\,G34.26$+$0.15 1 & 18:53:18.59 & $+$01:14:58.0 & $+0.60$ & $+0.30$ & $2502$ & $6636$ & $4\,200$ & dust+ff core\\ 
HMC\,G34.26$+$0.15 2 & 18:53:18.68 & $+$01:15:00.1 & $+1.95$ & $+2.40$ & $111$ & $145$ & $1\,400$ & dust+ff core\\ 
HMC\,G34.26$+$0.15 3 & 18:53:18.80 & $+$01:14:56.1 & $+3.75$ & $-1.65$ & $81$ & $100$ & $1\,200$ & dust+ff core\\ 
HMC\,G34.26$+$0.15 4 & 18:53:18.70 & $+$01:14:57.3 & $+2.25$ & $-0.45$ & $62$ & $89$ & $1\,200$ & dust core\\ 
HMC\,G34.26$+$0.15 5 & 18:53:18.72 & $+$01:14:55.6 & $+2.55$ & $-2.10$ & $35$ & $46$ & $1\,100$ & dust core\\ 
HMC\,G34.26$+$0.15 6 & 18:53:18.69 & $+$01:14:54.7 & $+2.10$ & $-3.00$ & $19$ & $14$ & $700$ & \\ 
HMC\,G34.26$+$0.15 7 & 18:53:18.22 & $+$01:15:00.6 & $-4.95$ & $+2.85$ & $18$ & $64$ & $1\,600$ & \\ 
\hline 
UCH{\sc ii}\,G10.30$-$0.15 1 & 18:08:55.96 & $-$20:05:55.5 & $+0.30$ & $-0.75$ & $18$ & $1006$ & $59\,300$ & cometary UCH{\sc ii} region\\ 
\hline 
UCH{\sc ii}\,G13.87$+$0.28 1 & 18:14:35.81 & $-$16:45:37.4 & $-2.10$ & $-0.75$ & $7.0$ & $1819$ & $66\,100$ & cometary UCH{\sc ii} region\\ 
\end{longtable}

\begin{longtable}{lrrr|rrrl}
\caption[Radial temperature profiles in the ALMA sample.]{Radial temperature profiles in the ALMA sample. The radial temperature profiles and power-law fit according to Eq. \eqref{eq:temperatureprofile} are shown in Figs. \ref{fig:ALMATrad} and \ref{fig:ALMATradapp}.}
\label{tab:ALMAradialtemp}
\\
\hline\hline
 & \multicolumn{3}{c}{HCN-to-HNC ratio} & \multicolumn{3}{c}{CH$_{3}$CN} & \\
 & $r_\mathrm{in}$ & $T_\mathrm{in}$ & $q$ & $r_\mathrm{in}$ & $T_\mathrm{in}$ & $q$ & classification\\
 & (au) & (K) & & (au) & (K) & & \\
\hline
\endfirsthead
\caption[]{continued.}\\
\hline\hline
 & \multicolumn{3}{c}{HCN-to-HNC ratio} & \multicolumn{3}{c}{CH$_{3}$CN} & \\
 & $r_\mathrm{in}$ & $T_\mathrm{in}$ & $q$ & $r_\mathrm{in}$ & $T_\mathrm{in}$ & $q$ & classification\\
 & (au) & (K) & & (au) & (K) & & \\
\hline
\endhead
\hline
\endfoot
IRDC\,G11.11$-$4 1 & \ldots & \ldots & \ldots & 2175 & 39.9$\pm$1.5 & 0.05$\pm$0.01 & dust core\\ 
IRDC\,18223$-$3 1 & 765 & 34.0$\pm$1.6 & 0.14$\pm$0.01 & \ldots & \ldots & \ldots & dust core\\ 
IRDC\,18223$-$3 2 & 765 & 25.2$\pm$0.6 & 0.11$\pm$0.01 & \ldots & \ldots & \ldots & dust core\\ 
IRDC\,18223$-$3 3 & \ldots & \ldots & \ldots & \ldots & \ldots & \ldots & dust core\\ 
IRDC\,18223$-$3 4 & \ldots & \ldots & \ldots & \ldots & \ldots & \ldots & dust core\\ 
IRDC\,18223$-$3 5 & 765 & 32.2$\pm$0.8 & 0.17$\pm$0.01 & \ldots & \ldots & \ldots & dust core\\ 
IRDC\,18223$-$3 6 & 2295 & 30.0$\pm$1.1 & 0.11$\pm$0.01 & \ldots & \ldots & \ldots & dust core\\ 
IRDC\,18310$-$4 1 & \ldots & \ldots & \ldots & \ldots & \ldots & \ldots & dust core\\ 
HMPO\,IRAS\,18089 1 & \ldots & \ldots & \ldots & 5625 & 149.5$\pm$11.4 & 0.76$\pm$0.1 & dust core\\ 
HMPO\,IRAS\,18089 2 & \ldots & \ldots & \ldots & \ldots & \ldots & \ldots & dust core\\ 
HMPO\,IRAS\,18089 3 & \ldots & \ldots & \ldots & 5625 & 118.6$\pm$35.9 & 0.75$\pm$0.14 & dust core\\ 
HMPO\,IRAS\,18089 4 & \ldots & \ldots & \ldots & \ldots & \ldots & \ldots & dust core\\ 
HMPO\,IRAS\,18182 1 & 3172 & 26.4$\pm$1.0 & 0.07$\pm$0.01 & 3172 & 107.7$\pm$30.8 & 0.68$\pm$0.07 & dust core\\ 
HMPO\,IRAS\,18182 2 & \ldots & \ldots & \ldots & \ldots & \ldots & \ldots & dust core\\ 
HMPO\,IRAS\,18182 3 & 3172 & 25.7$\pm$1.5 & 0.11$\pm$0.04 & \ldots & \ldots & \ldots & dust core\\ 
HMPO\,IRAS\,18182 4 & \ldots & \ldots & \ldots & \ldots & \ldots & \ldots & dust core\\ 
HMPO\,IRAS\,18264 1 & 2228 & 32.0$\pm$1.5 & 0.19$\pm$0.01 & 2228 & 122.9$\pm$3.0 & 0.38$\pm$0.03 & dust core\\ 
HMPO\,IRAS\,18264 2 & 2228 & 38.9$\pm$2.2 & 0.22$\pm$0.01 & 2228 & 121.1$\pm$6.1 & 0.22$\pm$0.02 & dust core\\ 
HMPO\,IRAS\,18264 3 & 2228 & 34.7$\pm$1.8 & 0.11$\pm$0.02 & \ldots & \ldots & \ldots & dust core\\ 
HMPO\,IRAS\,18264 4 & \ldots & \ldots & \ldots & \ldots & \ldots & \ldots & dust core\\ 
HMPO\,IRAS\,18264 5 & 2228 & 36.9$\pm$3.6 & 0.21$\pm$0.01 & 742 & 129.8$\pm$2.9 & 0.49$\pm$0.04 & dust core\\ 
HMPO\,IRAS\,18264 6 & \ldots & \ldots & \ldots & \ldots & \ldots & \ldots & dust core\\ 
HMPO\,IRAS\,18264 7 & \ldots & \ldots & \ldots & \ldots & \ldots & \ldots & dust core\\ 
HMPO\,IRAS\,18264 8 & 2228 & 33.1$\pm$1.9 & 0.22$\pm$0.01 & 742 & 111.7$\pm$5.9 & 0.4$\pm$0.03 & dust core\\ 
HMPO\,IRAS\,18264 9 & \ldots & \ldots & \ldots & \ldots & \ldots & \ldots & dust core\\ 
HMPO\,IRAS\,18264 10 & \ldots & \ldots & \ldots & \ldots & \ldots & \ldots & dust core\\ 
HMPO\,IRAS\,18264 11 & 2228 & 43.7$\pm$1.7 & 0.31$\pm$0.02 & \ldots & \ldots & \ldots & dust core\\ 
HMC\,G9.62$+$0.19 1 & \ldots & \ldots & \ldots & 11700 & 93.5$\pm$46.6 & 0.67$\pm$0.09 & dust+ff core\\ 
HMC\,G9.62$+$0.19 2 & \ldots & \ldots & \ldots & 11700 & 129.8$\pm$21.7 & 0.73$\pm$0.17 & dust+ff core\\ 
HMC\,G9.62$+$0.19 3 & \ldots & \ldots & \ldots & 11700 & 157.4$\pm$23.0 & 0.6$\pm$0.09 & dust core\\ 
HMC\,G9.62$+$0.19 4 & \ldots & \ldots & \ldots & \ldots & \ldots & \ldots & dust core\\ 
HMC\,G9.62$+$0.19 5 & \ldots & \ldots & \ldots & 11700 & 130.7$\pm$11.8 & 0.7$\pm$0.06 & dust core\\ 
HMC\,G9.62$+$0.19 6 & \ldots & \ldots & \ldots & \ldots & \ldots & \ldots & dust+ff core\\ 
HMC\,G9.62$+$0.19 7 & \ldots & \ldots & \ldots & 11700 & 114.6$\pm$9.9 & 0.64$\pm$0.06 & dust core\\ 
HMC\,G9.62$+$0.19 8 & \ldots & \ldots & \ldots & 11700 & 105.6$\pm$8.1 & 0.34$\pm$0.08 & dust core\\ 
HMC\,G9.62$+$0.19 9 & \ldots & \ldots & \ldots & \ldots & \ldots & \ldots & cometary UCH{\sc ii} region\\ 
HMC\,G9.62$+$0.19 10 & \ldots & \ldots & \ldots & \ldots & \ldots & \ldots & dust core\\ 
HMC\,G9.62$+$0.19 11 & \ldots & \ldots & \ldots & \ldots & \ldots & \ldots & dust core\\ 
HMC\,G10.47$+$0.03 1 & \ldots & \ldots & \ldots & 19350 & 210.4$\pm$31.9 & 0.62$\pm$0.13 & dust+ff core\\ 
HMC\,G10.47$+$0.03 2 & 6450 & 21.6$\pm$0.7 & 0.16$\pm$0.02 & 19350 & 143.2$\pm$4.3 & 0.24$\pm$0.06 & dust+ff core\\ 
HMC\,G10.47$+$0.03 3 & \ldots & \ldots & \ldots & 19350 & 167.1$\pm$23.5 & 0.27$\pm$0.19 & dust core\\ 
HMC\,G34.26$+$0.15 1 & \ldots & \ldots & \ldots & 1080 & 245.8$\pm$63.2 & 0.68$\pm$0.04 & dust+ff core\\ 
HMC\,G34.26$+$0.15 2 & \ldots & \ldots & \ldots & 1080 & 128.1$\pm$5.5 & 0.16$\pm$0.04 & dust+ff core\\ 
HMC\,G34.26$+$0.15 3 & \ldots & \ldots & \ldots & \ldots & \ldots & \ldots & dust+ff core\\ 
HMC\,G34.26$+$0.15 4 & \ldots & \ldots & \ldots & 360 & 246.6$\pm$8.0 & 0.22$\pm$0.03 & dust core\\ 
HMC\,G34.26$+$0.15 5 & \ldots & \ldots & \ldots & 360 & 233.5$\pm$5.4 & 0.24$\pm$0.0 & dust core\\ 
UCH{\sc ii}\,G10.30$-$0.15 1 & \ldots & \ldots & \ldots & \ldots & \ldots & \ldots & cometary UCH{\sc ii} region\\ 
UCH{\sc ii}\,G13.87$+$0.28 1 & \ldots & \ldots & \ldots & 2925 & 36.2$\pm$1.4 & 0.06$\pm$0.03 & cometary UCH{\sc ii} region\\ 
\end{longtable}

\setlength{\tabcolsep}{3pt}
\begin{longtable}{lrrrrrrr}
\caption[H$_{2}$ column density and mass in the ALMA sample.]{H$_{2}$ column density $N$(H$_{2}$) and mass $M$ in the ALMA sample. The beam-averaged temperature $\overline T$ are computed from the temperature maps (Sect. \ref{sec:ALMAtemperature}). The dust continuum optical depth $\tau^\mathrm{cont}_\nu$, H$_{2}$ column density $N$(H$_{2}$), and mass $M$ are calculated according to Eqs. \eqref{eq:opticaldepth}, \eqref{eq:H2calc}, and \eqref{eq:Mcalc}, respectively.}
\label{tab:ALMAbeamavgtempNM}
\\
\hline\hline
 & HCN/HNC & CH$_{3}$CN & CH$_{3}^{13}$CN & & & & \\
 & $\overline T$ & $\overline T$ & $\overline T$ & $\tau^\mathrm{cont}_\nu$ &$N$(H$_{2}$) & $M$ & classification\\
 & (K) & (K) & (K) & & (cm$^{-2}$) & ($M_\odot$) & \\ 
\hline
\endfirsthead
\caption[]{continued.}\\
\hline\hline
 & HCN/HNC & CH$_{3}$CN & CH$_{3}^{13}$CN & & & & \\
 & $\overline T$ & $\overline T$ & $\overline T$ & $\tau^\mathrm{cont}_\nu$ &$N$(H$_{2}$) & $M$ & classification\\
 & (K) & (K) & (K) & & (cm$^{-2}$) & ($M_\odot$) & \\ 
\hline
\endhead
\hline
\multicolumn{8}{l}{Notes. The optical depth $\tau^\mathrm{cont}_\nu$ and column density $N$ are given as a(b) meaning a$\times10^{\mathrm{b}}$.}
\endfoot
IRDC\,G11.11$-$4 1 & \ldots & 39.9$\pm$0.9 & \ldots & 5.9($-$3) & 1.1(24)$\pm$2.2(23) & 0.82$\pm$0.16 & dust core\\ 
IRDC\,18223$-$3 1 & 34.0$\pm$1.5 & 42.4$\pm$5.2 & \ldots & 1.3($-$2) & 2.4(24)$\pm$5.8(23) & 1.21$\pm$0.29 & dust core\\ 
IRDC\,18223$-$3 2 & 25.2$\pm$0.4 & \ldots & \ldots & 9.0($-$3) & 1.7(24)$\pm$3.4(23) & 0.60$\pm$0.12 & dust core\\ 
IRDC\,18223$-$3 3 & \ldots & \ldots & \ldots & \ldots & \ldots & \ldots & dust core\\ 
IRDC\,18223$-$3 4 & \ldots & \ldots & \ldots & \ldots & \ldots & \ldots & dust core\\ 
IRDC\,18223$-$3 5 & 32.2$\pm$0.6 & \ldots & \ldots & 3.4($-$3) & 6.4(23)$\pm$1.3(23) & 0.67$\pm$0.14 & dust core\\ 
IRDC\,18223$-$3 6 & 33.5$\pm$1.2 & \ldots & \ldots & 2.9($-$3) & 5.5(23)$\pm$1.1(23) & 0.36$\pm$0.07 & dust core\\ 
IRDC\,18310$-$4 1 & \ldots & \ldots & \ldots & \ldots & \ldots & \ldots & dust core\\ 
HMPO\,IRAS\,18089 1 & 19.6$\pm$0.4 & 179.0$\pm$1.0 & 96.0$\pm$1.2 & 6.0($-$2) & 1.1(25)$\pm$2.2(24) & 3.20$\pm$0.64 & dust core\\ 
HMPO\,IRAS\,18089 2 & \ldots & \ldots & \ldots & \ldots & \ldots & \ldots & dust core\\ 
HMPO\,IRAS\,18089 3 & \ldots & 151.6$\pm$7.0 & \ldots & 2.1($-$3) & 4.0(23)$\pm$8.3(22) & 0.35$\pm$0.07 & dust core\\ 
HMPO\,IRAS\,18089 4 & \ldots & \ldots & \ldots & \ldots & \ldots & \ldots & dust core\\ 
HMPO\,IRAS\,18182 1 & 29.0$\pm$0.5 & 101.6$\pm$2.4 & \ldots & 2.1($-$2) & 3.9(24)$\pm$7.9(23) & 6.08$\pm$1.22 & dust core\\ 
HMPO\,IRAS\,18182 2 & 22.0$\pm$0.7 & 33.1$\pm$2.2 & \ldots & 1.1($-$2) & 2.1(24)$\pm$4.5(23) & 1.28$\pm$0.27 & dust core\\ 
HMPO\,IRAS\,18182 3 & 26.1$\pm$0.2 & \ldots & \ldots & 1.4($-$2) & 2.6(24)$\pm$5.3(23) & 1.44$\pm$0.29 & dust core\\ 
HMPO\,IRAS\,18182 4 & 18.5$\pm$0.1 & 51.1$\pm$11.3 & \ldots & 1.5($-$3) & 2.8(23)$\pm$8.4(22) & 0.56$\pm$0.17 & dust core\\ 
HMPO\,IRAS\,18264 1 & 35.6$\pm$0.7 & 130.0$\pm$1.1 & \ldots & 1.4($-$2) & 2.6(24)$\pm$5.2(23) & 1.18$\pm$0.24 & dust core\\ 
HMPO\,IRAS\,18264 2 & 42.6$\pm$0.5 & 136.7$\pm$2.7 & \ldots & 1.0($-$2) & 1.9(24)$\pm$3.9(23) & 1.14$\pm$0.23 & dust core\\ 
HMPO\,IRAS\,18264 3 & 36.2$\pm$0.7 & \ldots & \ldots & 2.1($-$2) & 4.0(24)$\pm$8.1(23) & 0.84$\pm$0.17 & dust core\\ 
HMPO\,IRAS\,18264 4 & 20.8$\pm$1.1 & \ldots & \ldots & 3.1($-$2) & 5.7(24)$\pm$1.2(24) & 2.48$\pm$0.52 & dust core\\ 
HMPO\,IRAS\,18264 5 & 39.5$\pm$1.0 & 129.8$\pm$14.3 & \ldots & 3.4($-$3) & 6.4(23)$\pm$1.5(23) & 0.45$\pm$0.10 & dust core\\ 
HMPO\,IRAS\,18264 6 & 22.5$\pm$0.3 & 26.2$\pm$6.4 & \ldots & 1.5($-$2) & 2.7(24)$\pm$9.1(23) & 1.12$\pm$0.37 & dust core\\ 
HMPO\,IRAS\,18264 7 & 22.1$\pm$0.4 & \ldots & \ldots & 1.5($-$2) & 2.9(24)$\pm$5.8(23) & 0.99$\pm$0.20 & dust core\\ 
HMPO\,IRAS\,18264 8 & 37.6$\pm$1.3 & 111.7$\pm$20.9 & \ldots & 2.3($-$3) & 4.3(23)$\pm$1.2(23) & 0.21$\pm$0.06 & dust core\\ 
HMPO\,IRAS\,18264 9 & 18.2$\pm$0.1 & \ldots & \ldots & 1.4($-$2) & 2.7(24)$\pm$5.3(23) & 2.25$\pm$0.45 & dust core\\ 
HMPO\,IRAS\,18264 10 & 19.8$\pm$0.5 & 23.8$\pm$4.7 & \ldots & 4.1($-$3) & 7.7(23)$\pm$2.3(23) & 0.23$\pm$0.07 & dust core\\ 
HMPO\,IRAS\,18264 11 & 43.6$\pm$0.2 & \ldots & \ldots & 2.1($-$3) & 3.9(23)$\pm$7.8(22) & 0.19$\pm$0.04 & dust core\\ 
HMC\,G9.62$+$0.19 1 & \ldots & 72.2$\pm$1.9 & \ldots & 6.4($-$2) & 1.2(25)$\pm$2.4(24) & 6.44$\pm$1.30 & dust+ff core\\ 
HMC\,G9.62$+$0.19 2 & 17.8$\pm$0.2 & 195.0$\pm$29.7 & 124.4$\pm$3.6 & 3.1($-$2) & 5.7(24)$\pm$1.1(24) & 3.69$\pm$0.75 & dust+ff core\\ 
HMC\,G9.62$+$0.19 3 & \ldots & 161.9$\pm$6.2 & 120.8$\pm$7.9 & 1.8($-$2) & 3.3(24)$\pm$7.0(23) & 3.55$\pm$0.75 & dust core\\ 
HMC\,G9.62$+$0.19 4 & \ldots & 97.8$\pm$6.3 & \ldots & 8.9($-$3) & 1.7(24)$\pm$3.5(23) & 1.72$\pm$0.36 & dust core\\ 
HMC\,G9.62$+$0.19 5 & 15.2$\pm$0.0 & 160.2$\pm$36.9 & \ldots & 5.3($-$3) & 1.0(24)$\pm$3.1(23) & 0.90$\pm$0.28 & dust core\\ 
HMC\,G9.62$+$0.19 6 & 24.2$\pm$0.7 & 36.2$\pm$1.2 & \ldots & 8.2($-$4) & 1.5(23)$\pm$3.1(22) & 0.47$\pm$0.10 & dust+ff core\\ 
HMC\,G9.62$+$0.19 7 & \ldots & 109.5$\pm$17.7 & \ldots & 6.2($-$3) & 1.2(24)$\pm$3.0(23) & 1.57$\pm$0.41 & dust core\\ 
HMC\,G9.62$+$0.19 8 & \ldots & 108.0$\pm$1.6 & \ldots & 3.8($-$3) & 7.2(23)$\pm$1.5(23) & 1.17$\pm$0.23 & dust core\\ 
HMC\,G9.62$+$0.19 9 & \ldots & \ldots & \ldots & \ldots & \ldots & \ldots & cometary UCH{\sc ii} region\\ 
HMC\,G9.62$+$0.19 10 & \ldots & 80.6$\pm$4.0 & \ldots & 2.1($-$3) & 3.9(23)$\pm$8.0(22) & 1.47$\pm$0.30 & dust core\\ 
HMC\,G9.62$+$0.19 11 & \ldots & 36.1$\pm$3.0 & \ldots & 4.2($-$3) & 7.9(23)$\pm$1.7(23) & 1.26$\pm$0.28 & dust core\\ 
HMC\,G10.47$+$0.03 1 & \ldots & 89.0$\pm$32.6 & 148.9$\pm$10.2 & 5.0($-$1) & 7.4(25)$\pm$1.6(25) & 150.61$\pm$31.89 & dust+ff core\\ 
HMC\,G10.47$+$0.03 2 & 21.6$\pm$0.5 & 192.8$\pm$21.5 & \ldots & 7.7($-$4) & 1.5(23)$\pm$3.3(22) & 0.40$\pm$0.09 & dust+ff core\\ 
HMC\,G10.47$+$0.03 3 & 16.1$\pm$0.6 & 216.5$\pm$13.6 & 143.4$\pm$5.1 & 9.1($-$3) & 1.7(24)$\pm$3.5(23) & 3.90$\pm$0.79 & dust core\\ 
HMC\,G34.26$+$0.15 1 & \ldots & 249.8$\pm$0.2 & 153.6$\pm$7.1 & \ldots & 2.2(26)$\pm$4.6(25) & 17.84$\pm$3.66 & dust+ff core\\ 
HMC\,G34.26$+$0.15 2 & \ldots & 157.4$\pm$21.6 & \ldots & 3.7($-$2) & 6.8(24)$\pm$1.7(24) & 0.27$\pm$0.07 & dust+ff core\\ 
HMC\,G34.26$+$0.15 3 & \ldots & 112.9$\pm$10.3 & \ldots & 1.8($-$2) & 3.5(24)$\pm$7.6(23) & 0.13$\pm$0.03 & dust+ff core\\ 
HMC\,G34.26$+$0.15 4 & 38.1$\pm$1.2 & 246.6$\pm$3.6 & 173.5$\pm$13.6 & 8.6($-$2) & 1.6(25)$\pm$3.3(24) & 0.68$\pm$0.15 & dust core\\ 
HMC\,G34.26$+$0.15 5 & 21.5$\pm$0.4 & 233.5$\pm$2.5 & 179.6$\pm$14.0 & 4.7($-$2) & 8.6(24)$\pm$1.9(24) & 0.34$\pm$0.07 & dust core\\ 
UCH{\sc ii}\,G10.30$-$0.15 1 & 16.5$\pm$0.1 & 31.4$\pm$1.3 & \ldots & 1.4($-$2) & 2.6(24)$\pm$5.3(23) & 22.71$\pm$4.65 & cometary UCH{\sc ii} region\\ 
UCH{\sc ii}\,G13.87$+$0.28 1 & \ldots & 36.2$\pm$1.6 & \ldots & 8.7($-$3) & 1.6(24)$\pm$3.4(23) & 76.30$\pm$15.68 & cometary UCH{\sc ii} region\\ 
\end{longtable}

\begin{longtable}{lrrrrr}
\caption{Visibility and density profiles in the ALMA sample. The visibility profiles are shown in Figs. \ref{fig:ALMAvisibilityprofile} (HMC\,G9.62$+$0.19 2) and \ref{fig:ALMAvisibilityprofileapp} (all remaining sources). Two power-law profiles were fitted to the observed visibility profile with $\alpha_1$ and $\alpha_2$ tracing roughly the clump and core scales, respectively. The corresponding density power-law indices $p_1$ and $p_1$ (Eq. \ref{eq:densityprofile}) are then estimated according to Eq. \eqref{eq:uvanalysis} considering the observed temperature power-law index $q$ (Table \ref{tab:ALMAradialtemp}).}
\label{tab:ALMAdens}
\\
\hline\hline
 & $\alpha_{1}$ & $p_{1}$ & $\alpha_{2}$ & $p_{2}$ & classification\\
\hline
\endfirsthead
\caption[]{continued.}\\
\hline\hline
 & $\alpha_{1}$ & $p_{1}$ & $\alpha_{2}$ & $p_{2}$ & classification\\
\hline
\endhead
\hline
\endfoot
IRDC\,G11.11$-$4 1 & $-0.42\pm0.06$ & $2.53\pm0.06$ & $-0.06\pm0.02$ & $2.89\pm0.02$ & dust core\\ 
IRDC\,18223$-$3 1 & $-0.60\pm0.05$ & $2.26\pm0.05$ & $-0.03\pm0.01$ & $2.83\pm0.01$ & dust core\\ 
IRDC\,18223$-$3 2 & $-0.61\pm0.06$ & $2.28\pm0.06$ & $-0.01\pm0.01$ & $2.88\pm0.01$ & dust core\\ 
IRDC\,18223$-$3 3 & $-0.61\pm0.06$ & \ldots & $-0.02\pm0.01$ & \ldots & dust core\\ 
IRDC\,18223$-$3 4 & $-0.61\pm0.06$ & \ldots & $-0.00\pm0.01$ & \ldots & dust core\\ 
IRDC\,18223$-$3 5 & $-0.61\pm0.06$ & $2.22\pm0.06$ & $-0.00\pm0.01$ & $2.83\pm0.01$ & dust core\\ 
IRDC\,18223$-$3 6 & $-0.61\pm0.06$ & $2.28\pm0.06$ & $-0.00\pm0.01$ & $2.89\pm0.01$ & dust core\\ 
IRDC\,18310$-$4 1 & $-0.58\pm0.07$ & \ldots & $-0.00\pm0.01$ & \ldots & dust core\\ 
HMPO\,IRAS\,18089 1 & $-0.49\pm0.02$ & $1.75\pm0.10$ & $-0.49\pm0.03$ & $1.75\pm0.10$ & dust core\\ 
HMPO\,IRAS\,18089 2 & $-0.68\pm0.02$ & \ldots & $-0.11\pm0.02$ & \ldots & dust core\\ 
HMPO\,IRAS\,18089 3 & $-0.66\pm0.02$ & $1.59\pm0.14$ & $-0.12\pm0.04$ & $2.13\pm0.15$ & dust core\\ 
HMPO\,IRAS\,18089 4 & $-0.68\pm0.02$ & \ldots & $-0.07\pm0.01$ & \ldots & dust core\\ 
HMPO\,IRAS\,18182 1 & $-0.42\pm0.02$ & $1.90\pm0.07$ & $-0.43\pm0.07$ & $1.89\pm0.10$ & dust core\\ 
HMPO\,IRAS\,18182 2 & $-0.42\pm0.01$ & \ldots & $-0.46\pm0.06$ & \ldots & dust core\\ 
HMPO\,IRAS\,18182 3 & $-0.45\pm0.01$ & $2.44\pm0.04$ & $-0.39\pm0.07$ & $2.50\pm0.08$ & dust core\\ 
HMPO\,IRAS\,18182 4 & $-0.45\pm0.02$ & \ldots & $-0.39\pm0.07$ & \ldots & dust core\\ 
HMPO\,IRAS\,18264 1 & $-0.72\pm0.02$ & $1.90\pm0.04$ & $-0.19\pm0.02$ & $2.43\pm0.04$ & dust core\\ 
HMPO\,IRAS\,18264 2 & $-0.76\pm0.03$ & $2.02\pm0.04$ & $-0.17\pm0.02$ & $2.61\pm0.03$ & dust core\\ 
HMPO\,IRAS\,18264 3 & $-0.85\pm0.05$ & $2.04\pm0.05$ & $-0.13\pm0.01$ & $2.76\pm0.02$ & dust core\\ 
HMPO\,IRAS\,18264 4 & $-0.84\pm0.05$ & \ldots & $-0.14\pm0.02$ & \ldots & dust core\\ 
HMPO\,IRAS\,18264 5 & $-0.82\pm0.04$ & $1.69\pm0.06$ & $-0.09\pm0.01$ & $2.42\pm0.04$ & dust core\\ 
HMPO\,IRAS\,18264 6 & $-0.81\pm0.04$ & \ldots & $-0.10\pm0.01$ & \ldots & dust core\\ 
HMPO\,IRAS\,18264 7 & $-0.81\pm0.03$ & \ldots & $-0.10\pm0.02$ & \ldots & dust core\\ 
HMPO\,IRAS\,18264 8 & $-0.80\pm0.04$ & $1.80\pm0.05$ & $-0.11\pm0.02$ & $2.49\pm0.04$ & dust core\\ 
HMPO\,IRAS\,18264 9 & $-0.83\pm0.05$ & \ldots & $-0.13\pm0.02$ & \ldots & dust core\\ 
HMPO\,IRAS\,18264 10 & $-0.80\pm0.04$ & \ldots & $-0.11\pm0.01$ & \ldots & dust core\\ 
HMPO\,IRAS\,18264 11 & $-0.80\pm0.04$ & $1.89\pm0.04$ & $-0.10\pm0.01$ & $2.59\pm0.02$ & dust core\\ 
HMC\,G9.62$+$0.19 1 & $-1.30\pm0.12$ & $1.03\pm0.15$ & $-0.40\pm0.04$ & $1.93\pm0.10$ & dust+ff core\\ 
HMC\,G9.62$+$0.19 2 & $-1.32\pm0.10$ & $0.95\pm0.20$ & $-0.28\pm0.03$ & $1.99\pm0.17$ & dust+ff core\\ 
HMC\,G9.62$+$0.19 3 & $-1.36\pm0.07$ & $1.04\pm0.11$ & $-0.22\pm0.06$ & $2.18\pm0.11$ & dust core\\ 
HMC\,G9.62$+$0.19 4 & $-1.30\pm0.07$ & \ldots & $-0.09\pm0.05$ & \ldots & dust core\\ 
HMC\,G9.62$+$0.19 5 & $-1.30\pm0.07$ & $1.00\pm0.09$ & $-0.08\pm0.05$ & $2.22\pm0.08$ & dust core\\ 
HMC\,G9.62$+$0.19 6 & $-1.26\pm0.07$ & \ldots & $-0.18\pm0.05$ & \ldots & dust+ff core\\ 
HMC\,G9.62$+$0.19 7 & $-1.27\pm0.07$ & $1.09\pm0.09$ & $-0.09\pm0.06$ & $2.27\pm0.08$ & dust core\\ 
HMC\,G9.62$+$0.19 8 & $-1.28\pm0.07$ & $1.38\pm0.11$ & $-0.10\pm0.05$ & $2.56\pm0.09$ & dust core\\ 
HMC\,G9.62$+$0.19 9 & $-1.29\pm0.07$ & \ldots & $-0.17\pm0.04$ & \ldots & cometary UCH{\sc ii} region\\ 
HMC\,G9.62$+$0.19 10 & $-1.29\pm0.07$ & \ldots & $-0.09\pm0.05$ & \ldots & dust core\\ 
HMC\,G9.62$+$0.19 11 & $-1.29\pm0.07$ & \ldots & $-0.09\pm0.05$ & \ldots & dust core\\ 
HMC\,G10.47$+$0.03 1 & $-0.19\pm0.02$ & $2.19\pm0.13$ & $-0.64\pm0.02$ & $1.74\pm0.13$ & dust+ff core\\ 
HMC\,G10.47$+$0.03 2 & $-0.23\pm0.02$ & $2.53\pm0.06$ & $-0.61\pm0.06$ & $2.15\pm0.08$ & dust+ff core\\ 
HMC\,G10.47$+$0.03 3 & $-0.23\pm0.02$ & $2.50\pm0.19$ & $-0.56\pm0.06$ & $2.17\pm0.20$ & dust core\\ 
HMC\,G34.26$+$0.15 1 & $-0.13\pm0.03$ & $2.19\pm0.05$ & $-0.85\pm0.06$ & $1.47\pm0.07$ & dust+ff core\\ 
HMC\,G34.26$+$0.15 2 & $-0.18\pm0.02$ & $2.66\pm0.04$ & $-0.52\pm0.05$ & $2.32\pm0.06$ & dust+ff core\\ 
HMC\,G34.26$+$0.15 3 & $-0.18\pm0.02$ & \ldots & $-0.52\pm0.06$ & \ldots & dust+ff core\\ 
HMC\,G34.26$+$0.15 4 & $-0.17\pm0.02$ & $2.61\pm0.04$ & $-0.53\pm0.06$ & $2.25\pm0.07$ & dust core\\ 
HMC\,G34.26$+$0.15 5 & $-0.17\pm0.02$ & $2.59\pm0.02$ & $-0.52\pm0.06$ & $2.24\pm0.06$ & dust core\\ 
UCH{\sc ii}\,G10.30$-$0.15 1 & $-0.49\pm0.06$ & \ldots & $-1.06\pm0.08$ & \ldots & cometary UCH{\sc ii} region\\ 
UCH{\sc ii}\,G13.87$+$0.28 1 & $-1.80\pm0.14$ & $1.14\pm0.14$ & $-0.32\pm0.07$ & $2.62\pm0.08$ & cometary UCH{\sc ii} region\\ 
\end{longtable}

\section{Free-free contribution}

\begin{figure}[!htb]
\includegraphics[width=0.94\textwidth]{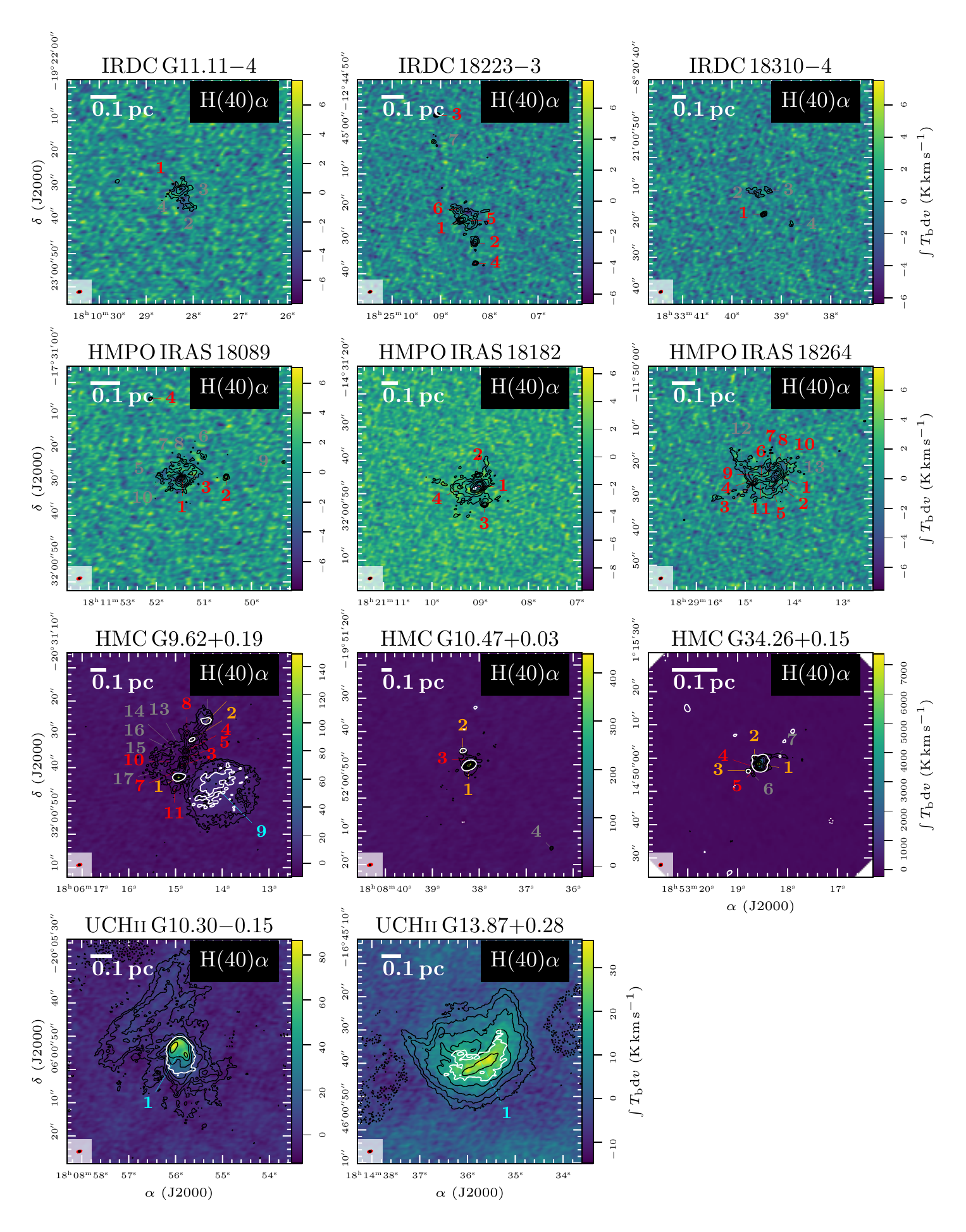}
\caption[Integrated intensity maps of H(40)$\alpha$.]{Integrated intensity maps of H(40)$\alpha$. In each panel, the H(40)$\alpha$ integrated intensity of the region is shown in color and white contours (dotted: $-5\sigma$ and solid: $+5\sigma$). The 3\,mm continuum data is shown by black contours. The dotted black contour marks the $-5\sigma_\mathrm{cont}$ level. The solid black contours start at $5\sigma_\mathrm{cont}$ and contour steps increase by a factor of 2 (e.g., 5, 10, 20, $40\sigma_\mathrm{cont}$). The synthesized beam of the continuum (black) and line (red) data is shown in the bottom left corner. The continuum fragments are classified into dust cores (red), dust+ff cores (orange), cometary UCH{\sc ii} regions (cyan), further explained in Sect. \ref{sec:ALMAfrag}. Fragments with $S$/$N < 15$ are not analyzed in this study and are labeled in grey.}
\label{fig:Halpha_moment0}
\end{figure}

	Figure \ref{fig:Halpha_moment0} show the H(40)$\alpha$ recombination line integrated intensity map of all regions. In Sect. \ref{sec:ALMAffemission} we estimate the contribution of free-free emission at 3\,mm wavelengths based on the H(40)$\alpha$ recombination line (Fig. \ref{fig:freefree}) using Eq. \eqref{eq:ff}.

\begin{figure*}[!htb]
\centering
\includegraphics[width=0.9\textwidth]{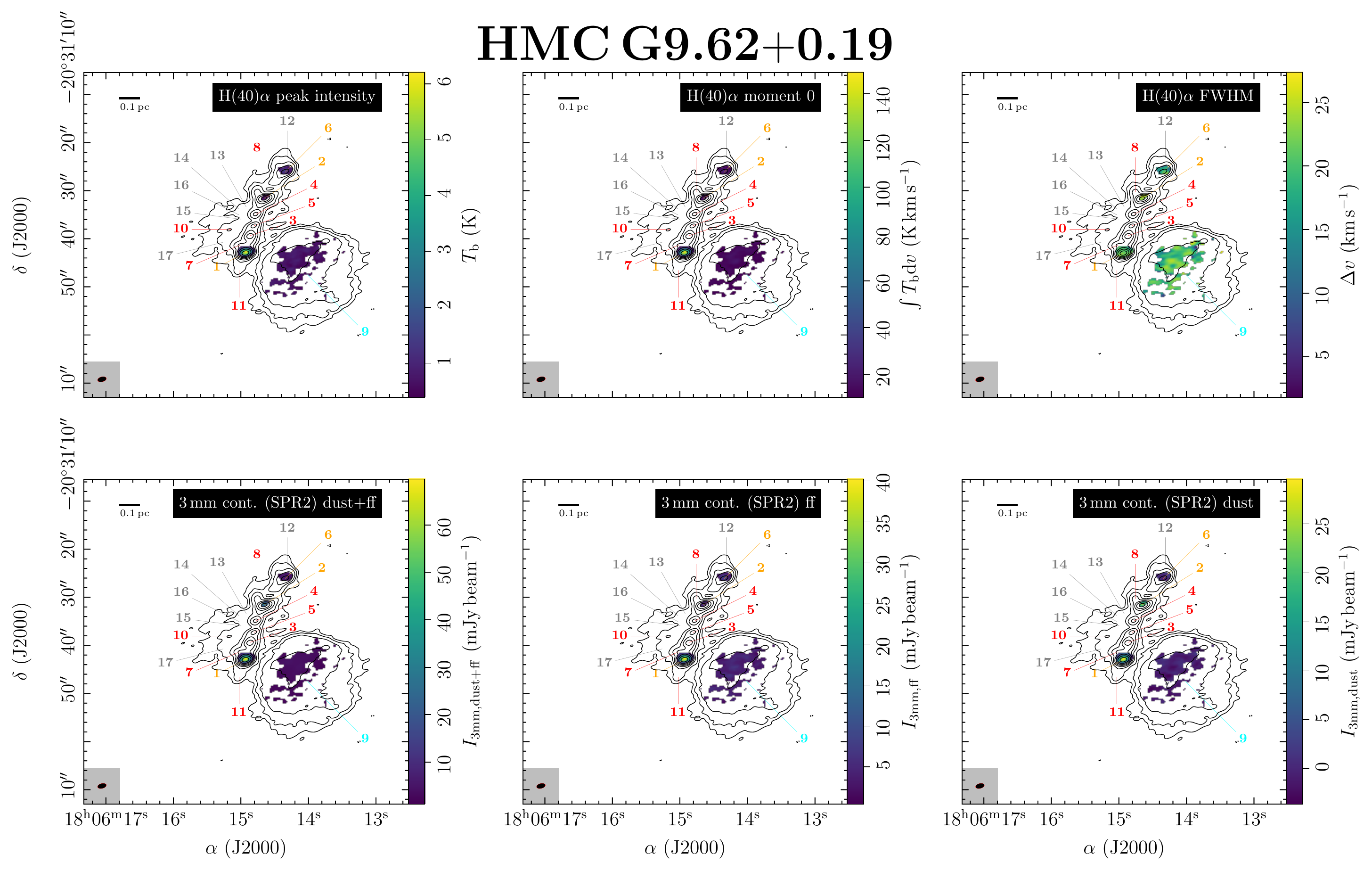}\\
\includegraphics[width=0.9\textwidth]{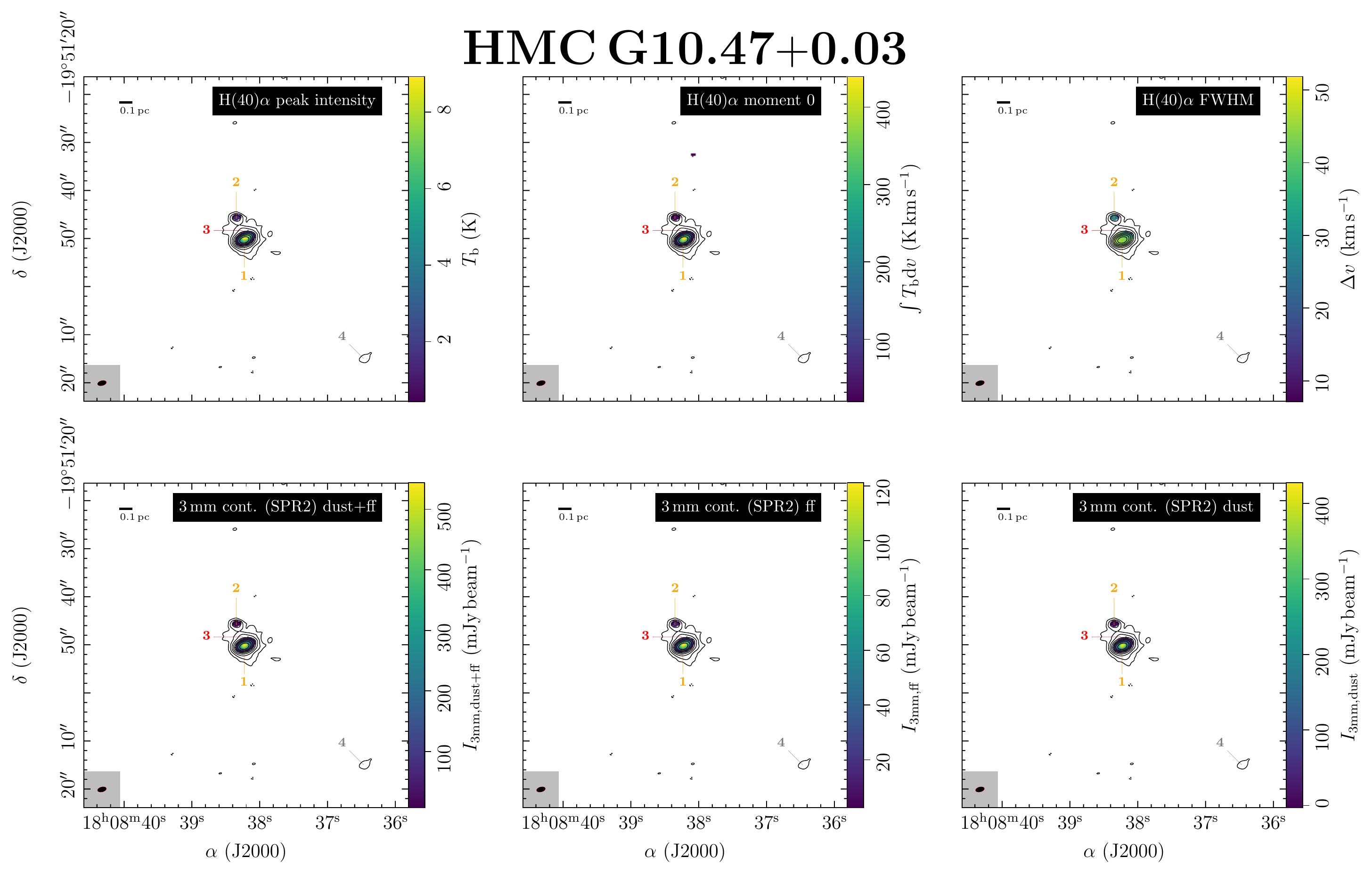}\\
\caption{Estimate of the free-free contribution at 3\,mm (Eq. \ref{eq:ff}). In all panels, the SPR2 3\,mm continuum is shown in contours. The dotted black contour marks the $-5\sigma_\mathrm{cont}$ level. The solid black contours start at $5\sigma_\mathrm{cont}$ and contour steps increase by a factor of 2 (e.g., 5, 10, 20, $40\sigma_\mathrm{cont}$). The top panels show in color the H(40)$\alpha$ peak intensity, integrated intensity, and line width derived from the 2nd moment. The bottom panels show in color the SPR2 3\,mm continuum (dust + ff), 3\,mm ff continuum (estimated using Eq. \ref{eq:ff}), and 3\,mm dust continuum. The synthesized beam size of the continuum and line data is shown in the bottom left corner in black and red, respectively. In all panels, the area where the H(40)$\alpha$ integrated intensity and SPR2 continuum have a $S$/$N < 5$ are masked. The continuum fragments are classified into dust cores (red), dust+ff cores (orange), cometary UCH{\sc ii} regions (cyan), further explained in Sect. \ref{sec:ALMAfrag}. Fragments with $S$/$N < 15$ are not analyzed in this study and are labeled in grey.}
\label{fig:freefree}
\end{figure*}

\begin{figure*}
\ContinuedFloat
\captionsetup{list=off,format=cont}
\includegraphics[width=0.9\textwidth]{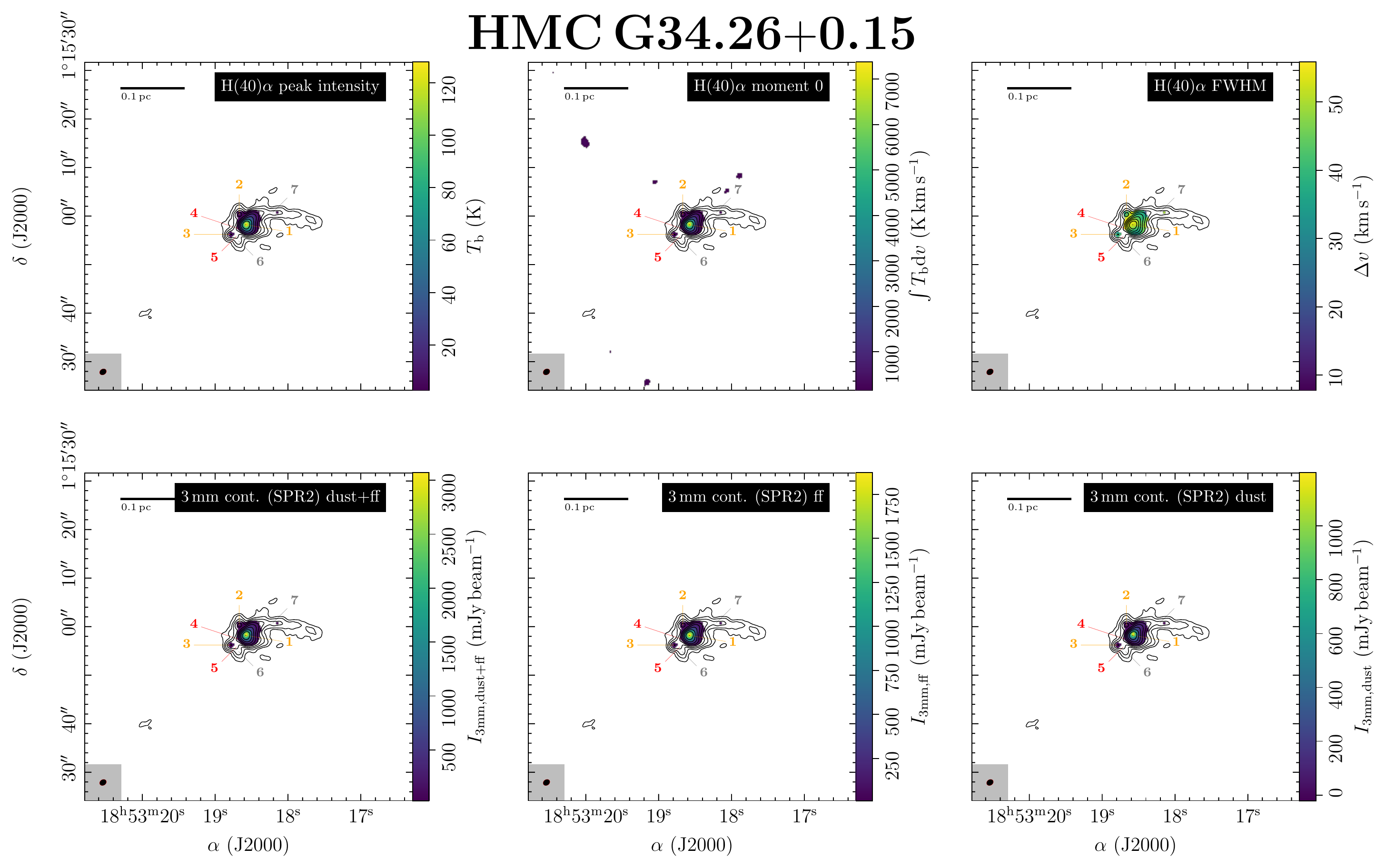}\\
\includegraphics[width=0.9\textwidth]{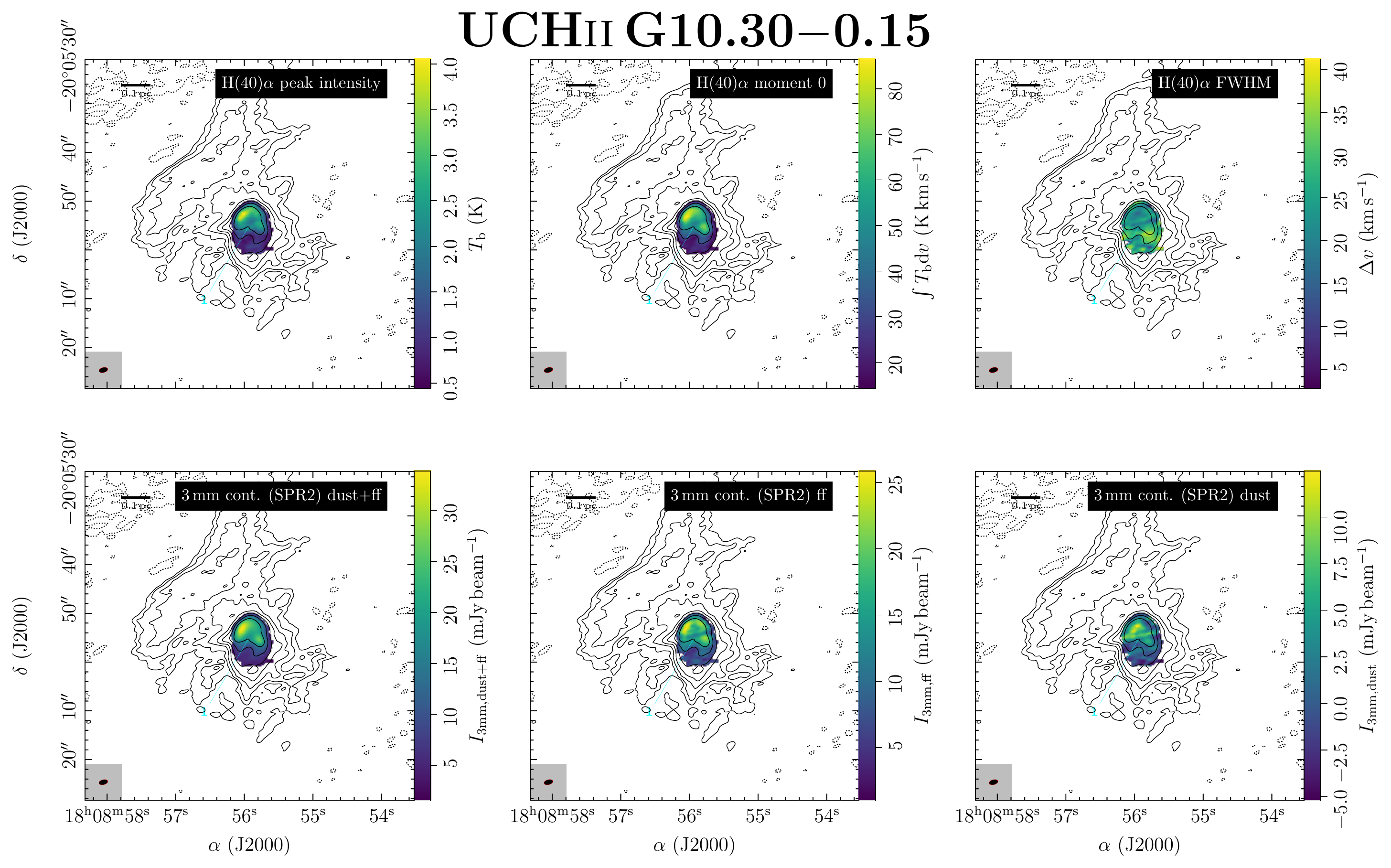}
\caption{Estimate of the free-free contribution at 3\,mm (Eq. \ref{eq:ff}). In all panels, the SPR2 3\,mm continuum is shown in contours. The dotted black contour marks the $-5\sigma_\mathrm{cont}$ level. The solid black contours start at $5\sigma_\mathrm{cont}$ and contour steps increase by a factor of 2 (e.g., 5, 10, 20, $40\sigma_\mathrm{cont}$). The top panels show in color the H(40)$\alpha$ peak intensity, integrated intensity, and line width derived from the 2nd moment. The bottom panels show in color the SPR2 3\,mm continuum (dust + ff), 3\,mm ff continuum (estimated using Eq. \ref{eq:ff}), and 3\,mm dust continuum. The synthesized beam size of the continuum and line data is shown in the bottom left corner in black and red, respectively. In all panels, the area where the H(40)$\alpha$ integrated intensity and SPR2 continuum have a $S$/$N < 5$ are masked. The continuum fragments are classified into dust cores (red), dust+ff cores (orange), cometary UCH{\sc ii} regions (cyan), further explained in Sect. \ref{sec:ALMAfrag}. Fragments with $S$/$N < 15$ are not analyzed in this study and are labeled in grey.}
\end{figure*}

\begin{figure*}
\ContinuedFloat
\captionsetup{list=off,format=cont}
\includegraphics[width=0.99\textwidth]{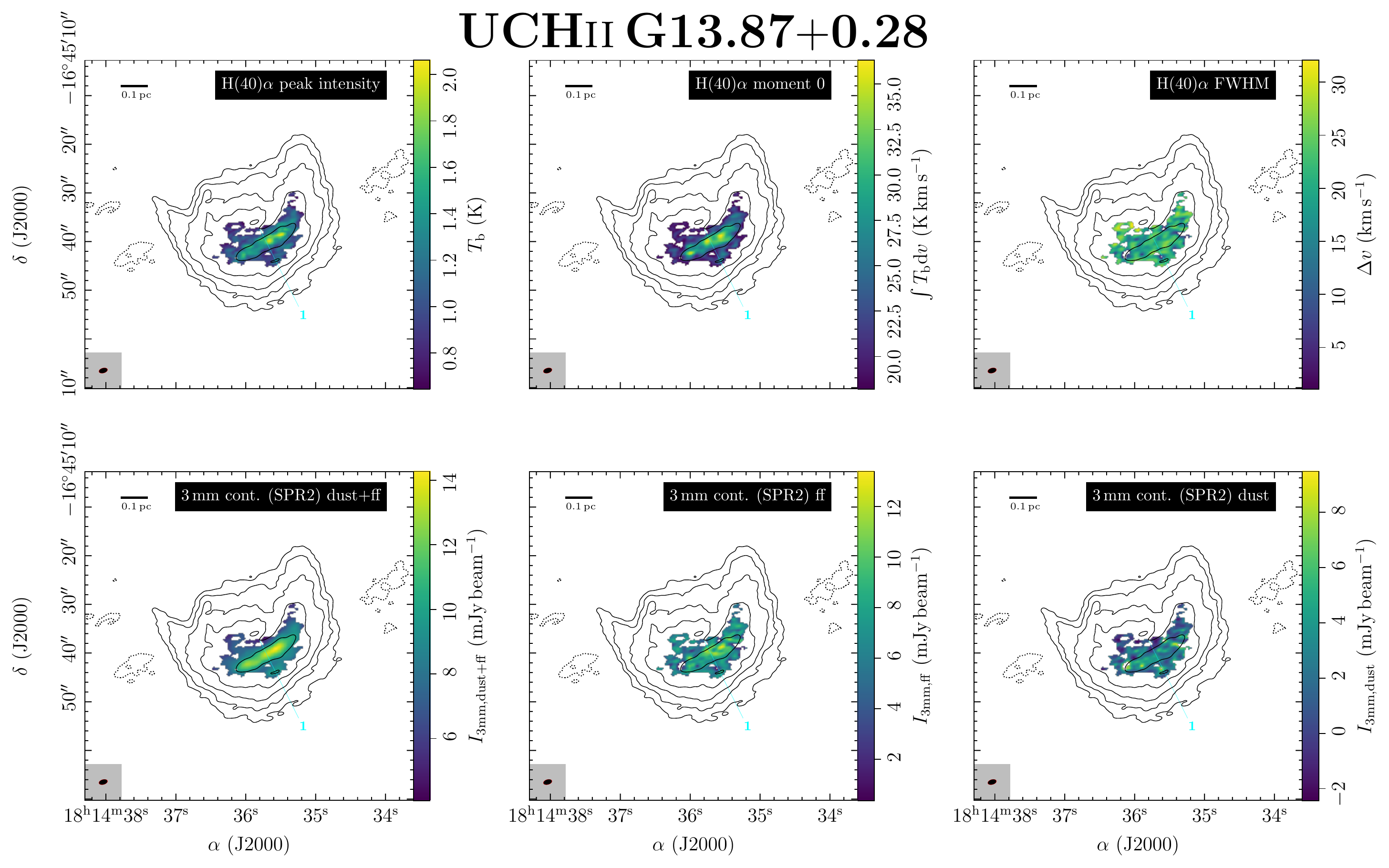}
\caption{Estimate of the free-free contribution at 3\,mm (Eq. \ref{eq:ff}). In all panels, the SPR2 3\,mm continuum is shown in contours. The dotted black contour marks the $-5\sigma_\mathrm{cont}$ level. The solid black contours start at $5\sigma_\mathrm{cont}$ and contour steps increase by a factor of 2 (e.g., 5, 10, 20, $40\sigma_\mathrm{cont}$). The top panels show in color the H(40)$\alpha$ peak intensity, integrated intensity, and line width derived from the 2nd moment. The bottom panels show in color the SPR2 3\,mm continuum (dust + ff), 3\,mm ff continuum (estimated using Eq. \ref{eq:ff}), and 3\,mm dust continuum. The synthesized beam size of the continuum and line data is shown in the bottom left corner in black and red, respectively. In all panels, the area where the H(40)$\alpha$ integrated intensity and SPR2 continuum have a $S$/$N < 5$ are masked. The continuum fragments are classified into dust cores (red), dust+ff cores (orange), cometary UCH{\sc ii} regions (cyan), further explained in Sect. \ref{sec:ALMAfrag}. Fragments with $S$/$N < 15$ are not analyzed in this study and are labeled in grey.}
\end{figure*}

\end{appendix}

\end{document}